Flemming Nielson · Hanne Riis Nielson

# Program Analysis

## *An Appetizer*

*December 2020*

# Preface

This book is an introduction to program analysis that is meant to be considerably more elementary than our advanced book *Principles of Program Analysis* (Springer, 2005). Rather than using flow charts as the model of programs, the book follows our introductory book *Formal Methods – An Appetizer* (Springer, 2019) using program graphs as the model of programs. In our experience this makes the underlying ideas more accessible to our computer science and computer engineering students on the master course *02242: Program Analysis* at The Technical University of Denmark. Here we have gradually replaced our use of the more elementary parts of *Principles of Program Analysis* with material from the current book.

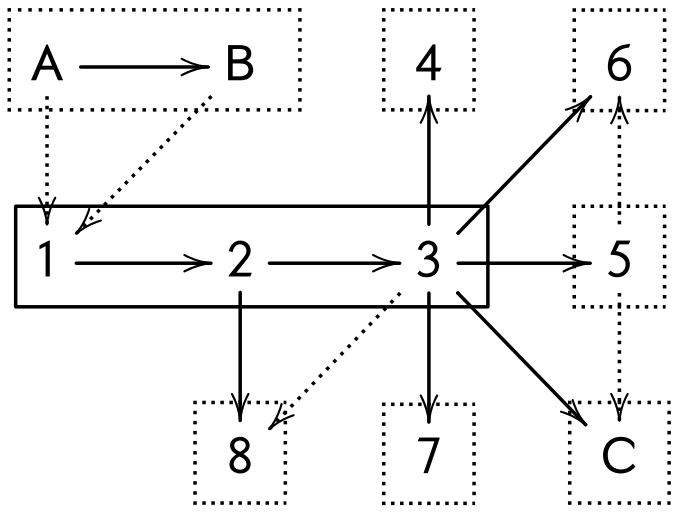

Dependencies between chapters.

Chapter 1 presents the basic model of program graphs, borrowing from our treatment in *Formal Methods – An Appetizer*, without introducing any specific programming language. Appendices A and B present two prototypical programming languages, based on *Guarded Commands* and a subset of C, and discuss how to generate program graphs from programs.

Chapters 2 and 3 present the basic ideas of intra-procedural program analysis. First the four classical bit-vector analyses are considered and then the monotone framework is developed. We aim at showing that the distinction between forward and backward analyses corresponds to reversing the edges in the program graphs, and that the distinction between least and greatest solutions corresponds to dualising the partial order.

Chapter 4 goes beyond the simple non-deterministic iteration algorithms used in previous chapters and presents some structure on the worklists, including reverse postorder, strong components and natural loops (for reducible program graphs).

Chapter 5 studies the analysis of mathematical integers: detection of signs, constant propagation and interval analysis are studies as independent attribute analyses.

Chapter 6 introduces abstract interpretation: abstraction and concretisation functions and their relationship as well as widenings; some knowledge of Section 5.1 is required and a few examples require knowledge of Section 5.3. It concludes in Section 6.4 with a treatment of a relational detection of signs analysis generalising the detection of signs analysis of Section 5.1.

Chapter 7 studies how to develop information flow analyses for confidentiality and integrity using type systems. This requires paying attention to the syntax of the programming language and we base the development on Guarded Commands of Appendix A. It both presents a type system for measuring leakage and one for avoiding leakage.

Chapter 8 studies how to use Datalog to express and efficiently solve a number of program analyses. Essentially all powerset based analyses covered in Chapters 2 and 3 can be dealt with and we also show how to deal with some more advanced analyses.

Appendix C discusses the projects that we have used in our master course *02242: Program Analysis* and the systems available at `FormalMethods.dk` allow to experiment with several of the developments covered in this book.

We should like to thank René Rydhof Hansen, Salvador Jacobi, Mike Castro Lundin, Emad Jacob Maroun, Magnus Gether Sørensen, Martin Obel Thomsen, Panagiotis Vasilikos, for having provided feed-back on earlier versions.

Denmark, December 2020
Flemming Nielson
Hanne Riis Nielson

# Contents

# Chapter 1

# Programs as Graphs

A *program graph* is a graphical representation of the *control structure* of a program or system. A program graph is a directed graph so it has a set of *nodes* and a set of *edges*. However, there is a little more structure to a program graph. It has an *initial node* that intuitively represents the starting point of the executions modelled by the program graph and by symmetry it has a *final node* that represents a point where the execution will have terminated. Each of the edges is labelled with an *action* modelling the computation to take place when control follows the edge. An action may for example be an assignment or a test as illustrated on the program graph of Figure 1.1.

EXAMPLE 1.1: Figure 1.1 shows a program graph with five nodes: $q_\triangleright$ is the initial node, $q_\blacktriangleleft$ is the final node and then there are three additional nodes $q_1, q_2$ and $q_3$. The edges are labelled with the associated actions; as an example the action $\texttt{y} := \texttt{x} * \texttt{y}$ is associated with the edge with source $q_2$ and target $q_3$. The five actions $\texttt{y} := 1, \texttt{x} > 0, \texttt{x} \leq 0, \texttt{y} := \texttt{x} * \texttt{y}$ and $\texttt{x} := \texttt{x} - 1$ have been chosen so as to be suggestive of the meaning that we will be giving them when we come to the semantics – it should be no surprise that the program graph is intended to compute the *factorial* function: upon termination of the program, the value of the variable y should be equal to the factorial of the initial value of x.

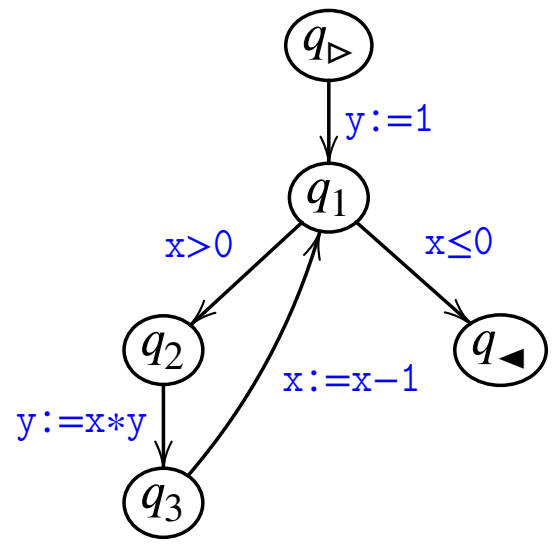

Figure 1.1: Program graph for the factorial function.

Program graphs can be automatically constructed from programs. In Appendix A we give systematic algorithms for this for a language of Guarded Commands, and Appendix B presents a series of tasks for doing so for MicroC, a tiny C-like language. For the factorial function in Figure 1.1 take a look at Figures A.1 and B.1. In this book we will develop the analyses for the program graphs – so we shall not be concerned about whether the program graph has been constructed from Guarded Commands, MicroC, or some other programming language.

## 1.1 Actions

The actions of the program graphs represent the basic computations of the underlying programming language. An *assignment* takes the form $x := a$ where $x$ is a variable and $a$ is an arithmetic expression. In addition to simple variables (as $x$) we shall also have *arrays* and the action $A[a_1] := a_2$ represents an assignment to an entry in the array named $A$. On top of this we shall have input and output actions. An *input* action has the form $c?x$ where $c$ is the name of the channel (or file) from which we read a value, and the idea is that the value read on $c$ will be assigned to the variable $x$; an input action may also take the form $c?A[a]$ and the idea is here that the value read on $c$ will be assigned to the specified entry of the array. The *output* action has the form $c!a$ where again $c$ is the name of the channel (or file) on which the value of the arithmetic expression $a$ is now being written. A *test* $b$ is simply a boolean expression, and we consider it an action to pass the test.

To summarise, the *actions* $\alpha$ of the program graphs take one of the following forms:

$$\alpha \quad ::= \quad x := a \mid A[a_1] := a_2 \mid c?x \mid c?A[a] \mid c!a \mid b \mid \texttt{skip}$$

We make use of arithmetic expressions $a$ and boolean expressions $b$ given by

$$a \quad ::= \quad n \mid x \mid A[a_0] \mid a_1 \; op_a \; a_2 \mid -a_0 \mid A\#$$
$$b \quad ::= \quad \texttt{true} \mid \texttt{false} \mid a_1 \; op_r \; a_2 \mid b_1 \; op_b \; b_2 \mid \neg b_0$$

where we shall insert parentheses ( and ) to disambiguate the syntax whenever needed. We shall assume that $op_a \in \{+, -, *, /, \%, \ldots\}$, $op_r \in \{=, >, \geq, \ldots\}$ and $op_b \in \{\wedge, \vee, \ldots\}$. The syntax of variables $x$, array names $A$, channel names $c$ and numbers $n$ is left unspecified.

The values assigned to the variables will always be integers. An array $A$ has elements $A[0], A[1], \cdots, A[\mathsf{length}(A)-1]$ where $\mathsf{length}(A)$ is a fixed *positive* number giving the length; the values of the elements will always be integers. The arithmetic expression $A[a]$ can only be evaluated if the value of $a$ is a number between 0 and $\mathsf{length}(A)-1$; otherwise the result is undefined. The channels (or files) will contain sequences of integers; an input action is successful as long as it is possible to read from the channel whereas an output action always is possible.

The arithmetic operators $+$, $-$ and $*$ are the standard operations for addition, subtraction and multiplication of integers. We write $/$ and $\%$ for the division and modulo operations on integers; in the case the second argument is 0 these operations are undefined. Otherwise, given two integers $n$ and $d$ ($\neq 0$), the following equation will hold

$$n = (n \,/\, d) * d + (n \,\%\, d)$$

Somewhat surprisingly, it turns out that different programming languages may produce different results when one or both arguments are negative. We shall therefore define $n \,/\, d$ as the result of truncating the real number obtained by the division $n/d$ towards 0, and let $n \,\%\, d$ be the remainder required to make the above equation hold. So $5/(-3)$ evaluates to $-1$ and $5\%(-3)$ evaluates to 2.

The meaning of the comparison operators $=$, $>$ and $\geq$ are the standard operations for comparing integers. Similarly, the boolean operators $\wedge$ and $\vee$ are the standard logical operators for conjunction and disjunction.

## 1.2 Memories

In order to describe the meaning of the actions we shall introduce the notion of a memory. It will describe the values of the variables, the contents of the arrays, and the data available on the channels. In our case the values will simply be integers and we shall write **Int** for the set of integers. Let us consider the three components one by one.

**Variables.** The values of the variables are modelled by a mapping

$$\sigma_{\mathsf{V}} : \textbf{Var} \rightarrow \textbf{Int}$$

where **Var** is a *finite* set of variables of interest; this set is called the domain of $\sigma_{\mathsf{V}}$. The mapping $\sigma_{\mathsf{V}}$ determines the value of each of the variables of **Var** and we shall write $\sigma_{\mathsf{V}}(x)$ for the value of $x \in$ **Var**; the notation $\sigma_{\mathsf{V}}(x)$ is undefined in case $x \notin$ **Var**. Figure 1.2 shows a memory $\sigma_{\mathsf{V}} : \{\mathtt{x}, \mathtt{y}\} \rightarrow$ **Int** with just two variables $\mathtt{x}$ and $\mathtt{y}$.

| | |
|---|---|
| x | 7 |
| y | 5 |

Figure 1.2: Example memory: $\sigma_{\mathsf{V}}$ with $\sigma_{\mathsf{V}}(\mathtt{x}) = 7$ and $\sigma_{\mathsf{V}}(\mathtt{y}) = 5$.

An assignment results in an *update* of the memory and we shall introduce the notation $\sigma_{\mathsf{V}}[x \mapsto v]$ for the memory that is as $\sigma_{\mathsf{V}}$ except that the value of the variable $x$ (of **Var**) now has the value $v$; once again $\sigma_{\mathsf{V}}[x \mapsto v]$ is undefined in case $x \notin$ **Var**. With this notation we may for example write $\sigma_{\mathsf{V}}[\mathtt{x} \mapsto 13]$ for the memory of Figure 1.3 obtained by updating the memory $\sigma_{\mathsf{V}}$ of Figure 1.2. On the other hand $\sigma_{\mathsf{V}}[\mathtt{z} \mapsto 0]$ is undefined because we are attempting to update an unknown variable (as $\mathtt{z} \notin \{\mathtt{x}, \mathtt{y}\}$).

| | |
|---|---|
| x | 13 |
| y | 5 |

Figure 1.3: Updated example memory: $\sigma_{\mathsf{V}}[\mathtt{x} \mapsto 13]$.

**Arrays.** The values of arrays are modelled by a mapping

$$\sigma_{\mathsf{A}} : \textbf{Arr} \rightarrow \textbf{Int}^*$$

where **Arr** is a *finite* set of array names of interest (being the domain of $\sigma_{\mathsf{A}}$) and where **Int**$^*$ denotes *lists* of values. The idea is that when $\sigma_{\mathsf{A}}(A) = [v_0, v_1, \cdots, v_{k-1}]$ then the size of the array $A$ is $k$, written $\mathsf{length}(\sigma_{\mathsf{A}}(A)) = k$, and the $i$'th element of the array is $v_i$; this is sometimes called indexing with origin 0. We shall occasionally write $\sigma_{\mathsf{A}}(A)_i$ for $v_i$ provided $\sigma_{\mathsf{A}}(A) = [v_0, v_1, \cdots, v_{k-1}]$ and $0 \leq i < k$; if $A \notin$ **Arr** or $i < 0$ or $i \geq k$ then $\sigma_A(A)_i$ is undefined. Figure 1.4 shows a memory $\sigma_A : \{A, B\} \to \mathbf{Int}^*$ with just two arrays A and B of length 3 and 4, respectively.

| | |
|---|---|
| A | $[4,3,0]$ |
| B | $[2,-3,7,10]$ |

Figure 1.4: Example memory $\sigma_A$; as an example $\sigma_A(B)_1 = -3$.

| | |
|---|---|
| A | $[4,3,0]$ |
| B | $[2,5,7,10]$ |

Figure 1.5: Updated example memory $\sigma_A[B[1] \mapsto 5]$.

An assignment to an array entry gives rise to an *update* of the memory and for this we introduce the notation $\sigma_A[A[i] \mapsto v]$. This is the memory that is as $\sigma_A$ except that the array $A$ now has its $i$'th position modified to $v$, provided that $0 \leq i < k$; if this is not the case then $\sigma_A[A[i] \mapsto v]$ is undefined. As an example, we may write $\sigma_A[B[1] \mapsto 5]$ for the memory that obtained by updating the memory $\sigma_A$ of Figure 1.4; this is shown in Figure 1.5. On the other hand, the memory $\sigma_A[A[3] \mapsto 5]$ is undefined because we are attempting to update the array A outside its bounds, and also $\sigma_A[C[3] \mapsto 5]$ is undefined because C is not a known array.

**Channels.** The final component of the memory takes care of the channels. We shall model channels by a mapping

$$\sigma_C : \mathbf{Chan} \to \mathbf{Int}^*$$

where **Chan** is a *finite* set of channel names; this set is the domain of $\sigma_C$. The idea is that $\sigma_C(c) = [v_0, v_1, \cdots, v_n]$ is the sequence of values available on the channel $c$; this notation is undefined if $c \notin \mathbf{Chan}$. This sequence may be empty and this is written [ ] (corresponding to $n < 0$); there are no limits to the length of the sequence. Figure 1.6 shows a memory $\sigma_C : \{in, out\} \to \mathbf{Int}^*$ with just two channels in and out; the channel in contains three values while the channel out is empty.

| | |
|---|---|
| in | $[7,9,13]$ |
| out | $[]$ |

Figure 1.6: Example memory $\sigma_C$ with two channels in and out.

Now assume that $\sigma_C(c) = [v_0, v_1, \cdots, v_n]$. It is only possible to perform an input from $c$ if $n \geq 0$ and we will get the value $v_0$ while at the same time the memory is modified to become $\sigma_C[c \mapsto [v_1, \cdots, v_n]]$ – thus we are reading from the beginning of the sequence and values can only be read once. It is always possible to perform an output of a value $v$ to $c$ and this will cause the memory to become $\sigma_C[c \mapsto [v_0, v_1, \cdots, v_n, v]]$ – thus we are writing to the end of the sequence. As an example, we may input a value from the channel in and output the value 2 to the channel out and the memory of Figure 1.6 will be modified to become that of Figure 1.7. An attempt to input from the channel out in case the memory is as in Figure 1.6 will fail.

| | |
|---|---|
| in | $[9,13]$ |
| out | $[2]$ |

Figure 1.7: Updated example memory after input and output.

To summarise, a *memory* has three components and is an element of the set:

$$\mathbf{Mem} = (\mathbf{Var} \to \mathbf{Int}) \times (\mathbf{Arr} \to \mathbf{Int}^*) \times (\mathbf{Chan} \to \mathbf{Int}^*)$$

A memory $\sigma$ determines the values of the variables ($\sigma_V$), the values of the entries of the arrays ($\sigma_A$) and the sequence of values present on the channels ($\sigma_C$). When we do not want to differentiate between the three parts of the memory we shall simply write $\sigma$; the different components can then be obtained using subscripts V, A and C – as for example in $\sigma = (\sigma_V, \sigma_A, \sigma_C)$.

Example 1.2: Figure 1.8 gives an example of a memory $\sigma$ specifying the values of four variables x, y, i and n together with two arrays A and B that

$$\left( \begin{array}{|c|c|} \hline \mathtt{x} & 9 \\ \hline \mathtt{y} & 7 \\ \hline \mathtt{i} & 3 \\ \hline \mathtt{n} & 8 \\ \hline \end{array}, \begin{array}{|c|c|} \hline \mathtt{A} & [1,2,4,6,7,12,4,19] \\ \hline \mathtt{B} & [7,9,13,17,0,4,3,13] \\ \hline \end{array}, \begin{array}{|c|c|} \hline \mathtt{in} & [4,12,19,7,0] \\ \hline \mathtt{out} & [\,] \\ \hline \end{array} \right)$$

Figure 1.8: Example memory with four variables, two arrays and two channels.

both contain eight elements, and finally the memory also specifies the contents of two channels, `in` and `out` where the latter is empty.

## 1.3 Semantics

Our goal in this section is to describe the semantics of the actions introduced in Section 1.1. We shall first specify how to determine the values of the arithmetic expressions using the memories of the previous section. Based on that we shall specify how to determine the truth values of the boolean expressions. And finally we show how each of the actions give rise to a transformation on memories.

**Arithmetic expressions.** An arithmetic expression was defined to be one of $n$, $x$, $A[a_0]$, $a_1\ op_a\ a_2$ and $-a_0$ where $a_0$, $a_1$ and $a_2$ are themselves arithmetic expressions and $op_a \in \{+, -, *, /, \%, \ldots\}$. This is a recursive definition of how to form an *abstract syntax tree* for an arithmetic expression and we are going to exploit that when defining the semantics.

The value of an arithmetic expression will be an integer, that is, an element of **Int**. In order to determine this value we need information about the memory. We shall therefore define a semantic function

$$\mathcal{A}[\![a]\!] : \mathbf{Mem} \hookrightarrow \mathbf{Int}$$

for each arithmetic expression $a$. It is a partial function and the idea is that $\mathcal{A}[\![a]\!]\sigma$ is the value of $a$ in the memory $\sigma$ if it can be computed.

The definition of these functions is recursive and mimics the form of the abstract syntax trees for the arithmetic expressions. For a constant $n$ the value is given directly, whereas for a variable $x$ the value is determined by consulting the memory. For the composite expressions we first determine the values of the subcomponents and next we can determine the value of the expression itself: for $A[a_0]$ this amounts to a lookup in the memory, whereas for the other expressions the corresponding arithmetic operations must be performed. This is summarised as follows:

$$\mathcal{A}[\![n]\!]\sigma = n$$

$$\mathcal{A}[\![x]\!]\sigma = \sigma_{\mathsf{V}}(x)$$

$$\mathcal{A}[\![A[a_0]]\!]\sigma = \begin{cases} v_i & \text{if } i = \mathcal{A}[\![a_0]\!]\sigma,\ \sigma_{\mathsf{A}}(A) = [v_0, v_1, \cdots, v_{k-1}] \\ & \text{and } 0 \leq i < k \\ \text{undefined} & \text{otherwise} \end{cases}$$

$$\mathcal{A}[\![a_1\ op_a\ a_2]\!]\sigma = \begin{cases} v & \text{if } v_1 = \mathcal{A}[\![a_1]\!]\sigma,\ v_2 = \mathcal{A}[\![a_2]\!]\sigma \\ & \text{and } v = v_1\ op_a\ v_2 \\ \text{undefined} & \text{otherwise} \end{cases}$$

$$\mathcal{A}[\![-a_0]\!]\sigma = \begin{cases} -v & \text{if } v = \mathcal{A}[\![a_0]\!]\sigma \\ \text{undefined} & \text{otherwise} \end{cases}$$

$$\mathcal{A}[\![A\#]\!]\sigma = \begin{cases} \text{length}(\sigma_{\mathsf{A}}(A)) & \text{if } A \in \textbf{Arr} \\ \text{undefined} & \text{otherwise} \end{cases}$$

Here we allow to write merely $n$ for the integer denoted by the number (or numeral) $n$, and similarly we omit distinguishing between the syntactic operations and their mathematical counterparts and simply write $op_a$ in both cases; as discussed in Section 1.1, the operations $+$, $-$ and $*$ are the addition, subtraction and multiplication on integers whereas $/$ and $\%$ are the division and modulo operations on integers. When we write $v = v_1\ op_a\ v_2$ it means that the operation $op_a$ is defined on the arguments $v_1$ and $v_2$ and that it gives the integer $v$. Similarly, when we write $v = \mathcal{A}[\![a]\!]\sigma$ it means that $\mathcal{A}[\![a]\!]\sigma$ is defined and that it gives the integer $v$.

$$\begin{aligned} & \mathcal{A}[\![\mathtt{x} * 3 - 5]\!]\sigma \\ & = \mathcal{A}[\![(\mathtt{x} * 3) - 5]\!]\sigma \\ & = \mathcal{A}[\![\mathtt{x} * 3]\!]\sigma - \mathcal{A}[\![5]\!]\sigma \\ & = (\mathcal{A}[\![\mathtt{x}]\!]\sigma * \mathcal{A}[\![3]\!]\sigma) - \mathcal{A}[\![5]\!]\sigma \\ & = (\sigma_{\mathsf{V}}(\mathtt{x}) * 3) - 5 \\ & = (7 * 3) - 5 \\ & = 16 \end{aligned}$$

Figure 1.9: Calculating the value of $\mathtt{x} * 3 - 5$ in a memory with $\sigma_{\mathsf{V}}(\mathtt{x}) = 7$.

EXAMPLE 1.3: Figure 1.9 shows in detail how to use the semantics to compute the value of $\mathtt{x} * 3 - 5$ using the usual precedence of operators and a memory $\sigma$ where $\sigma_{\mathsf{V}}(\mathtt{x}) = 7$.

TRY IT OUT 1.4: Perform calculations similar to those of Figure 1.9 to determine the value of the arithmetic expression `A[x]/x` in a memory $\sigma$ where $\sigma_{\mathsf{V}}(\mathtt{x}) = 2$ and $\sigma_{\mathsf{A}}(\mathtt{A}) = [7, 9, 13, 17]$.

Modify $\sigma$ to have $\sigma_{\mathsf{V}}(\mathtt{x}) = 4$ and repeat the calculations; be careful to point out exactly where the computation fails. Repeat the calculations for the case where $\sigma_{\mathsf{V}}(\mathtt{x}) = 0$ and pinpoint again exactly where the computation fails. □

**Boolean expressions.** The boolean expressions were defined to have one of the forms `true`, `false`, $a_1\ op_r\ a_2$, $b_1\ op_b\ b_2$ or $\neg\, b_0$, where $b_0$, $b_1$ and $b_2$ are themselves boolean expressions and $a_1$ and $a_2$ are arithmetic expressions. The relational operators $op_r \in \{=, >, \geq, \ldots\}$ allow us to compare values and the boolean operators $op_b \in \{\wedge, \vee, \ldots\}$ allow us to combine boolean conditions.

The value of a boolean expression is a truth value and we shall write **Bool** for the set $\{\text{true}, \text{false}\}$ of truth values. We shall therefore define a semantic function

$$\mathcal{B}[\![b]\!] : \textbf{Mem} \hookrightarrow \textbf{Bool}$$

for each boolean expression $b$. The idea is that $\mathcal{B}[\![b]\!]\sigma$ is the truth value of $b$ in the memory $\sigma$ provided that it can be computed.

Our approach for determining the value of a boolean expression is very similar to the one we used for arithmetic expressions. In particular, we shall need information about the memory in order to determine the values of the arithmetic expressions occurring inside the boolean expressions. Also we shall exploit that we have a recursive definition of how the abstract syntax trees for boolean expressions are formed.

$$\mathcal{B}[\![\texttt{true}]\!]\sigma = \text{true}$$

$$\mathcal{B}[\![\texttt{false}]\!]\sigma = \text{false}$$

$$\mathcal{B}[\![a_1\ op_r\ a_2]\!]\sigma = \begin{cases} t & \text{if } v_1 = \mathcal{A}[\![a_1]\!]\sigma,\ v_2 = \mathcal{A}[\![a_2]\!]\sigma \\ & \text{and } t = v_1\ op_r\ v_2 \\ \text{undefined} & \text{otherwise} \end{cases}$$

$$\mathcal{B}[\![b_1\ op_b\ b_2]\!]\sigma = \begin{cases} t & \text{if } t_1 = \mathcal{B}[\![b_1]\!]\sigma,\ t_2 = \mathcal{B}[\![b_2]\!]\sigma \\ & \text{and } t = t_1\ op_b\ t_2 \\ \text{undefined} & \text{otherwise} \end{cases}$$

$$\mathcal{B}[\![\neg b_0]\!]\sigma = \begin{cases} \neg t & \text{if } t = \mathcal{B}[\![b_0]\!]\sigma \\ \text{undefined} & \text{otherwise} \end{cases}$$

Also here we do not distinguish between the syntactic operations and their mathematical and logical counterparts. In analogy with before whenever we write $t = \mathcal{B}[\![b]\!]\sigma$ it means that $\mathcal{B}[\![b]\!]\sigma$ is defined and equals the truth value $t$. Note that for $b_1\ op_b\ b_2$ we require that both arguments are defined in order to determine the truth value of the expression.

Try It Out 1.5: Show that the boolean expression $(\texttt{y} > 0) \wedge (\texttt{x}/\texttt{y} > 0)$ evaluates to true in a memory $\sigma$ with $\sigma_V(\texttt{x}) = 5$ and $\sigma_V(\texttt{y}) = 3$. What is the result if we modify $\sigma$ to have $\sigma_V(\texttt{y}) = 0$? □

Exercise 1.6: Let us extend the set of boolean operators to include *short-circuit*

versions && and || of conjunction and disjunction. The idea is here that if the result of the operation can be determined from the first argument alone then it is not necessary to evaluate the second argument. Extend the above definition of $\mathcal{B}[\![b]\!]$ to cater for this and repeat Try It Out 1.5 but for the boolean expression $(\mathrm{y} > 0)\,\&\&\,(\mathrm{x}/\mathrm{y} > 0)$. □

**Actions.** We shall now define the semantics for the actions. Recall that a memory describes the values of the variables (in **Var**), the contents of the arrays (in **Arr**) and the data available on the channels (in **Chan**). An action will make use of this information when computing the values of arithmetic and boolean expressions; and the action may change the value of one of the variables, it may update an array entry, or it may read from one of the channels or write to one of the channels.

We shall view an action as a transformation on the memory and therefore the meaning of the actions is defined by a semantic function

$$\mathcal{S}[\![\alpha]\!] : \textbf{Mem} \hookrightarrow \textbf{Mem}$$

for each action $\alpha$. The idea is that if $\sigma$ is the memory before $\alpha$ is executed then $\mathcal{S}[\![\alpha]\!]\sigma$ is the memory afterwards – provided that the action was successful.

Let us first consider the two assignment actions: given a memory $\sigma$ they will transform it into another memory. The action $x := a$ will update $\sigma_{\mathsf{V}}$ to have a new value for $x$ namely the value $v$ that $a$ evaluates to; in Section 1.2 we introduced the notation $\sigma_{\mathsf{V}}[x \mapsto v]$ for this. The only caveat is that $x$ must be one of the variables that $\sigma_{\mathsf{V}}$ knows about so it must be the case that $x \in \textbf{Var}$; if not, the action will fail.

In a similar way, the action $A[a_1] := a_2$ will modify $\sigma_{\mathsf{A}}$ to have a new value for one of the indices of $A$. The new value $v$ is the one computed for $a_2$. The index $i$ that is modified is the value of $a_1$ and we shall insist that it is within the bounds of the array $A$ as specified by $\sigma_{\mathsf{A}}(A)$ – the notation $\sigma_{\mathsf{A}}[A[i] \mapsto v]$ introduced in Section 1.2 will make a check of this, and if successful it will return an updated memory. If we are attempting to index out of bounds or if the array $A$ is unknown (so $A \notin \textbf{Arr}$) then the action fails.

The semantics of the two assignment actions are given by:

$$\mathcal{S}[\![x := a]\!]\sigma = \begin{cases} (\sigma_{\mathsf{V}}[x \mapsto v], \sigma_{\mathsf{A}}, \sigma_{\mathsf{C}}) & \\ \qquad\qquad \text{if } v = \mathcal{A}[\![a]\!]\sigma \text{ and } x \in \textbf{Var} \\ \text{undefined} \quad \text{otherwise} \end{cases}$$

$$\mathcal{S}[\![A[a_1] := a_2]\!]\sigma = \begin{cases} (\sigma_V, \sigma_A[A[i] \mapsto v], \sigma_C) & \\ & \text{if } i = \mathcal{A}[\![a_1]\!]\sigma,\ v = \mathcal{A}[\![a_2]\!]\sigma \\ & \text{and } A \in \textbf{Arr},\ 0 \le i < \text{length}(\sigma_A(A)) \\ \text{undefined} & \text{otherwise} \end{cases}$$

EXAMPLE 1.7: Consider the action B[i] := B[i] + B[i]/10 and its semantics in the memory $\sigma$ of Figure 1.8. We have $\mathcal{A}[\![\texttt{i}]\!]\sigma = 3$ and $\mathcal{A}[\![\texttt{B[i]} + \texttt{B[i]}/10]\!]\sigma = 18$ so the new memory will update index 3 of the array B to have the value 18 as shown in Figure 1.10; the two other components of the memory are unchanged by the action so they are as in Figure 1.8.

| | |
|---|---|
| A | $[1, 2, 4, 6, 7, 12, 4, 19]$ |
| B | $[7, 9, 13, 18, 0, 4, 3, 13]$ |

Figure 1.10: The array memory obtained from updating B[i] from $\sigma$ in Figure 1.8.

Next let us consider the three actions making use of the channels. The input action $c?x$ makes a modification of $\sigma_V$ as well as $\sigma_C$: we are reading a value $v_0$ from the channel $c$ so $\sigma_C$ is modified to reflect that, and we are updating the value of $x$ to have the value $v_0$ so $\sigma_V$ will also be modified. The channel must contain at least one value in order for the input action to be successful, so we require that $\sigma_C(c) = [v_0, v_1, \cdots, v_n]$ where $n \ge 0$, and as a result of the input we update $\sigma_C$ to record that only the values $[v_1, \cdots, v_n]$ are available for further inputs on $c$. The update of $\sigma_V$ is analogous to an assignment $x := v_0$.

The input action $c?A[a]$ makes modifications of $\sigma_A$ and $\sigma_C$. It is basically a combination of what we have seen above: we are reading a value $v_0$ from the channel $c$ so $\sigma_C$ is updated exactly as explained above. This time we update an entry of the array $A$ in $\sigma_A$ and this is analogous to an assignment $A[a] := v_0$.

The output action $c!a$ only modifies $\sigma_C$. We determine the value of the arithmetic expression $a$ and its value is written to the channel $c$.

The semantics of the three actions involving channels are given by:

$$\mathcal{S}[\![c?x]\!]\sigma = \begin{cases} (\sigma_V[x \mapsto v_0], \sigma_A, \sigma_C[c \mapsto [v_1, \cdots, v_n]]) & \\ & \text{if } \sigma_C(c) = [v_0, v_1, \cdots, v_n],\ n \ge 0 \\ & \text{and } x \in \textbf{Var}, c \in \textbf{Chan} \\ \text{undefined} & \text{otherwise} \end{cases}$$

$$\mathcal{S}[\![c?A[a]]\!]\sigma = \begin{cases} (\sigma_V, \sigma_A[A[i] \mapsto v_0], \sigma_C[c \mapsto [v_1, \cdots, v_n]]) & \\ & \text{if } i = \mathcal{A}[\![a]\!]\sigma, \sigma_C(c) = [v_0, v_1, \cdots, v_n], n \ge 0 \\ & \text{and } A \in \textbf{Arr},\ 0 \le i < \text{length}(\sigma_A(A)), c \in \textbf{Chan} \\ \text{undefined} & \text{otherwise} \end{cases}$$

$$\mathcal{S}[\![c!a]\!]\sigma = \begin{cases} (\sigma_V, \sigma_A, \sigma_C[c \mapsto [v_0, v_1, \cdots, v_n, v]]) & \\ & \text{if } v = \mathcal{A}[\![a]\!]\sigma \\ & \text{and } \sigma_C(c) = [v_0, v_1, \cdots, v_n], c \in \textbf{Chan} \\ \text{undefined} & \text{otherwise} \end{cases}$$

| | |
|---|---|
| x | 4 |
| y | 7 |
| i | 3 |
| n | 8 |
| in | [12, 19, 7, 0] |
| out | [ ] |

Figure 1.11: The variable and channel memories obtained by executing in?x from $\sigma$ in Figure 1.8.

| | |
|---|---|
| in | [4, 12, 19, 7, 0] |
| out | [7] |

Figure 1.12: The channel memory obtained by executing out!y from $\sigma$ in Figure 1.8.

EXAMPLE 1.8: Consider the action in?x and its semantics in the memory $\sigma$ of Figure 1.8. We will read the first value on the channel in and update the variable component and the channel component of the memory as shown in Figure 1.11. The array component is unchanged so it is as in Figure 1.8.

The semantics of the action out!y in the memory $\sigma$ of Figure 1.8 will only update the channel part of the memory as shown in Figure 1.12; the other two components of the memory are unchanged so they are as in Figure 1.8.

The remaining two actions are tests and skip and they both leave the memory unchanged. The action $b$ inspects the memory and leaves it unchanged provided that the test evaluates to true. If the test evaluates to false or its value is undefined then the action fails. The action skip does not even inspect the memory and it will always succeed (so it will have the same effect as the test true).

The semantics of the tests and skip actions are given by:

$$\mathcal{S}[\![b]\!]\sigma = \begin{cases} \sigma & \text{if } \mathcal{B}[\![b]\!]\sigma = \text{true} \\ \text{undefined} & \text{otherwise} \end{cases}$$

$$\mathcal{S}[\![\texttt{skip}]\!]\sigma = \sigma$$

## 1.4 Program Graphs

As already explained in the beginning of this chapter, a program graph is a graphical representation of a program. In Figure 1.1 we saw an example of a program graph: it has a set of nodes, including an initial node $q_\triangleright$ and a final node $q_\blacktriangleleft$, and it has a set of directed edges labelled by actions. Our formal requirements to a program graph are summarised in the following definition.

DEFINITION 1.9: A *program graph* **PG** consists of the following:

- **Q**: a finite set of *nodes*
- $q_\triangleright, q_\blacktriangleleft \in \mathbf{Q}$: two nodes called the *initial node* and the *final node*, respectively
- **Act**: a finite set of *actions*
- $\mathbf{E} \subseteq \mathbf{Q} \times \mathbf{Act} \times \mathbf{Q}$: a finite set of *edges*

An edge $(q_\circ, \alpha, q_\bullet)$ has *source* node $q_\circ$, *target* node $q_\bullet$, and it is labelled with the *action* $\alpha$.

We shall *require* that the initial and final nodes are distinct, that all nodes are reachable from $q_{\triangleright}$, and that $q_{\blacktriangleleft}$ is reachable from all nodes.

It is possible to weaken the assumption that all nodes are reachable from $q_{\triangleright}$ and that $q_{\blacktriangleleft}$ is reachable from all nodes by complicating parts of the development.

EXAMPLE 1.10: The program graph of Figure 1.1 can be formally represented by the following sets of nodes and edges:

$$\begin{aligned} \mathbf{Q} &= \{q_{\triangleright}, q_1, q_2, q_3, q_{\blacktriangleleft}\} \\ \mathbf{E} &= \{(q_{\triangleright}, \texttt{y := 1}, q_1), (q_1, \texttt{x > 0}, q_2), (q_1, \texttt{x} \le \texttt{0}, q_{\blacktriangleleft}), \\ &\quad (q_2, \texttt{y := x * y}, q_3), (q_3, \texttt{x := x - 1}, q_1)\} \end{aligned}$$

The actions are $\mathbf{Act} = \{\texttt{y := 1}, \texttt{x > 0}, \texttt{x} \le \texttt{0}, \texttt{y := x * y}, \texttt{x := x - 1}\}$.

EXAMPLE 1.11: The program graph of Figure 1.13 makes use of all the different kinds of actions presented in the previous sections. It uses ordinary assignments as well as assignments to array entries, it uses channels for input as well as output, and it uses tests as well as the `skip` action. The program graph has a total of 13 nodes and a closer inspection reveals that it contains two loops, an outer loop starting at the node $q_1$ and an inner loop starting at $q_4$. Note that all nodes are reachable from $q_{\triangleright}$ and that all nodes can reach $q_{\blacktriangleleft}$.

The program graph represents a function that inputs a value from the channel `in` and searches for it in the array `A`. If successful, the corresponding entry in the array `B` will be updated and the resulting value will be output on the channel `out`. If not successful, the value -1 will be output on the channel `out`. The program terminates when a negative value is provided as input.

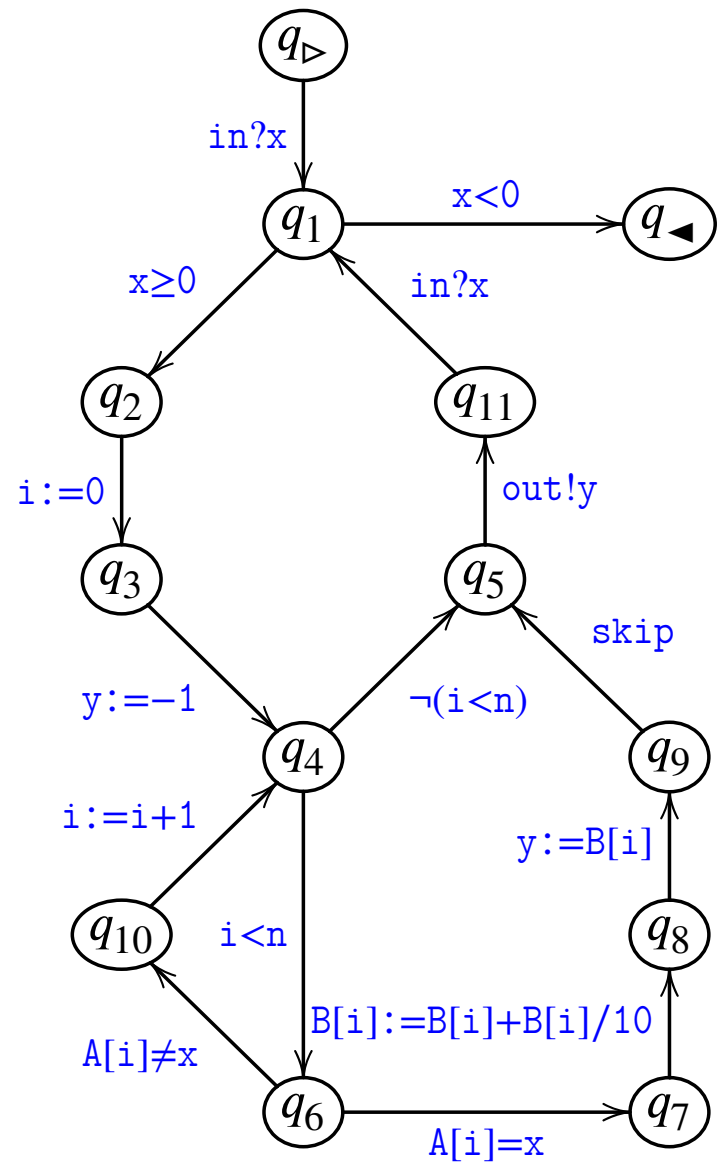

Figure 1.13: Program graph for searching and updating array entries.

EXERCISE 1.12: Construct program graphs for the following functions:

(a) A function computing the Fibonacci function; you may assume that it inputs a value from the channel `in` and outputs the result to the channel `out`.

(b) A function that sorts an array `A` with `n` elements; you may model various sorting algorithms as for example insertion sort and bubble sort.

(c) A function for adding two matrices `A` and `B` both of size $\texttt{n} \times \texttt{m}$; here you will need to represent the matrices as arrays and perform the appropriate calculations in order to index into the matrices.

(d) A function for multiplying two matrices `A` and `B` of size $\texttt{n} \times \texttt{m}$ and $\texttt{m} \times \texttt{p}$, respectively; as above the matrices should be represented as arrays. □

**Paths.** We shall conclude by introducing a notion of path in a program graph:

A *path* $\pi$ in a program graph has the form $q_0, \alpha_1, q_1, \alpha_2, q_2, \cdots, q_{n-1}, \alpha_n, q_n$ where $n \geq 0$ and $(q_{i-1}, \alpha_i, q_i)$ (for $1 \leq i \leq n$) is an edge in the program graph.

EXAMPLE 1.13: Returning to the program graph of Figure 1.1 we have, among others, the following paths:

(a) $q_\triangleright$

(b) $q_\triangleright, \texttt{y} := \texttt{1}, q_1, \texttt{x} \leq \texttt{0}, q_\blacktriangleleft$

(c) $q_1, \texttt{x} > \texttt{0}, q_2, \texttt{y} := \texttt{x} * \texttt{y}, q_3$

(d) $q_\triangleright, \texttt{y} := \texttt{1}, q_1, \texttt{x} > \texttt{0}, q_2, \texttt{y} := \texttt{x} * \texttt{y}, q_3, \texttt{x} := \texttt{x} - \texttt{1}, q_1, \texttt{x} > \texttt{0}, q_2$

## 1.5 Execution Sequences

The semantics of the program graphs will explain how the memory is changed as the actions labelling the edges are executed one by one. The semantics of the actions given in Section 1.3 describes how a *single* action modifies the memory. We shall be interested in executing *sequences* of actions corresponding to paths in the program graph; the execution sequences introduced below capture that idea.

A *configuration* $\langle q; \sigma \rangle$ is a node $q \in \mathbf{Q}$ together with a memory $\sigma \in \mathbf{Mem}$. It is called an *initial configuration* when $q = q_\triangleright$ and a *final configuration* when $q = q_\blacktriangleleft$.

EXAMPLE 1.14: Let us once again consider the program graph of Figure 1.1 and let us write $\sigma_{uv}$ for the memory that maps the variable $\texttt{x}$ to the value $u$ and the variable $\texttt{y}$ to the value $v$. The configuration $\langle q_\triangleright; \sigma_{37} \rangle$ is then an initial configuration stating that $\texttt{x}$ has the value 3 and $\texttt{y}$ the value 7. Similarly, $\langle q_\blacktriangleleft; \sigma_{06} \rangle$ is a final configuration where $\texttt{x}$ has the value 0 and $\texttt{y}$ the value 6.

**Execution steps.** We shall use the semantics to explain how to move between configurations. Whenever we have an edge $(q_\circ, \alpha, q_\bullet)$ in the program graph and a configuration $\langle q_\circ; \sigma \rangle$ we have the *potential* of modifying the configuration to become $\langle q_\bullet; \sigma' \rangle$ for some memory $\sigma'$. Whether or not this is indeed possible depends on the action $\alpha$ as well as the memory $\sigma$. If the semantics $\mathcal{S}[\![\alpha]\!]\sigma$ as specified in Section 1.3 is defined, that is $\mathcal{S}[\![\alpha]\!]\sigma = \sigma'$ for some $\sigma'$, then we can move from the configuration $\langle q_\circ; \sigma \rangle$ to the configuration $\langle q_\bullet; \sigma' \rangle$. However, if $\mathcal{S}[\![\alpha]\!]\sigma$ is undefined then it will not be possible to make a move following the edge $(q_\circ, \alpha, q_\bullet)$.

Whenever $(q_\circ, \alpha, q_\bullet)$ is an edge in the program graph we have an *execution step*

$$\langle q_\circ; \sigma\rangle \overset{\alpha}{\Longrightarrow} \langle q_\bullet; \sigma'\rangle \quad \text{if } \mathcal{S}[\![\alpha]\!]\sigma = \sigma'$$

Note that $\mathcal{S}[\![\alpha]\!]\sigma = \sigma'$ means that $\sigma$ is in the domain of $\mathcal{S}[\![\alpha]\!]$ and that the result of the function is $\sigma'$; if this is not the case then there will be no execution step.

A configuration $\langle q; \sigma\rangle$ is called *stuck* if it is not a final configuration and if there is no execution step of the form $\langle q; \sigma\rangle \overset{\alpha}{\Longrightarrow} \langle q'; \sigma'\rangle$ for any action $\alpha$ and node $q'$ of the program graph.

EXAMPLE 1.15: Continuing Example 1.14 we have the execution step

$$\langle q_\triangleright; \sigma_{37}\rangle \overset{\texttt{y:=1}}{\Longrightarrow} \langle q_1; \sigma_{31}\rangle$$

where $\sigma_{31}$ reflects that the memory is updated so that $\texttt{y}$ has the value 1. At this point we have potentially two ways to proceed. One possibility is to follow the edge $(q_1, \texttt{x} > 0, q_2)$; this is possible since $\mathcal{S}[\![\texttt{x} > 0]\!]\sigma_{31} = \sigma_{31}$ and we get the execution step

$$\langle q_1; \sigma_{31}\rangle \overset{\texttt{x>0}}{\Longrightarrow} \langle q_2; \sigma_{31}\rangle$$

Another possibility is to try to follow the edge $(q_1, \texttt{x} \leq 0, q_\blacktriangleleft)$; however, this will fail since $\mathcal{S}[\![\texttt{x} \leq 0]\!]\sigma_{31}$ is undefined. Thus in this case there is only one possible next step.

In general, there may be no next step (this happens if we replace the test $\texttt{x} > 0$ in Figure 1.1 with $\texttt{x} > 3$), or there may be exactly one next step (as in the case above), or there may be several next steps (this happens if we replaced the test $\texttt{x} \leq 0$ in Figure 1.1 with $\texttt{x} \leq 3$) in which case we have a free choice as to which edge to follow.

**Execution sequences.** An execution step corresponds to following a single edge of the program graph; an execution sequence corresponds to following a *path* in the program graph.

An *execution sequence* following the path $q_0, \alpha_1, q_1, \alpha_2, q_2, \cdots, q_{n-1}, \alpha_n, q_n$ (where $n \geq 0$) takes the form

$$\langle q_0; \sigma_0\rangle \overset{\alpha_1}{\Longrightarrow} \langle q_1; \sigma_1\rangle \overset{\alpha_2}{\Longrightarrow} \cdots \overset{\alpha_n}{\Longrightarrow} \langle q_n; \sigma_n\rangle$$

provided that each $\langle q_{i-1}; \sigma_{i-1}\rangle \overset{\alpha_i}{\Longrightarrow} \langle q_i; \sigma_i\rangle$ is an execution step.

We shall be most interested in execution sequences that start with the initial node

(so they have $q_0 = q_\triangleright$) or that end in the final node (so they have $q_n = q_\blacktriangleleft$); a *complete execution sequence* is one where both $q_0 = q_\triangleright$ and $q_n = q_\blacktriangleleft$.

It is important to notice that not all paths have execution sequences following them.

A path $\pi = q_0, \alpha_1, q_1, \alpha_2, q_2, \cdots, q_{n-1}, \alpha_n, q_n$ in a program graph is a *realisable path* whenever there is an execution sequence following it.

$$
\begin{array}{lcl}
\langle q_\triangleright; \sigma_{37}\rangle & \overset{\texttt{y:=1}}{\Longrightarrow} & \langle q_1; \sigma_{31}\rangle \\
& \overset{\texttt{x>0}}{\Longrightarrow} & \langle q_2; \sigma_{31}\rangle \\
& \overset{\texttt{y:=x*y}}{\Longrightarrow} & \langle q_3; \sigma_{33}\rangle \\
& \overset{\texttt{x:=x-1}}{\Longrightarrow} & \langle q_1; \sigma_{23}\rangle \\
& \overset{\texttt{x>0}}{\Longrightarrow} & \langle q_2; \sigma_{23}\rangle \\
& \overset{\texttt{y:=x*y}}{\Longrightarrow} & \langle q_3; \sigma_{26}\rangle \\
& \overset{\texttt{x:=x-1}}{\Longrightarrow} & \langle q_1; \sigma_{16}\rangle \\
& \overset{\texttt{x>0}}{\Longrightarrow} & \langle q_2; \sigma_{16}\rangle \\
& \overset{\texttt{y:=x*y}}{\Longrightarrow} & \langle q_3; \sigma_{16}\rangle \\
& \overset{\texttt{x:=x-1}}{\Longrightarrow} & \langle q_1; \sigma_{06}\rangle \\
& \overset{\texttt{x}\leq\texttt{0}}{\Longrightarrow} & \langle q_\blacktriangleleft; \sigma_{06}\rangle
\end{array}
$$

Figure 1.14: Execution sequence for the factorial function.

Example 1.16: In Figure 1.14 we give a complete execution sequence for the program graph of Figure 1.1; as in Example 1.14 we write $\sigma_{uv}$ for the memory where x has the value $u$ and y has the value $v$. The corresponding realisable path is

$$
\begin{array}{l}
q_\triangleright, \texttt{y := 1}, q_1, \texttt{x > 0}, q_2, \texttt{y := x * y}, q_3, \texttt{x := x - 1}, \\
\qquad q_1, \texttt{x > 0}, q_2, \texttt{y := x * y}, q_3, \texttt{x := x - 1}, q_1, \texttt{x} \leq \texttt{0}, q_\blacktriangleleft
\end{array}
$$

Stronger notions of being a realisable path would result by not only insisting that there is an execution sequence following it but also requiring that $q_0 = q_\triangleright$ and that the execution sequence starts in a state $\sigma_0$ being an element of some designated set $\mathbf{Mem}_\triangleright$.

## 1.6 The Nature of Approximation

The aim of program analysis is to *statically* predict properties of the *dynamic* behaviour of the program.

This is non-trivial because for most questions of interest it is *undecidable* to find the correct answer. Recall that undecidable questions are the ones where there does not exist an algorithm that always terminates and that always provides the correct answer; the existence of undecidable questions is related to the *halting problem*.

The approach of program analysis is to overcome undecidability by providing approximate answers. We shall insist that one of the answers ('yes' or 'no', respectively) can be truly trusted whereas the complementary answer might be wrong ('no' or 'yes', respectively). Sometimes it is useful to use an exclamation mark '!' to indicate that an answer can be truly trusted, and a question mark '?' to indicate that the answer might be wrong; note that negation changes question marks into exclamation marks and vice versa. Whenever the result of the program analysis is used, we need to ensure that the answer relied upon can indeed be fully trusted.

As an example, consider the question of whether or not a node $q$ in the program graph is *reachable* from the initial node $q_\triangleright$ along some path in the program graph. If we insist on dealing only with realisable paths the question is undecidable. If we allow to consider all paths, whether realisable or not, the question can be answered by a simple algorithm that gradually builds the sets of reachable nodes by starting with $\{q_{\triangleright}\}$ and inspecting the edges of nodes found to be reachable. In this case the positive answer (namely that $q$ is reachable) might be wrong (so we could write 'reachable?'). However, the negative answer (namely $q$ is not reachable) can be fully trusted (so we could write 'unreachable!'). The reason is that all the realisable paths are indeed contained in the set of all paths.

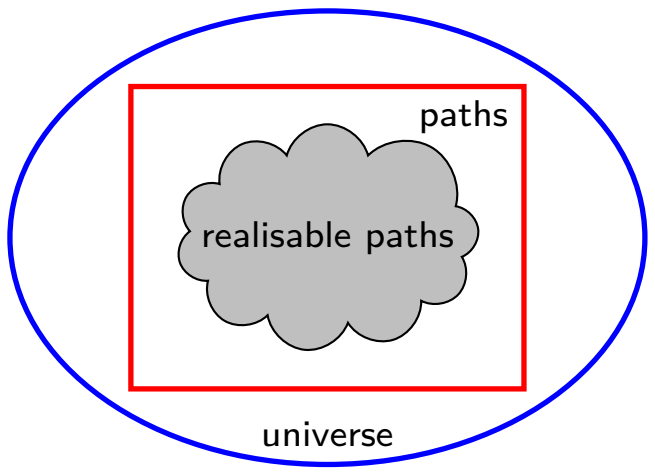

Figure 1.15: Paths and realisable paths within the universe.

Figure 1.15 illustrates how the realisable paths fit within all the paths that again fit within the universe of all alternating sequences of nodes and actions. For most program analyses we want to show that the set of realisable paths satisfy some property $\Phi$ (or '$\Phi$!' using the notation explained above). We do so by showing that the larger set of all paths have an empty overlap with the paths *not* satisfying $\Phi$. This idea is illustrated on Figure 1.16 where the property $\Phi$ divides the universe in two parts: the property holds to the right of the green line and it does not hold to the left of it. The yellow area of the figure is then the set of paths that does *not* satisfy $\Phi$ and if we can show that this set is empty then it follows that all the realisable paths will indeed satisfy the property $\Phi$.

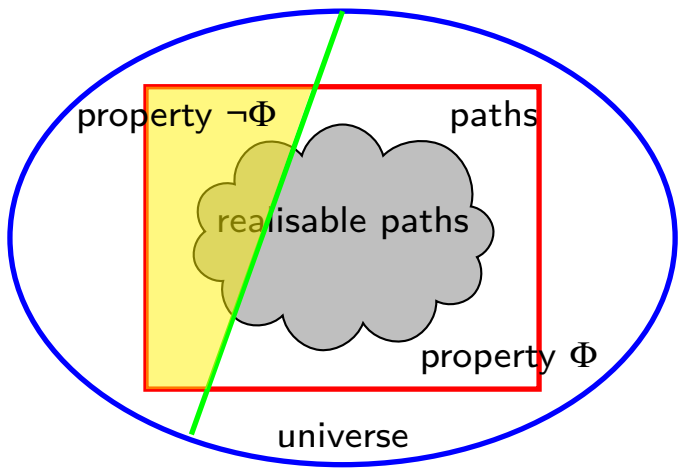

Figure 1.16: The property $\Phi$.

Another reason for the approximative nature of program analysis, beyond the need to overcome undecidability, is to obtain analyses that can be computed reasonably efficiently, meaning that answers can be found by algorithms that have acceptable time- and space-complexity. The art of program analysis is to find the right trade-off between the time- and space-complexity of the analyses and the benefits of using the results of the analyses.

# Chapter 2

# The Bit-Vector Framework

In this chapter we introduce four program analyses. The first is the *Reaching Definitions* analysis; it gives information about where variables and arrays have most recently obtained their values. The second is the *Live Variables* analysis that gives information about whether or not the values of variables and arrays might still be used before being redefined (or before the program ends). Then we introduce the *Available Expressions* analysis; it gives information about which expressions that have previously been computed and that will evaluate to the same value if recomputed now. The final analysis is the *Very Busy Expressions* analysis that gives information about which expressions that will be computed before the program ends and that would have the same value if they were computed already now. These four analyses constitute the classical examples of the so-called *Bit-Vector Framework* that constitutes a simple example of *Data Flow Analysis*.

## 2.1 Reaching Definitions

Reaching Definitions analysis aims at finding out for each node of a program graph where a variable or array might have been previously defined (or modified) the last time; if a variable might have retained its initial value we also want to know this. As an example for the program graph of Figure 2.1 the analysis should tell us that at the node $q_1$ there are two possibilities for the variable y: either it was modified using the action from $q_▷$ to $q_1$ or the one from $q_2$ to $q_3$. For the variables x there are also two possibilities: either it was modified by the action from $q_3$ to $q_1$ or it has not been modified at all and hence it has retained its original value.

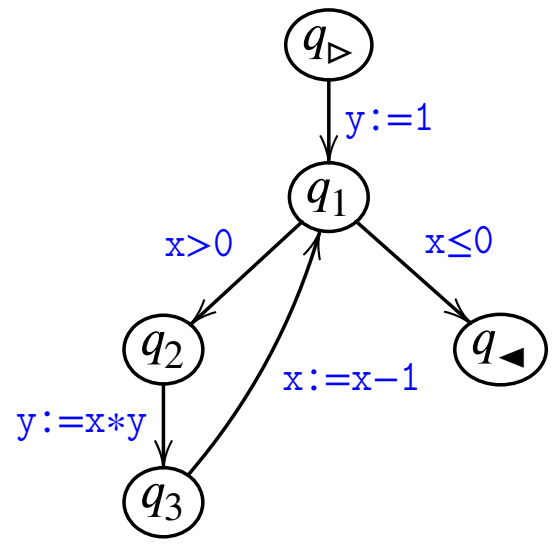

Figure 2.1: Program graph for the factorial function.

The information provided by the Reaching Definitions analysis is useful in compiler optimisation and program understanding where it, for example, is used to link the places where variables are modified to where their values are used. The analysis is also useful for validating security properties; it will for example help spotting situations where variables are not properly initialised and hence the program might be vulnerable to certain forms of attacks.

**Analysis assignments.** We shall use triples of the form $(x, q_\circ, q_\bullet)$ and $(A, q_\circ, q_\bullet)$ to express the information of interest. If $q_\circ$ and $q_\bullet$ are both in **Q** this intends to say that the variable $x$ or array $A$ might have been last defined (or modified) by the action $\alpha$ on an edge $(q_\circ, \alpha, q_\bullet)$ in the program graph. We shall allow $q_\circ$ to be the special symbol '$?$' to indicate that $x$ or $A$ might have retained its initial value.

EXAMPLE 2.1: For the program graph of Figure 2.1 the triple $(\mathtt{y}, q_2, q_3)$ tells us that the variable $\mathtt{y}$ is modified by the action labelling the edge from $q_2$ to $q_3$. The triple $(\mathtt{y}, ?, q_\triangleright)$ tells us that $\mathtt{y}$ has not been modified and hence has its initial value.

For the node $q_1$ we can use the set $\{(\mathtt{y}, ?, q_\triangleright), (\mathtt{y}, q_\triangleright, q_1), (\mathtt{y}, q_2, q_3)\}$ to describe where $\mathtt{y}$ was last defined. Clearly, it will be more informative to use a smaller set and the analysis to be presented shortly will compute the smallest sets possible.

The result of a *Reaching Definitions* analysis is given by an *analysis assignment* RD that maps each node in the program graph to a set of triples of the form $(x, q_\circ, q_\bullet)$ or $(A, q_\circ, q_\bullet)$ where $q_\bullet$ is in **Q** and $q_\circ$ is in either the special symbol '$?$' or is in **Q**:

$$\mathsf{RD} : \mathbf{Q} \to \mathsf{PowerSet}(\,(\mathbf{Var} \cup \mathbf{Arr}) \times \mathbf{Q}_? \times \mathbf{Q}\,)$$

where $\mathbf{Q}_? = \{?\} \cup \mathbf{Q}$.

Recall that an element of the *powerset* $\mathsf{PowerSet}(\,(\mathbf{Var} \cup \mathbf{Arr}) \times \mathbf{Q}_? \times \mathbf{Q}\,)$ is merely a subset of $(\mathbf{Var} \cup \mathbf{Arr}) \times \mathbf{Q}_? \times \mathbf{Q}$.

(If one is concerned about parallel edges with different actions one could use $\mathsf{RD} : \mathbf{Q} \to \mathsf{PowerSet}(\,(\mathbf{Var} \cup \mathbf{Arr}) \times \mathbf{Q}_? \times \mathbf{Act} \times \mathbf{Q}\,)$ instead.)

EXAMPLE 2.2: Let us reconsider the program graph shown in Figure 2.1. Here we would like that an analysis assignment RD satisfies the following conditions:

$$\begin{array}{lcl}
\mathsf{RD}(q_\triangleright) & \supseteq & \{(\mathtt{x}, ?, q_\triangleright), (\mathtt{y}, ?, q_\triangleright)\} \\
\mathsf{RD}(q_1) & \supseteq & \{(\mathtt{x}, ?, q_\triangleright), (\mathtt{x}, q_3, q_1), (\mathtt{y}, q_\triangleright, q_1), (\mathtt{y}, q_2, q_3)\} \\
\mathsf{RD}(q_2) & \supseteq & \{(\mathtt{x}, ?, q_\triangleright), (\mathtt{x}, q_3, q_1), (\mathtt{y}, q_\triangleright, q_1), (\mathtt{y}, q_2, q_3)\} \\
\mathsf{RD}(q_3) & \supseteq & \{(\mathtt{x}, ?, q_\triangleright), (\mathtt{x}, q_3, q_1), (\mathtt{y}, q_2, q_3)\} \\
\mathsf{RD}(q_\blacktriangleleft) & \supseteq & \{(\mathtt{x}, ?, q_\triangleright), (\mathtt{x}, q_3, q_1), (\mathtt{y}, q_\triangleright, q_1), (\mathtt{y}, q_2, q_3)\}
\end{array}$$

If equalities hold in all cases we would intuitively think that RD is the best analysis assignment that can be hoped for.

**What is defined on a path.** Let us be more precise about what we mean by being defined. For each action $\alpha$ we shall specify the set of variables and arrays that are being defined (or modified) by it as follows:

$$
\begin{array}{lcl}
\mathsf{Def}(x := a) & = & \{x\} \\
\mathsf{Def}(A[a_1] := a_2) & = & \{A\} \\
\mathsf{Def}(c?x) & = & \{x\} \\
\mathsf{Def}(c?A[a]) & = & \{A\} \\
\mathsf{Def}(c!a) & = & \{\,\} \\
\mathsf{Def}(b) & = & \{\,\} \\
\mathsf{Def}(\mathtt{skip}) & = & \{\,\}
\end{array}
$$

Note that for the assignment $A[a_1] := a_2$ we just record that the array $A$ is being modified and in particular we have decided not to record a specific entry into the array.

For a path $\pi = q_0, \alpha_1, q_1, \cdots, q_{n-1}, \alpha_n, q_n$ with $n \geq 0$ and a variable $x$ we define

$$
\mathsf{Def}(\pi, x) = \begin{cases} (x, q_{i-1}, q_i) & \text{if } x \in \mathsf{Def}(\alpha_i) \text{ and } \forall j > i : x \notin \mathsf{Def}(\alpha_j) \\ (x, ?, q_0) & \text{if } \forall j : x \notin \mathsf{Def}(\alpha_j) \end{cases}
$$

because we only record the *last* occurrence of a definition of $x$. However, for a path $\pi$ as above and an array $A$ we define

$$
\mathsf{Def}(\pi, A) = \{(A, ?, q_0)\} \cup \{(A, q_{i-1}, q_i) \mid A \in \mathsf{Def}(\alpha_i)\}
$$

because we cannot be sure that a redefinition of $A$ overwrites all previous definitions.

Example 2.3: Consider the path $\pi$ of Example 1.16:

$$
\begin{array}{l}
q_\triangleright, \mathtt{y} := 1, q_1, \mathtt{x} > 0, q_2, \mathtt{y} := \mathtt{x} * \mathtt{y}, q_3, \mathtt{x} := \mathtt{x} - 1, \\
\qquad q_1, \mathtt{x} > 0, q_2, \mathtt{y} := \mathtt{x} * \mathtt{y}, q_3, \mathtt{x} := \mathtt{x} - 1, q_1, \mathtt{x} \leq 0, q_\blacktriangleleft
\end{array}
$$

Here we have $\mathsf{Def}(\pi, \mathtt{x}) = (\mathtt{x}, q_3, q_1)$ and $\mathsf{Def}(\pi, \mathtt{y}) = (\mathtt{y}, q_2, q_3)$.

Putting this together we define:

$$
\mathsf{Def}(\pi) = \{\mathsf{Def}(\pi, x) \mid x \in \mathbf{Var}\} \cup \bigcup_{A \in \mathbf{Arr}} \mathsf{Def}(\pi, A)
$$

Try It Out 2.4: Determine the set $\mathsf{Def}(\pi)$ for the paths $\pi$ of Example 1.13. □

We can now reformulate the aim of the Reaching Definitions analysis as follows: for each node $q$ of a program graph we want to determine the set of all possible definitions that may arise for paths starting at $q_\triangleright$ and ending at $q$. This then gives rise to a notion of correctness of a Reaching Definitions analysis given by:

DEFINITION 2.5: An analysis assignment RD *summarises the paths* of a program graph provided that $\text{Def}(\pi) \subseteq \text{RD}(q)$ for every path $\pi$ from $q_\triangleright$ to $q$.

The *least* analysis assignment that summarises the paths of a program graph is sometimes called the *MOP solution* (which we may read as Merge Over Paths solution).

TRY IT OUT 2.6: Argue that an analysis assignment RD summarises the paths for the program graph of Figure 2.1 exactly when the conditions of Example 2.2 hold. □

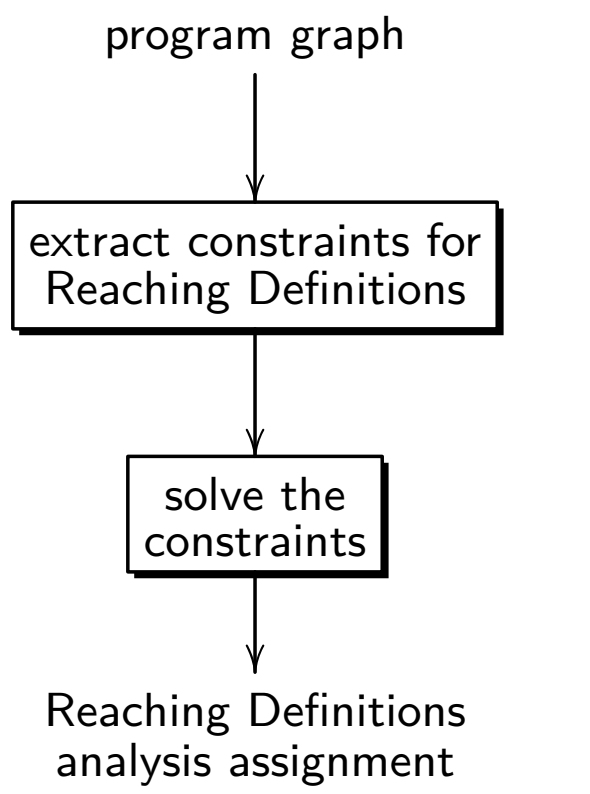

Figure 2.2: Computing the analysis assignment.

**Constraints for Reaching Definitions.** The above treatment does not present an effective way of calculating a correct Reaching Definitions analysis. To remedy this we shall now formulate a system of constraints that the Reaching Definitions assignment must satisfy and we shall present an algorithm for solving these constraints. The overall process is illustrated on Figure 2.2.

The first step is to define sets of so-called killed and generated definitions for each edge $(q_\circ, \alpha, q_\bullet)$ in the program graph. Let us consider an assignment $x := a$; here we will *generate* the definition $(x, q_\circ, q_\bullet)$ because $x$ is being defined by the action. All previous definitions of $x$ will be *killed* as they are no longer of interest; they are all included in the set $\{x\} \times \mathbf{Q}_? \times \mathbf{Q}$ as recorded in the table below:

| $\alpha$ | $\text{kill}_{\text{RD}}(q_\circ, \alpha, q_\bullet)$ | $\text{gen}_{\text{RD}}(q_\circ, \alpha, q_\bullet)$ |
|---|---|---|
| $x := a$ | $\{x\} \times \mathbf{Q}_? \times \mathbf{Q}$ | $\{(x, q_\circ, q_\bullet)\}$ |
| $A[a_1] := a_2$ | $\{\}$ | $\{(A, q_\circ, q_\bullet)\}$ |
| $c?x$ | $\{x\} \times \mathbf{Q}_? \times \mathbf{Q}$ | $\{(x, q_\circ, q_\bullet)\}$ |
| $c?A[a]$ | $\{\}$ | $\{(A, q_\circ, q_\bullet)\}$ |
| $c!a$ | $\{\}$ | $\{\}$ |
| $b$ | $\{\}$ | $\{\}$ |
| skip | $\{\}$ | $\{\}$ |

In general a triple $(x, q_\circ, q_\bullet)$ or $(A, q_\circ, q_\bullet)$ is generated whenever the action $\alpha$ might define (or modify) the variable $x$ or array $A$. However, a triple $(x, q_\circ, q_\bullet)$ or $(A, q_\circ, q_\bullet)$ is killed only when we can be sure that a redefinition overwrites all previous definitions; this will be the case for variables but not for arrays. Note that the two cases of input on channels resemble those of assignments and that for output, test and skip we neither generate nor kill any definitions.

TRY IT OUT 2.7: For the edge $(q_\triangleright, \mathtt{y} := 1, q_1)$ of Figure 2.1 we get

$$\begin{array}{rcl} \text{kill}_{\text{RD}}(q_\triangleright, \mathtt{y} := 1, q_1) & = & \{(\mathtt{y}, q, q') \mid q \in \mathbf{Q}_?, q' \in \mathbf{Q}\} \\ \text{gen}_{\text{RD}}(q_\triangleright, \mathtt{y} := 1, q_1) & = & \{(\mathtt{y}, q_\triangleright, q_1)\} \end{array}$$

Construct the sets of killed and generated definitions for the remaining edges of Figure 2.1. □

EXERCISE 2.8: Consider the program graph of Figure 1.13 and construct the sets of killed and generated definitions for each of the edges. □

EXERCISE 2.9: Suppose that the array $A$ in question has length 1. Is it possible to improve upon the definition of $\mathsf{kill}_{\mathsf{RD}}$? What are the consequences for the definition of Def? □

The next step is to impose constraints on the analysis assignment RD. For each edge $(q_\circ, \alpha, q_\bullet)$ in the program graph we shall require that:

$$\big(\mathsf{RD}(q_\circ) \setminus \mathsf{kill}_{\mathsf{RD}}(q_\circ, \alpha, q_\bullet)\big) \cup \mathsf{gen}_{\mathsf{RD}}(q_\circ, \alpha, q_\bullet) \subseteq \mathsf{RD}(q_\bullet)$$

For the entry node we shall additionally require that:

$$(\mathbf{Var} \cup \mathbf{Arr}) \times \{?\} \times \{q_\triangleright\} \subseteq \mathsf{RD}(q_\triangleright)$$

The latter constraint simply expresses that at the initial node all variables and arrays have their initial value.

EXAMPLE 2.10: Using that $\mathbf{Var} = \{\mathtt{x}, \mathtt{y}\}$ and $\mathbf{Arr} = \{\,\}$ and transposing the subset inclusions to superset inclusions the constraints obtained for the program graph of Figure 2.1 are:

$$\begin{array}{lcl}
\mathsf{RD}(q_\triangleright) & \supseteq & \{(\mathtt{x}, ?, q_\triangleright), (\mathtt{y}, ?, q_\triangleright)\} \\
\mathsf{RD}(q_1) & \supseteq & (\mathsf{RD}(q_\triangleright) \setminus (\{\mathtt{y}\} \times \{?, q_\triangleright, q_1, \cdots, q_\blacktriangleleft\} \times \{q_\triangleright, q_1, \cdots, q_\blacktriangleleft\})) \cup \{(\mathtt{y}, q_\triangleright, q_1)\} \\
\mathsf{RD}(q_1) & \supseteq & (\mathsf{RD}(q_3) \setminus (\{\mathtt{x}\} \times \{?, q_\triangleright, q_1, \cdots, q_\blacktriangleleft\} \times \{q_\triangleright, q_1, \cdots, q_\blacktriangleleft\})) \cup \{(\mathtt{x}, q_3, q_1)\} \\
\mathsf{RD}(q_2) & \supseteq & (\mathsf{RD}(q_1) \setminus \{\,\}) \cup \{\,\} \\
\mathsf{RD}(q_3) & \supseteq & (\mathsf{RD}(q_2) \setminus (\{\mathtt{y}\} \times \{?, q_\triangleright, q_1, \cdots, q_\blacktriangleleft\} \times \{q_\triangleright, q_1, \cdots, q_\blacktriangleleft\})) \cup \{(\mathtt{y}, q_2, q_3)\} \\
\mathsf{RD}(q_\blacktriangleleft) & \supseteq & (\mathsf{RD}(q_1) \setminus \{\,\}) \cup \{\,\}
\end{array}$$

DEFINITION 2.11: An analysis assignment RD *solves the constraints* whenever it satisfies all the constraints constructed for the Reaching Definitions analysis.

TRY IT OUT 2.12: Returning to Example 2.2 consider the analysis assignment RD given by

$$\begin{array}{lcl}
\mathsf{RD}(q_\triangleright) & = & \{(\mathtt{x}, ?, q_\triangleright), (\mathtt{y}, ?, q_\triangleright)\} \\
\mathsf{RD}(q_1) & = & \{(\mathtt{x}, ?, q_\triangleright), (\mathtt{x}, q_3, q_1), (\mathtt{y}, q_\triangleright, q_1), (\mathtt{y}, q_2, q_3)\} \\
\mathsf{RD}(q_2) & = & \{(\mathtt{x}, ?, q_\triangleright), (\mathtt{x}, q_3, q_1), (\mathtt{y}, q_\triangleright, q_1), (\mathtt{y}, q_2, q_3)\} \\
\mathsf{RD}(q_3) & = & \{(\mathtt{x}, ?, q_\triangleright), (\mathtt{x}, q_3, q_1), (\mathtt{y}, q_2, q_3)\} \\
\mathsf{RD}(q_\blacktriangleleft) & = & \{(\mathtt{x}, ?, q_\triangleright), (\mathtt{x}, q_3, q_1), (\mathtt{y}, q_\triangleright, q_1), (\mathtt{y}, q_2, q_3)\}
\end{array}$$

and show that it solves the constraints of Example 2.10. □

EXERCISE 2.13: Rather than amalgamating all the array entries $A[0]$, ..., $A[k-1]$ into the array name $A$ we might consider to deal with them individually. Then the analysis assignments would then be mappings

$$\mathbf{Q} \to \mathsf{PowerSet}(\,(\mathbf{Var} \cup \{A[i] \mid A \in \mathbf{Arr}, 0 \leq i < \mathsf{length}(A)\})) \times \mathbf{Q}_? \times \mathbf{Q}\,)$$

The additional constraint for the entry node would then be

$$(\mathbf{Var} \cup \{A[i] \mid A \in \mathbf{Arr}, 0 \leq i < \mathsf{length}(A)\}) \times \{?\} \times \{q_\triangleright\} \subseteq \mathsf{RD}(q_\triangleright)$$

Redefine the sets Def, $\mathsf{kill}_{\mathsf{RD}}$, and $\mathsf{gen}_{\mathsf{RD}}$ according to this idea. Recalling the discussion in Section 1.6: Is it worth the trouble? □

**Solving the constraints.** It is fairly direct to sketch an algorithm for solving the constraints of the Reaching Definitions analysis.

| | |
|---|---|
| INPUT | a program graph with nodes $\mathbf{Q}$, initial node $q_\triangleright$ and edges $\mathbf{E}$ |
| OUTPUT | RD: an analysis assignment for Reaching Definitions analysis |
| METHOD | forall $q \in \mathbf{Q} \setminus \{q_\triangleright\}$ do $\mathsf{RD}(q) := \{\,\}$ ;<br>$\mathsf{RD}(q_\triangleright) := (\mathbf{Var} \cup \mathbf{Arr}) \times \{?\} \times \{q_\triangleright\}$ ;<br>while there exists an edge $(q_\circ, \alpha, q_\bullet) \in \mathbf{E}$<br>such that $(\mathsf{RD}(q_\circ) \setminus \mathsf{kill}_{\mathsf{RD}}(q_\circ, \alpha, q_\bullet)) \cup \mathsf{gen}_{\mathsf{RD}}(q_\circ, \alpha, q_\bullet) \nsubseteq \mathsf{RD}(q_\bullet)$<br>do $\mathsf{RD}(q_\bullet) := \mathsf{RD}(q_\bullet) \cup (\mathsf{RD}(q_\circ) \setminus \mathsf{kill}_{\mathsf{RD}}(q_\circ, \alpha, q_\bullet)) \cup \mathsf{gen}_{\mathsf{RD}}(q_\circ, \alpha, q_\bullet)$ |

The idea is to initialise the analysis assignment to record no information (that is, $\{\,\}$) for all the nodes except the initial node where we record that the variables and arrays all have their initial value. This is then followed by an iteration step where the analysis assignment is extended with more and more information in order to ensure that the constraints are satisfied for all the edges of the program graph. We shall study more efficient algorithms in Chapter 4.

TRY IT OUT 2.14: Returning to the program graph of Figure 2.1 let $\mathbf{Var} = \{\mathtt{x}, \mathtt{y}\}$ and $\mathbf{Arr} = \{\,\}$ and use the algorithm to compute the analysis assignment RD of Example 2.12. (Hint: Take a look at Example 2.10.) □

EXERCISE 2.15: Returning to the program graph of Figure 1.13 use the algorithm to compute an analysis assignment RD; as part of this you may reuse your definitions of the killed and generated definitions of Exercise 2.8. □

EXERCISE 2.16: Perform the Reaching Definitions analysis on the program graphs constructed in Exercise 1.12. □

PROPOSITION 2.17: The algorithm terminates and produces an analysis assignment that solves the constraints for the Reaching Definitions analysis.

PROOF: First we show that, if the algorithm terminates, then the analysis assignment solves the constraints. For the constraints generated from the edges this follows directly from the condition of the while loop. For the additional constraint generated for the entry node we observe that it is established already in the initialisation phase and that $\mathsf{RD}(q_{\triangleright})$ never decreases thereafter.

To show that the algorithm terminates note that each iteration of the while loop makes some $\mathsf{RD}(q_{\bullet})$ larger while not decreasing any of the other $\mathsf{RD}(q)$. Since the sets **Var**, **Arr**, $\mathbf{Q}_?$ and **Q** are finite this cannot continue forever. □

EXERCISE 2.18: Prove that if RD is the analysis assignment produced by the algorithm, and if RD′ is some other analysis assignment that also solves the constraints, then $\mathsf{RD}(q) \subseteq \mathsf{RD}'(q)$ holds for all nodes $q$ in **Q**. □

Exercise 2.18 shows that the algorithm produces the least analysis assignment that solves the constraints, and this solution is sometimes called the *MFP solution* (which we may read as Minimal Fixed Point solution in this case).

**Solutions summarise the paths.** It remains to relate the two notions of "solving the constraints" and "summarising the paths".

PROPOSITION 2.19: Whenever an analysis assignment solves the constraints of the Reaching Definitions analysis it also summarises the paths.

PROOF: Suppose that the analysis assignment RD solves the constraints. It suffices to consider an arbitrary path $\pi = q_0, \alpha_1, q_1, \cdots, q_{n-1}, \alpha_n, q_n$ with $q_0 = q_{\triangleright}$ and an arbitrary variable $x \in \mathbf{Var}$ and prove that $\mathsf{Def}(\pi, x) \in \mathsf{RD}(q_n)$ as the case for an array $A \in \mathbf{Arr}$ is mostly analogous.

If $\mathsf{Def}(\pi, x) = (x, ?, q_{\triangleright})$ we know that $\forall j : x \notin \mathsf{Def}(\alpha_j)$. It follows that $\forall j : (x, ?, q_{\triangleright}) \notin \mathsf{kill}_{\mathsf{RD}}(\alpha_j)$ and since $(x, ?, q_{\triangleright}) \in \mathsf{RD}(q_{\triangleright})$ it is immediate that $(x, ?, q_{\triangleright}) \in \mathsf{RD}(q_n)$.

If $\mathsf{Def}(\pi, x) = (x, q_{i-1}, q_i)$ we know that $x \in \mathsf{Def}(\alpha_i)$ and that $\forall j > i : x \notin \mathsf{Def}(\alpha_j)$. It follows that $(x, q_{i-1}, q_i) \in \mathsf{gen}_{\mathsf{RD}}(q_{i-1}, \alpha_i, q_i)$ and that $\forall j > i : (x, q_{i-1}, q_i) \notin \mathsf{kill}_{\mathsf{RD}}(\alpha_j)$. From the first we get $(x, q_{i-1}, q_i) \in \mathsf{RD}(q_i)$ and from the second we get $(x, q_{i-1}, q_i) \in \mathsf{RD}(q_n)$. □

EXERCISE 2.20: Conduct the proof of Proposition 2.19 for the case of an array $A \in \mathbf{Arr}$. □

TEASER 2.21: Is it possible to find an analysis assignment that summarises the paths but that does not solve the constraints? □

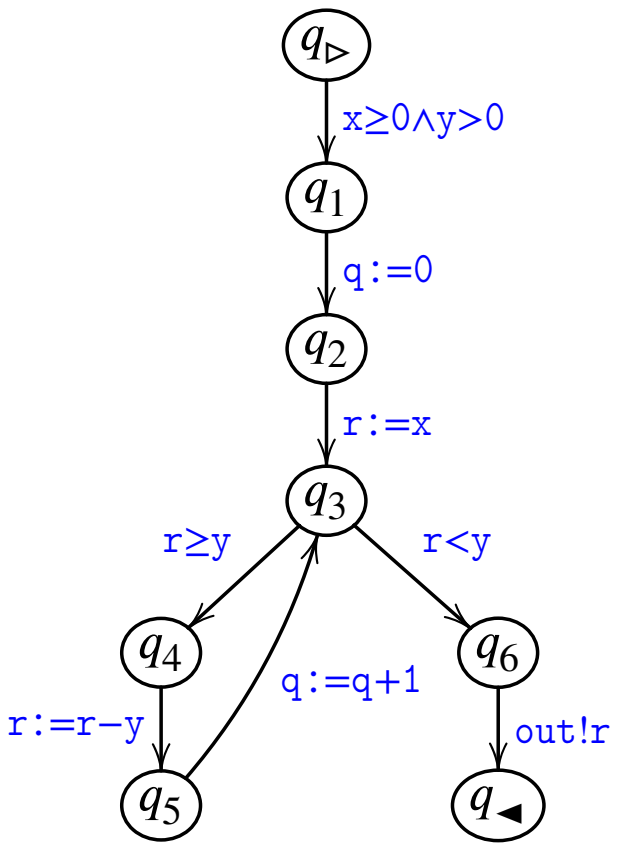

Figure 2.3: Program graph for the modulo function.

## 2.2 Live Variables

Live Variables analysis aims a finding out at every node of a program graph which variables or arrays might be used before being redefined (or the program terminates) – these variables are often called *live* variables in contrast to *dead* variables. As an example for the program graph of Figure 2.3 the analysis should tell us that once computation reaches the node $q_3$ there will no further use of the variable x and at the node $q_6$ only the value of r will be needed for the rest of the computation.

The information provided by the Live Variables analysis is useful in compiler construction where it for example will help us to determine when a register designated to hold the value of a variable may be reused for another purpose. The information is also useful for spotting parts of the program that perform computations whose result will never be needed, which can be exploited in compiler construction as well as in program understanding, and may arise when programs are continuously being modified by different teams of programmers. With respect to security the information may indicate information flow violations in case a variable is still used after its last intended use.

**Analysis assignments.** The Live Variables analysis will compute with *sets* of variables and array names. The idea is that if a variable $x$ is included in the analysis information for some node it means that the value of the variable might be used in the future computations before it is redefined (or the program terminates). Similarly, if an array name $A$ is included in the analysis result of a node then it means that one of the entries in the array might be used before being redefined.

Example 2.22: For the node $q_3$ of the program graph of Figure 2.3 we may use the set $\{x, y, q, r\}$ to tell us which variables *might* be used in the future computations. Clearly it will be more informative to use a smaller set – in this case we may remove x from the set but it will not be correct to remove any of the other variables from the set.

The result of a *Live Variables* analysis will be given by an *analysis assignment* LV that maps each node in the program graph to a set of variable and array names:

$$\mathsf{LV} : \mathbf{Q} \to \mathsf{PowerSet}(\,\mathbf{Var} \cup \mathbf{Arr}\,)$$

Example 2.23: Consider the program graph for the *modulo* function of Figure 2.3. Here we would like that an analysis assignment LV satisfies the following

conditions:

$$\begin{array}{lcllcl} \mathsf{LV}(q_\triangleright) & \supseteq & \{x, y\} & \mathsf{LV}(q_4) & \supseteq & \{y, q, r\} \\ \mathsf{LV}(q_1) & \supseteq & \{x, y\} & \mathsf{LV}(q_5) & \supseteq & \{y, q, r\} \\ \mathsf{LV}(q_2) & \supseteq & \{x, y, q\} & \mathsf{LV}(q_6) & \supseteq & \{r\} \\ \mathsf{LV}(q_3) & \supseteq & \{y, q, r\} & \mathsf{LV}(q_\blacktriangleleft) & \supseteq & \{\,\} \end{array}$$

If equalities hold in all cases we would intuitively think that LV is the best analysis assignment that can be hoped for.

**What is used on a path.** Let us be more precise about what we mean by being used. For each action $\alpha$ we shall be specifying the set of variables and arrays that are being used by it as follows:

$$\begin{array}{lcl} \mathsf{Use}(x := a) & = & \mathbf{fv}(a) \\ \mathsf{Use}(A[a_1] := a_2) & = & \mathbf{fv}(a_1) \cup \mathbf{fv}(a_2) \\ \mathsf{Use}(c?x) & = & \{\,\} \\ \mathsf{Use}(c?A[a]) & = & \mathbf{fv}(a) \\ \mathsf{Use}(c!a) & = & \mathbf{fv}(a) \\ \mathsf{Use}(b) & = & \mathbf{fv}(b) \\ \mathsf{Use}(\mathtt{skip}) & = & \{\,\} \end{array}$$

For the assignment $x := a$ we will need the set of variables (and arrays) mentioned in $a$ and we shall write $\mathbf{fv}(a)$ for this set; note that we will only need $x$ if it occurs within $a$. For the assignment $A[a_1] := a_2$ we will need the variables (and arrays) mentioned in $a_1$ as well as $a_2$ because we need to compute the index of the array as well as its new value. For the test $b$ we will need the values of all the variables (and arrays) occurring within the boolean expression; this set is denoted $\mathbf{fv}(b)$.

More formally $\mathbf{fv}(\cdots)$ denotes the sets of (so-called free) variables and arrays of a syntactic construct; in the case of arithmetic expressions and boolean expressions it is defined as follows:

$$\begin{array}{rcl} \mathbf{fv}(n) & = & \{\,\} \\ \mathbf{fv}(x) & = & \{x\} \\ \mathbf{fv}(A[a_0]) & = & \{A\} \cup \mathbf{fv}(a_0) \\ \mathbf{fv}(a_1\ op_a\ a_2) & = & \mathbf{fv}(a_1) \cup \mathbf{fv}(a_2) \\ \mathbf{fv}(-a_0) & = & \mathbf{fv}(a_0) \end{array} \qquad \begin{array}{rcl} \mathbf{fv}(\mathtt{true}) & = & \{\,\} \\ \mathbf{fv}(\mathtt{false}) & = & \{\,\} \\ \mathbf{fv}(a_1\ op_r\ a_2) & = & \mathbf{fv}(a_1) \cup \mathbf{fv}(a_2) \\ \mathbf{fv}(b_1\ op_b\ b_2) & = & \mathbf{fv}(b_1) \cup \mathbf{fv}(b_2) \\ \mathbf{fv}(\neg b_0) & = & \mathbf{fv}(b_0) \end{array}$$

TRY IT OUT 2.24: Use the above definitions to determine the sets $\mathbf{fv}(\mathtt{x} * \mathtt{y})$ and $\mathbf{fv}(\mathtt{A[i]} = \mathtt{x})$. □

For a path $\pi = q_0, \alpha_1, q_1, \cdots, q_{n-1}, \alpha_n, q_n$ with $n \geq 0$ and a variable $x$ we define

$$\mathsf{Use}(\pi, x) = \left(\exists i : x \in \mathsf{Use}(\alpha_i) \wedge \forall j < i : x \notin \mathsf{Def}(\alpha_j)\right)$$

However, for a path $\pi$ as above and an array $A$ we define

$$\mathsf{Use}(\pi, A) = \left(\exists i : A \in \mathsf{Use}(\alpha_i)\right)$$

because we cannot be sure that a redefinition of $A$ overwrites all previous definitions.

EXAMPLE 2.25: Let us write $\pi$ for the path

$$q_\triangleright, \texttt{x} \geq 0 \wedge \texttt{y} > 0, q_1, \texttt{q} := 0, q_2, \texttt{r} := \texttt{x}, q_3, \texttt{r} < \texttt{y}, q_6, \texttt{out!r}, q_\blacktriangleleft$$

of Figure 2.3. Then $\mathsf{Use}(\pi, \texttt{x})$ and $\mathsf{Use}(\pi, \texttt{y})$ are true whereas $\mathsf{Use}(\pi, \texttt{q})$ and $\mathsf{Use}(\pi, \texttt{r})$ are false.

Putting this together we define:

$$\mathsf{Use}(\pi) = \{x \in \textbf{Var} \mid \mathsf{Use}(\pi, x) \text{ is true}\} \cup \{A \in \textbf{Arr} \mid \mathsf{Use}(\pi, A) \text{ is true}\}$$

TRY IT OUT 2.26: Determine the sets $\mathsf{Use}(\pi)$ for the following paths $\pi$ of the program graph of Figure 2.3:

(a) $q_3, \texttt{r} < \texttt{y}, q_6, \texttt{out!r}, q_\blacktriangleleft$

(b) $q_3, \texttt{r} \geq \texttt{y}, q_4, \texttt{r} := \texttt{r} - \texttt{y}, q_5, \texttt{q} := \texttt{q} + 1, q_3, \texttt{r} < \texttt{y}, q_6, \texttt{out!r}, q_\blacktriangleleft$ □

We can now reformulate the aim of the Live Variables analysis as follows: for each node $q$ of a program graph we want to determine the set of all variables and arrays that possibly might be used in the future, that is, on paths starting at $q$ and ending at $q_\blacktriangleleft$. This then gives rise to the following notion of correctness of a Live Variables analysis:

DEFINITION 2.27: An analysis assignment LV *summarises the paths* of a program graph provided that $\mathsf{Use}(\pi) \subseteq \mathsf{LV}(q)$ for every path $\pi$ from $q$ to $q_\blacktriangleleft$.

The *least* analysis assignment that summarises the paths of a program graph is sometimes called the *MOP solution* (which we may read as Merge Over Paths solution).

TRY IT OUT 2.28: Argue that an analysis assignment LV summarises the paths for the program graph of Figure 2.3 exactly when the conditions of Example 2.23 hold. □

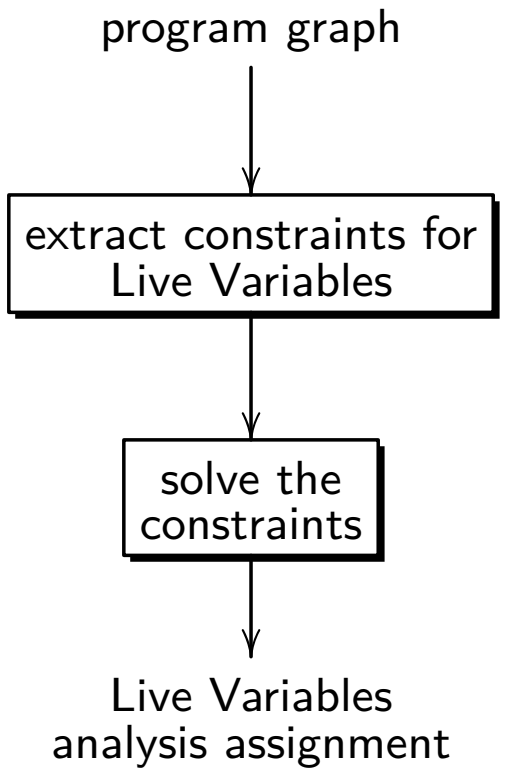

Figure 2.4: Computing the analysis assignment.

**Constraints for Live Variables.** The above treatment does not present an effective way of calculating a correct Live Variables analysis. We shall now show how to formulate a system of constraints that the Live Variables assignment must satisfy and we shall give an algorithm for solving these constraints. This is illustrated on Figure 2.4. Our approach is indeed very similar to the one for Reaching Definitions; however the form of the constraints will be different.

The first step is to define sets of so-called killed and generated variables and arrays for each edge $(q_\circ, \alpha, q_\bullet)$ of the program graph. First consider the assignment $x := a$;

it gives a new value to the variable $x$ so the previous value of $x$ is no longer of interest and hence it is *killed*. On the other hand we need the values of the variables occurring in $a$ in order to determine the new value of $x$; these variables are the ones of the set $\mathbf{fv}(a)$ so this set will be *generated*. This is recorded in the following table:

| $\alpha$ | $\mathsf{kill}_{\mathsf{LV}}(q_\circ, \alpha, q_\bullet)$ | $\mathsf{gen}_{\mathsf{LV}}(q_\circ, \alpha, q_\bullet)$ |
|---|---|---|
| $x$ := $a$ | $\{x\}$ | $\mathbf{fv}(a)$ |
| $A[a_1]$ := $a_2$ | $\{\,\}$ | $\mathbf{fv}(a_1) \cup \mathbf{fv}(a_2)$ |
| $c?x$ | $\{x\}$ | $\{\,\}$ |
| $c?A[a]$ | $\{\,\}$ | $\mathbf{fv}(a)$ |
| $c!a$ | $\{\,\}$ | $\mathbf{fv}(a)$ |
| $b$ | $\{\,\}$ | $\mathbf{fv}(b)$ |
| `skip` | $\{\,\}$ | $\{\,\}$ |

In general, a variable $x$ or array $A$ is generated whenever the action $\alpha$ might use the variable $x$ or array $A$. However, a variable $x$ or array $A$ is killed only when we can be sure that a redefinition overwrites all previous definitions; this will generally be the case for variables but not for arrays.

Try It Out 2.29: For the edge ($q_4$, `r := r - y`, $q_5$) of Figure 2.3 we get

$$\begin{array}{rcl} \mathsf{kill}_{\mathsf{LV}}(q_4, \texttt{r := r - y}, q_5) & = & \{\texttt{r}\} \\ \mathsf{gen}_{\mathsf{LV}}(q_4, \texttt{r := r - y}, q_5) & = & \{\texttt{r}, \texttt{y}\} \end{array}$$

Construct the sets of killed and generated variables for the remaining edges of Figure 2.3. □

Exercise 2.30: Consider the program graph of Figure 1.13 and construct the sets of killed and generated variables and arrays for each of the edges. □

Exercise 2.31: Suppose that the array $A$ in question has length 1. Is it possible to improve upon the definition of $\mathsf{kill}_{\mathsf{LV}}$? What are the consequences for the definition of Use? □

The next step is to impose constraints on the analysis assignment LV. For each edge $(q_\circ, \alpha, q_\bullet)$ in the program graph we shall require that:

$$\big(\mathsf{LV}(q_\bullet) \setminus \mathsf{kill}_{\mathsf{LV}}(q_\circ, \alpha, q_\bullet)\big) \cup \mathsf{gen}_{\mathsf{LV}}(q_\circ, \alpha, q_\bullet) \subseteq \mathsf{LV}(q_\circ)$$

For the exit node we shall additionally require that:

$$\{\,\} \subseteq \mathsf{LV}(q_\blacktriangleleft)$$

The last constraint is clearly superfluous; however, we find it instructive to highlight the fact that one should consider what to demand at the end of the program. For the factorial function of Figure 2.1 we may, for example, consider introducing the constraint with $\{\texttt{y}\} \subseteq \mathsf{LV}(q_\blacktriangleleft)$ for the final node in order to express that we are interested in the value of `y` when the program terminates.

TRY IT OUT 2.32: Construct the set of constraints obtained for the program graph of Figure 2.3 (in the manner of Example 2.10 and taking **Var** $= \{\mathtt{x}, \mathtt{y}, \mathtt{q}, \mathtt{r}\}$ and **Arr** $= \{\ \}$). □

DEFINITION 2.33: An analysis assignment LV *solves the constraints* whenever it satisfies all the constraints constructed for the Live Variables analysis.

TRY IT OUT 2.34: Returning to Example 2.23 consider the analysis assignment LV given by

$$
\begin{array}{lcl@{\qquad}lcl}
\mathsf{LV}(q_{\triangleright}) & = & \{\mathtt{x}, \mathtt{y}\} & \mathsf{LV}(q_4) & = & \{\mathtt{y}, \mathtt{q}, \mathtt{r}\} \\
\mathsf{LV}(q_1) & = & \{\mathtt{x}, \mathtt{y}\} & \mathsf{LV}(q_5) & = & \{\mathtt{y}, \mathtt{q}, \mathtt{r}\} \\
\mathsf{LV}(q_2) & = & \{\mathtt{x}, \mathtt{y}, \mathtt{q}\} & \mathsf{LV}(q_6) & = & \{\mathtt{r}\} \\
\mathsf{LV}(q_3) & = & \{\mathtt{y}, \mathtt{q}, \mathtt{r}\} & \mathsf{LV}(q_{\blacktriangleleft}) & = & \{\ \}
\end{array}
$$

and show that is solves the constraints of Try It Out 2.32. □

EXERCISE 2.35: Rather than amalgamating all the array entries $A[0]$, ..., $A[n-1]$ into the array name $A$ we might consider to deal with them individually. Redefine Use, $\mathsf{kill}_{\mathsf{LV}}$, and $\mathsf{gen}_{\mathsf{LV}}$ according to this idea. Recalling the discussion in Section 1.6: Is it worth the trouble? □

**Solving the constraints.** It is fairly direct to sketch an algorithm for solving the constraints of the Live Variables analysis.

| | |
|---|---|
| INPUT | a program graph with nodes **Q**, final node $q_{\blacktriangleleft}$ and edges **E** |
| OUTPUT | LV: an analysis assignment for Live Variables analysis |
| METHOD | forall $q \in \mathbf{Q}$ do $\mathsf{LV}(q) := \{\ \}$ ; |
| | while there exists an edge $(q_\circ, \alpha, q_\bullet) \in \mathbf{E}$<br>such that $(\mathsf{LV}(q_\bullet) \setminus \mathsf{kill}_{\mathsf{LV}}(q_\circ, \alpha, q_\bullet)) \cup \mathsf{gen}_{\mathsf{LV}}(q_\circ, \alpha, q_\bullet) \not\subseteq \mathsf{LV}(q_\circ)$<br>do $\mathsf{LV}(q_\circ) := \mathsf{LV}(q_\circ) \cup (\mathsf{LV}(q_\bullet) \setminus \mathsf{kill}_{\mathsf{LV}}(q_\circ, \alpha, q_\bullet)) \cup \mathsf{gen}_{\mathsf{LV}}(q_\circ, \alpha, q_\bullet)$ |

The algorithm is very similar to the one we presented for the Reaching Definitions analysis in the previous section. It consists of an initialisation step followed by an iteration step where the analysis assignment is extended to ensure that all constraints are satisfied. We shall study more efficient algorithms in Chapter 4.

TRY IT OUT 2.36: Returning to the program graph of Figure 2.3 let **Var** $= \{\mathtt{x}, \mathtt{y}, \mathtt{q}, \mathtt{r}\}$ and **Arr** $= \{\ \}$ and use the algorithm to construct the the analysis assignment LV of Try It Out 2.34. (Hint: take a look at Try It Out 2.32.) □

EXERCISE 2.37: Returning to the program graph of Figure 1.13 use the algorithm to compute an analysis assignment LV; as part of this you may reuse your definitions of the killed and generated variables and arrays of Exercise 2.30. □

Exercise 2.38: Perform the Live Variables analysis on the program graphs constructed in Exercise 1.12. □

The following result is analogous to Proposition 2.17 and the proof is similar:

> Proposition 2.39: The algorithm terminates and produces an analysis assignment that solves the constraints for the Live Variables analysis.

Exercise 2.40: Prove that if LV is the analysis assignment produced by the algorithm, and if LV′ is some other analysis assignment that also solves the constraints, then $\mathsf{LV}(q) \subseteq \mathsf{LV}'(q)$ holds for all nodes $q$ in $\mathbf{Q}$. □

Exercise 2.40 shows that the algorithm produces the least analysis assignment that solves the constraints, and this solution is sometimes called the *MFP solution* (which we may read as Minimal Fixed Point solution in this case).

**Solutions summarise the paths.** It remains to relate the two notions of "solving the constraints" and "summarising the paths". The following result is established in the same way as Proposition 2.19:

> Proposition 2.41: Whenever an analysis assignment solves the constraints of the Live Variables analysis it also summarises the paths.

Teaser 2.42: Is it possible to find an analysis assignment that summarises the paths but that does not solve the constraints? □

## 2.3 Available Expressions

An expression is said to be *available* at a node of a program graph if its value has been computed previously and if a recomputation at the current node will give the same value. Available Expressions analysis then aims to determine for each node of a program graph which (non-trivial) expressions are available. As an example for the program graph of Figure 2.5 the analysis could tell us that at node $q_6$ the expressions $\mathtt{i * m}$ and $\mathtt{i * m + j}$ are available; however, they are not available at the node $q_3$.

The information provided by the Available Expressions analysis is useful for compiler construction because it opens up for improvements (or 'optimisations') of the code where time is saved by fetching a previously computed value rather than recomputing it. As an example the optimisation may modify the program graph of Figure 2.5 to have an assignment $\mathtt{z := i * m}$ before entering the loop at node $q_3$ and then the value of $\mathtt{u}$ can be computed using an assignment $\mathtt{u := z + j}$.

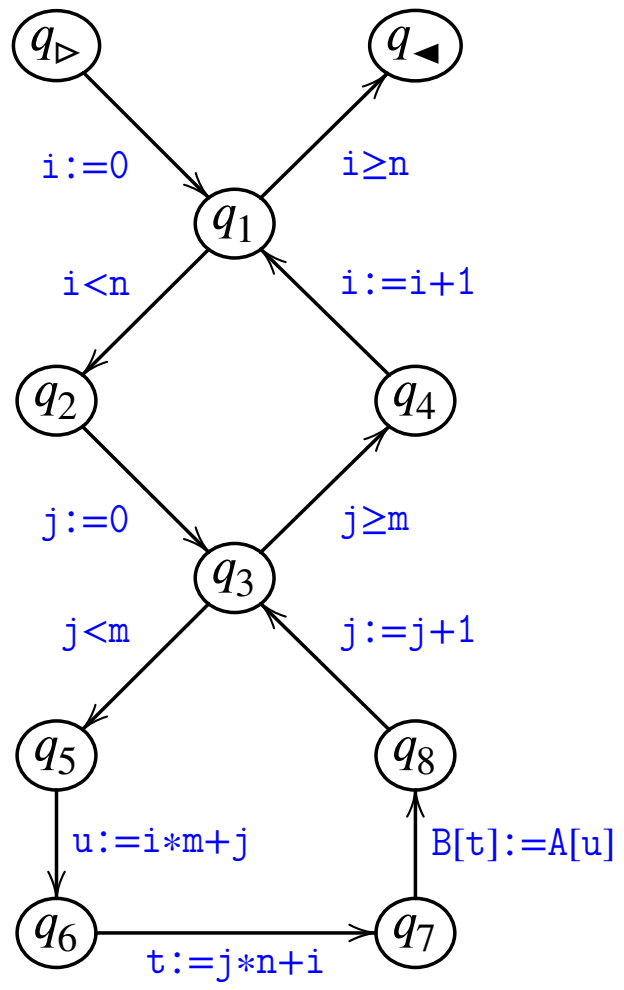

Figure 2.5: Program graph for matrix transpose.

**Analysis assignments.** The analysis is only concerned about *non-trivial* expressions, that is, arithmetic expressions involving at least one operator. So $a_1\ op_a\ a_2$ is a non-trivial expression (for all choices of $op_a$) and so are $A[a]$ and $-a$; however, $n$ and $x$ are not. We shall write **AExp** for the set of non-trivial expressions in the program graph of interest; this includes all the non-trivial subexpressions occurring in the arithmetic and boolean expressions of the program graph.

EXAMPLE 2.43: Consider the program graph of Figure 2.5. Here we have

$$\textbf{AExp} = \{\texttt{i}+\texttt{1}, \texttt{j}+\texttt{1}, \texttt{i}*\texttt{m}, \texttt{i}*\texttt{m}+\texttt{j}, \texttt{j}*\texttt{n}, \texttt{j}*\texttt{n}+\texttt{i}, \texttt{A[u]}\}$$

Note that in addition to the arithmetic expressions $\texttt{i}*\texttt{m}+\texttt{j}$ and $\texttt{j}*\texttt{m}+\texttt{i}$ that directly occur in the actions we also include the two subexpressions $\texttt{i}*\texttt{m}$ and $\texttt{j}*\texttt{n}$ whose value will be computed as part of computing the value of the more complex expressions.

For the node $q_6$ we may use the set $\{\texttt{i}*\texttt{m}+\texttt{j}, \texttt{i}*\texttt{m}\}$ to tell which expressions are available. Clearly it will be more informative to use a *larger* set – in this case we may remove $\texttt{i}*\texttt{m}$ from the set but it will not be correct to include $\texttt{j}*\texttt{n}$ in the set. Thus, unlike the analyses we have seen so far, it is better to include too few possibilities than to include too many.

The result of an *Available Expressions* analysis will be given by an analysis assignment AE that maps each node in the program graph to a set of non-trivial expressions:

$$\mathsf{AE} : \mathbf{Q} \rightarrow \mathsf{PowerSet}(\,\textbf{AExp}\,)$$

EXAMPLE 2.44: Returning to the program graph of Figure 2.5 we would like that the analysis assignment AE satisfies the following conditions:

$$\begin{array}{lcl@{\qquad}lcl}
\mathsf{AE}(q_\triangleright) & \subseteq & \{\,\} & \mathsf{AE}(q_5) & \subseteq & \{\,\} \\
\mathsf{AE}(q_1) & \subseteq & \{\,\} & \mathsf{AE}(q_6) & \subseteq & \{\texttt{i}*\texttt{m}, \texttt{i}*\texttt{m}+\texttt{j}\} \\
\mathsf{AE}(q_2) & \subseteq & \{\,\} & \mathsf{AE}(q_7) & \subseteq & \{\texttt{i}*\texttt{m}, \texttt{i}*\texttt{m}+\texttt{j}, \texttt{j}*\texttt{n}, \texttt{j}*\texttt{n}+\texttt{i}\} \\
\mathsf{AE}(q_3) & \subseteq & \{\,\} & \mathsf{AE}(q_8) & \subseteq & \{\texttt{i}*\texttt{m}, \texttt{i}*\texttt{m}+\texttt{j}, \texttt{j}*\texttt{n}, \texttt{j}*\texttt{n}+\texttt{i}, \texttt{A[u]}\} \\
\mathsf{AE}(q_4) & \subseteq & \{\,\} & \mathsf{AE}(q_\blacktriangleleft) & \subseteq & \{\,\}
\end{array}$$

If equalities hold in all cases we would intuitively think that AE is the best analysis assignment that can be hoped for.

**What is available on a path.** Following the approach of the previous sections we shall start by being precise about the information to be determined by the analysis. For each action $\alpha$ we define $\mathsf{AExp}(\alpha)$ to be the set of non-trivial expressions occurring

in $\alpha$:

$$
\begin{array}{rcl}
\mathsf{AExp}(x \; \mathtt{:=} \; a) & = & \mathbf{ae}(a) \\
\mathsf{AExp}(A\mathtt{[}a_1\mathtt{]} \; \mathtt{:=} \; a_2) & = & \mathbf{ae}(a_1) \cup \mathbf{ae}(a_2) \\
\mathsf{AExp}(c\mathtt{?}x) & = & \{\,\} \\
\mathsf{AExp}(c\mathtt{?}A\mathtt{[}a\mathtt{]}) & = & \mathbf{ae}(a) \\
\mathsf{AExp}(c\mathtt{!}a) & = & \mathbf{ae}(a) \\
\mathsf{AExp}(b) & = & \mathbf{ae}(b) \\
\mathsf{AExp}(\mathtt{skip}) & = & \{\,\}
\end{array}
$$

Basically this amounts to inspecting all the arithmetic and boolean expressions of the action and determining their non-trivial subexpressions. This is formalised by the following definitions of $\mathbf{ae}(a)$ and $\mathbf{ae}(b)$:

$$
\begin{array}{rcl}
\mathbf{ae}(n) & = & \{\,\} \\
\mathbf{ae}(x) & = & \{\,\} \\
\mathbf{ae}(A\mathtt{[}a_0\mathtt{]}) & = & \mathbf{ae}(a_0) \cup \{A\mathtt{[}a_0\mathtt{]}\} \\
\mathbf{ae}(a_1 \; op_a \; a_2) & = & \mathbf{ae}(a_1) \cup \mathbf{ae}(a_2) \\
 & \cup & \{a_1 \; op_a \; a_2\} \\
\mathbf{ae}(-a_0) & = & \{-a_0\} \cup \mathbf{ae}(a_0)
\end{array}
\qquad
\begin{array}{rcl}
\mathbf{ae}(\mathtt{true}) & = & \{\,\} \\
\mathbf{ae}(\mathtt{false}) & = & \{\,\} \\
\mathbf{ae}(a_1 \; op_r \; a_2) & = & \mathbf{ae}(a_1) \cup \mathbf{ae}(a_2) \\
\mathbf{ae}(b_1 \; op_b \; b_2) & = & \mathbf{ae}(b_1) \cup \mathbf{ae}(b_2) \\
\mathbf{ae}(\neg b_0) & = & \mathbf{ae}(b_0)
\end{array}
$$

Note that for the composite arithmetic expressions $A\mathtt{[}a_0\mathtt{]}$ and $a_1 \; op_a \; a_2$ we include the non-trivial expressions of the subexpressions as well as the expression itself.

TRY IT OUT 2.45: Use the above definitions to determine the sets $\mathbf{ae}(\mathtt{x} * \mathtt{y} * \mathtt{z})$ and $\mathbf{ae}(\mathtt{B[i]} + \mathtt{B[i]/10})$. □

For a path $\pi = q_0, \alpha_1, q_1, \cdots, q_{n-1}, \alpha_n, q_n$ with $n \geq 0$ we define

$$
\mathsf{AExp}(\pi) = \{a \in \mathbf{AExp} \mid \exists i \leq n : a \in \mathsf{AExp}(\alpha_i) \wedge \forall j \geq i : \mathbf{fv}(a) \cap \mathsf{Def}(\alpha_j) = \emptyset\}
$$

This ensures that an available expression is disregarded as soon as one of the variables or arrays occurring in it is redefined – it could be in the same action since we take $j \geq i$ rather than $j > i$. In this way we capture that in this case the value of the expression might change so it will no longer be available.

EXAMPLE 2.46: Returning to the program graph of Figure 2.5 let us consider the following paths that both start in $q_\triangleright$ and end in $q_3$:

$$
\begin{array}{rcl}
\pi_1 & = & q_\triangleright, \mathtt{i} := \mathtt{0}, q_1, \mathtt{i} < \mathtt{n}, q_2, \mathtt{j} := \mathtt{0}, q_3 \\
\pi_2 & = & q_\triangleright, \mathtt{i} := \mathtt{0}, q_1, \mathtt{i} < \mathtt{n}, q_2, \mathtt{j} := \mathtt{0}, q_3, \mathtt{j} < \mathtt{m}, q_5, \mathtt{u} := \mathtt{i} * \mathtt{m} + \mathtt{j}, q_6, \\
 & & \qquad \mathtt{t} := \mathtt{j} * \mathtt{n} + \mathtt{i}, q_7, \mathtt{B[t]} := \mathtt{A[u]}, q_8, \mathtt{j} := \mathtt{j} + \mathtt{1}, q_3
\end{array}
$$

We have $\mathsf{AExp}(\pi_1) = \{\,\}$ and $\mathsf{AExp}(\pi_2) = \{\mathtt{i} * \mathtt{m}, \mathtt{A[u]}, \mathtt{j} + \mathtt{1}\}$. The expressions $\mathtt{i} * \mathtt{m} + \mathtt{j}$ and $\mathtt{j} * \mathtt{n} + \mathtt{i}$ are not included in the latter since $\mathtt{j}$ is updated in the last action of $\pi_2$.

We can now reformulate the aim of the Available Expressions analysis as follows: For each node $q$ of a program graph we want to determine the set of non-trivial expressions that are available on *all* paths starting at $q_\triangleright$ and ending at $q$. This then gives rise to the following notion of correctness of an Available Expressions analysis:

DEFINITION 2.47: An analysis assignment AE *summarises the paths* of a program graph provided that $\mathsf{AExp}(\pi) \supseteq \mathsf{AE}(q)$ for every path $\pi$ from $q_\triangleright$ to $q$.

The *greatest* analysis assignment that summarises the paths of a program graph is sometimes called the *MOP solution* (which we may read as Merge Over Paths solution).

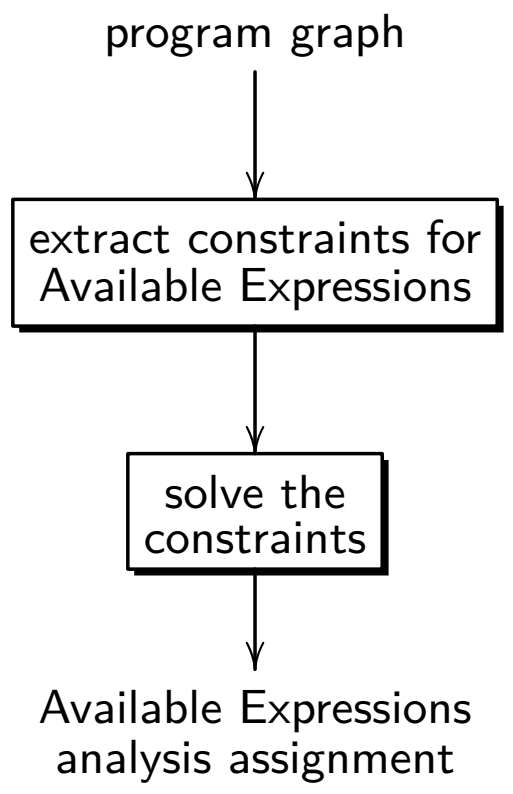

Figure 2.6: Computing the analysis assignment.

TRY IT OUT 2.48: Argue that an analysis assignment AE summarises the paths for the program graph of Example 2.5 exactly when the conditions of Example 2.43 hold. □

**Constraints for Available Expressions.** The above treatment does not present an effective way of calculating a correct Available Expressions analysis. Following the approach of the previous analyses we shall therefore formulate a system of constraints that the Available Expressions assignment must satisfy and next we shall present an algorithm solving these constraints. This is illustrated on Figure 2.6.

Our construction of the constraints follow the same overall approach as in the previous sections. So the first step is to define sets of so-called killed and generated available expressions for each edge $(q_\circ, \alpha, q_\bullet)$ in the program graph:

| $\alpha$ | $\mathsf{kill}_{\mathsf{AE}}(q_\circ, \alpha, q_\bullet)$ | $\mathsf{gen}_{\mathsf{AE}}(q_\circ, \alpha, q_\bullet)$ |
|---|---|---|
| $x$ := $a$ | $\{a' \in \mathbf{AExp} \mid x \in \mathbf{fv}(a')\}$ | $\{a' \in \mathbf{ae}(a) \mid x \notin \mathbf{fv}(a')\}$ |
| $A$[$a_1$] := $a_2$ | $\{a' \in \mathbf{AExp} \mid A \in \mathbf{fv}(a')\}$ | $\{a' \in \mathbf{ae}(a_1) \cup \mathbf{ae}(a_2) \mid A \notin \mathbf{fv}(a')\}$ |
| $c$?$x$ | $\{a' \in \mathbf{AExp} \mid x \in \mathbf{fv}(a')\}$ | $\{\,\}$ |
| $c$?$A$[$a$] | $\{a' \in \mathbf{AExp} \mid A \in \mathbf{fv}(a')\}$ | $\{a' \in \mathbf{ae}(a) \mid A \notin \mathbf{fv}(a')\}$ |
| $c$!$a$ | $\{\,\}$ | $\mathbf{ae}(a)$ |
| $b$ | $\{\,\}$ | $\mathbf{ae}(b)$ |
| skip | $\{\,\}$ | $\{\,\}$ |

The idea is that a non-trivial expression $a'$ is *killed* by an action whenever the action defines any variable or array used in $a'$ because then the value of $a'$ might change. A non-trivial expression $a'$ is *generated* by an action whenever it is computed by the action and it does not make use of a variable or an array that is killed by the action; again the reason is that a recomputation of $a'$ then may give another value.

TRY IT OUT 2.49: For the edge $(q_8, \texttt{j := j + 1}, q_3)$ of Figure 2.5 we then get

$$\mathsf{kill}_{\mathsf{AE}}(q_8, \texttt{j := j + 1}, q_3) = \{\texttt{j + 1}, \texttt{i * m + j}, \texttt{j * n}, \texttt{j * n + i}\}$$
$$\mathsf{gen}_{\mathsf{AE}}(q_8, \texttt{j := j + 1}, q_3) = \{\,\}$$

where we recall that the set **AExp** of non-trivial expressions is as in Example 2.43. Construct the sets of killed and generated variables for the remaining edges of Figure 2.5. □

EXERCISE 2.50: Consider the program graph of Figure 1.13 and construct the sets of killed and generated variables and arrays for each of the edges. □

The next step is to impose constraints on the analysis assignment AE. For each edge $(q_\circ, \alpha, q_\bullet)$ in the program graph we shall require that

$$\big(\mathsf{AE}(q_\circ) \setminus \mathsf{kill}_{\mathsf{AE}}(q_\circ, \alpha, q_\bullet)\big) \cup \mathsf{gen}_{\mathsf{AE}}(q_\circ, \alpha, q_\bullet) \supseteq \mathsf{AE}(q_\bullet)$$

For the entry node we shall additionally require that

$$\{\,\} \supseteq \mathsf{AE}(q_\triangleright)$$

Note that the last constraint is *not* superfluous. It is equivalent to $\mathsf{AE}(q_\triangleright) = \{\,\}$ and thus captures that no expressions are available when the computation starts.

TRY IT OUT 2.51: Construct the set of constraints obtained for the program graph of Figure 2.5 (in the manner of Example 2.10 and when **AExp** is as in Example 2.43). □

DEFINITION 2.52: An analysis assignment AE *solves the constraints* whenever it satisfies all the constraints constructed for the Available Expressions analysis.

TRY IT OUT 2.53: Returning to Example 2.43 consider the analysis assignment AE given by

$$\begin{array}{lcllcl}
\mathsf{AE}(q_\triangleright) & = & \{\,\} & \mathsf{AE}(q_5) & = & \{\,\} \\
\mathsf{AE}(q_1) & = & \{\,\} & \mathsf{AE}(q_6) & = & \{\mathtt{i}*\mathtt{m}, \mathtt{i}*\mathtt{m}+\mathtt{j}\} \\
\mathsf{AE}(q_2) & = & \{\,\} & \mathsf{AE}(q_7) & = & \{\mathtt{i}*\mathtt{m}, \mathtt{i}*\mathtt{m}+\mathtt{j}, \mathtt{j}*\mathtt{n}, \mathtt{j}*\mathtt{n}+\mathtt{i}\} \\
\mathsf{AE}(q_3) & = & \{\,\} & \mathsf{AE}(q_8) & = & \{\mathtt{i}*\mathtt{m}, \mathtt{i}*\mathtt{m}+\mathtt{j}, \mathtt{j}*\mathtt{n}, \mathtt{j}*\mathtt{n}+\mathtt{i}, \mathtt{A[u]}\} \\
\mathsf{AE}(q_4) & = & \{\,\} & \mathsf{AE}(q_\blacktriangleleft) & = & \{\,\}
\end{array}$$

and show that is solves the constraints of Try It Out 2.51. □

**Solving the constraints.** As in the previous sections it is fairly direct to sketch an algorithm for solving the constraints of the Available Expressions analysis.

| | |
|---|---|
| INPUT | a program graph with nodes **Q**, initial node $q_{\triangleright}$ and edges **E** |
| OUTPUT | AE: an analysis assignment for Available Expressions analysis |
| METHOD | forall $q \in \mathbf{Q} \setminus \{q_{\triangleright}\}$ do $\mathsf{AE}(q) := \mathbf{AExp}$ ;<br>$\mathsf{AE}(q_{\triangleright}) := \{\,\}$ ;<br>while there exists an edge $(q_\circ, \alpha, q_\bullet) \in \mathbf{E}$<br>such that $(\mathsf{AE}(q_\circ) \setminus \mathsf{kill}_{\mathsf{AE}}(q_\circ, \alpha, q_\bullet)) \cup \mathsf{gen}_{\mathsf{AE}}(q_\circ, \alpha, q_\bullet) \not\supseteq \mathsf{AE}(q_\bullet)$<br>do $\mathsf{AE}(q_\bullet) := \mathsf{AE}(q_\bullet) \cap \big((\mathsf{AE}(q_\circ) \setminus \mathsf{kill}_{\mathsf{AE}}(q_\circ, \alpha, q_\bullet)) \cup \mathsf{gen}_{\mathsf{AE}}(q_\circ, \alpha, q_\bullet)\big)$ |

Again the algorithm consists of an initialisation step followed by an iteration step where the analysis assignment is extended until all constraints are satisfied. We shall study more efficient algorithms in Chapter 4.

Try It Out 2.54: Returning to the program graph of Figure 2.5 let

$$\mathbf{AExp} = \{\mathtt{i+1}, \mathtt{j+1}, \mathtt{i*m}, \mathtt{i*m+j}, \mathtt{j*n}, \mathtt{j*n+i}, \mathtt{A[u]}\}$$

and use the algorithm to construct the analysis assignment AE of Example 2.44. □

Exercise 2.55: Returning to the program graph of Figure 1.13 use the algorithm to compute an analysis assignment AE; as part of this you may reuse your definitions of the killed and generated available expressions of Exercise 2.50. □

Exercise 2.56: Perform the Available Expressions analysis on the program graphs constructed in Exercise 1.12. □

Corresponding to Proposition 2.17 and 2.39 we have:

Proposition 2.57: The algorithm terminates and produces an analysis assignment that solves the constraints of the Available Expressions analysis.

Exercise 2.58: Prove that if AE is the analysis assignment produced by the algorithm, and if AE′ is some other analysis assignment that also solves the constraints, then $\mathsf{AE}(q) \supseteq \mathsf{AE}'(q)$ holds for all nodes $q$ in **Q**. □

Exercise 2.58 shows that the algorithm produces the greatest analysis assignment that solves the constraints, and this solution is sometimes called the *MFP solution* (which we may read as Maximal Fixed Point solution in this case).

**Solutions summarise the paths.** It remains to relate the notion of "solving the constraints" and "summarising the paths". Corresponding to Proposition 2.19 and 2.41 we have:

PROPOSITION 2.59: Whenever an analysis assignment solves the constraints it also summarises the paths.

TEASER 2.60: Is it possible to find an analysis assignment that summarises the paths but that does not solve the constraints? □

Let us conclude with a more detailed discussion of what it means for an expression to be available. An expression $a$ will *become* available as a result of executing an action on an edge $(q_\circ, \alpha, q_\bullet)$ and the analysis will then record that $a$ is available at $q_\bullet$. The expression may *continue* to be available at some of the subsequent nodes; if it is also available at $q$ it means that the evaluation of $a$ at $q_\bullet$ and $q$ gives the same result. This holds independently of whether the evaluation gives a proper value or it fails.

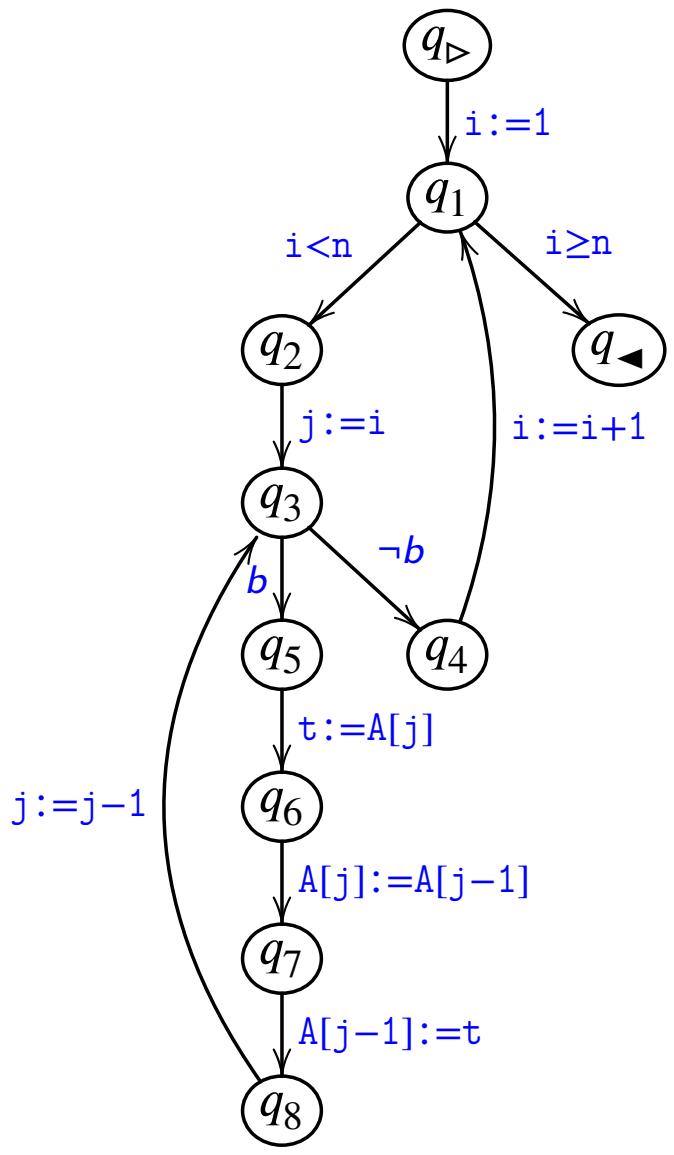

Figure 2.7: Program graph for insertion sort; writing $b$ for the test `j>0 && A[j-1]>A[j]`.

In Exercise 1.6 we introduced short-circuit versions of conjunction and disjunction; for the program graph of Figure 2.7 it ensures that the test `A[j-1]>A[j]` only will be performed when the test $\texttt{j} > 0$ evaluates to true. A direct application of the Available Expressions analysis as presented above will determine that `A[j-1]` is available at $q_4$ as we have taken $\mathbf{ae}(b_1\ op_b\ b_2) = \mathbf{ae}(b_1) \cup \mathbf{ae}(b_2)$ for all boolean operators $op_b$ (including && and ||). At first sight this may be a bit surprising but indeed the expression is available in the sense that it evaluates to the same result at $q_3$ and $q_4$.

TEASER 2.61: To ensure that an expression is only available if it has indeed been evaluated we shall in this Teaser investigate a couple of suggestions for how to modify the analysis.

One possibility is to redefine $\mathbf{ae}(b_1\ op_b\ b_2)$ in the case where $op_b \in \{\&\&, ||\}$ and let $\mathbf{ae}(b_1\ op_b\ b_2) = \mathbf{ae}(b_1)$ in order to reflect that we might not evaluate the second argument.

Another possibility is to introduce two versions of $\mathbf{ae}(b)$, one corresponding to the case where we want $b$ to evaluate to true (denoted $\mathsf{ae_t}(b)$), and another where we want it to evaluate to false (denoted $\mathsf{ae_f}(b)$). Some of the clauses are:

$$
\begin{array}{rclcrcl}
\mathsf{ae_t}(b_1\ op_b\ b_2) & = & \mathsf{ae_t}(b_1) \cup \mathsf{ae_t}(b_2) & \quad & \mathsf{ae_f}(b_1\ op_b\ b_2) & = & \mathsf{ae_f}(b_1) \cup \mathsf{ae_f}(b_2) \\
\mathsf{ae_t}(b_1\ \&\&\ b_2) & = & \mathsf{ae_t}(b_1) \cup \mathsf{ae_t}(b_2) & & \mathsf{ae_f}(b_1\ \&\&\ b_2) & = & \mathsf{ae_f}(b_1) \\
\mathsf{ae_t}(b_1\ ||\ b_2) & = & \mathsf{ae_t}(b_1) & & \mathsf{ae_f}(b_1\ ||\ b_2) & = & \mathsf{ae_f}(b_1) \cup \mathsf{ae_f}(b_2) \\
\mathsf{ae_t}(\neg b_0) & = & \mathsf{ae_f}(b_0) & & \mathsf{ae_f}(\neg b_0) & = & \mathsf{ae_t}(b_0)
\end{array}
$$

where $op_b$ stands for the binary boolean operators with the exception of && and ||.

Discuss the two alternatives and the extent to which they live up to our expectations. Modify the analysis to use your preferred choice and apply it to the program graph of Figure 2.7. □

## 2.4 Very Busy Expressions

An expression is said to be *very busy* if its value will be computed before the program ends and furthermore its value would be same if computed at the current node. Very Busy Expressions analysis then aims to determine for each node of the program graph which (non-trivial) expressions are very busy. As an example for the program graph of Figure 2.8 the analysis could tell us that at the node $q_4$ the expressions A[i] and i + 1 are very busy; however, only i + 1 is very busy at the node $q_6$ and none of the two expressions are very busy at $q_3$.

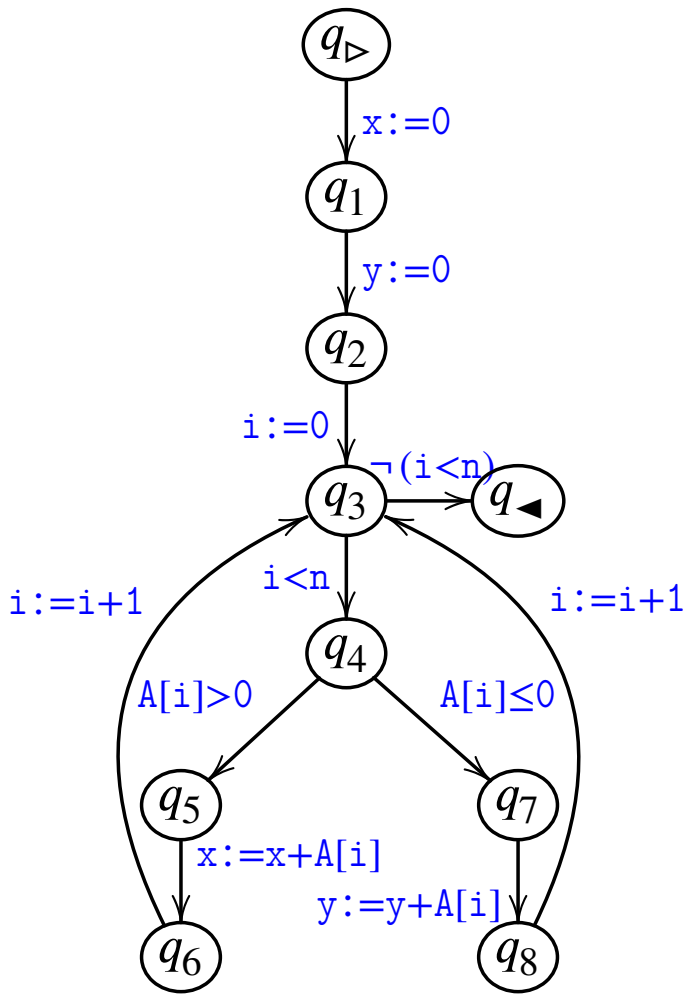

Figure 2.8: Program graph for adding positive and negative entries of an array.

The information provided by the Very Busy Expressions analysis is useful in compiler optimisation where it opens up for improvements to the code; in particular the code can be shortened by only generating code for evaluating the expression once. For the program graph of Figure 2.8 we may modify the program graph to have an assignment z := A[i] before the node $q_3$ and then make use of z when updating the values of x and y.

**Analysis assignments.** The Very Busy Expressions analysis will only be concerned with non-trivial expressions as was the case for the Available Expressions analysis. Following Section 2.3 we shall write **AExp** for this set; it will include all the non-trivial subexpressions occurring in the arithmetic and boolean expressions of the program graph.

EXAMPLE 2.62: For the program graph of Figure 2.8 we have

$$\mathbf{AExp} = \{\texttt{A[i]}, \texttt{x} + \texttt{A[i]}, \texttt{y} + \texttt{A[i]}, \texttt{i} + 1\}$$

For the node $q_4$ we may use the set {A[i], i + 1} to tell which expressions are very busy. It is more informative to have a *larger* set – in this case we may remove i + 1 from the set but it will not be correct to include x + A[i] in the set. Thus, as for the Available Expressions analysis (and unlike the Reaching Definitions analysis and the Live Variables analysis), it is better to include too few possibilites than to include too many.

The result of a *Very Busy Expressions* analysis will be given by an analysis assignment VB that maps each node in the program graph to a set of non-trivial arithmetic expressions:

$$\mathsf{VB} : \mathbf{Q} \to \mathsf{PowerSet}(\mathbf{AExp})$$

EXAMPLE 2.63: Returning to the program graph of Figure 2.8 we would like the analysis assignment VB to satisfy the following conditions:

$$
\begin{array}{rclrcl}
\mathsf{VB}(q_{\triangleright}) & \subseteq & \{\,\} & \mathsf{VB}(q_5) & \subseteq & \{\mathtt{A[i]}, \mathtt{i}+1, \mathtt{x}+\mathtt{A[i]}\} \\
\mathsf{VB}(q_1) & \subseteq & \{\,\} & \mathsf{VB}(q_6) & \subseteq & \{\mathtt{i}+1\} \\
\mathsf{VB}(q_2) & \subseteq & \{\,\} & \mathsf{VB}(q_7) & \subseteq & \{\mathtt{A[i]}, \mathtt{i}+1, \mathtt{y}+\mathtt{A[i]}\} \\
\mathsf{VB}(q_3) & \subseteq & \{\,\} & \mathsf{VB}(q_8) & \subseteq & \{\mathtt{i}+1\} \\
\mathsf{VB}(q_4) & \subseteq & \{\mathtt{A[i]}, \mathtt{i}+1\} & \mathsf{VB}(q_{\blacktriangleleft}) & \subseteq & \{\,\}
\end{array}
$$

If equalities hold in all cases we would intuitively think that VB is the best analysis assignment that can be hoped for.

**What is busy on a path.** As in Section 2.3 we write $\mathsf{ae}(a)$ for set of non-trivial arithmetic expressions occurring in $a$, $\mathsf{ae}(b)$ for the set of non-trivial arithmetic expressions occurring in $b$, and $\mathsf{AExp}(\alpha)$ for the set of non-trivial arithmetic expressions occurring in $\alpha$.

For a path $\pi = q_0, \alpha_1, q_1, \cdots, q_{n-1}, \alpha_n, q_n$ with $n \geq 0$ we define

$$\mathsf{VBExp}(\pi) = \{a \in \mathbf{AExp} \mid \exists i \leq n : a \in \mathsf{AExp}(\alpha_i) \wedge \forall j < i : \mathbf{fv}(a) \cap \mathsf{Def}(\alpha_j) = \emptyset\}$$

This ensures that a very busy expression is disregarded as soon as one of its constituent variables or arrays is being defined, so that the value of the expression might change.

EXAMPLE 2.64: Returning to the program graph of Figure 2.8 let us consider the following paths starting in $q_3$ and ending in $q_{\blacktriangleleft}$:

$$
\begin{array}{rcl}
\pi_1 & = & q_3, \neg(\mathtt{i} < \mathtt{n}), q_{\blacktriangleleft} \\
\pi_2 & = & q_3, \mathtt{i} < \mathtt{n}, q_4, \mathtt{A[i]} > 0, q_5, \mathtt{x} := \mathtt{x} + \mathtt{A[i]}, q_6, \mathtt{i} := \mathtt{i} + 1, q_3, \neg(\mathtt{i} < \mathtt{n}), q_{\blacktriangleleft}
\end{array}
$$

Then $\mathsf{VBExp}(\pi_1) = \{\,\}$ whereas $\mathsf{VBExp}(\pi_2) = \{\mathtt{A[i]}, \mathtt{x} + \mathtt{A[i]}, \mathtt{i} + 1\}$.

TRY IT OUT 2.65: Discuss whether or not we should use $j < i$ or $j \leq i$ in the definition of $\mathsf{VBExp}(\pi)$. □

We are now ready to reformulate the aim of the Very Busy Expressions analysis: For each node $q$ of a program graph we want to determine the set of non-trivial expressions that need to be evaluated on *all* paths starting in $q$ and ending at $q_{\blacktriangleleft}$. This then gives rise to the following notion of correctness of a Very Busy Expressions analysis:

DEFINITION 2.66: An analysis assignment VB *summarises the paths* of a program graph provided that $\mathsf{VBExp}(\pi) \supseteq \mathsf{VB}(q)$ for every path $\pi$ from $q$ to $q_{\blacktriangleleft}$.

The *greatest* analysis assignment that summarises the paths of a program graph is sometimes called the *MOP solution* (which we may read as Merge Over Paths solution).

Try It Out 2.67: Try to argue that an analysis assignment VB summarises the paths for the program graph of Figure 2.8 exactly when the conditions of Example 2.63 hold. □

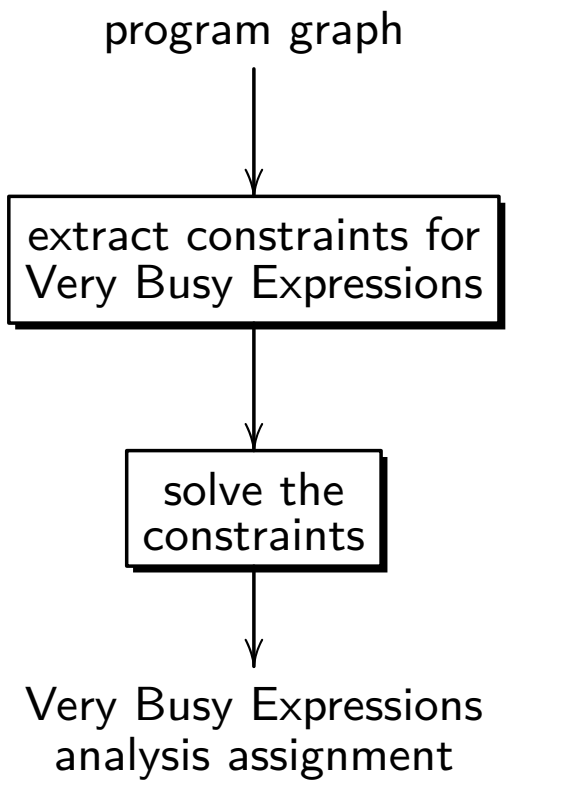

Figure 2.9: Computing the analysis assignment.

**Constraints for Very Busy Expressions.** The above treatment does not present an effective way of calculating a correct Very Busy Expressions analysis. Following the approach of the previous sections we shall now formulate a system of constraints that the Very Busy Expressions analysis assignment must satisfy and next we shall adapt the algorithms we have seen to solve these constraints. This is illustrated on Figure 2.9.

In order to specify the constraints we shall first define sets of so-called killed and generated expressions for each edge $(q_\circ, \alpha, q_\bullet)$ in the program graph:

| $\alpha$ | $\mathsf{kill}_{\mathsf{VB}}(q_\circ, \alpha, q_\bullet)$ | $\mathsf{gen}_{\mathsf{VB}}(q_\circ, \alpha, q_\bullet)$ |
|---|---|---|
| $x$ := $a$ | $\{a' \in \mathbf{AExp} \mid x \in \mathbf{fv}(a')\}$ | $\mathsf{ae}(a)$ |
| $A[a_1]$ := $a_2$ | $\{a' \in \mathbf{AExp} \mid A \in \mathbf{fv}(a')\}$ | $\mathsf{ae}(a_1) \cup \mathsf{ae}(a_2)$ |
| $c?x$ | $\{a' \in \mathbf{AExp} \mid x \in \mathbf{fv}(a')\}$ | $\{\,\}$ |
| $c?A[a]$ | $\{a' \in \mathbf{AExp} \mid A \in \mathbf{fv}(a')\}$ | $\mathsf{ae}(a)$ |
| $c!a$ | $\{\,\}$ | $\mathsf{ae}(a)$ |
| $b$ | $\{\,\}$ | $\mathsf{ae}(b)$ |
| skip | $\{\,\}$ | $\{\,\}$ |

The idea is that a non-trivial expression $a$ is *generated* whenever the action $\alpha$ computes it because it obviously will be needed in the rest of the computation. On the other hand, a non-trivial expression $a$ is *killed* whenever the action $\alpha$ defines any variable or array used in $a$ because then the value of $a$ might change – indeed this definition is exactly as for the Available Expressions analysis in the previous section.

Try It Out 2.68: For the edge $(q_5, \texttt{x := x + A[i]}, q_6)$ of Figure 2.8 we have

$$\begin{aligned}\mathsf{kill}_{\mathsf{VB}}(q_5, \texttt{x := x + A[i]}, q_6) &= \{\texttt{x + A[i]}\}\\ \mathsf{gen}_{\mathsf{VB}}(q_5, \texttt{x := x + A[i]}, q_6) &= \{\texttt{A[i]}, \texttt{x + A[i]}\}\end{aligned}$$

where we recall that the set **AExp** of non-trivial expressions is as in Example 2.63. Construct the sets of killed and generated variables for the remaining edges of Figure 2.8. □

Exercise 2.69: Consider the program graph of Figure 1.13 and construct the sets of killed and generated variables and arrays for each of the edges. □

The next step is to impose constraints for each edge $(q_\circ, \alpha, q_\bullet)$ in the program graph:

$$(\mathsf{VB}(q_\bullet) \backslash \mathsf{kill}_{\mathsf{VB}}(q_\circ, \alpha, q_\bullet)) \cup \mathsf{gen}_{\mathsf{VB}}(q_\circ, \alpha, q_\bullet) \supseteq \mathsf{VB}(q_\circ)$$

Furthermore, we impose an additional constraint for the final node:

$$\{\ \} \supseteq \mathsf{VB}(q_\blacktriangleleft)$$

Note that the last constraint is *not* superfluous and is equivalent to $\mathsf{VB}(q_\blacktriangleleft) = \{\ \}$.

TRY IT OUT 2.70: Construct the set of constraints obtained for the program graph of Figure 2.8 (in the manner of Example 2.10). □

DEFINITION 2.71: An analysis assignment VB *solves the constraints* whenever it satisfies the constraints constructed for the Very Busy Expressions analysis.

TRY IT OUT 2.72: Returning to Example 2.62 consider the analysis assignment VB given by

$$\begin{array}{lcl@{\qquad}lcl}
\mathsf{VB}(q_\triangleright) & = & \{\ \} & \mathsf{VB}(q_5) & = & \{\mathtt{A[i]}, \mathtt{i}+1, \mathtt{x}+\mathtt{A[i]}\} \\
\mathsf{VB}(q_1) & = & \{\ \} & \mathsf{VB}(q_6) & = & \{\mathtt{i}+1\} \\
\mathsf{VB}(q_2) & = & \{\ \} & \mathsf{VB}(q_7) & = & \{\mathtt{A[i]}, \mathtt{i}+1, \mathtt{y}+\mathtt{A[i]}\} \\
\mathsf{VB}(q_3) & = & \{\ \} & \mathsf{VB}(q_8) & = & \{\mathtt{i}+1\} \\
\mathsf{VB}(q_4) & = & \{\mathtt{A[i]}, \mathtt{i}+1\} & \mathsf{VB}(q_\blacktriangleleft) & = q & \{\ \}
\end{array}$$

and show that it solves the constraints of Try It Out 2.70. □

**Solving the constraints.** The following algorithm solves the constraints of the Very Busy Expressions analysis.

| | |
|---|---|
| INPUT | a program graph with **Q**, $q_\triangleright$, **E** |
| OUTPUT | VB: an analysis assignment |
| METHOD | forall $q \in \mathbf{Q} \setminus \{q_\blacktriangleleft\}$ do $\mathsf{VB}(q) := \mathbf{AExp}$ ;<br>$\mathsf{VB}(q_\blacktriangleleft) := \{\ \}$ ;<br>while there exists an edge $(q_\circ, \alpha, q_\bullet) \in \mathbf{E}$<br>such that $(\mathsf{VB}(q_\bullet) \backslash \mathsf{kill}_{\mathsf{VB}}(q_\circ, \alpha, q_\bullet)) \cup \mathsf{gen}_{\mathsf{VB}}(q_\circ, \alpha, q_\bullet) \not\supseteq \mathsf{VB}(q_\circ)$<br>do $\mathsf{VB}(q_\circ) := \mathsf{VB}(q_\circ) \cap \big((\mathsf{VB}(q_\bullet) \backslash \mathsf{kill}_{\mathsf{VB}}(q_\circ, \alpha, q_\bullet)) \cup \mathsf{gen}_{\mathsf{VB}}(q_\circ, \alpha, q_\bullet)\big)$ |

The algorithm has the same overall structure as the previous algorithms of this chapter; we shall study more efficient algorithms in Chapter 4.

TRY IT OUT 2.73: Returning to the program graph of Figure 2.8 let

$$\mathbf{AExp} = \{\mathtt{A[i]}, \mathtt{x}+\mathtt{A[i]}, \mathtt{y}+\mathtt{A[i]}, \mathtt{i}+1\}$$

and use the algorithm to construct the the analysis assignment VB of Example 2.72. □

EXERCISE 2.74: Returning to the program graph of Figure 1.13 use the algorithm to compute an analysis assignment VB; as part of this you may reuse your definitions of the killed and generated available expressions of Exercise 2.69. □

EXERCISE 2.75: Perform the Very Busy Expressions analysis on the program graphs constructed in Exercise 1.12. □

In analogy with the results of the previous sections we have:

PROPOSITION 2.76: The algorithm terminates and produces an analysis assignment that solves the constraints of the Very Busy Expressions analysis.

EXERCISE 2.77: Prove that if VB is the analysis assignment produced by the algorithm, and if VB′ is some other analysis assignment that also solves the constraints, then $\mathsf{VB}(q) \supseteq \mathsf{VB}'(q)$ holds for all nodes $q$ in $\mathbf{Q}$. □

Exercise 2.77 shows that the algorithm produces the greatest analysis assignment that solves the constraints, and this solution is sometimes called the *MFP solution* (which we may read as Maximal Fixed Point solution in this case).

**Solutions summarise the paths.** It remains to relate the notion of "solving the constraints" and "summarising the paths". As in the previous sections we have:

PROPOSITION 2.78: Whenever an analysis assignment solves the constraints it also summarises the paths.

TEASER 2.79: Is it possible to find an analysis assignment that summarises the paths but that does not solve the constraints? □

As for the Available Expressions analysis we shall conclude by a discussion of what it means for an expression to be very busy. An expression $a$ *becomes* very busy as a result of executing an action on an edge $(q_\circ, \alpha, q_\bullet)$ and the analysis will associate this information with $q_\circ$. The expression may then *continue* to be very busy at some previous nodes; if it is very busy at $q$ then it means that evaluation of $a$ at $q$ and $q_\circ$ gives the same result. This holds independently of whether the evaluation gives a proper value or it fails.

# Chapter 3

# The Monotone Framework

In Chapter 2 we considered a number of program analyses where the analysis domains were powersets and where the analysis functions were expressed using 'kill' and 'gen' functions. Sometimes it is advantageous to allow for more general forms of analysis domains and analysis functions and we perform the necessary developments in the present chapter. This gives rise to a general approach known as *Monotone Framework* that constitutes a robust framework for *Data Flow Analysis*.

We begin by recasting the Reaching Definitions analysis from Section 2.1 in a slightly different form in order to illustrate the approach. Then we formalise the requirements on the analysis domains and on the analysis functions as needed for the Monotone Framework to work out. We conclude by showing how all of the analyses of Chapter 2 can be recast in the Monotone Framework as developed here.

## 3.1 Beyond the Bit-Vector Framework

In this section we will be presenting the *Reaching Definitions* analysis of Section 2.1 in a different form. Recall that a set of reaching definitions $R$ is a subset of $(\mathbf{Var} \cup \mathbf{Arr}) \times \mathbf{Q}_? \times \mathbf{Q}$ or equivalently an element of

$$\text{PowerSet}(\,(\mathbf{Var} \cup \mathbf{Arr}) \times \mathbf{Q}_? \times \mathbf{Q}\,)$$

This means for each variable $x$ (and similarly array $A$) that we can obtain a set of pairs of nodes $\{(q_1, q_2) \mid (x, q_1, q_2) \in R\}$ telling more directly where the variable $x$ might have been defined.

**Analysis domains.** To represent this more directly we could decide to represent reaching definitions as a mapping

$$(\mathbf{Var} \cup \mathbf{Arr}) \to \text{PowerSet}(\,\mathbf{Q}_? \times \mathbf{Q}\,)$$

$$R = \{(\mathtt{x}, ?, q_{\triangleright}), (\mathtt{x}, q_3, q_1), (\mathtt{y}, q_{\triangleright}, q_1), (\mathtt{y}, q_2, q_3)\}$$

$$R' = \begin{bmatrix} \mathtt{x} \mapsto \{(?, q_{\triangleright}), (q_3, q_1)\} \\ \mathtt{y} \mapsto \{(q_{\triangleright}, q_1), (q_2, q_3)\} \end{bmatrix}$$

Figure 3.1: Reaching definitions as a set ($R$) and as a mapping ($R'$).

The two representations are entirely isomorphic, meaning that we can pass freely between them without introducing any errors. Given $R \in \text{PowerSet}(\,(\mathbf{Var} \cup \mathbf{Arr}) \times \mathbf{Q}_? \times \mathbf{Q}\,)$ in the set-based representation, we can construct $R' : (\mathbf{Var} \cup \mathbf{Arr}) \to \text{PowerSet}(\,\mathbf{Q}_? \times \mathbf{Q}\,)$ in the mapping-based representation, by setting

$$\begin{array}{rcll} R'(x) & = & \{(q_1, q_2) \mid (x, q_1, q_2) \in R\} & \text{for all } x \in \mathbf{Var} \\ R'(A) & = & \{(q_1, q_2) \mid (A, q_1, q_2) \in R\} & \text{for all } A \in \mathbf{Arr} \end{array}$$

This is illustrated on Figure 3.1 for a simple example.

Similarly, given $R' : (\mathbf{Var} \cup \mathbf{Arr}) \to \text{PowerSet}(\,\mathbf{Q}_? \times \mathbf{Q}\,)$ in the mapping-based representation, we can construct $R \in \text{PowerSet}(\,(\mathbf{Var} \cup \mathbf{Arr}) \times \mathbf{Q}_? \times \mathbf{Q}\,)$ in the set-based representation, as

$$\begin{array}{rcl} R & = & \{(x, q_1, q_2) \mid (q_1, q_2) \in R'(x) \wedge x \in \mathbf{Var}\} \\ & \cup & \{(A, q_1, q_2) \mid (q_1, q_2) \in R'(A) \wedge A \in \mathbf{Arr}\} \end{array}$$

Furthermore, going from the set-based representation to the mapping-based representation and back gives us the same set that we started with. Similarly, going from the mapping-based representation to the set-based representation and back again gives us the same mapping that we started with.

**Analysis functions.** When we modify the analysis domain to use the mapping representation we also need to modify the form of the constraints of the Reaching Definitions analysis. In order to do so we shall introduce notation for updating mappings $R' : (\mathbf{Var} \cup \mathbf{Arr}) \to \text{PowerSet}(\,\mathbf{Q}_? \times \mathbf{Q}\,)$.

We shall define $R'[x \mapsto S]$ to be the mapping that is as $R'$ except that it maps $x$ to $S$ rather than $R'(x)$, and similarly we define $R'[A \mapsto S]$ to be the mapping that is as $R'$ except that it maps $A$ to $S$ rather than $R'(A)$. Finally, write $[\longmapsto S]$ for the mapping that maps every variable or array of $\mathbf{Var} \cup \mathbf{Arr}$ to $S$.

This allows us to define analysis functions on mappings that somehow mimick what we did in Section 2.1.

| $\alpha$ | $\widehat{S}_{\mathsf{RD}'}[\![q_\circ, \alpha, q_\bullet]\!](R')$ |
|---|---|
| $x$ `:=` $a$ | $R'[x \mapsto \{(q_\circ, q_\bullet)\}]$ |
| $A$`[`$a_1$`]` `:=` $a_2$ | $R'[A \mapsto R'(A) \cup \{(q_\circ, q_\bullet)\}]$ |
| $c$`?`$x$ | $R'[x \mapsto \{(q_\circ, q_\bullet)\}]$ |
| $c$`?`$A$`[`$a$`]` | $R'[A \mapsto R'(A) \cup \{(q_\circ, q_\bullet)\}]$ |
| $c$`!`$a$ | $R'$ |
| $b$ | $R'$ |
| `skip` | $R'$ |

For $x$ `:=` $a$ the mapping $R'$ is simply updated to record that $x$ is mapped to the set $\{(q_\circ, q_\bullet)\}$ thereby capturing not only that this pair is generated but also that all previous pairs associated with $x$ in $R'$ are killed. For the assignment $A$`[`$a_1$`]` `:=` $a_2$ we need to capture that we cannot kill the previous information, so $A$ needs to be associated with the old information $R'(A)$ as well as the new information $\{(q_\circ, q_\bullet)\}$. The remaining clauses follow the same overall idea.

EXERCISE 3.1: Show that if $R'$ is the mapping-based representation of $R$ then

$$\widehat{S}_{\mathsf{RD}'}[\![q_\circ, \alpha, q_\bullet]\!]\big(R'\big)$$

is the mapping-based representation of

$$\big(R \setminus \mathsf{kill}_{\mathsf{RD}}(q_\circ, \alpha, q_\bullet)\big) \cup \mathsf{gen}_{\mathsf{RD}}(q_\circ, \alpha, q_\bullet)$$

by considering each of the cases of $\alpha$ in turn. □

We can extend the definition of $\widehat{S}_{\mathsf{RD}'}[\![\cdots]\!]$ from edges to paths as follows. Consider a path $\pi = q_0, \alpha_1, q_1, \cdots, q_{n-1}, \alpha_n, q_n$ with $n \geq 0$ and define

$$\widehat{S}_{\mathsf{RD}'}[\![\pi]\!](R') = \widehat{S}_{\mathsf{RD}'}[\![q_{n-1}, \alpha_n, q_n]\!]\Big(\cdots \widehat{S}_{\mathsf{RD}'}[\![q_0, \alpha_1, q_1]\!](R') \cdots\Big)$$

which means $\widehat{S}_{\mathsf{RD}'}[\![\pi]\!](R') = R'$ in case $n = 0$.

**Analysis constraints.** The next step is to rephrase the constraints of the Reaching Definitions analysis. When presenting the constraints in Section 2.1 we used the subset $\subseteq$ relation freely to express how some $\mathsf{RD}(q)$ should contain some $R$: the general format was $R \subseteq \mathsf{RD}(q)$ and in the case of an edge $(q_\circ, x := a, q_\bullet)$ it specialised to

$$\big(\mathsf{RD}(q_\circ) \setminus \{(x, q, q') \mid q \in \mathbf{Q}_?, q' \in \mathbf{Q}\}\big) \cup \{(x, q_\circ, q_\bullet)\} \subseteq \mathsf{RD}(q_\bullet)$$

Turning to the mapping-based representation we would like to be able to express how some $\mathsf{RD}'(q)$ should contain some $R'$.

We shall do so by defining a relation $\sqsubseteq$ that somehow mimicks the subset relation $\subseteq$. This would allow us to replace the above constraint for assignment with a constraint of the form

$$\mathsf{RD}'(q_\circ)[x \mapsto \{(x, q_\circ, q_\bullet)\}] \sqsubseteq \mathsf{RD}'(q_\bullet)$$

We define $R'_1 \sqsubseteq R'_2$ to mean that for all variables $x$ we have $R'_1(x) \subseteq R'_2(x)$ and that for all arrays $A$ we have $R'_1(A) \subseteq R'_2(A)$. Note that we have defined this relation such that $R'_1 \sqsubseteq R'_2$ is equivalent to $R_1 \subseteq R_2$ in case $R'_1$ is the mapping-based representation of $R_1$ and similarly $R'_2$ for $R_2$.

This allows us to reformulate the constraints of Section 2.1 in a mapping-based manner. So for an analysis assignment $\mathsf{RD}'$ we require that for each edge $(q_\circ, \alpha, q_\bullet)$ in the program graph:

$$\widehat{S}_{\mathsf{RD}'}[\![q_\circ, \alpha, q_\bullet]\!]\big(\mathsf{RD}'(q_\circ)\big) \sqsubseteq \mathsf{RD}'(q_\bullet)$$

For the entry node we additionally require that:

$$[\longmapsto \{(?, q_\triangleright)\}] \sqsubseteq \mathsf{RD}'(q_\triangleright)$$

TRY IT OUT 3.2: Construct the set of constraints for the program graph of Figure 2.1 using the mapping-based representation. □

EXERCISE 3.3: Consider a set-based analysis assignment RD and a mapping-based analysis assignment RD′ such that each $\mathsf{RD}'(q)$ is the mapping-based representation of $\mathsf{RD}(q)$. Argue that RD satisfies the constraints of Section 2.1 exactly when RD′ satisfies the constraints listed above. □

**Solving the constraints.** The next step is to adapt the algorithm of Section 2.1 to solve the constraints expressed using the mapping-based representation. In addition to the subset ordering ⊆ we also used the empty set { } and set union ∪ to express the algorithm. Turning to the mapping-based representation we would like to have similar notation.

INPUT a program graph with **Q**, $q_\triangleright$, **E**
OUTPUT RD′: an analysis assignment

METHOD forall $q \in \mathbf{Q} \setminus \{q_\triangleright\}$ do $\mathsf{RD}'(q) := \bot$ ;
$\mathsf{RD}'(q_\triangleright) := [\longmapsto \{(?, q_\triangleright)\}]$ ;

while there exists an edge $(q_\circ, \alpha, q_\bullet) \in \mathbf{E}$
such that $\widehat{S}_{\mathsf{RD}'}[\![q_\circ, \alpha, q_\bullet]\!](\mathsf{RD}'(q_\circ)) \not\sqsubseteq \mathsf{RD}'(q_\bullet)$
do $\mathsf{RD}'(q_\bullet) := \mathsf{RD}'(q_\bullet) \sqcup \widehat{S}_{\mathsf{RD}'}[\![q_\circ, \alpha, q_\bullet]\!](\mathsf{RD}'(q_\circ))$

For this define ⊥ to be the mapping that maps each variable to the empty set and each array to the empty set; in other words, $\bot = [\longmapsto \{\,\}]$. Then it should be clear that ⊥ is the mapping-based representation of { }. Next define $R'_1 \sqcup R'_2$ to be the mapping that maps each variable $x$ to $R'_1(x) \cup R'_2(x)$ and each array $A$ to $R'_1(A) \cup R'_2(A)$. Then it should be clear that $R'_1 \sqcup R'_2$ is the mapping-based representation of $R_1 \cup R_2$ in case $R'_1$ is the mapping-based representation of $R_1$ and similarly $R'_2$ for $R_2$.

TRY IT OUT 3.4: Use the above algorithm to solve the set of constraints constructed in Try It Out 3.2. □

EXERCISE 3.5: Show that the above algorithm is equivalent to the algorithm of Section 2.1 in that each test and update in the above algorithm is the mapping-based analogue of the similar test and update in the algorithm from Section 2.1. □

It follows that the algorithm terminates and produces an analysis assignment RD′ that solves the constraints for Reaching Definitions. As before, this analysis assignment is in fact the least analysis assignment that solves the constraints. Finally, whenever RD′ is an analysis assignment that solves the constraints, and $\pi$ is a path from $q_\triangleright$ to $q$, then $\widehat{S}_{\mathsf{RD}'}[\![\pi]\!]([\longmapsto \{(?, q_\triangleright)\}]) \sqsubseteq \mathsf{RD}'(q)$.

## 3.2 Analysis Domains

The purpose of a program analysis is to obtain an *analysis assignment* of the form

$$\mathsf{AA} : \mathbf{Q} \to \widehat{\mathbf{D}}$$

where $\widehat{\mathbf{D}}$ is a suitable analysis domain. In Section 3.1 we considered two versions of $\widehat{\mathbf{D}}$; one was PowerSet( $(\mathbf{Var} \cup \mathbf{Arr}) \times \mathbf{Q}_? \times \mathbf{Q}$ ) and the other was $(\mathbf{Var} \cup \mathbf{Arr}) \to$ PowerSet( $\mathbf{Q}_? \times \mathbf{Q}$ ).

**Partially ordered sets.** Both of the choices for $\widehat{\mathbf{D}}$ considered in Section 3.1 were equipped with a relation for expressing that some $\hat{d}$ ($\in \widehat{\mathbf{D}}$) is contained in some AA($q$). For PowerSet( $(\mathbf{Var} \cup \mathbf{Arr}) \times \mathbf{Q}_? \times \mathbf{Q}$ ) we simply used subset $\subseteq$ and for $(\mathbf{Var} \cup \mathbf{Arr}) \to$ PowerSet( $\mathbf{Q}_? \times \mathbf{Q}$ ) we defined a relation $\sqsubseteq$ that mimicked it. This suggests that every analysis domain should be equipped with such a relation and we should therefore consider what properties we would require it to have.

It is quite natural to demand that $\hat{d} \sqsubseteq \hat{d}$ is always the case; this merely says that $\sqsubseteq$ should be a *reflexive* relation. It is also quite natural to demand that if $\hat{d}_1 \sqsubseteq \hat{d}_2$ and $\hat{d}_2 \sqsubseteq \hat{d}_3$ then also $\hat{d}_1 \sqsubseteq \hat{d}_3$; this merely says that $\sqsubseteq$ should be a *transitive* relation. It is often considered natural to demand that if $\hat{d}_1 \sqsubseteq \hat{d}_2$ and $\hat{d}_2 \sqsubseteq \hat{d}_1$ then in fact $\hat{d}_1 = \hat{d}_2$; this merely says $\sqsubseteq$ should be an *anti-symmetric* relation. Putting it all together this requires $\sqsubseteq$ to be a *partial order*. We then say that $(\widehat{\mathbf{D}}, \sqsubseteq)$ is a *partially ordered set*.

TRY IT OUT 3.6: Show that the subset relation $\subseteq$ used for PowerSet( $(\mathbf{Var} \cup \mathbf{Arr}) \times \mathbf{Q}_? \times \mathbf{Q}$ ) is a partial order. Also show that the relation $\sqsubseteq$ used for $(\mathbf{Var} \cup \mathbf{Arr}) \to$ PowerSet( $\mathbf{Q}_? \times \mathbf{Q}$ ) is a partial order. □

Given a partially ordered set there are systematic ways for constructing new partially ordered sets. As an example, once we have argued that PowerSet( $\mathbf{Q}_? \times \mathbf{Q}$ ) with the relation $\subseteq$ is a partially ordered set then it follows that $(\mathbf{Var} \cup \mathbf{Arr}) \to$ PowerSet( $\mathbf{Q}_? \times \mathbf{Q}$ ) is also a partially ordered set provided that we define the relation $\sqsubseteq$ in a particular manner as explored in the following exercise:

EXERCISE 3.7: Suppose that $(D, \sqsubseteq_D)$ is a partially ordered set and define the analysis domain $\widehat{\mathbf{D}} = (\mathbf{Var} \cup \mathbf{Arr}) \to D$ to consist of all mappings from variables and array names to $D$. Define the relation $\sqsubseteq$ by $\hat{d}_1 \sqsubseteq \hat{d}_2$ whenever $\forall x \in \mathbf{Var} : \hat{d}_1(x) \sqsubseteq_D \hat{d}_2(x)$ and $\forall A \in \mathbf{Arr} : \hat{d}_1(A) \sqsubseteq_D \hat{d}_2(A)$. Show that $(\widehat{\mathbf{D}}, \sqsubseteq)$ is a partially ordered set. □

It is fairly easy to come up with example analysis domains $\widehat{\mathbf{D}}$ beyond the ones considered so far. Recall that a *directed graph* $(D, \to)$ is a set $D$ together with an edge relation $\to \subseteq D \times D$. Define $\sqsubseteq_D$ as the reflexive and transitive closure $\to^*$ of the edge relation; to be precise, let $d_0 \sqsubseteq_D d_n$ whenever there is a sequence of edges $d_0 \to d_1$, ..., $d_{n-1} \to d_n$ in the graph for $n \geq 0$. It is then immediate that $\sqsubseteq_D$ is a partial order whenever the directed graph is *acyclic* (meaning that $d \to d' \to^* d$ is never possible for any $d'$). Thus every acyclic directed graph $D$ will be a candidate for an analysis domain $\widehat{\mathbf{D}}$.

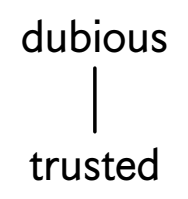

Figure 3.2: Hasse diagram for an integrity analysis.

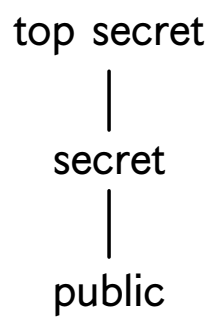

Figure 3.3: Hasse diagram for a confidentiality analysis.

Finite acyclic directed graphs can be drawn very simply as so-called *Hasse diagrams*. A Hasse diagram is essentially a graph drawn in such a way that there are no horisontal edges and all remaining edges are directed in the 'upwards slanting direction'. In Figure 3.2 we see a simple example of this; here the intention is that trusted $\sqsubseteq$ trusted, trusted $\sqsubseteq$ dubious, and dubious $\sqsubseteq$ dubious but that dubious $\not\sqsubseteq$ trusted. It expresses that dubious data should not flow to trusted data. Since we are only interested in the transitive closure of the edges relation, or equivalently, the partial order induced by the edges, it is customary not to draw edges that do not contribute to the transitive closure. Figure 3.3 presents an example expressing that public data can be freely used, secret data should not flow to public data, and finally that top secret data should only flow to top secret data. Here we do not add the 'superfluous edge' from public to top secret.

TEASER 3.8: Let $(D, \sqsubseteq)$ be a finite partially ordered set and show that it can be expressed as a Hasse diagram. You can do so in two steps. First define $d \to d'$ whenever $d \sqsubseteq d' \land d \neq d'$; it is immediate that the partial order $\sqsubseteq$ coincides with the reflexive and transitive closure $\to^*$ of the newly defined edges relation. Next shrink $\to$ by removing $d_1 \to d_3$ whenever $d_1 \to d_2$ and $d_2 \to d_3$ are already present; it is immediate that the partial order $\sqsubseteq$ continues to coincide with the reflexive and transitive closure $\to^*$ of the edges relation. It remains to show that this non-deterministic algorithm always gives the same result in the end. □

**Pointed semi-lattices.** The development so far suffices for sketching the constraints we generate for our analyses.

For each edge $(q_\circ, \alpha, q_\bullet)$ in the program graph we impose the constraint:

$$\widehat{S}[\![q_\circ, \alpha, q_\bullet]\!]\big(\mathsf{AA}(q_\circ)\big) \sqsubseteq \mathsf{AA}(q_\bullet)$$

Furthermore, we impose an additional constraint for the entry node:

$$\hat{d}_\diamond \sqsubseteq \mathsf{AA}(q_\triangleright)$$

Here we assume that the analysis specifies appropriate analysis functions $\widehat{S}[\![q_\circ, \alpha, q_\bullet]\!] : \widehat{\mathbf{D}} \to \widehat{\mathbf{D}}$ and an appropriate *initial element* $\hat{d}_\diamond \in \widehat{\mathbf{D}}$. We shall impose conditions on the functions $\widehat{S}[\![q_\circ, \alpha, q_\bullet]\!]$ in the next section.

EXAMPLE 3.9: Let us return to the Reaching Definitions analysis of Section 2.1. It makes use of the analysis domain $\widehat{\mathbf{D}} = \mathsf{PowerSet}((\mathbf{Var} \cup \mathbf{Arr}) \times \mathbf{Q}_? \times \mathbf{Q})$ and the analysis functions $\widehat{\mathcal{S}}_{\mathsf{RD}}[\![q_\circ, \alpha, q_\bullet]\!]$ will now be defined by

$$\widehat{\mathcal{S}}_{\mathsf{RD}}[\![q_\circ, \alpha, q_\bullet]\!](R) = (R \setminus \mathsf{kill}_{\mathsf{RD}}(q_\circ, \alpha, q_\bullet)) \cup \mathsf{gen}_{\mathsf{RD}}(q_\circ, \alpha, q_\bullet)$$

where $R \subseteq (\mathbf{Var} \cup \mathbf{Arr}) \times \mathbf{Q}_? \times \mathbf{Q}$ and the sets $\mathsf{kill}_{\mathsf{RD}}(q_\circ, \alpha, q_\bullet)$ and $\mathsf{gen}_{\mathsf{RD}}(q_\circ, \alpha, q_\bullet)$ are as in Section 2.1. The initial element $\hat{d}_\diamond$ is $(\mathbf{Var} \cup \mathbf{Arr}) \times \{?\} \times \{q_\triangleright\}$.

To actually solve the constraints we need to stipulate a suitable element $\bot \in \widehat{\mathbf{D}}$ and a suitable function $\sqcup : \widehat{\mathbf{D}} \times \widehat{\mathbf{D}}$ mimicking what was done in Section 3.1. For $\bot$ we shall assume that there is an element $\bot \in \widehat{\mathbf{D}}$ such that

$$\forall \hat{d} \in \widehat{\mathbf{D}} : \bot \sqsubseteq \hat{d}$$

It is straightforward to show that condition is satisfied in the case of $\{\ \}$ as used in Section 2.1 as well as in the case of $\bot$ as defined in Section 3.1. The element $\bot$ is usually read as '*bottom*' and it is the least element of $\widehat{\mathbf{D}}$.

For $\sqcup$ we shall assume that there is a function $\sqcup : \widehat{\mathbf{D}} \times \widehat{\mathbf{D}} \to \widehat{\mathbf{D}}$ such that

$$\forall \hat{d}, \hat{d}_1, \hat{d}_2 : \left(\hat{d}_1 \sqsubseteq \hat{d} \ \wedge \ \hat{d}_2 \sqsubseteq \hat{d} \quad \text{if and only if} \quad \hat{d}_1 \sqcup \hat{d}_2 \sqsubseteq \hat{d}\right)$$

Here we are using infix notation $\hat{d}_1 \sqcup \hat{d}_2$ rather than the prefix notation $\sqcup(\hat{d}_1, \hat{d}_2)$. The function $\sqcup$ is usually read as '*join*', '*lub*' or *least upper bound*.

EXERCISE 3.10: Show that the set union operator $\cup$ as used in Section 2.1 satisfies this condition. Show that the operator $\sqcup$ as defined in Section 3.1 satisfies this condition. □

A partially ordered set equipped with a least element ($\bot$ satisfying $\bot \sqsubseteq \hat{d}$ for all $\hat{d}$) and a join operation ($\sqcup$ satisfying $\hat{d}_1 \sqsubseteq \hat{d} \wedge \hat{d}_2 \sqsubseteq \hat{d} \Leftrightarrow \hat{d}_1 \sqcup \hat{d}_2 \sqsubseteq \hat{d}$) is called a *pointed semi-lattice*.

TRY IT OUT 3.11: Show that in a pointed semi-lattice we have

- $\hat{d} \sqcup \hat{d} = \hat{d}$,
- $\hat{d} \sqcup \bot = \hat{d}$,
- $\hat{d}_1 \sqcup \hat{d}_2 = \hat{d}_2 \sqcup \hat{d}_1$,
- $\hat{d}_1 \sqcup (\hat{d}_2 \sqcup \hat{d}_3) = (\hat{d}_1 \sqcup \hat{d}_2) \sqcup \hat{d}_3$

Furthermore, show that $\hat{d}_1 \sqsubseteq \hat{d}_2$ is equivalent to $\hat{d}_1 \sqcup \hat{d}_2 = \hat{d}_2$ □

This gives rise to the following (by now familiar) algorithm (leaving more efficient algorithms to Chapter 4).

| | |
|---|---|
| INPUT | a program graph with $\mathbf{Q}$, $q_\triangleright$, $\mathbf{E}$<br>an analysis specification with $\widehat{\mathbf{D}}$, $\widehat{\mathcal{S}}[\![q_\circ, \alpha, q_\bullet]\!]$ and $\hat{d}_\diamond$ |
| OUTPUT | AA: an analysis assignment |
| METHOD | forall $q \in \mathbf{Q} \setminus \{q_\triangleright\}$ do AA$(q) := \bot$ ;<br>AA$(q_\triangleright) := \hat{d}_\diamond$ ;<br>while there exists an edge $(q_\circ, \alpha, q_\bullet) \in \mathbf{E}$<br>such that $\widehat{\mathcal{S}}[\![q_\circ, \alpha, q_\bullet]\!](\mathsf{AA}(q_\circ)) \not\sqsubseteq \mathsf{AA}(q_\bullet)$<br>do AA$(q_\bullet) :=$ AA$(q_\bullet) \sqcup \widehat{\mathcal{S}}[\![q_\circ, \alpha, q_\bullet]\!](\mathsf{AA}(q_\circ))$ |

PROPOSITION 3.12: If the algorithm terminates it finishes with an analysis assignment AA that solves the constraints.

PROOF: On termination of the algorithm we have that $\widehat{\mathcal{S}}[\![q_\circ, \alpha, q_\bullet]\!](\mathsf{AA}(q_\circ)) \sqsubseteq \mathsf{AA}(q_\bullet)$ for all edges $(q_\circ, \alpha, q_\bullet) \in \mathbf{E}$. After the initialisation we have that $\hat{d}_\diamond \sqsubseteq \mathsf{AA}(q_\triangleright)$ and this is maintained throughout the algorithm. □

EXERCISE 3.13: Present an example of an analysis domain, analysis functions, and program graph such that the algorithm never terminates. (Hint: It suffices to choose a program graph with just one loop and to use PowerSet( **Int** ) as analysis domain.) □

TEASER 3.14: Present a (possibly contrived) example of an analysis domain, analysis functions, and program graph such that the analysis assignment computed by the algorithm is *not* the least one. (Hint: It suffices to choose a program graph with just one edge and to use the Hasse diagram of Figure 3.3 as analysis domain.) □

**Ascending chain condition.** We shall need an extra assumption on our analysis domain $\widehat{\mathbf{D}}$ beyond being a pointed semi-lattice in order to ensure that the above algoritm always terminates. To express this it is helpful to define the relation $\sqsubset$ by $\hat{d}_1 \sqsubset \hat{d}_2$ whenever $\hat{d}_1 \sqsubseteq \hat{d}_2$ and $\hat{d}_1 \neq \hat{d}_2$.

An analysis domain $\widehat{\mathbf{D}}$ satisfies the *ascending chain condition* whenever it is *impossible* to find a countably infinite sequence of elements $\hat{d}_i \in \widehat{\mathbf{D}}$ such that $\hat{d}_1 \sqsubset \hat{d}_2 \sqsubset \cdots \sqsubset \hat{d}_n \sqsubset \cdots$.

An analysis domain $\widehat{\mathbf{D}}$ has *finite height h* whenever it is *impossible* to find a sequence of elements $\hat{d}_i \in \widehat{\mathbf{D}}$ such that $\hat{d}_1 \sqsubset \hat{d}_2 \sqsubset \cdots \sqsubset \hat{d}_n$ with $n > h$.

Clearly an analysis domain satisfies the ascending chain condition whenever it has finite height (for some $h$), and has finite height (for some $h$) whenever it is finite.

EXERCISE 3.15: An analysis domain for Constant Propagation of integers has as elements all the integers together with elements $\bot$ and $\top$. The partial order is defined by $\bot \sqsubseteq z \sqsubseteq \top$ for all integers $z$ and is illustrated in Figure 3.4. Show that this is a pointed semi-lattice and that it satisfies the ascending chain condition (in spite of being infinite). □

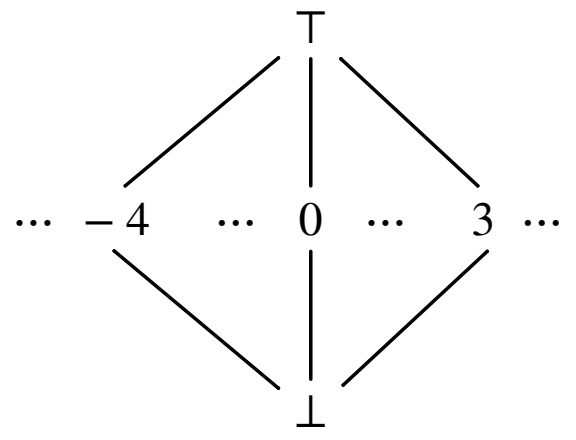

Figure 3.4: Analysis domain for Constant Propagation analysis.

PROPOSITION 3.16: If the analysis domain satisfies the ascending chain condition then the algorithm terminates.

PROOF: Consider the analysis domain $\mathbf{Q} \to \widehat{\mathbf{D}}$ which may be partially ordered by setting $\mathsf{AA}_1 \sqsubseteq' \mathsf{AA}_2$ whenever $\forall x \in \mathbf{Var} : \mathsf{AA}_1(x) \sqsubseteq \mathsf{AA}_2(x)$ and $\forall A \in \mathbf{Arr} : \mathsf{AA}_1(A) \sqsubseteq \mathsf{AA}_2(A)$. Next define $\bot'$ by setting setting $\bot'(x) = \bot$ and $\bot'(A) = \bot$ for all $x \in \mathbf{Var}$ and $A \in \mathbf{Arr}$, and define $\sqcup'$ by setting $(\mathsf{AA}_1 \sqcup' \mathsf{AA}_2)(x) = (\mathsf{AA}_1(x)) \sqcup (\mathsf{AA}_2(x))$ and $(\mathsf{AA}_1 \sqcup' \mathsf{AA}_2)(A) = (\mathsf{AA}_1(A)) \sqcup (\mathsf{AA}_2(A))$ for all $x \in \mathbf{Var}$ and $A \in \mathbf{Arr}$. Since $\widehat{\mathbf{D}}$ is a pointed semi-lattice it follows that also $\mathbf{Q} \to \widehat{\mathbf{D}}$ is a pointed semi-lattice. Recall by Definition 1.9 that $\mathbf{Q}$ is a finite (and non-empty) set of nodes. Since $\widehat{\mathbf{D}}$ satisfies the ascending chain condition it follows that also $\mathbf{Q} \to \widehat{\mathbf{D}}$ does.

Next write $\mathsf{AA}_i$ for the value of AA just before doing the $i$'th iteration of the loop, so that $\mathsf{AA}_0$ is the result of performing the initialisation. It is straightforward to prove that $\mathsf{AA}_0 \sqsubset' \mathsf{AA}_1 \sqsubset' \cdots \sqsubset' \mathsf{AA}_n \sqsubset' \cdots$ and this shows that the algorithm must terminate. □

ESSENTIAL EXERCISE 3.17: Show that a *pointed semi-lattice* is the same as a partially ordered set $(L, \sqsubseteq)$ where for every *finite* subset $Y \subseteq L$ there is an element $\bigsqcup Y$ such that $\forall y \in Y : y \sqsubseteq l$ is equivalent to $\bigsqcup Y \sqsubseteq l$ for all choices of $l \in L$.

TEASER 3.18: A *complete lattice* is a partially ordered set $(L, \sqsubseteq)$ such that for every subset $Y \subseteq L$ there is an element $\bigsqcup Y$ such that $\forall y \in Y : y \sqsubseteq l$ is equivalent to $\bigsqcup Y \sqsubseteq l$ for all choices of $l \in L$.

Clearly a complete lattice is a pointed semi-lattice: define $\bot = \bigsqcup\{\ \}$ and $l_1 \sqcup l_2 = \bigsqcup\{l_1, l_2\}$. Give an example of a partially ordered set that is a pointed semi-lattice but not a complete lattice.

Next show that if $(L, \sqsubseteq)$ is a pointed semi-lattice satisfying the ascending chain condition then $(L, \sqsubseteq)$ is also a complete lattice. (Hint: This is easiest if the elements of $L$ can be enumerated as in $L = \{\mathsf{l}_0, \mathsf{l}_1, \cdots, \mathsf{l}_n, \cdots\}$.) □

**Summary of demands on analysis domains.** Let us conclude by summarising our demands on an analysis domain – leaving our demands on the analysis functions to the next section.

DEFINITION 3.19: An analysis domain $\widehat{\mathbf{D}}$ should be a partially ordered set (expressed by $\sqsubseteq$), that is a pointed semi-lattice (and so has $\bot$ and $\sqcup$), and that satisfies the ascending chain condition.

These demands suffice for Propositions 3.12 and 3.16.

ESSENTIAL EXERCISE 3.20: Show that if $\widehat{\mathbf{D}}_1$ and $\widehat{\mathbf{D}}_2$ are analysis domains, then so is $\widehat{\mathbf{D}}_1 \times \widehat{\mathbf{D}}_2$; for this you should first define the partial order and next check the conditions.

Futhermore, show that if $\widehat{\mathbf{D}}$ is an analysis domain and $\mathbf{V}$ is a finite (and non-empty) set, then $\mathbf{V} \to \widehat{\mathbf{D}}$ is an analysis domain; again you should first defining the partial order and next check the conditions. (Hint: See the proof of Proposition 3.16.)

## 3.3 Analysis Functions

In the previous section we established in Propositions 3.12 and 3.16 a number of properties of our algorithm for solving the constraints. However, unlike what we did for Reaching Definitions in Section 3.1 (and in Chapter 2) we did not claim that the algorithm computed the least solution nor that the solution summarises all the paths. This is because it requires some additional assumptions on the analysis functions for this to hold.

**Monotone functions.** It is essential for our development that the analysis functions preserve the partial ordering of its arguments as expressed by:

An analysis function $f : \widehat{\mathbf{D}} \to \widehat{\mathbf{D}}$ is *monotone* whenever $\hat{d}_1 \sqsubseteq \hat{d}_2$ implies that $f(\hat{d}_1) \sqsubseteq f(\hat{d}_2)$ (for all $\hat{d}_1$ and $\hat{d}_2$ of $\widehat{\mathbf{D}}$).

TRY IT OUT 3.21: The Reaching Definitions analysis of Section 2.1 makes use of analysis functions $f$ of the form

$$f(R) = \big(R \setminus \mathsf{kill}_{\mathsf{RD}}(q_\circ, \alpha, q_\bullet)\big) \cup \mathsf{gen}_{\mathsf{RD}}(q_\circ, \alpha, q_\bullet)$$

To show that $f$ is monotone let $R_1$ and $R_2$ be elements of the analysis domain $\mathsf{PowerSet}((\mathbf{Var} \cup \mathbf{Arr}) \times \mathbf{Q}_? \times \mathbf{Q})$ and assume that $R_1 \subseteq R_2$. Show that $f(R_1) \subseteq f(R_2)$ for all choices of $\alpha$. □

EXERCISE 3.22: Show that the analysis function $\widehat{S}_{\mathsf{RD'}}[\![q_\circ, \alpha, q_\bullet]\!]$ introduced in Section 3.1 for the mapping-based version of Reaching Definitions is monotone for all choices of $\alpha$. □

EXAMPLE 3.23: Reaching Definitions can be used to indicate points in the program where a variable might be *dangerous* in the sense that its value might depend on the initial memory. It is possible to define an analysis called *Dangerous Variables* for more directly obtaining this information. As analysis domain we shall take the set PowerSet( **Var** ∪ **Arr** ) of all subsets of variables and arrays. We then define the analysis functions as follows:

| $\alpha$ | $\widehat{S}_{\mathsf{DV}}[\![q_\circ, \alpha, q_\bullet]\!](R)$ |
|---|---|
| $x$ := $a$ | $\begin{cases} R \setminus \{x\} & \text{if } \mathbf{fv}(a) \cap R = \{\,\} \\ R \cup \{x\} & \text{if } \mathbf{fv}(a) \cap R \neq \{\,\} \end{cases}$ |
| $A[a_1]$ := $a_2$ | $\begin{cases} R & \text{if } (\mathbf{fv}(a_1) \cup \mathbf{fv}(a_2)) \cap R = \{\,\} \\ R \cup \{A\} & \text{if } (\mathbf{fv}(a_1) \cup \mathbf{fv}(a_2)) \cap R \neq \{\,\} \end{cases}$ |
| $c?x$ | $R \setminus \{x\}$ |
| $c?A[a]$ | $\begin{cases} R & \text{if } \mathbf{fv}(a) \cap R = \{\,\} \\ R \cup \{A\} & \text{if } \mathbf{fv}(a) \cap R \neq \{\,\} \end{cases}$ |
| $c!a$ | $R$ |
| $b$ | $R$ |
| skip | $R$ |

For each edge $(q_\circ, \alpha, q_\bullet)$ in the program graph we impose the constraint:

$$\widehat{S}_{\mathsf{DV}}[\![q_\circ, \alpha, q_\bullet]\!](\mathsf{DV}(q_\circ)) \sqsubseteq \mathsf{DV}(q_\bullet)$$

For the entry node we additionally require:

$$\mathbf{Var} \cup \mathbf{Arr} \sqsubseteq \mathsf{DV}(q_\triangleright)$$

Note that it would be impossible to specify the analysis in the Bit-Vector Framework as considered in Chapter 2.

TRY IT OUT 3.24: Explain why the analysis of Example 3.23 does what it is supposed to do. □

TRY IT OUT 3.25: Construct the set of constraints for Dangerous Variables analysis for the program graph of Figure 2.1 and use the above algorithm to compute the analysis assignment DV. □

EXERCISE 3.26: Perform the Dangerous Variables analysis on the program graphs constructed in Exercise 1.12. □

EXERCISE 3.27: Show that the analysis function $\widehat{S}_{\mathsf{DV}}[\![q_\circ, \alpha, q_\bullet]\!]$ from Example 3.23 is monotone for all choices of $\alpha$. □

> PROPOSITION 3.28: If the analysis functions are monotone then the algorithm computes the least solution to the constraints.

PROOF: We must prove that if AA is the analysis assignment produced by the algorithm, and if AA′ is some other analysis assignment that also solves the constraints, then $\forall q \in \mathbf{Q} : \mathsf{AA}(q) \sqsubseteq \mathsf{AA}'(q)$. We do so by induction on the number of times the loop is executed.

For the base case we observe that the condition holds immediately after the initialisation phase of the algorithm.

For the induction step we consider some edge $(q_\circ, \alpha, q_\bullet) \in \mathbf{E}$. Then by monotonicity

$$\widehat{\mathcal{S}}[\![q_\circ, \alpha, q_\bullet]\!]\big(\mathsf{AA}(q_\circ)\big) \sqsubseteq \widehat{\mathcal{S}}[\![q_\circ, \alpha, q_\bullet]\!]\big(\mathsf{AA}'(q_\circ)\big)$$

and since AA′ solves the constraints we get

$$\widehat{\mathcal{S}}[\![q_\circ, \alpha, q_\bullet]\!]\big(\mathsf{AA}(q_\circ)\big) \sqsubseteq \mathsf{AA}'(q_\bullet)$$

and hence

$$\mathsf{AA}(q_\bullet) \sqcup \widehat{\mathcal{S}}[\![q_\circ, \alpha, q_\bullet]\!]\big(\mathsf{AA}(q_\circ)\big) \sqsubseteq \mathsf{AA}'(q_\bullet)$$

which suffices for showing that the property is maintained after the assignment to $\mathsf{AA}(q_\bullet)$. □

> PROPOSITION 3.29: If the analysis functions are monotone and AA solves the constraints then AA summarises the paths.

PROOF: Suppose that the analysis assignment AA solves the constraints. We must prove that if $\pi = q_0, \alpha_1, q_1, \cdots, q_{n-1}, \alpha_n, q_n$ with $q_0 = q_\triangleright$ is a path from $q_\triangleright$ to $q_n$ then $\widehat{\mathcal{S}}[\![\pi]\!]\big(\hat{d}_\diamond\big) \sqsubseteq \mathsf{AA}(q_n)$. We do so by induction on the length of the path $\pi$.

If the path $\pi$ is empty (so $\pi = q_0 = q_n = q_\triangleright$) then $\widehat{\mathcal{S}}[\![\pi]\!]\big(\hat{d}_\diamond\big) = \hat{d}_\diamond$ and $\hat{d}_\diamond \sqsubseteq \mathsf{AA}(q_n)$ because AA solves the constraints.

If the path $\pi$ is non-empty it follows by the induction hypothesis that

$$\widehat{\mathcal{S}}[\![\pi']\!]\big(\hat{d}_\diamond\big) \sqsubseteq \mathsf{AA}(q_{n-1})$$

where $\pi' = q_0, \alpha_1, q_1, \cdots, q_{n-1}$. Hence by monotonicity

$$\widehat{\mathcal{S}}[\![q_{n-1}, \alpha, q_n]\!]\Big(\widehat{\mathcal{S}}[\![\pi']\!]\big(\hat{d}_\diamond\big)\Big) \sqsubseteq \widehat{\mathcal{S}}[\![q_{n-1}, \alpha, q_n]\!]\big(\mathsf{AA}(q_{n-1})\big)$$

and since AA solves the constraints we have

$$\begin{array}{rcl} \widehat{\mathcal{S}}[\![\pi]\!]\big(\hat{d}_\diamond\big) & = & \widehat{\mathcal{S}}[\![q_{n-1}, \alpha, q_n]\!]\Big(\widehat{\mathcal{S}}[\![\pi']\!]\big(\hat{d}_\diamond\big)\Big) \\ & \sqsubseteq & \widehat{\mathcal{S}}[\![q_{n-1}, \alpha, q_n]\!]\big(\mathsf{AA}(q_{n-1})\big) \\ & \sqsubseteq & \mathsf{AA}(q_n) \end{array}$$

as was to be shown. □

**Summary of demands on analysis functions.** We are now ready to summarise our conditions on the specification of analyses.

DEFINITION 3.30: A *specification* of an analysis in the *monotone framework* is given by:

- an analysis domain $\widehat{\mathbf{D}}$ (with ordering $\sqsubseteq$) that is a pointed semi-lattice satisfying the ascending chain condition,
- monotone analysis functions $\widehat{\mathcal{S}}[\![q_\circ, \alpha, q_\bullet]\!] : \widehat{\mathbf{D}} \to \widehat{\mathbf{D}}$ for each choice of action $\alpha$ and nodes $q_\circ$ and $q_\bullet$, and
- an initial element $\hat{d}_\diamond \in \widehat{\mathbf{D}}$ (sometimes written $\hat{d}_\triangleright$ or $\hat{d}_\blacktriangleleft$).

These demands suffice for Propositions 3.12, 3.16, 3.28 and 3.29.

ESSENTIAL EXERCISE 3.31: Show that if $f_1 : \widehat{\mathbf{D}}_1 \to \widehat{\mathbf{D}}_1$ and $f_2 : \widehat{\mathbf{D}}_2 \to \widehat{\mathbf{D}}_2$ are monotone analysis functions then the analysis function $f : \widehat{\mathbf{D}}_1 \times \widehat{\mathbf{D}}_2 \to \widehat{\mathbf{D}}_1 \times \widehat{\mathbf{D}}_2$ defined by

$$f(\hat{d}_1, \hat{d}_2) = (f_1(\hat{d}_1), f_2(\hat{d}_2))$$

is also monotone.

Furthermore, show that if $f : \widehat{\mathbf{D}} \to \widehat{\mathbf{D}}$ is a monotone analysis function and $X$ is a set then the analysis function $f' : (X \to \widehat{\mathbf{D}}) \to (X \to \widehat{\mathbf{D}})$ defined by

$$f'(g)(x) = f(g(x))$$

is also monotone.

**Distributive functions.** Recall from Chapter 2 that the least analysis assignment solving the constraints is sometimes called the *MFP solution* and the least analysis assignment that summarises the paths is sometimes called the *MOP solution*. It follows from Proposition 3.29 that for an analysis in the monotone framework the MOP solution is always less than or equal to the MFP solution. We conclude by considering a sufficient condition for when the MOP and MFP solutions agree.

An analysis function $f : \widehat{\mathbf{D}} \to \widehat{\mathbf{D}}$ is *distributive* whenever $f(\hat{d}_1 \sqcup \hat{d}_2) = f(\hat{d}_1) \sqcup f(\hat{d}_2)$ (for all $\hat{d}_1$ and $\hat{d}_2$ of $\widehat{\mathbf{D}}$).

TRY IT OUT 3.32: Show that a distributive analysis function is also a monotone function. □

EXERCISE 3.33: Show that the analysis function $f$ defined by

$$f(R) = \left(R \setminus \mathsf{kill}_{\mathsf{RD}}(q_\circ, \alpha, q_\bullet)\right) \cup \mathsf{gen}_{\mathsf{RD}}(q_\circ, \alpha, q_\bullet)$$

is distributive (using $\subseteq$ for the partial order $\sqsubseteq$) for all choices of $\alpha$. Also show that the analysis function $\widehat{S}_{\mathsf{RD}'}[\![q_\circ, \alpha, q_\bullet]\!]$ from Section 3.1 is distributive for all choices of $\alpha$. □

EXERCISE 3.34: Show that the analysis functions $\widehat{S}_{\mathsf{DV}}[\![q_\circ, \alpha, q_\bullet]\!]$ of the Dangerous Variables analysis of Example 3.23 are distributive for all choices of $\alpha$. □

PROPOSITION 3.35: Assume that the analysis functions are distributive, and that the analysis domain is a pointed semi-lattice that satisfies the ascending chain condition. Then the least analysis assignment that summarises the paths also solves the constraints.

PROOF: Let AA be the least analysis assignment that summarises the paths, that is, the least analysis assignment AA such that $\widehat{S}[\![\pi]\!](\hat{d}_\diamond) \sqsubseteq \mathsf{AA}(q)$ whenever $\pi$ is a path from $q_\triangleright$ to some node $q$.

This holds for the empty path ($\pi = q_\triangleright$) so we have that $\hat{d}_\diamond \sqsubseteq \mathsf{AA}(q_\triangleright)$.

Next consider an edge $(q_\circ, \alpha, q_\bullet)$ in the program graph. By our requirements on program graphs in Definition 1.9 there is at least one path from $q_\triangleright$ to $q_\circ$. Hence we can number the paths from $q_\triangleright$ to $q_\bullet$ as $\pi_1$, $\pi_2$, and so on, where we number a path more than once in case there are only finitely many paths from $q_\triangleright$ to $q_\circ$. Since AA summarises the paths we have that $\widehat{S}[\![\pi_i]\!](\hat{d}_\diamond) \sqsubseteq \mathsf{AA}(q_\circ)$ for all $i$.

Since $\widehat{\mathbf{D}}$ satisfies the ascending chain condition, and AA is the least analysis assignment that summarises the paths, we can find some number $N > 0$ such that

$$\mathsf{AA}(q_\circ) = \widehat{S}[\![\pi_1]\!](\hat{d}_\diamond) \sqcup \cdots \sqcup \widehat{S}[\![\pi_N]\!](\hat{d}_\diamond)$$

as otherwise the chain (writing $d_i = \widehat{S}[\![\pi_1]\!](\hat{d}_\diamond)$)

$$d_0 \sqsubseteq d_0 \sqcup d_1 \sqsubseteq \cdots \sqsubseteq (d_0 \sqcup d_1 \sqcup \cdots \sqcup d_i) \sqsubseteq \cdots$$

would contain a sub-chain violating the ascending chain condition or the leastness of AA.

By the distributivity of $\widehat{S}[\![q_\circ, \alpha, q_\bullet]\!]$ we then have that

$$\widehat{S}[\![q_\circ, \alpha, q_\bullet]\!](\mathsf{AA}(q_\circ)) = \widehat{S}[\![\pi_1, \alpha, q_\bullet]\!](\hat{d}_\diamond) \sqcup \cdots \sqcup \widehat{S}[\![\pi_N, \alpha, q_\bullet]\!](\hat{d}_\diamond)$$

and since AA summarises all the paths from $q_\triangleright$ to $q_\bullet$ it summarises all the paths $\pi_i, \alpha, q_\bullet$ and hence

$$\widehat{S}[\![q_\circ, \alpha, q_\bullet]\!](\mathsf{AA}(q_\circ)) \sqsubseteq \mathsf{AA}(q_\bullet)$$

showing that AA solves the constraints. □

Distributive functions are those monotone functions that preserve the join operator of the pointed semi-lattice. Those functions that preserve the least element (i.e. $f(\bot) = \bot$, are said to be *strict*. Only few of the analysis functions considered so far satisfy this condition.

## 3.4 Duality and Reversal

In Chapter 2 we presented four program analyses that clearly have similarities but also important differences. We have shown that one of these, Reaching Definitions analysis, fits within the Monotone Framework developed in this chapter, and in this section we are going to show that this also holds for the remaining three analyses.

**Reversal.** Two of the analyses from Chapter 2, Reaching Definitions and Available Expressions are so-called *forward analyses* because they analyse the program graphs by traversing the edges in the same direction as when executing the program graph. The remaining two analyses, Live Variables and Very Busy Expressions are so-called *backward analyses* because they analyse the program graphs by traversing the eges in the opposite direction. Our treatment so far in this chapter might seem to deal only with forward analyses.

Rather than redoing the entire development for backward analyses we shall employ a simple trick. By merely reversing the direction of the edges in a program graph we shift between forward and backward analyses.

Recall that a program graph **PG** as in Definition 1.9 consists of a set of nodes **Q**, an initial node $q_{\triangleright}$, a final node $q_{\blacktriangleleft}$, a set of actions **Act**, and a set of edges **E**. It is sometimes helpful to write $\mathbf{PG} = (\mathbf{Q}, q_{\triangleright}, q_{\blacktriangleleft}, \mathbf{Act}, \mathbf{E})$ to summarise this.

Given a program graph $\mathbf{PG} = (\mathbf{Q}, q_{\triangleright}, q_{\blacktriangleleft}, \mathbf{Act}, \mathbf{E})$ the *reverse program graph* $\mathbf{PG}^R$ is given by

$$\mathbf{PG}^R = (\mathbf{Q}, q_{\blacktriangleleft}, q_{\triangleright}, \mathbf{Act}, \mathbf{E}^R)$$

where

$$\mathbf{E}^R = \{(q_\bullet, \alpha, q_\circ) \mid (q_\circ, \alpha, q_\bullet) \in \mathbf{E}\}$$

Note how we simply change the role of initial and final nodes and the source and targets of edges and that we indeed obtain a program graph as required in Definition 1.9. Hence we can apply our generation of constraints, and our algorithm for solving them, also on reverse program graphs.

EXAMPLE 3.36: Let us consider once again the *Live Variables* analysis from Section 2.2 as it applies to a program graph **PG**. First we express it as a specification of an analysis in the monotone framework of Definition 3.30.

- The analysis domain $\widehat{\mathbf{D}}$ is PowerSet( **Var** ∪ **Arr** ) with the subset ordering.
- The analysis functions are given by

$$\widehat{\mathcal{S}}_{\mathsf{LV}}[\![q_\bullet, \alpha, q_\circ]\!](L) = \big(L \setminus \mathsf{kill}_{\mathsf{LV}}(q_\circ, \alpha, q_\bullet)\big) \cup \mathsf{gen}_{\mathsf{LV}}(q_\circ, \alpha, q_\bullet)$$

- The initial element $\hat{d}_\diamond$ is $\{\,\}$.

It is straightforward to check that the analysis functions are monotone (as in Exercise 3.22).

Finally we perform the constraint generation, and the computation of the least solution, on the reverse program graph $\mathbf{PG}^R$ and this then gives exactly the same results as in Section 2.2.

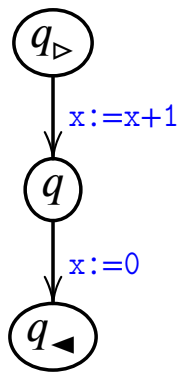

Figure 3.5: Program graph illustrating Live Variables and Faint Variables analysis.

Try It Out 3.37: Construct the reverse program graph corresponding to Figure 2.3. Use the approach of Example 3.36 to construct the corresponding set of constraints and compare it with the result of Try It Out 2.32. □

Example 3.38: Consider the program graph of Figure 3.5; the Live Variables analysis records that `x` is live at node $q_\triangleright$ as it is used to compute a new value of `x` in the next action. However, the analysis also tells us that `x` is dead at the node $q$. We shall therefore be interested in a variation of Live Variables analysis called *Faint Variables* analysis where we do not consider a variable to be live if it is only used to compute a variable that is dead.

As in Example 3.36 we take the analysis domain $\widehat{\mathbf{D}}$ to be PowerSet( **Var** ∪ **Arr** ) (with the subset ordering) and the initial element $\hat{d}_\diamond$ to be $\{\,\}$ and the analysis is performed on the reverse program graph. We then define the analysis functions as follows:

| $\alpha$ | $\widehat{S}_{\mathsf{FV}}[\![q_\bullet, \alpha, q_\circ]\!](L)$ |
|---|---|
| $x$ := $a$ | $\begin{cases} (L \setminus \{x\}) \cup \mathbf{fv}(a) & \text{if } x \in L \\ L & \text{if } x \notin L \end{cases}$ |
| $A[a_1]$ := $a_2$ | $\begin{cases} L \cup \mathbf{fv}(a_1) \cup \mathbf{fv}(a_2) & \text{if } A \in L \\ L & \text{if } A \notin L \end{cases}$ |
| $c?x$ | $L \setminus \{x\}$ |
| $c?A[a]$ | $\begin{cases} L \cup \mathbf{fv}(a) & \text{if } A \in L \\ L & \text{if } A \notin L \end{cases}$ |
| $c!a$ | $L \cup \mathbf{fv}(a)$ |
| $b$ | $L \cup \mathbf{fv}(b)$ |
| `skip` | $L$ |

The variables of the set $L$ are sometimes called the strongly live variables and the complement is then the set of faint variables.

Try It Out 3.39: Construct the set of constraints obtained for the Faint Variables analysis of the program graph of Figure 3.5. □

Exercise 3.40: Perform the Faint Variables analysis on (variations of) the program graphs constructed in Exercise 1.12. □

EXERCISE 3.41: The analysis functions $\widehat{S}_{\mathsf{LV}}[\![q_\bullet, \alpha, q_\circ]\!]$ and $\widehat{S}_{\mathsf{FV}}[\![q_\bullet, \alpha, q_\circ]\!]$ are defined in rather different formats. Show how to define $\widehat{S}_{\mathsf{LV}}[\![q_\bullet, \alpha, q_\circ]\!]$ in the same format as used for $\widehat{S}_{\mathsf{FV}}[\![q_\bullet, \alpha, q_\circ]\!]$. Argue that it would be impossible to specify the Faint Variables analysis in the Bit-Vector Framework as considered in Chapter 2. □

EXERCISE 3.42: Show that the analysis functions $\widehat{S}_{\mathsf{LV}}[\![q_\bullet, \alpha, q_\circ]\!]$ of the Faint Variables analysis of Example 3.38 are distributive for all choices of $\alpha$. □

**Duality** Changing the perspective, two of the analyses, Reaching Definitions and Live Variables operate on sets and we are interested in finding the *least solution*. The remaining two analyses, Available Expressions and Very Busy Expressions also operate on sets, but we are interested in finding the *greatest solution*, and the constraints use $\supseteq$ rather than $\subseteq$. So far, our treatment in this chapter might seem only to deal with least solutions. Rather than redoing the entire development for greatest solutions we shall again employ a simple trick.

Whenever we have a powerset like PowerSet($X$) it is customary to think of the partial order being subset $\subseteq$ and to write (PowerSet($X$), $\subseteq$) for the partially ordered set. This partially ordered set is depicted in Figure 3.6 and we use $\subseteq$ for the constraints and we want the least solution. Formally, when $\sqsubseteq$ is $\subseteq$ we have $\bot = \{\,\}$ and $\hat{d}_1 \sqcup \hat{d}_2 = \hat{d}_1 \cup \hat{d}_2$. The partially ordered set (PowerSet($X$), $\sqsubseteq$) = (PowerSet($X$), $\subseteq$) is a pointed semi-lattice, which satisfies the ascending chain condition when $X$ is finite, and hence can be used as an analysis domain.

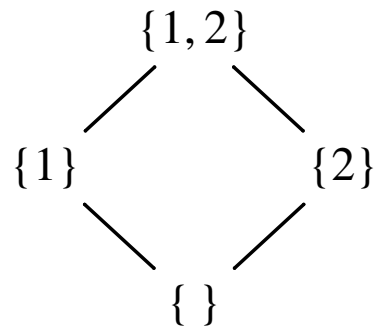

Figure 3.6: Hasse diagram for (PowerSet({1,2}), $\sqsubseteq$) when $\sqsubseteq$ is $\subseteq$.

However, it is equally possible to think of the partial order being superset $\supseteq$ and to write (PowerSet($X$), $\supseteq$) for the partially ordered set. This partially ordered set is depicted in Figure 3.7 and we use $\supseteq$ for the constraints and we want the greatest solution. Formally, when $\sqsubseteq$ is $\supseteq$ we have $\bot = X$ and $\hat{d}_1 \sqcup \hat{d}_2 = \hat{d}_1 \cap \hat{d}_2$. The partially ordered set (PowerSet($X$), $\sqsubseteq$) = (PowerSet($X$), $\supseteq$) is a pointed semi-lattice, which satisfies the ascending chain condition when $X$ is finite, and hence can be used as an analysis domain.

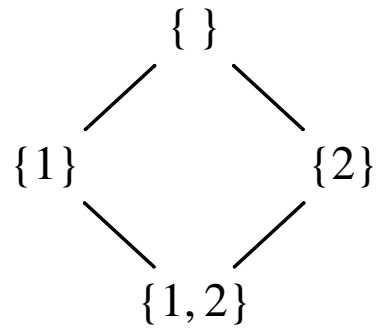

Figure 3.7: Hasse diagram for (PowerSet({1,2}), $\sqsubseteq$) when $\sqsubseteq$ is $\supseteq$.

EXAMPLE 3.43: Let us consider once again the *Available Expressions* analysis from Section 2.3 as it applies to a program graph **PG**. First we express it as a specification of an analysis in the monotone framework of Definition 3.30.

- The analysis domain $\widehat{\mathbf{D}}$ is taken to be PowerSet( **AExp** ) with the superset ordering.
- The analysis functions are given by

$$\widehat{S}_{\mathsf{AE}}[\![q_\circ, \alpha, q_\bullet]\!](A) = \big(A \setminus \mathsf{kill}_{\mathsf{AE}}(q_\circ, \alpha, q_\bullet)\big) \cup \mathsf{gen}_{\mathsf{AE}}(q_\circ, \alpha, q_\bullet)$$

- The initial element $\hat{d}_\diamond$ is $\{\,\}$.

It is straightforward to check that the analysis functions are monotone (as in Exercise 3.22).

Finally we perform the constraint generation, and the computation of the least solution, on the program graph **PG** and this then gives exactly the same results as in Section 2.3.

EXERCISE 3.44: Consider the *Very Busy Expressions* analysis from Section 2.4 as it applies to a program graph **PG**. Explain how to specify it as an analysis in the monotone framework of Definition 3.30 and on which program graph the analysis should be performed. □

TEASER 3.45: The observation that a powerset can be partially ordered by subset and superset gives rise to a general construction on partially ordered sets.

> Given a partially ordered set $(D, \sqsubseteq)$ define the *dual partially ordered set* $(D, \sqsubseteq)^{op}$ by $(D, \sqsubseteq)^{op} = (D, \sqsupseteq)$.

It is immediate that if $(D, \sqsubseteq)$ is a partially ordered set then so is $(D, \sqsupseteq)$; however, that $(D, \sqsubseteq)$ is a pointed semi-lattice (or satisfies the ascending chain condition) does not suffice for $(D, \sqsupseteq)^{op}$ being a pointed semi-lattice (or satisfying the ascending chain condition).

Returning to Teaser 3.18, show that if $(L, \sqsubseteq)$ is a complete lattice then so is $(L, \sqsubseteq)^{op}$.

(Hint: This is equivalent to showing that if $(L, \sqsubseteq)$ is a complete lattice, then for every subset $Y \subseteq L$ there is an element $\bigsqcap Y$, such that $\forall y \in Y : y \sqsupseteq l$ is equivalent to $\bigsqcap Y \sqsupseteq l$ for all choices of $l \in L$.) □

To informally summarise this chapter, we can deal with all the analyses considered so far by focusing on a partial order $\sqsubseteq$ (that can be chosen to be $\subseteq$ or $\supseteq$), and by focusing on directed edges (that can be chosen to be those of the program graph or the reversal of those of the program graph). In all cases our focus is on the least solution to the constrains expressed using $\sqsubseteq$, which turns out to be the largest solutions in case $\sqsubseteq = \supseteq$. This is very helpful for the development of algorithms for solving program analyses.

# Chapter 4

# Advanced Algorithms

In Chapter 2 we have seen specialised algorithms for solving the various bit-vector problems and they were generalised in Section 3.2 to cater for arbitrary analyses. The *Chaotic Iteration* algorithm may be summarised as in Figure 4.1 and in this chapter we consider advanced worklist algorithms for solving the analysis problems more efficiently. We focus on forward analyses and least solutions but in Section 4.4 we show how to adapt the development to arbitrary analyses and solutions.

## 4.1 The Worklist Algorithm

The algorithm of Figure 4.1 is inherently non-deterministic: the main loop calls for selecting an edge in the program graph for which the analysis constraint is not satisfied, and it will then update the analysis assignment; the algorithm terminates when this is no longer possible. To resolve most of this non-determinism we shall introduce a so-called *worklist* that will keep a record of the nodes of the program graph we need to reconsider.

| | |
|---|---|
| INPUT | a program graph $\mathbf{PG} = (\mathbf{Q}, q_\triangleright, q_\blacktriangleleft, \mathbf{Act}, \mathbf{E})$<br>an analysis specification with $\widehat{\mathbf{D}}$, $\widehat{\mathcal{S}}[\![q_\circ, \alpha, q_\bullet]\!]$ and $\hat{d}_\diamond$ |
| OUTPUT | the least solution AA to the forward analysis problem |
| METHOD | forall $q \in \mathbf{Q} \setminus \{q_\triangleright\}$ do $\mathsf{AA}(q) := \bot$ ;<br>$\mathsf{AA}(q_\triangleright) := \hat{d}_\diamond$ ;<br>while there exists an edge $(q_\circ, \alpha, q_\bullet) \in \mathbf{E}$<br>  such that $\widehat{\mathcal{S}}[\![q_\circ, \alpha, q_\bullet]\!](\mathsf{AA}(q_\circ)) \not\sqsubseteq \mathsf{AA}(q_\bullet)$<br>do $\mathsf{AA}(q_\bullet) := \mathsf{AA}(q_\bullet) \sqcup \widehat{\mathcal{S}}[\![q_\circ, \alpha, q_\bullet]\!](\mathsf{AA}(q_\circ))$ |

Figure 4.1: Chaotic Iteration (as considered in Chapters 2 and 3) for producing the least solution to a forward analysis problem.

| | |
|---|---|
| INPUT | a program graph $\mathbf{PG} = (\mathbf{Q}, q_\triangleright, q_\blacktriangleleft, \mathbf{Act}, \mathbf{E})$<br>an analysis specification with $\widehat{\mathbf{D}}$, $\widehat{\mathcal{S}}[\![q_\circ, \alpha, q_\bullet]\!]$ and $\hat{d}_\diamond$ |
| OUTPUT | the least solution AA to the forward analysis problem |
| METHOD | $\mathsf{W} := \mathsf{empty}$ ;<br>forall $q \in \mathbf{Q}$ do $\mathsf{AA}(q) := \bot$ ;<br>$\quad\quad\quad\quad \mathsf{W} := \mathsf{insert}(q, \mathsf{W})$ ;<br>$\mathsf{AA}(q_\triangleright) := \hat{d}_\diamond$ ;<br>while $\mathsf{W} \neq \mathsf{empty}$ do<br>$\quad (q_\circ, \mathsf{W}) := \mathsf{extract}(\mathsf{W})$ ;<br>$\quad$ forall $(q_\circ, \alpha, q_\bullet) \in \mathbf{E}$ do<br>$\quad\quad$ if $\widehat{\mathcal{S}}[\![q_\circ, \alpha, q_\bullet]\!](\mathsf{AA}(q_\circ)) \not\sqsubseteq \mathsf{AA}(q_\bullet)$<br>$\quad\quad$ then $\mathsf{AA}(q_\bullet) := \mathsf{AA}(q_\bullet) \sqcup \widehat{\mathcal{S}}[\![q_\circ, \alpha, q_\bullet]\!](\mathsf{AA}(q_\circ))$ ;<br>$\quad\quad\quad\quad \mathsf{W} := \mathsf{insert}(q_\bullet, \mathsf{W})$ |

Figure 4.2: Worklist Algorithm parameterised on a representation of the worklist.

The general form of the worklist algorithm is shown in Figure 4.2. Here we are not explicit about how the worklist W is organised; we merely assume that it contains nodes from the program graph and that we have the following three operations on the worklist:

- empty: this gives the empty worklist containing no nodes;
- insert$(q, \mathsf{W})$: this operation inserts a node $q$ in the worklist W and returns the updated worklist; and
- extract(W): this operation returns a pair $(q, \mathsf{W}')$ consisting of a node $q$ from the worklist W together with an updated worklist W′ where the node $q$ has been removed.

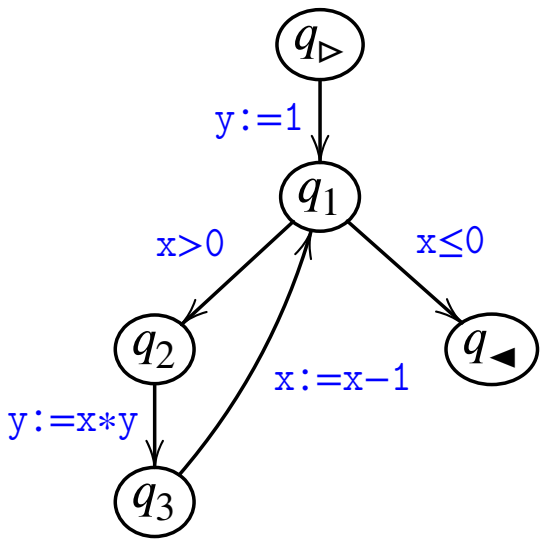

Figure 4.3: Program graph used in Figure 4.4.

The algorithm of Figure 4.2 has an initialisation phase and an iteration phase just as the algorithm of Figure 4.1. The initialisation phase ensures that the worklist W contains all the nodes of the program graph; starting from the empty worklist empty this is achieved by inserting all the nodes of the program graph using the operation insert$(\cdot, \cdot)$. In the iteration phase we select a node from the worklist using the extract$(\cdot)$ operation. Then all the edges with this node as the source node are inspected, and the analysis assignment will be updated when required, and if this happens the target nodes are added to the worklist using the insert$(\cdot, \cdot)$ operation.

| $(q, \mathsf{W}) = $ extract$(\cdot)$ | | edge | RD($q_\triangleright$) | RD($q_1$) | RD($q_2$) | RD($q_3$) | RD($q_\blacktriangleleft$) | insert$(\cdot,\cdot)$ |
|---|---|---|---|---|---|---|---|---|
| $\{q_\triangleright, q_1, q_2, q_3, q_\blacktriangleleft\}$ | | | $\{(x,?,q_\triangleright), (y,?,q_\triangleright)\}$ | $\{\,\}$ | $\{\,\}$ | $\{\,\}$ | $\{\,\}$ | |
| $q_\triangleright$ | $\{q_1, q_2, q_3, q_\blacktriangleleft\}$ | $(q_\triangleright, y := 1, q_1)$ | $\{(x,?,q_\triangleright), (y,?,q_\triangleright)\}$ | $\{(x,?,q_\triangleright), (y,q_\triangleright,q_1)\}$ | $\{\,\}$ | $\{\,\}$ | $\{\,\}$ | $q_1$ |
| $q_1$ | $\{q_2, q_3, q_\blacktriangleleft\}$ | $(q_1, x > 0, q_2)$ | $\{(x,?,q_\triangleright), (y,?,q_\triangleright)\}$ | $\{(x,?,q_\triangleright), (y,q_\triangleright,q_1)\}$ | $\{(x,?,q_\triangleright), (y,q_\triangleright,q_1)\}$ | $\{\,\}$ | $\{\,\}$ | $q_2$ |
| | | $(q_1, x \leq 0, q_\blacktriangleleft)$ | $\{(x,?,q_\triangleright), (y,?,q_\triangleright)\}$ | $\{(x,?,q_\triangleright), (y,q_\triangleright,q_1)\}$ | $\{(x,?,q_\triangleright), (y,q_\triangleright,q_1)\}$ | $\{\,\}$ | $\{(x,?,q_\triangleright), (y,q_\triangleright,q_1)\}$ | $q_\blacktriangleleft$ |
| $q_2$ | $\{q_3, q_\blacktriangleleft\}$ | $(q_2, y := x * y, q_3)$ | $\{(x,?,q_\triangleright), (y,?,q_\triangleright)\}$ | $\{(x,?,q_\triangleright), (y,q_\triangleright,q_1)\}$ | $\{(x,?,q_\triangleright), (y,q_\triangleright,q_1)\}$ | $\{(x,?,q_\triangleright), (y,q_2,q_3)\}$ | $\{(x,?,q_\triangleright), (y,q_\triangleright,q_1)\}$ | $q_3$ |
| $q_3$ | $\{q_\blacktriangleleft\}$ | $(q_3, x := x - 1, q_1)$ | $\{(x,?,q_\triangleright), (y,?,q_\triangleright)\}$ | $\{(x,?,q_\triangleright), (y,q_\triangleright,q_1), (x,q_3,q_1), (y,q_2,q_3)\}$ | $\{(x,?,q_\triangleright), (y,q_\triangleright,q_1)\}$ | $\{(x,?,q_\triangleright), (y,q_2,q_3)\}$ | $\{(x,?,q_\triangleright), (y,q_\triangleright,q_1)\}$ | $q_1$ |
| $q_1$ | $\{q_\blacktriangleleft\}$ | $(q_1, x > 0, q_2)$ | $\{(x,?,q_\triangleright), (y,?,q_\triangleright)\}$ | $\{(x,?,q_\triangleright), (y,q_\triangleright,q_1), (x,q_3,q_1), (y,q_2,q_3)\}$ | $\{(x,?,q_\triangleright), (y,q_\triangleright,q_1), (x,q_3,q_1), (y,q_2,q_3)\}$ | $\{(x,?,q_\triangleright), (y,q_2,q_3)\}$ | $\{(x,?,q_\triangleright), (y,q_\triangleright,q_1)\}$ | $q_2$ |
| | | $(q_1, x \leq 0, q_\blacktriangleleft)$ | $\{(x,?,q_\triangleright), (y,?,q_\triangleright)\}$ | $\{(x,?,q_\triangleright), (y,q_\triangleright,q_1), (x,q_3,q_1), (y,q_2,q_3)\}$ | $\{(x,?,q_\triangleright), (y,q_\triangleright,q_1), (x,q_3,q_1), (y,q_2,q_3)\}$ | $\{(x,?,q_\triangleright), (y,q_2,q_3)\}$ | $\{(x,?,q_\triangleright), (y,q_\triangleright,q_1), (x,q_3,q_1), (y,q_2,q_3)\}$ | $q_\blacktriangleleft$ |
| $q_2$ | $\{q_\blacktriangleleft\}$ | $(q_2, y := x * y, q_3)$ | $\{(x,?,q_\triangleright), (y,?,q_\triangleright)\}$ | $\{(x,?,q_\triangleright), (y,q_\triangleright,q_1), (x,q_3,q_1), (y,q_2,q_3)\}$ | $\{(x,?,q_\triangleright), (y,q_\triangleright,q_1), (x,q_3,q_1), (y,q_2,q_3)\}$ | $\{(x,?,q_\triangleright), (y,q_2,q_3), (x,q_3,q_1)\}$ | $\{(x,?,q_\triangleright), (y,q_\triangleright,q_1), (x,q_3,q_1), (y,q_2,q_3)\}$ | $q_3$ |
| $q_3$ | $\{q_\blacktriangleleft\}$ | $(q_3, x := x - 1, q_1)$ | $\{(x,?,q_\triangleright), (y,?,q_\triangleright)\}$ | $\{(x,?,q_\triangleright), (y,q_\triangleright,q_1), (x,q_3,q_1), (y,q_2,q_3)\}$ | $\{(x,?,q_\triangleright), (y,q_\triangleright,q_1), (x,q_3,q_1), (y,q_2,q_3)\}$ | $\{(x,?,q_\triangleright), (y,q_2,q_3), (x,q_3,q_1)\}$ | $\{(x,?,q_\triangleright), (y,q_\triangleright,q_1), (x,q_3,q_1), (y,q_2,q_3)\}$ | |
| $q_\blacktriangleleft$ | $\{\,\}$ | | $\{(x,?,q_\triangleright), (y,?,q_\triangleright)\}$ | $\{(x,?,q_\triangleright), (y,q_\triangleright,q_1), (x,q_3,q_1), (y,q_2,q_3)\}$ | $\{(x,?,q_\triangleright), (y,q_\triangleright,q_1), (x,q_3,q_1), (y,q_2,q_3)\}$ | $\{(x,?,q_\triangleright), (y,q_2,q_3), (x,q_3,q_1)\}$ | $\{(x,?,q_\triangleright), (y,q_\triangleright,q_1), (x,q_3,q_1), (y,q_2,q_3)\}$ | |

Figure 4.4: Computing the Reaching Definitions analysis assignment for the program graph of Figure 4.3 using the worklist representation of Figure 4.5.

EXAMPLE 4.1: In a simple scenario the worklist is just a set of nodes so $\mathsf{W} \subseteq \mathbf{Q}$ and the three operations can then be specified as in Figure 4.5. Here the operation extract(W) non-deterministically selects one of the nodes of the set and returns it together with the remainder of the set.

This form of the worklist algorithm is very close to the Chaotic Iteration of Figure 4.1. Figure 4.4 shows one choice of the iteration steps in the computation of the Reaching Definitions analysis of the program graph of Figure 4.3. The first line of Figure 4.4 shows the result after the initialisation step and each of the following lines corresponds to the (non-deterministic) extraction of a node from the worklist and processing of one of its edges. The first two columns show the node extracted from the set and the remaining set as obtained using

empty $= \{\,\}$

insert$(q, \mathsf{W}) = \mathsf{W} \cup \{q\}$

extract$(\mathsf{W}) = (q, \mathsf{W}')$
  where $q \in \mathsf{W}$
  and $\mathsf{W}' = \mathsf{W} \setminus \{q\}$

Figure 4.5: Worklist as a set.

the extract$(\cdot)$ operation. The third column shows the edge under consideration in the inner loop and the next four columns show the updated values of the analysis assignment. The final column shows the node to be inserted in the set using the insert$(\cdot,\cdot)$ operation (if any).

An alternative is to organise the worklist as a list – as the name suggests. There are two obvious ways of doing so as explored in the following examples.

empty = nil

insert$(q, \mathsf{W})$ = cons$(q, \mathsf{W})$

extract$(\mathsf{W}) = (q, \mathsf{W}')$
 where $q$ = head(W)
 and $\mathsf{W}'$ = tail(W)

Figure 4.6: LIFO: Worklist as a stack.

EXAMPLE 4.2: The first corresponds to a stack – so nodes are inserted and removed according to the last-in first-out (*LIFO*) principle. This is captured by the definitions of the worklist operations in Figure 4.6. The underlying data structure can simply be a list with nil being the empty list and cons, head and tail being the standard operations for respectively inserting an element in the front of a list and returning head and the tail of a list. These operations will all be constant time operations.

EXERCISE 4.3: Show how to compute the analysis result for the Reaching Definitions analysis of the program graph of Figure 4.3 using the LIFO worklist representation. The resulting analysis assignment should be as in Figure 4.5. □

empty = nil

insert$(q, \mathsf{W})$ = extend$(\mathsf{W}, q)$

extract$(\mathsf{W}) = (q, \mathsf{W}')$
 where $q$ = head(W)
 and $\mathsf{W}'$ = tail(W)

Figure 4.7: FIFO: Worklist as a queue.

EXAMPLE 4.4: An alternative is to use a queue – so nodes are inserted and removed according to the first-in first-out (*FIFO*) principle. This is captured by the definition of the worklist operations in Figure 4.7. In order to get constant time operations we can use a linked list as the underlying data structure and then the extend operation will append a node to the end of the list while the operations nil, head and tail perform the same operations as above but using this data structure.

EXERCISE 4.5: Show how to compute the analysis assignment for the Reaching Definitions analysis of the program graph of Figure 4.3 using the FIFO worklist representation. □

HANDS ON 4.6: Use the http://www.formalmethods.dk/pa4fun/ tool to perform a more detailed comparison of the iteration steps needed to compute the Reaching Definitions analysis assignments using the different worklist representations of Figures 4.5, 4.6 and 4.7. □

The following result establishes a number of correctness results corresponding to those for the Chaotic Iteration algorithm of Chapter 3. It follows that the worklist algorithm produces the same result as the Chaotic Iteration algorithm – but hopefully more efficiently.

PROPOSITION 4.7: If the analysis domain is a pointed semi-lattice that satisfies the ascending chain condition then the algorithm of Figure 4.2 terminates with an analysis assignment AA that solves the constraints of the analysis problem.

If additionally the analysis functions are monotone then the algorithm computes the least solution to the constraints and the analysis assignment AA summarises the paths.

Proof: The first statement is analogous to those of Propositions 3.12 and 3.16. The key observation when adapting the proofs of these results is the observation that whenever $q_\circ$ is *not* in the worklist during the iteration phase then indeed $\widehat{S}[\![q_\circ, \alpha, q_\bullet]\!](\mathsf{AA}(q_\circ)) \sqsubseteq \mathsf{AA}(q_\bullet)$ holds for all edges $(q_\circ, \alpha, q_\bullet)$ with source $q_\circ$. The second result is analogous to those of Propositions 3.28 and 3.29. □

If the program graph has $e$ edges and the analysis domain has finite height $h$ then the worklist algorithm only performs $\mathcal{O}(e \cdot h)$ basic operations. To see this observe that a node $q$ is only placed on the worklist $\mathcal{O}(h)$ times and each time some number $e_q$ of edges is considered. This accounts for $\mathcal{O}(\Sigma_{q\in\mathbf{Q}}\, h \cdot e_q)$ basic operations and since $\Sigma_{q\in\mathbf{Q}}\, e_q = e$ this establishes the complexity result.

Exercise 4.8: Consider a modification of Algorithm 4.2 where only $q_\triangleright$ is initially placed on the worklist and show that the algorithm does not always compute the least solution. (Hint: Find a program graph and analysis specification exhibiting the problem.) □

## 4.2 Reverse Postorder

For programs not containing loops it is possible to organise the worklist such that each node only needs to be considered once. This is done by iterating through the nodes in a special order known as *reverse postorder* and we shall explain it in this section. For programs containing loops one usually needs to consider nodes more than once, but the more efficient algorithms still use reverse postorder as part of their operation as we shall see in Section 4.3.

**Depth-First Spanning Tree** The reverse postorder of the nodes is obtained as part of constructing a depth-first spanning tree for the program graph. For this consider the algorithm in Figure 4.8. It takes as input a program graph with nodes **Q**, edges **E** and initial node $q_\triangleright$. It produces a *depth-first spanning tree* **T** (initially empty) and a so-called *reverse postorder* numbering **rP**. It makes use of a current number k (initially the number of nodes) and a set **V** of visited nodes (initially empty). The key part of the algorithm is the procedure DFS($q$) that traverses all outgoing edges (pointing to nodes that have not previously been traversed) in a depth-first manner while building the tree along the way. The procedure assigns the node its reverse postorder number just before returning.

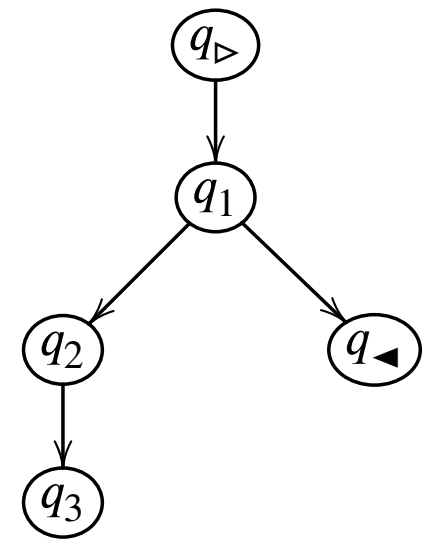

Figure 4.9: Depth-first spanning tree for the program graph of Figure 4.3.

Try It Out 4.9: Show that the algorithm of Figure 4.8 applied to the program graph of Figure 4.3 may give rise to the depth-first spanning tree of Figure 4.9. Is it

| | |
|---|---|
| INPUT | a program graph $\mathbf{PG} = (\mathbf{Q}, q_\triangleright, q_\blacktriangleleft, \mathbf{Act}, \mathbf{E})$ |
| OUTPUT | $\mathbf{T}$: a set of edges in a depth-first spanning tree<br>$\mathbf{rP}$: a reverse postorder numbering of the nodes in $\mathbf{PG}$ |
| DATA | $\mathbf{T} \subseteq \mathbf{Q} \times \mathbf{Q}$: the tree constructed so far<br>$\mathbf{V} \subseteq \mathbf{Q}$: the nodes visited so far<br>$\mathsf{k} \in \mathbf{Int}$: the current number<br>$\mathbf{rP} : \mathbf{Q} \to \mathbf{Int}$: the reverse postorder numbering |
| ALGORITHM | $\mathbf{T} := \{\ \}$;<br>$\mathbf{V} := \{\ \}$;<br>k := $\lvert\mathbf{Q}\rvert$;<br>DFS($q_\triangleright$) |
| PROCEDURE | DFS($q_\circ$) is defined by<br>$\mathbf{V} := \mathbf{V} \cup \{q_\circ\}$;<br>while there exists an edge $(q_\circ, \alpha, q_\bullet) \in \mathbf{E}$<br>such that $q_\bullet \notin \mathbf{V}$<br>do $\mathbf{T} := \mathbf{T} \cup \{(q_\circ, q_\bullet)\}$; DFS($q_\bullet$);<br>$\mathbf{rP}[q_\circ]$ := k; k:= k-1; |

Figure 4.8: Depth-First Spanning Tree and Reverse Postorder.

possible to construct other depth-first spanning trees? What are the possible reverse postorders? □

Exercise 4.10: Use the algorithm of Figure 4.8 to construct a depth-first spanning tree for the insertion sort program graph of Figure 2.7 and determine the associated reverse postorder numbering. How many different reverse postorders may arise from different ways of running the algorithm? □

**Classification of Edges** A main purpose of a depth-first spanning tree and the associated reverse postorder is to be able to classify the edges in the program graph. For this it will be helpful to write $\mathbf{T}^*$ for the reflexive and transitive closure of $\mathbf{T}$ (that is, following zero or more edges in $\mathbf{T}$), $\mathbf{T}^+$ for the transitive closure of $\mathbf{T}$ (that is, following one or more edges in $\mathbf{T}$), and $\mathbf{T}^{++}$ for $\mathbf{T}^+\mathbf{T}$ (that is, following two or more edges in $\mathbf{T}$).

Definition 4.11: The edges in $\mathbf{E}$ can be classified with respect to $\mathbf{T}$ into one of four types:

- a *tree edge* is an edge $(q, \alpha, q') \in \mathbf{E}$ where $(q, q') \in \mathbf{T}$,
- a *forward edge* is an $(q, \alpha, q') \in \mathbf{E}$ where $(q, q') \in \mathbf{T}^{++}$, meaning that it points further down in the tree,

- a *back edge* is an edge $(q, \alpha, q') \in \mathbf{E}$ where $(q', q) \in \mathbf{T}^*$ (including $q = q'$), meaning that it points back up in the tree,
- a *cross edge* is an edge $(q, \alpha, q') \in \mathbf{E}$ where neither $(q, q')$ nor $(q', q)$ is in $\mathbf{T}^*$, meaning that neither $q$ nor $q'$ is an ancestor of the other.

EXAMPLE 4.12: The edges of the program graph of Figure 4.3 can be classified as follows with respect to the depth-first spanning tree of Figure 4.9: all edges except $(q_3, \texttt{x := x} - 1, q_1)$ are tree edges while $(q_3, \texttt{x := x} - 1, q_1)$ is a back edge. There are no forward edges and no cross edges.

In general, a program graph may have many different depth-first spanning trees and therefore the notions of tree edge, forward edge, back edge, and cross edge are not uniquely determined but are always relative to the depth-first spanning tree of interest. However, for a large class of so-called *reducible program graphs* the notion of back edge will be the same regardless of the depth-first spanning tree considered; this will be a consequence of Proposition 4.28 to be presented in Section 4.3. Most program graphs constructed from 'structured programming languages' will indeed produce reducible program graphs (and the Guarded Commands language is an example of this). Intuitively, a reducible program graph is one where paths from the initial node always enter a loop at the same node. A non-reducible program graph is shown in Figure 4.10.

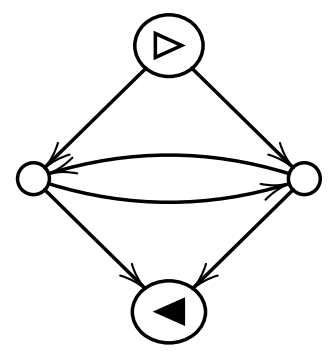

Figure 4.10: A non-reducible program graph.

EXERCISE 4.13: Returning to Exercise 4.10 determine whether or not the different reverse postorders of the various program graphs give rise to the same back edges. □

EXERCISE 4.14: Construct a depth-first spanning tree for the program graph of Figure 4.11 together with the reverse postorder numbering. Classify the edges as tree edges, forward edges, back edges and cross edges. □

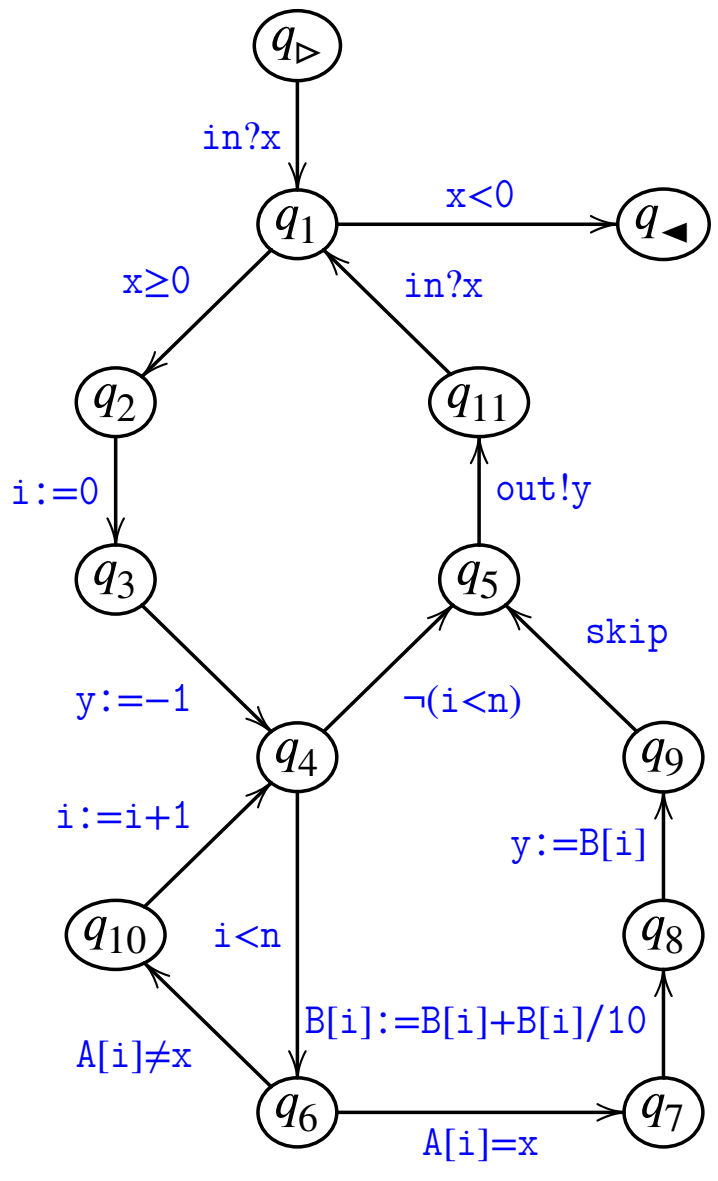

Figure 4.11: Program graph for searching and updating array entries.

PROPOSITION 4.15: An edge $(q, \alpha, q') \in \mathbf{E}$ is a *back edge* if and only if $\mathbf{rP}[q] \geq \mathbf{rP}[q']$.

PROOF: We first assume that $(q, \alpha, q') \in \mathbf{E}$ is a back edge and show that $\mathbf{rP}[q] \geq \mathbf{rP}[q']$. If $(q, \alpha, q')$ is additionally a self loop, i.e. $q = q'$, we have $\mathbf{rP}[q] = \mathbf{rP}[q']$ and the result is immediate. Otherwise $q'$ is a proper ancestor of $q$ in the depth-first spanning tree $\mathbf{T}$. Hence $q'$ is visited before $q$ in the depth-first traversal and the call DFS($q'$) is pending throughout the entire call of DFS($q$) meaning that $\mathbf{rP}[q] > \mathbf{rP}[q']$ and the result is immediate.

We next assume that $\mathbf{rP}[q] \geq \mathbf{rP}[q']$ and show that $(q, \alpha, q') \in \mathbf{E}$ is a back edge. If $\mathbf{rP}[q] = \mathbf{rP}[q']$ we have $q = q'$ and hence we have a self loop which is clearly a back edge. Otherwise $\mathbf{rP}[q] > \mathbf{rP}[q']$ showing that the call DFS($q$) is left before that of DFS($q'$). This means that the call of DFS($q'$) is either pending during the call of DFS($q$) or has not yet been initiated. If the call of DFS($q'$) is pending during the call of DFS($q$) then $q'$ is a proper ancestor of $q$ in the depth-first spanning tree $\mathbf{T}$ and the edge is a back edge. If the call of DFS($q'$) has not yet been initiated before the call of DFS($q$) is left we arrive at a contradiction: during the call of DFS($q$) we have $q' \notin \mathbf{V}$ and when we examine the edge $(q, \alpha, q')$ we will perform the call DFS($q'$), which contradicts our assumptions. □

| | |
|---|---|
| INPUT | a program graph $\mathbf{PG} = (\mathbf{Q}, q_{\triangleright}, q_{\blacktriangleleft}, \mathbf{Act}, \mathbf{E})$<br>an analysis specification with $\widehat{\mathbf{D}}$, $\widehat{\mathcal{S}}[\![q_\circ, \alpha, q_\bullet]\!]$ and $\hat{d}_\diamond$ |
| OUTPUT | the least solution AA to the forward analysis problem |
| ALGORITHM | forall $q \in \mathbf{Q} \setminus \{q_\triangleright\}$ do $\mathsf{AA}(q) := \bot$ ;<br>$\mathsf{AA}(q_\triangleright) := \hat{d}_\diamond$;<br>while there exists an edge $(q_\circ, \alpha, q_\bullet) \in \mathbf{E}$<br>such that $\widehat{\mathcal{S}}[\![q_\circ, \alpha, q_\bullet]\!](\mathsf{AA}(q_\circ)) \not\sqsubseteq \mathsf{AA}(q_\bullet)$<br>do forall $q_\circ \in \mathbf{Q}$ in reverse postorder<br>do forall edges $(q_\circ, \alpha, q_\bullet) \in \mathbf{E}$<br>do $\mathsf{AA}(q_\bullet) := \mathsf{AA}(q_\bullet) \sqcup \widehat{\mathcal{S}}[\![q_\circ, \alpha, q_\bullet]\!](\mathsf{AA}(q_\circ))$ |

Figure 4.12: Round Robin Iteration towards the solution.

**Proposition 4.16:** Each loop in $\mathbf{E}$ will involve at least one back edge.

Proof: We prove this by contradiction and so let us suppose that there is a loop $(q_0, \alpha_1, q_1), \cdots, (q_{n-1}, \alpha_n, q_n)$ (with $n > 0$ and $q_0 = q_n$) that involves no back edges. It then follows from Proposition 4.15 that $\mathbf{rP}[q_0] < \mathbf{rP}[q_n]$ but this is a contradiction since $\mathbf{rP}[q_0] = \mathbf{rP}[q_n]$. □

In the beginning of this section we claimed that for programs not containing loops it is possible to organise the worklist in reverse postorder so that each node only needs to be considered once. This follows from the two propositions just established.

**The Round Robin Algorithm** It is instructive to consider a very simple algorithm that is reasonably efficient due to the use of reverse postorder. To do so we shall change our mindset from the general Chaotic Iteration algorithm of Figure 4.1 that attempts to follow only edges along which propagations are needed. Instead we shall aim at propagating information along *all* edges is case we have not yet obtained the desired solution. An abstract version of this algorithm for forward analyses is shown in Figure 4.12 and is called the *Round Robin* algorithm.

Intuitively, the Round Robin algorithm works well because the number of iterations needed only depends on the nesting structure of the loops (and of the finite height of the analysis domain) and most programs do not have deeply nested loops. The reverse postorder organisation of the worklist ensures that information is propagated in a *forward* manner of the analysis and that a small number of rounds are needed in order to obtain the analysis assignment.

| $(q,(\mathsf{V},\mathsf{t})) = \mathsf{extract}(\cdot)$ | | edge | $\mathsf{RD}(q_\triangleright)$ | $\mathsf{RD}(q_1)$ | $\mathsf{RD}(q_2)$ | $\mathsf{RD}(q_3)$ | $\mathsf{RD}(q_\blacktriangleleft)$ | $\mathsf{insert}(\cdot,\cdot)$ |
|---|---|---|---|---|---|---|---|---|
| | $[\,]$ true | | $\{(\mathtt{x},?,q_\triangleright),(\mathtt{y},?,q_\triangleright)\}$ | $\{\}$ | $\{\}$ | $\{\}$ | $\{\}$ | |
| $q_\triangleright$ | $[q_1,q_2,q_3,q_\blacktriangleleft]$ false | $(q_\triangleright,\mathtt{y}:=1,q_1)$ | $\{(\mathtt{x},?,q_\triangleright),(\mathtt{y},?,q_\triangleright)\}$ | $\{(\mathtt{x},?,q_\triangleright),(\mathtt{y},q_\triangleright,q_1)\}$ | $\{\}$ | $\{\}$ | $\{\}$ | $q_1$ |
| $q_1$ | $[q_2,q_3,q_\blacktriangleleft]$ true | $(q_1,\mathtt{x}>0,q_2)$ | $\{(\mathtt{x},?,q_\triangleright),(\mathtt{y},?,q_\triangleright)\}$ | $\{(\mathtt{x},?,q_\triangleright),(\mathtt{y},q_\triangleright,q_1)\}$ | $\{(\mathtt{x},?,q_\triangleright),(\mathtt{y},q_\triangleright,q_1)\}$ | $\{\}$ | $\{\}$ | $q_2$ |
| | | $(q_1,\mathtt{x}\leq 0,q_\blacktriangleleft)$ | $\{(\mathtt{x},?,q_\triangleright),(\mathtt{y},?,q_\triangleright)\}$ | $\{(\mathtt{x},?,q_\triangleright),(\mathtt{y},q_\triangleright,q_1)\}$ | $\{(\mathtt{x},?,q_\triangleright),(\mathtt{y},q_\triangleright,q_1)\}$ | $\{\}$ | $\{(\mathtt{x},?,q_\triangleright),(\mathtt{y},q_\triangleright,q_1)\}$ | $q_\blacktriangleleft$ |
| $q_2$ | $[q_3,q_\blacktriangleleft]$ true | $(q_2,\mathtt{y}:=\mathtt{x}*\mathtt{y},q_3)$ | $\{(\mathtt{x},?,q_\triangleright),(\mathtt{y},?,q_\triangleright)\}$ | $\{(\mathtt{x},?,q_\triangleright),(\mathtt{y},q_\triangleright,q_1)\}$ | $\{(\mathtt{x},?,q_\triangleright),(\mathtt{y},q_\triangleright,q_1)\}$ | $\{(\mathtt{x},?,q_\triangleright),(\mathtt{y},q_2,q_3)\}$ | $\{(\mathtt{x},?,q_\triangleright),(\mathtt{y},q_\triangleright,q_1)\}$ | $q_3$ |
| $q_3$ | $[q_\blacktriangleleft]$ true | $(q_3,\mathtt{x}:=\mathtt{x}-1,q_1)$ | $\{(\mathtt{x},?,q_\triangleright),(\mathtt{y},?,q_\triangleright)\}$ | $\{(\mathtt{x},?,q_\triangleright),(\mathtt{y},q_\triangleright,q_1),(\mathtt{x},q_3,q_1),(\mathtt{y},q_2,q_3)\}$ | $\{(\mathtt{x},?,q_\triangleright),(\mathtt{y},q_\triangleright,q_1)\}$ | $\{(\mathtt{x},?,q_\triangleright),(\mathtt{y},q_2,q_3)\}$ | $\{(\mathtt{x},?,q_\triangleright),(\mathtt{y},q_\triangleright,q_1)\}$ | $q_3$ |
| $q_\blacktriangleleft$ | $[\,]$ true | | $\{(\mathtt{x},?,q_\triangleright),(\mathtt{y},?,q_\triangleright)\}$ | $\{(\mathtt{x},?,q_\triangleright),(\mathtt{y},q_\triangleright,q_1),(\mathtt{x},q_3,q_1),(\mathtt{y},q_2,q_3)\}$ | $\{(\mathtt{x},?,q_\triangleright),(\mathtt{y},q_\triangleright,q_1)\}$ | $\{(\mathtt{x},?,q_\triangleright),(\mathtt{y},q_2,q_3)\}$ | $\{(\mathtt{x},?,q_\triangleright),(\mathtt{y},q_\triangleright,q_1)\}$ | |

Figure 4.14: The first round in computing the Reaching Definitions analysis assignment for the program graph of Figure 4.3 using the Round Robin worklist representation of Figure 4.13.

EXAMPLE 4.17: The algorithm can also be obtained as an instantiation of the general worklist algorithm of Figure 4.2. To see this we shall represent the worklist W by a pair (V, t) where V is a list of nodes and t is a boolean indicating whether an extra round is needed. The operations on the worklist are then as in Figure 4.13. Here nil, head and tail are as in Examples 4.2 and 4.4; additionally we write $\mathsf{sort}_{\mathsf{rP}}$ for the function that sorts a set in reverse postorder.

Figure 4.14 shows the iteration steps of the first round needed for computing the Reaching Definitions analysis assignment for the program graph of Figure 4.3. We make use of the depth-first spanning tree of Figure 4.9 and the associated reverse post order $q_\triangleright, q_1, q_2, q_3, q_\blacktriangleleft$. There is a total of three rounds where the last simply amounts to checking that all the constraints are satisfied.

$\mathsf{empty} = (\mathsf{nil}, \mathsf{false})$

$\mathsf{insert}(q,(\mathsf{V},\mathsf{t})) = (\mathsf{V},\mathsf{true})$

$\mathsf{extract}(\mathsf{V},\mathsf{t}) =$
  if $\mathsf{V} = \mathsf{nil}$
  then $(q_\triangleright, (\mathsf{V}_{\mathsf{rP}}, \mathsf{false}))$
    where $\mathsf{V}_{\mathsf{rP}} =$
      $\mathsf{sort}_{\mathsf{rP}}(\mathbf{Q}\setminus\{q_\triangleright\})$
  else $(q, (\mathsf{V}', \mathsf{t}))$
    where $q = \mathsf{head}(\mathsf{V})$
    and $\mathsf{V}' = \mathsf{tail}(\mathsf{V})$

Figure 4.13: Worklist for Round Robin.

EXERCISE 4.18: Complete the iteration steps for the remaining two rounds of computing the Reaching Definitions analysis assignment of Example 4.17. □

**The Worklist Algorithm and Reverse Postorder** In the Round Robin algorithm of Figure 4.12 we iterated through *all* the nodes in reverse postorder whenever we had not yet obtained the desired solution. The worklist version of this approach is to only iterate through the *necessary* nodes in reverse postorder.

To faciliate this it is natural to let the worklist W have two components (V, P):

- a list V of *current* nodes to be visited in the order given, and
- a set (or list) P of *pending* nodes that need to be reconsidered.

The key point of the algorithm is to set the list of current nodes to be the set of pending nodes sorted in reverse postorder whenever the former becomes empty, and to terminate when both are empty.

$\mathsf{empty} = (\mathsf{nil}, \{\ \})$

$\mathsf{insert}(q, (\mathsf{V}, \mathsf{P})) =$
  if $q \notin \mathsf{V}$
  then $(\mathsf{V}, \{q\} \cup \mathsf{P})$
  else $(\mathsf{V}, \mathsf{P})$

$\mathsf{extract}(\mathsf{V}, \mathsf{P}) =$
  if $\mathsf{V} = \mathsf{nil}$
  then $(q, (\mathsf{V}', \{\ \}))$
    where $\mathsf{V}_{\mathsf{rP}} = \mathsf{sort}_{\mathsf{rP}}(\mathsf{P})$
    and $q = \mathsf{head}(\mathsf{V}_{\mathsf{rP}})$
    and $\mathsf{V}' = \mathsf{tail}(\mathsf{V}_{\mathsf{rP}})$
  else $(q, (\mathsf{V}', \mathsf{P}))$
    where $q = \mathsf{head}(\mathsf{V})$
    and $\mathsf{V}' = \mathsf{tail}(\mathsf{V})$

Figure 4.15: Worklist for iterating in reverse postorder.

The three operations on the worklist are presented in Figure 4.15; here $\mathsf{sort}_{\mathsf{rP}}$ once again is the function that sorts a set into a list in reverse postorder, and we use $q \in \mathsf{V}$ to indicate whether or not the node $q$ is a member of the list V. It is clear that nodes are always inserted into the set of pending nodes and are always extracted from the head of the list of current nodes.

EXAMPLE 4.19: We can now redo the computation of the analysis assignment for the Reaching Definitions analysis of the program graph of Figure 4.3 using the worklist representation of Figure 4.15. Initially the pending set will contain all the nodes, that is $\{q_\triangleright, q_1, q_2, q_3, q_\blacktriangleleft\}$, and in the first round these nodes will be sorted in reverse post-order and considered one by one. While doing so the pending set is updated; in most cases the node to be inserted is already in the current list and at the end of the first round the pending set will just contain the node $q_1$. Now follows a (short) round where $q_2$ and $q_\blacktriangleleft$ are inserted in the pending set. Yet another round is needed and here only $q_3$ enters the pending set and after yet another (very short) round the algorithm terminates.

HANDS ON 4.20: Use the http://www.formalmethods.dk/pa4fun/ tool to perform a more detailed comparison of the iteration steps outlined in Example 4.19 and Example 4.17. How big is the reduction in the number of iteration steps? Repeat the experiment for the program graph of Figure 4.11. □

## 4.3 Strong Components and Natural Loops

In the previous section we explained that programs without loops can by analysed efficiently by iterating in reverse postorder. For programs with loops this leaves the question of what strategy to use for dealing with the loops. We present two possibilities in this section.

**Strong Components** Two nodes $q'$ and $q''$ are *strongly connected* whenever there is a (possibly trivial) directed path from $q'$ to $q''$ as well as a (possibly trivial) directed path from $q''$ to $q'$. The consideration of possibly trivial directed paths ensures that a node is always strongly connected to itself and that *being strongly connected to* is an equivalence relation. A *strong component* therefore is a *maximal* set of nodes that are strongly connected to each other.

Strong components form the nodes of a so-called *reduced graph* that has an edge from one strong component $Q'$ to another strong component $Q''$ whenever $Q' \neq Q''$ and there is a node $q' \in Q'$ and a node $q'' \in Q''$ such that the program graph has an edge from $q'$ to $q''$. The reduced graph is a directed acyclic graph (DAG) and is topologically sorted by the reverse postorder.

| | |
|---|---|
| INPUT | a program graph $\mathbf{PG} = (\mathbf{Q}, q_\triangleright, q_\blacktriangleleft, \mathbf{Act}, \mathbf{E})$ and<br>a reverse postorder numbering **rP** of the nodes in **PG** |
| OUTPUT | a list **SClist** of strong components partitioning **Q** |
| DATA | **SC**: a strongly connected set of nodes<br>$\mathbf{V} \subseteq \mathbf{Q}$: the nodes in append(**SClist**, **SC**) |
| ALGORITHM | **SClist** := [ ]; **SC** := { }; **V** := { };<br>for $i = 1$ to $\lvert\mathbf{Q}\rvert$ do<br>    $q := \mathbf{rP}^{-1}[i]$;<br>    if $q \notin \mathbf{V}$ then<br>        Assign($q$);<br>        **SClist** := append(**SClist**, **SC**); **SC** := { } |
| PROCEDURE | Assign($q_\bullet$) is defined by<br>    $\mathbf{SC} := \mathbf{SC} \cup \{q_\bullet\}$; $\mathbf{V} := \mathbf{V} \cup \{q_\bullet\}$;<br>    for each $(q_\circ, \alpha, q_\bullet) \in \mathbf{E}$ do<br>        if $q_\circ \notin \mathbf{V}$ then Assign($q_\circ$) |

Figure 4.16: Strong components.

The strong components can be computed from the reverse postorder using *Kosaraju's algorithm* as shown in Figure 4.16.

We use $\mathbf{rP}^{-1}[i]$ to indicate the node $q$ such that $\mathbf{rP}[q] = i$, which makes sense because **rP** can be seen as a bijection between **Q** and $\{1, \cdots, \lvert\mathbf{Q}\rvert\}$; hence $\mathbf{rP}^{-1}[1] = q_\triangleright$ is the first node considered. We use append(**SClist**, **SC**) to express the result of appending **SC** as an element at the end of the list **SClist**; strong components are collected in this order so as to ensure that the list topologically sorts the reduced graph.

The procedure Assign($q$) recursively performs a *backward* traversal of the graph from $q$, adding the new nodes found to the strong component currently being constructed. It is the fact that we consider the nodes in increasing reverse postorder that ensures that we account for both forward and backward reachability as needed for **SC** being strongly connected.

EXAMPLE 4.21: Consider the program graph of Figure 4.3 together with the reverse postorder $q_\triangleright, q_1, q_2, q_3, q_\blacktriangleleft$ of the nodes. The algorithm of Figure 4.16 will give rise to the following list of strong components:

$$[\{q_\triangleright\}, \{q_1, q_2, q_3\}, \{q_\blacktriangleleft\}]$$

The associated reduced graph is shown in Figure 4.17. When computing the analysis assignment we will iterate through the strong components in this order:

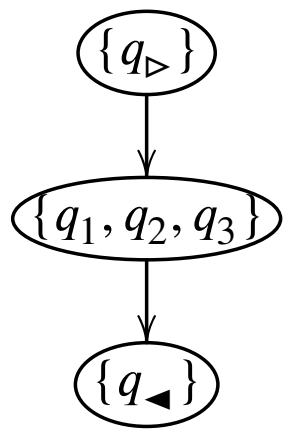

Figure 4.17: Reduced graph for the program graph of Figure 4.3.

first $\{q_{\triangleright}\}$, then $\{q_1, q_2, q_3\}$ and finally $\{q_{\blacktriangleleft}\}$.

Exercise 4.22: Continuing Exercise 4.14, determine the strong components and the associated reduced graph of the program graph based on the reverse postorder identified. □

Proposition 4.23: The algorithm of Figure 4.16 computes the strong components of the program graph.

Sketch of Proof: We shall argue that we have the following invariant: *(a)* all **SC** on **SClist** are indeed strong components, and *(b)* all **SC** on **SClist** occur in increasing order of their value of **rP** as defined by $\mathbf{rP}[\mathbf{SC}] = \min\{\mathbf{rP}[q] \mid q \in \mathbf{SC}\}$. This clearly holds initially and when the algorithm terminates all nodes of the graph have been included in a strong component.

Whenever during a *top level* call of Assign($q$) we encounter a call of Assign($q'$) we know $q$ is reachable from $q'$. We also know that $\mathbf{rP}[q'] > \mathbf{rP}[q]$ because $\mathbf{rP}[q'] = \mathbf{rP}[q]$ is prevented by updating **V** already after the first call to $q$, and $\mathbf{rP}[q'] < \mathbf{rP}[q]$ is prevented by having already constructed on **SClist** the strong component containing $q'$.

The call of DFS($q'$) therefore is left before that of DFS($q$). This means that the call of DFS($q$) is either pending during the call of DFS($q'$) or has not yet been initiated. If the call of DFS($q$) is pending during the call of DFS($q'$) this means that $q'$ is reachable from $q$ and hence it is correct to include $q'$ in the strong component being constructed.

If the call of DFS($q$) has not yet been initiated before the call of DFS($q'$) is left we arrive at a contradiction. Setting $q_0 = q'$ and $q_{n+1} = q$ there would be edges $(q_0, \alpha_0, q_1), (q_1, \alpha_1, q_2), \cdots, (q_n, \alpha_n, q_{n+1})$ (for $n \geq 0$) such that $\forall i : \mathbf{rP}[q_i] > \mathbf{rP}[q]$. Let $j$ be maximal such that the call of DFS($q$) has not been initiated before the call of DFS($q_j$) has been left. Hence the call DFS($q$) is pending during the call of DFS($q_{j+1}$), where we take a call to be pending during its own call, and therefore the call of DFS($q_{j+1}$) has not been initiated before the call of DFS($q_j$) has been left. During the call of DFS($q_j$) the edge $(q_j, \alpha_j, q_{j+1})$ is considered and would give rise to a call of DFS($q_{j+1}$). This establishes the desired contradiction. □

```
empty = (nil, { })

insert(q, (V, P)) =
  if q ∈ V
  then (V, P)
  else (V, {q} ∪ P)

extract(V, P) =
  if V = nil
  then (q, (V′, P′))
    where (S, P′) = scs_rP(P)
    and V_rP = sort_rP(S)
    and q = head(V_rP)
    and V′ = tail(V_rP)
  else (q, (V′, P))
    where q = head(V)
    and V′ = tail(V)
```

Figure 4.18: Worklist for iterating through strong components.

**Iterating through Strong Components** We shall iterate through the strong components in a top-down manner with respect to the reduced graph. In Figure 4.18 we present this idea using operations that operate on a worklist W with two components (V, P) as above. When the component V is exhausted, the set P of pending nodes is split into two (S, P′). The set S ⊆ P contains all the nodes of P belonging to a strong component where none of its ancestors in the reduced graph contains any nodes from P; this is called a *topmost* strong component. (If there is more than one candidate one might choose the one that has a node with the smallest reverse postorder among all nodes in the candidates.) The set $\mathsf{P}' = \mathsf{P} \setminus \mathsf{S}$ then contains the remaining nodes from P. We use the function $\mathsf{scs}_{\mathsf{rP}}$ to achieve this, and as above we write $\mathsf{sort}_{\mathsf{rP}}$ for the function that sorts a set into a list in reverse postorder.

EXAMPLE 4.24: We can now redo the computation of the analysis assignment for the Reaching Definitions analysis of the program graph of Figure 4.3. We will use the worklist represention of Figure 4.18 and the list $[\{q_\triangleright\}, \{q_1, q_2, q_3\}, \{q_\blacktriangleleft\}]$ of strong components identified in Example 4.21. After the initialisation the pending set will contain all the nodes, that is, $\{q_\triangleright, q_1, q_2, q_3, q_\blacktriangleleft\}$. We now extract the first strong component, that is, $\{q_\triangleright\}$, and note that the pending set contains the remaining nodes $\{q_1, q_2, q_3, q_\blacktriangleleft\}$. After processing the node $q_\triangleright$, the pending set is still $\{q_1, q_2, q_3, q_\blacktriangleleft\}$ so we extract the second strong component $\{q_1, q_2, q_3\}$ and leave $\{q_\blacktriangleleft\}$ in the pending set for later processing.

We now have a round where we process the nodes of $\{q_1, q_2, q_3\}$ in reverse post-order and as a result of this $q_1$ is added to the pending set. The process is repeated for $\{q_1\}$ and the pending set will grow from $\{q_\blacktriangleleft\}$ to $\{q_2, q_\blacktriangleleft\}$. There will be yet another round for $\{q_2\}$ and $q_3$ will be added to the pending set. And then we will have a round with $q_3$ but this time no further nodes are added to the pending set so we are left with the set $\{q_\blacktriangleleft\}$ which is a strong component and it is then processed in the last step.

HANDS ON 4.25: Use the http://www.formalmethods.dk/pa4fun/ tool to perform a more detailed comparison of the iteration steps outlined in Example 4.24 and Example 4.19. How big is the reduction in the number of iteration steps? Repeat the experiment for the program graph of Figure 4.11. □

TEASER 4.26: Let us consider an alternative definition of the operation $\mathsf{scs}_{\mathsf{rP}}(\mathsf{P})$: It will return a pair $(\mathsf{S}, \mathsf{P}')$ as in Figure 4.18 but will take S to contain *all* the nodes of the topmost strong component having a member in P; as before $\mathsf{P}' = \mathsf{P} \setminus \mathsf{S}$. Compare this alternative worklist representation with the one of Figure 4.18 for example by redoing computation of the analysis assignment in Example 4.24 (and Hands On 4.25). □

**Natural Loops** The strong components merely merge loops within loops and we next consider an approach that is less aggressive in merging loops. As an example the program graph of Figure 4.11 contains two loops but since one of them is within the other, they are in the same strong component.

Given a program graph $\mathbf{PG} = (\mathbf{Q}, q_\triangleright, q_\blacktriangleleft, \mathbf{Act}, \mathbf{E})$ a node $q_\bullet$ is said to *dominate* a node $q_\circ$ whenever any path from $q_\triangleright$ to $q_\circ$ also contains $q_\bullet$. An edge $(q_\circ, \alpha, q_\bullet)$ is said to be a *dominator edge* whenever the target $q_\bullet$ dominates the source $q_\circ$. The program graph **PG** is *reducible* whenever the subgraph obtained by removing all dominator edges is acyclic. The program graph in Figure 4.10 has no dominator edges and hence is non-reducible, although it has a back edge (because there is a loop) regardless of the choice of reverse postorder.

Example 4.27: In the program graph of Figure 4.3, the node $q_1$ is dominating the node $q_3$ and the edge $(q_3, \texttt{x := x} - 1, q_1)$ is a dominator edge. Indeed this is the only dominator edge and hence the program graph is reducible.

Proposition 4.28: All dominator edges are back edges, and a program graph is reducible exactly when dominator edges and back edges coincide (regardless of the choice of depth-first spanning tree and reverse postorder).

Sketch of Proof: We first show that all dominator edges will be classified as back edges. To see this consider a dominator edge $(q_\circ, \alpha, q_\bullet)$ and note that whenever the call DFS($q_\circ$) is performed there must be a pending call of DFS($q_\bullet$) as otherwise $q_\bullet$ would not dominate $q_\circ$.

For a reducible program graph all back edges are dominator edges as well. To see this note that the DFS algorithm will never construct a tree edge that is also a dominator edge. Hence the depth-first spanning tree constructed for **PG** could also be constructed for the subgraph obtained by removing all dominator edges. If the DFS algorithm was to encounter an edge in the subgraph that would be classified as a back edge then this would violate the acyclicity of the subgraph.

Finally, suppose that back edges and dominator edges coincide. As the program graph obtained by removing all back edges is acyclic this shows that the program graph is reducible. □

The *natural loop* induced by the dominator edge $(q_\circ, \alpha, q_\bullet)$ is the set of nodes that occur on a path from $q_\bullet$ to $q_\circ$ that does not contain $q_\bullet$ as an intermediate node. The *natural loop* headed at $q_\bullet$ is the union of all the natural loops induced by dominator edges with $q_\bullet$ as a target; all nodes in the set will be dominated by $q_\bullet$. Each natural loop will be a strongly connected set of nodes but may be properly contained in a strong component.

Two natural loops headed at distinct nodes $q_1$ and $q_2$ will either be disjoint or one will be properly contained in the other. To see this suppose that the natural loops headed at $q_1$ and $q_2$ have a node $q$ in common and consider a shortest path from $q_\triangleright$ to $q$. Both $q_1$ and $q_2$ must occur on this path so one must occur before the other and without loss of generality assume that $q_1$ occurs before $q_2$. Then the natural loop headed at $q_2$ cannot contain $q_1$ but all nodes in the natural loop headed at $q_2$ will also be in the natural loop headed at $q_1$. Hence the natural loop headed at $q_2$ is properly contained in the one headed at $q_1$.

Reverse postorder is useful for constructing natural loops for reducible program graphs. The algorithm of Figure 4.19 considers all back edges and adds their natural loops to the appropriate headers as represented in the data structure L. If it encounters a back edge that is not a dominator edge then the program graph is not reducible and the algorithm aborts.

| | |
|---|---|
| INPUT | a program graph $\mathbf{PG} = (\mathbf{Q}, q_{\triangleright}, q_{\blacktriangleleft}, \mathbf{Act}, \mathbf{E})$ and<br>a reverse postorder numbering $\mathbf{rP}$ of the nodes in $\mathbf{PG}$ |
| OUTPUT | a mapping such that $\mathsf{L}[q]$ is the natural loop headed at $q$ |
| ABORTS | whenever the program graph is non-reducible |
| ALGORITHM | for $q \in \mathbf{Q}$ do $\mathsf{L}[q] := \{\ \}$;<br>for all $(q_\circ, \alpha, q_\bullet) \in \mathbf{E}$ with $\mathbf{rP}[q_\bullet] \leq \mathbf{rP}[q_\circ]$ do<br>    $\mathsf{L}[q_\bullet] := \mathsf{L}[q_\bullet] \cup \{q_\bullet\}$; Build$(q_\circ,q_\bullet)$; |
| PROCEDURE | Build$(q_\circ,q_\bullet)$ is defined by<br>    if $\mathbf{rP}[q_\bullet] \leq \mathbf{rP}[q_\circ]$ and $q_\circ \notin \mathsf{L}[q_\bullet]$ then<br>        $\mathsf{L}[q_\bullet] := \mathsf{L}[q_\bullet] \cup \{q_\circ\}$;<br>        forall $(q, \alpha, q_\circ) \in \mathbf{E}$ do Build$(q,q_\bullet)$<br>    else if $\mathbf{rP}[q_\bullet] > \mathbf{rP}[q_\circ]$ then abort |

Figure 4.19: Natural loops.

EXAMPLE 4.29: The program graph of Figure 4.3 only has one natural loop consisting of the nodes $q_1$, $q_2$ and $q_3$ (and corresponding to the single back edge $(q_3, \texttt{x := x} - 1, q_1)$).

PROPOSITION 4.30: The algorithm of Figure 4.19 computes the natural loops for reducible program graphs, and aborts for non-reducible program graphs.

SKETCH OF PROOF: It is an invariant of the algorithm that $\mathsf{L}[q_\bullet]$ is a subset of $\{q \mid \mathbf{rP}[q_\bullet] \leq \mathbf{rP}[q]\}$. The algorithm considers all back edges $(q_\circ, \alpha, q_\bullet)$, which are the edges with $\mathbf{rP}[q_\bullet] \leq \mathbf{rP}[q_\circ]$, and attempts to build the natural loop induced by it. If dominator edges coincide with back edges this constructs the natural loops as desired.

If a back edge is encountered that is not a dominator edge, the algorithm aborts. This happens if we perform Build$(q_\circ,q_\bullet)$ and $\mathbf{rP}[q_\bullet] > \mathbf{rP}[q_\circ]$ as then the back edge to $q_\bullet$ initiating the topmost call of Build$(\cdot,q_\bullet)$ would not be a dominator edge. □

EXERCISE 4.31: Let us return to the program graph of Figure 4.11 and the reverse postorder constructed in Exercise 4.14. Use the algorithm of Figure 4.19 to show that there are two natural loops, namely $\mathsf{Q}_1 = \{q_1, q_2, q_3, q_4, q_5, q_6, q_7, q_8, q_9, q_{10}, q_{11}\}$ and $\mathsf{Q}_4 = \{q_4, q_6, q_{10}\}$. □

**Iterating through Natural Loops** There may be nodes in the program graph that do not occur in any natural loop (unlike what was the case for strong components above). We shall therefore define a *natural component* to be either a natural loop or a single node that is not a member of any natural loop. In this way all nodes are members of a natural component, and two natural components are either disjoint or one is contained in the other.

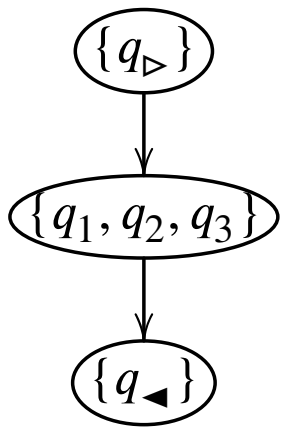

Figure 4.20: Graph of loops for the program graph of Figure 4.3.

We next construct a directed graph on the natural components. We have an edge from one natural component $Q'$ to another natural component $Q''$ whenever at least one of the following applies:

- $Q' \subset Q''$ (meaning $Q' \subseteq Q''$ and $Q' \neq Q''$), or
- $Q'$ and $Q''$ are disjoint and there is a node $q' \in Q'$ and a node $q'' \in Q''$ such that the program graph has an edge from $q'$ to $q''$.

The resulting directed graph will be called the *graph of loops*.

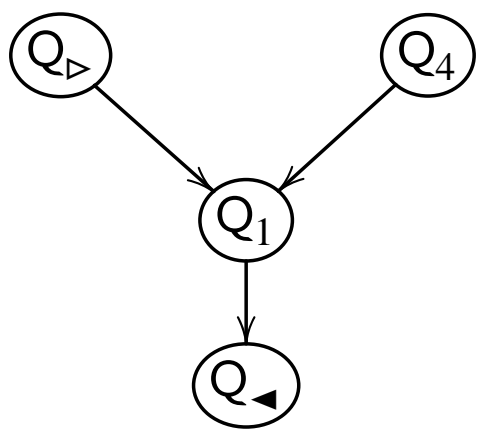

Figure 4.21: Graph of loops for the program graph of Figure 4.11.

Example 4.32: The graph of loops for the program graph of Figure 4.3 has three nodes as shown in Figure 4.20; only one of them is a natural loop.

Example 4.33: Continuing Example 4.31 let us write $\mathsf{Q}_{\triangleright} = \{q_{\triangleright}\}$ and $\mathsf{Q}_{\blacktriangleleft} = \{q_{\blacktriangleleft}\}$. Then the corresponding graph of loops is as shown in Figure 4.21.

The graph of loops is acyclic. To see this consider a shortest cycle $Q_1, \cdots, Q_k$ with $Q_1 = Q_k$ and let us arrive at a contradiction. Note that $k \geq 2$ by construction of the edges between natural components. It cannot be the case that all edges are of the first kind as $Q_1 \subset Q_k$ would contradict $Q_1 = Q_k$, so at least one edge must be of the second kind. It then cannot be the case that any edge is of the first kind as then an even shorter cycle could be constructed, so all natural components in the cycle must be disjoint. If there is an edge from a node $q_i \in Q_i$ to a node $q_{i+1} \in Q_{i+1}$ (for $1 \leq i < k$) then $Q_{i+1}$ cannot be a natural loop as otherwise $q_i$ would be included in the construction of the natural loop $Q_{i+1}$ as part of the backward traversal from $q_{i+1}$. Hence the cycle $Q_1, \cdots, Q_k$ consists of singleton sets of nodes not belonging to any natural loop. But at least one of these edges must be a back edge (by Proposition 4.16) and this violates our assumptions.

```
empty = (nil, { })

insert(q, (V, P)) =
  if q ∉ V
  then (V, {q} ∪ P)
  else (V, P)

extract(V, P) =
  if V = nil
  then (q, (V′, P′))
    where (S, P′) = nat_rP(P)
    and V_rP = sort_rP(S)
    and q = head(V_rP)
    and V′ = tail(V_rP)
  else (q, (V′, P))
    where q = head(V)
    and V′ = tail(V)
```

Figure 4.22: Worklist for iterating through natural loops.

Since we direct edges from inner natural loops to outer natural loops, as opposed to the other direction, it will not necessarily be the case that the graph of loops is topologically sorted by the reverse postorder of the headers of the loops. However, this direction of edges is necessary in order to give preference to iterating through inner loops before iterating through outer loops.

We shall iterate through the natural components in a top-down manner with respect to the graph of loops. In Figure 4.22 we present this idea using operations that operate on a worklist W with two components (V, P) as before. When the component V is exhausted, the set P of pending nodes is split into two (S, P′). The set $\mathsf{S} \subseteq \mathsf{P}$ contains all the nodes of P belonging to a natural component where none of its ancestors in the graph of loops contains any nodes from P; this is called a *topmost* natural component. (If there is more than one candidate one might choose the one that has a node with the smallest reverse postorder among all nodes in the candidates.) The set $P' = P \setminus S$ then contains the remaining nodes from P. We use the function $\mathsf{nat}_{\mathsf{rP}}$ to achieve this, and as above we write $\mathsf{sort}_{\mathsf{rP}}$ for the function that sorts a set into a list in reverse postorder.

EXAMPLE 4.34: We can now redo the computation of the analysis assignment for the Reaching Definitions analysis of the program graph of Figure 4.3 using the worklist represention of Figure 4.22. As the graph of loops in Figure 4.20 is equal to the reduced graph in Figure 4.17 the iteration steps are indeed similar.

HANDS ON 4.35: Based on the program graph of Figure 4.11 and the development of Example 4.33 construct the iteration sequences for the Reaching Definitions analysis using the worklist representation of Figure 4.22. Compare with the results obtained using the representation based on strong components as obtained in the http://www.formalmethods.dk/pa4fun/ tool. □

TEASER 4.36: Let us reconsider the definition of $\mathsf{nat}_{\mathsf{rP}}(\mathsf{P})$ used in Figure 4.22 and modify it to return a pair $(S, P')$ where now S contains all the nodes of a topmost natural component having members in P and, as before, $P' = P \setminus S$. Compare this alternative worklist representation with the one of Figure 4.22. □

## 4.4 Duality and Reversal

In the previous sections we have focussed on algorithms for forward analyses where we are interested in the least solution to the constraints. In this section we shall consider how to adapt this development to backward analyses and to analyses where the greatest solution of the constraints is desired. Rather than redoing the entire development we shall follow the approach of Section 3.4 and employ some simple transformations in order to generalise our approach.

**Reversal** Recall that in Section 3.4 we introduced the notion of a reversed program graph $\mathbf{PG}^R$ obtained from a program graph $\mathbf{PG} = (\mathbf{Q}, q_{\triangleright}, q_{\blacktriangleleft}, \mathbf{Act}, \mathbf{E})$ by simply reversing the edges; thus

$$\mathbf{PG}^R = (\mathbf{Q}, q_{\blacktriangleleft}, q_{\triangleright}, \mathbf{Act}, \mathbf{E}^R)$$

where

$$\mathbf{E}^R = \{(q_\bullet, \alpha, q_\circ) \mid (q_\circ, \alpha, q_\bullet) \in \mathbf{E}\}$$

The least solution to a backward analysis problem can then be computed using the algorithms of the previous sections using the reversed program graph. Working it out in detail, the algorithm of Figure 4.2 applied to the reversed program graph $\mathbf{PG}^R$ gives rise to the algorithm of Figure 4.23 applied to the original program graph $\mathbf{PG}$.

| | |
|---|---|
| INPUT | a program graph $\mathbf{PG} = (\mathbf{Q}, q_\triangleright, q_\blacktriangleleft, \mathbf{Act}, \mathbf{E})$<br>an analysis specification with $\widehat{\mathbf{D}}$, $\widehat{\mathcal{S}}[\![q_\bullet, \alpha, q_\circ]\!]$ and $\hat{d}_\diamond$ |
| OUTPUT | the least solution AA to the backward analysis problem |
| METHOD | W := empty ;<br>forall $q \in \mathbf{Q}$ do AA($q$) := $\bot$ ;<br>        W := insert($q$, W) ;<br>AA($q_\blacktriangleleft$) := $\hat{d}_\diamond$ ;<br>while W $\neq$ empty do<br>    ($q_\bullet$, W) := extract(W) ;<br>    forall $(q_\circ, \alpha, q_\bullet) \in \mathbf{E}$ do<br>        if $\widehat{\mathcal{S}}[\![q_\bullet, \alpha, q_\circ]\!](\mathsf{AA}(q_\bullet)) \not\sqsubseteq \mathsf{AA}(q_\circ)$<br>        then $\mathsf{AA}(q_\circ) := \mathsf{AA}(q_\circ) \sqcup \widehat{\mathcal{S}}[\![q_\bullet, \alpha, q_\circ]\!](\mathsf{AA}(q_\bullet))$ ;<br>            W := insert($q_\circ$, W) |

Figure 4.23: Adaptation of the Worklist Algorithm of Figure 4.2 to a backward analysis problem.

The algorithm of Figure 4.23 can directly be used with the worklist representations of Figures 4.5, 4.6 and 4.7.

Hands On 4.37: Use the http://www.formalmethods.dk/pa4fun/ tool to compute the Live Variables analysis result for the program graph of Figure 4.3. Compare the iteration sequences obtained when using the LIFO representation of Figure 4.6 and the FIFO worklist representation of Figure 4.7. □

In order to use the worklist representations exploiting reverse postorders we have to compute the reverse postorder numbering based on the reversed program graph, that is, we apply the algorithm of Figure 4.8 to $\mathbf{PG}^R$ rather than **PG**. We can then apply the algorithm of Figure 4.16 to the reversed program graph and its reverse postordering and obtain the associated list of strong components; similarly, we can apply the algorithm of Figure 4.19 to the reversed program graph and its reverse postordering and obtain the associated natural loops. With this modification the worklist representations of Figures 4.13, 4.15, 4.18 and 4.22 can be used together with the worklist algorithm of Figure 4.23.

Try It Out 4.38: Construct the reversed program graph for the program graph of Figure 4.3 and use the algorithm of Figure 4.8 to determine a reverse postordering for it. Show that $q_\blacktriangleleft, q_1, q_3, q_2, q_\triangleright$ is a possible reverse postorder of the nodes. □

Hands On 4.39: Once again use the http://www.formalmethods.dk/pa4fun/ tool to compute the Live Variables analysis result for the program graph of Figure 4.3. Compare the iteration sequences obtained using the worklist representations of the Figures 4.13, 4.15 and 4.18 and 4.22. □

| | |
|---|---|
| INPUT | a program graph $\mathbf{PG} = (\mathbf{Q}, q_\triangleright, q_\blacktriangleleft, \mathbf{Act}, \mathbf{E})$<br>an analysis specification with PowerSet($\mathbf{X}$), $\widehat{S}[\![q_\circ, \alpha, q_\bullet]\!]$ and $\hat{d}_\diamond$ |
| OUTPUT | the greatest solution AA to the forward analysis problem |
| METHOD | W := empty ;<br>forall $q \in \mathbf{Q}$ do AA($q$) := $\mathbf{X}$ ;<br>            W := insert($q$, W) ;<br>AA($q_\triangleright$) := $\hat{d}_\diamond$ ;<br>while W $\neq$ empty do<br>    ($q_\circ$, W) := extract(W) ;<br>    forall $(q_\circ, \alpha, q_\bullet) \in \mathbf{E}$ do<br>        if $\widehat{S}[\![q_\circ, \alpha, q_\bullet]\!]$(AA($q_\circ$)) $\not\supseteq$ AA($q_\bullet$)<br>        then AA($q_\bullet$) := AA($q_\bullet$) $\cap$ $\widehat{S}[\![q_\circ, \alpha, q_\bullet]\!]$(AA($q_\circ$)) ;<br>            W := insert($q_\bullet$, W) |

Figure 4.24: Adaptation of the Worklist Algorithm of Figure 4.2 to a forward analysis problem computing the greatest solution over a powerset.

**Duality** Most of the analyses we have considered require that we compute the *least* solution to the constraint system; however, we have also seen examples of analyses where we are interested in the *greatest* solution, most noticeably Available Expressions analysis and Very Busy Expressions analysis. So far, our treatment in this chapter might seem only to deal with least solutions. Rather than redoing the entire development for greatest solutions we shall again employ a simple transformation.

In Section 3.4 we observed that an analysis domain (PowerSet($X$), $\sqsubseteq$) may have $\sqsubseteq$ to be either $\subseteq$ or $\supseteq$ but that this affects the definition of least element and least upper bound:

| analysis domain | ordering $\sqsubseteq$ | least element $\bot$ | least upper bound $\sqcup$ |
|---|---|---|---|
| (PowerSet($\mathbf{X}$), $\subseteq$) | $\subseteq$ | { } | $\cup$ |
| (PowerSet($\mathbf{X}$), $\supseteq$) | $\supseteq$ | $\mathbf{X}$ | $\cap$ |

With this in mind we can adapt the worklist algorithm of Figure 4.2 to forward analysis problems requiring the computation of greatest solutions over powersets; it is shown in Figure 4.24. The algorithm can be combined with any of the worklist representations considered in the previous sections.

EXERCISE 4.40: Perform a similar adaption of the worklist algorithm of Figure 4.2 to solve backward analysis problems requiring the greatest solutions. □

To informally summarise this chapter, our advanced worklist algorithms apply to all the analyses considered so far, by focusing on a partial order $\sqsubseteq$ (that can be chosen to be $\subseteq$ or $\supseteq$), and by focusing on directed edges (that can be chosen to be those of the program graph or the reversal of those of the program graph). In all cases our focus is on the least solution to the constrains expressed using $\sqsubseteq$, which turns out to be the largest solutions in case $\sqsubseteq = \sqsupseteq$. With respect to the use of reverse postorder for backward analyses it is important that the reverse postorder is only computed *after* the edges have been reversed.

# Chapter 5

# Analysis of Integers

In this chapter we present three analyses of integers. They all make use of analysis domains that record information about the potential values of the variables and array entries. The analyses differ in the kind of information they record. The analysis of Section 5.1 is a *Detection of Signs* analysis. It records the signs (that is, negative, zero or positive) of the potential values. The analysis of Section 5.2 is a *Constant Propagation* analysis. It aims to determine whether or not a variable always will evaluate to a constant value and if so, it will record that constant value. The final analysis presented in Section 5.3 is an *Interval analysis*. This analysis goes one step further than the previous ones and aims to determine intervals bounding the values, that is, it determines a lower bound and an upper bound on the value.

## 5.1 Detection of Signs Analysis

The aim of the *Detection of Signs* analysis is to determine, for each node of the program graph, what the possible signs of the variables and arrays could be. We shall write

$$\textbf{Sign} = \{-, 0, +\}$$

for the three signs and the analysis computes with *sets* of signs, that is, elements of PowerSet(**Sign**). For a program graph as the one of Figure 5.1 we can then express that z can have any sign at the node $q_2$ by using the set $\{-, 0, +\}$ and we can express that z is positive at the node $q_5$ by using the singleton set $\{+\}$.

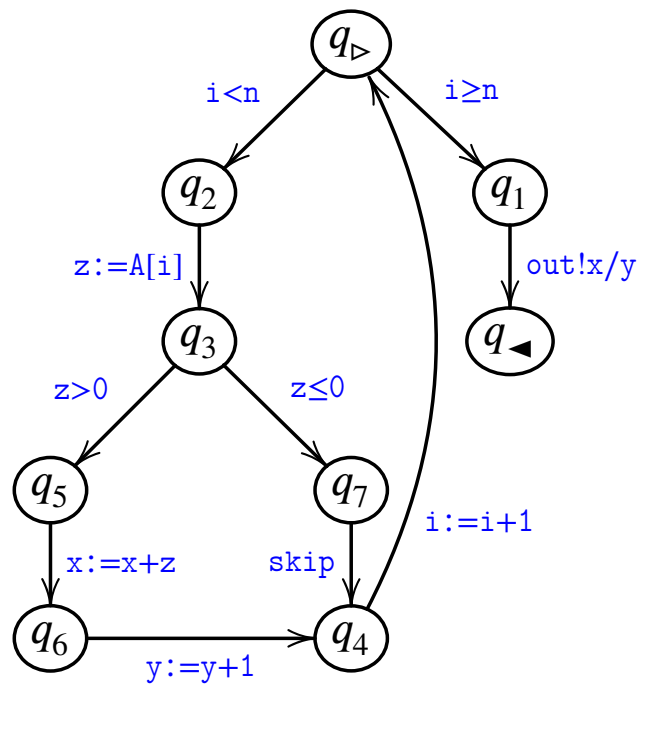

Figure 5.1: Computing the average of some of the elements of an array.

The information provided by the Detection of Signs analysis can be used for checking whether a division by zero may happen. For arithmetic expressions of the form $a_1/a_2$ (division) and $a_1 \% a_2$ (modulo) we can inspect whether or not the potential signs of $a_2$ include $0$; if it does *not* then we can guarantee that a division by zero never will happen. The analysis can also be used for checking whether we index outside the lower bound of arrays. For occurrences of $A[a]$ we can inspect whether or not the potential signs of $a$ include $-$; if it does *not* then we can guarantee that we never index outside the lower bound of the array – obviously we cannot provide any guarantees for the upper bound.

**Analysis assignments.** The analysis assignments of the analysis are going to be mappings

$$\mathsf{DS} : \mathbf{Q} \to \widehat{\mathbf{Mem}}_{\mathsf{DS}}$$

where $\widehat{\mathbf{Mem}}_{\mathsf{DS}}$ is the analysis domain of abstract memories. An abstract memory will then for each variable and array tell us what its signs might be; we shall write this as

$$\widehat{\mathbf{Mem}}_{\mathsf{DS}} = (\mathbf{Var} \to \mathsf{PowerSet}(\mathbf{Sign})) \times (\mathbf{Arr} \to \mathsf{PowerSet}(\mathbf{Sign}))$$

| | |
|---|---|
| x | $\{0\}$ |
| y | $\{0\}$ |
| z | $\{-,0,+\}$ |
| i | $\{0\}$ |
| n | $\{+\}$ |
| A | $\{0,+\}$ |

Figure 5.2: Example abstract memory for five variables and a single array.

Thus an abstract memory has the form $\hat{\sigma} = (\hat{\sigma}_{\mathsf{V}}, \hat{\sigma}_{\mathsf{A}})$, and it will determine a *set* of signs $\hat{\sigma}_{\mathsf{V}}(x)$ for each variable $x$, and it will determine a *set* of signs $\hat{\sigma}_{\mathsf{A}}(A)$ for *all* the entries for each array $A$. Figure 5.2 gives an example of an abstract memory where the signs of x, y and i are zero, z can have any sign whereas the signs of n is positive as recorded by the respective sets of signs. For the array A the abstract memory records that its entries are non-negative, that is, some of the entries may be 0, some may be positive but the array will have no negative entries.

**Computing with signs** Let us start by introducing a function $\mathsf{sign} : \mathbf{Int} \to \mathbf{Sign}$ that returns the sign of an integer:

$$\mathsf{sign}(n) = \begin{cases} - & \text{if } n < 0 \\ 0 & \text{if } n = 0 \\ + & \text{if } n > 0 \end{cases}$$

The analysis will compute with signs meaning that all the arithmetic operators will be given signs as arguments rather than integers and they will return signs as result. We will have abstract versions $\tilde{op}_a$ of the arithmetic operators $op_a$ that are given two signs as argument and returns a set of signs for the result – the abstract addition operator $\tilde{+}$ is specified in the leftmost table of Figure 5.3.

The abstract operators return a *set* of signs as a single sign will not suffice. It is essential that if $v = v_1 \ op_a \ v_2$ then $\mathsf{sign}(v)$ is recorded as a possible result when applying $\tilde{op}_a$ to $\mathsf{sign}(v_1)$ and $\mathsf{sign}(v_2)$; formally, $\mathsf{sign}(v) \in \mathsf{sign}(v_1) \ \tilde{op}_a \ \mathsf{sign}(v_2)$.

TRY IT OUT 5.1: Specify abstract versions of the arithmetic operators for subtraction ($-$) and multiplication ($*$) and argue that they correctly capture the possible signs of the result. □

| $s_1 \,\tilde{+}\, s_2$ | $-$ | $0$ | $+$ |
|---|---|---|---|
| $-$ | $\{-\}$ | $\{-\}$ | $\{-,0,+\}$ |
| $0$ | $\{-\}$ | $\{0\}$ | $\{+\}$ |
| $+$ | $\{-,0,+\}$ | $\{+\}$ | $\{+\}$ |

| $s_1 \,\tilde{<}\, s_2$ | $-$ | $0$ | $+$ |
|---|---|---|---|
| $-$ | {tt, ff} | {tt} | {tt} |
| $0$ | {ff} | {ff} | {tt} |
| $+$ | {ff} | {ff} | {tt, ff} |

Figure 5.3: The abstract operations $\tilde{+}$ and $\tilde{<}$ for Detection of Signs Analysis.

EXERCISE 5.2: Continuing Try It Out 5.1 specify abstract versions of the division (/) and modulo (%) operations and argue that they correctly capture the possible signs of the results. Discuss how to handle division and modulo in case the second argument is 0. □

Similarly we have abstract versions $\tilde{op}_r$ of the relational operators $op_r$. These operators will get two signs as arguments and they will return a *set* of boolean values; again a single result will not suffice. The rightmost table of Figure 5.3 defines the abstract comparison operator $\tilde{<}$ corresponding to the less-than operator <.

TRY IT OUT 5.3: Specify abstract versions of the relational operators for equality (=) and less-than-or-equal (≤). □

Later we are going to use these operations in a context where they are given *sets* of signs as arguments so it is convenient to lift the operations to such sets. We therefore define

$$S_1 \;\widehat{op}\; S_2 = \bigcup_{s_1 \in S_1, s_2 \in S_2} s_1 \;\tilde{op}\; s_2$$

corresponding to the pointwise application of the operator. As an example we have $\{0,+\}\widehat{+}\{0,+\} = \{0,+\}$ reflecting that the addition of two non-negative numbers is a non-negative number. Note that $\{\ \}\;\widehat{op}\;S = S\;\widehat{op}\;\{\ \} = \{\ \}$ meaning that if one of the arguments of $\widehat{op}$ does not have any signs then the result has no sign either.

**The analysis domain** Recall that we defined the abstract memories to be pairs of mappings

$$\widehat{\mathbf{Mem}}_{\mathsf{DS}} = (\mathbf{Var} \to \mathsf{PowerSet}(\mathbf{Sign})) \times (\mathbf{Arr} \to \mathsf{PowerSet}(\mathbf{Sign}))$$

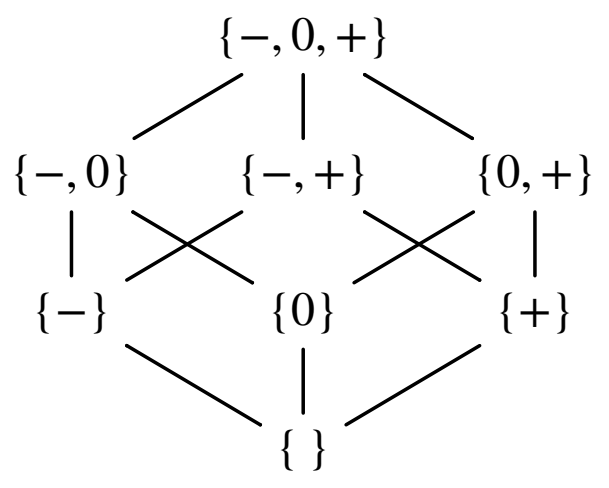

Figure 5.4: The analysis domain PowerSet(**Sign**) with the ordering ⊆.

We shall now provide the additional structure turning $\widehat{\mathbf{Mem}}_{\mathsf{DS}}$ into an analysis domain. Thus we shall provide it with a partial ordering $\sqsubseteq_{\mathsf{DS}}$, a least element $\bot_{\mathsf{DS}}$ and a join operator $\sqcup_{\mathsf{DS}}$ and we shall check that it satisfies the ascending chain condition.

The starting point is that PowerSet(**Sign**) with the subset ordering ⊆ already is an analysis domain with the empty set { } being its least element and set union ∪ being the join operation; Figure 5.4 illustrates the ordering.

In the manner of Essential Exercise 3.20, the ordering $\sqsubseteq_{\mathsf{DS}}$ on $\widehat{\mathbf{Mem}}_{\mathsf{DS}}$ is the pointwise lifting of the subset ordering $\subseteq$ on **Sign**; this amounts to

$$\hat{\sigma}_1 \sqsubseteq_{\mathsf{DS}} \hat{\sigma}_2 \quad \text{if and only if} \quad \begin{array}{l} \hat{\sigma}_{1\mathsf{V}}(x) \subseteq \hat{\sigma}_{2\mathsf{V}}(x) \text{ for all } x \in \mathbf{Var} \text{ and} \\ \hat{\sigma}_{1\mathsf{A}}(A) \subseteq \hat{\sigma}_{2\mathsf{A}}(A) \text{ for all } A \in \mathbf{Arr} \end{array}$$

We shall define $\bot_{\mathsf{DS}}$ to be the abstract memory that maps all variables $x \in \mathbf{Var}$ to the empty set of signs $\{\,\}$ and similarly all arrays $A \in \mathbf{Arr}$ are mapped to $\{\,\}$. The join operation will be the pointwise extension of the union of signs; formally $\hat{\sigma}_1 \sqcup_{\mathsf{DS}} \hat{\sigma}_2$ is the abstract memory $\hat{\sigma}$ given by $\hat{\sigma}_{\mathsf{V}}(x) = \hat{\sigma}_{1\mathsf{V}}(x) \cup \hat{\sigma}_{2\mathsf{V}}(x)$ for all variables $x$ and similarly $\hat{\sigma}(A) = \hat{\sigma}_{1\mathsf{A}}(A) \cup \hat{\sigma}_{2\mathsf{A}}(A)$ for all arrays $A$.

TRY IT OUT 5.4: Check that the above definitions of $\sqsubseteq_{\mathsf{DS}}$, $\bot_{\mathsf{DS}}$ and $\sqcup_{\mathsf{DS}}$ satisfy the requirements put forward in Section 3.2 for $\widehat{\mathbf{Mem}}_{\mathsf{DS}}$ being a pointed semi-lattice.□

We also have to argue that the ascending chain condition is satisfied for $\widehat{\mathbf{Mem}}_{\mathsf{DS}}$. Here we proceed by contradition. So let us assume that there is an infinite sequence $\hat{\sigma}_0 \sqsubset_{\mathsf{DS}} \hat{\sigma}_1 \sqsubset_{\mathsf{DS}} \cdots \sqsubset_{\mathsf{DS}} \hat{\sigma}_n \sqsubset_{\mathsf{DS}} \cdots$ of abstract memories from $\widehat{\mathbf{Mem}}_{\mathsf{DS}}$. We now observe that whenever $\hat{\sigma}_j \sqsubset_{\mathsf{DS}} \hat{\sigma}_{j+1}$ there must be at least one of the variables and arrays that will be mapped to a strictly larger set of signs by $\hat{\sigma}_{j+1}$ than by $\hat{\sigma}_j$ – since otherwise the two abstract memories would be equal. Since we have a finite set of variables, a finite set of arrays and a finite set of signs clearly this cannot happen an infinite number of times, so we have the desired contradiction.

**Analysing arithmetic and boolean expressions** Before defining the analysis functions for the actions we shall specify how to analyse the arithmetic and boolean expressions. For this we have functions

$$\begin{array}{rcl} \hat{\mathcal{A}}_{\mathsf{DS}}[\![a]\!] & : & \widehat{\mathbf{Mem}}_{\mathsf{DS}} \to \mathsf{PowerSet}(\mathbf{Sign}) \\ \hat{\mathcal{B}}_{\mathsf{DS}}[\![b]\!] & : & \widehat{\mathbf{Mem}}_{\mathsf{DS}} \to \mathsf{PowerSet}(\mathbf{Bool}) \end{array}$$

For an arithmetic expression we obtain a set of signs that might be obtained in the given abstract memory, and for a boolean expression we obtain a set of boolean values that might be obtained in the given abstract memory.

For arithmetic expressions we define:

$$\begin{array}{rcl} \hat{\mathcal{A}}_{\mathsf{DS}}[\![n]\!]\hat{\sigma} & = & \{\mathsf{sign}(n)\} \\ \hat{\mathcal{A}}_{\mathsf{DS}}[\![x]\!]\hat{\sigma} & = & \hat{\sigma}_{\mathsf{V}}(x) \\ \hat{\mathcal{A}}_{\mathsf{DS}}[\![A[a]]\!]\hat{\sigma} & = & \begin{cases} \hat{\sigma}_{\mathsf{A}}(A) & \text{if } \hat{\mathcal{A}}_{\mathsf{DS}}[\![a]\!]\hat{\sigma} \cap \{0,+\} \neq \{\,\} \\ \{\,\} & \text{otherwise} \end{cases} \\ \hat{\mathcal{A}}_{\mathsf{DS}}[\![a_1\ op_a\ a_2]\!]\hat{\sigma} & = & \hat{\mathcal{A}}_{\mathsf{DS}}[\![a_1]\!]\hat{\sigma}\ \widehat{op_a}\ \hat{\mathcal{A}}_{\mathsf{DS}}[\![a_2]\!]\hat{\sigma} \\ \hat{\mathcal{A}}_{\mathsf{DS}}[\![-a]\!]\hat{\sigma} & = & \hat{-}\ \hat{\mathcal{A}}_{\mathsf{DS}}[\![a]\!]\hat{\sigma} \end{array}$$

Note that for indexing into an array we only give a non-trivial result in case the index might be non-negative. If the index is known to be negative then we are sure that an error will occur when executing the program graph, and we use the empty set $\{\ \}$ to record that no value will be available in the continuation. As an example if $\widehat{\sigma}_{\mathsf{V}}(i) = \{-, 0\}$ then $\widehat{\mathcal{A}}_{\mathsf{DS}}[\![A[i]]\!]\widehat{\sigma}$ equals $\widehat{\sigma}_{\mathsf{A}}(A)$ as the lookup might succeed whereas if $\widehat{\sigma}_{\mathsf{V}}(i) = \{-\}$ then it equals $\{\ \}$ as we know that the lookup will fail (and we can issue a warning to the programmer).

In the analysis the binary and unary operators $op_a$ are replaced by their abstract versions defined above; however, note that we are using $\widehat{op}_a$ (rather than $\tilde{op}_a$) as we are operating on *sets* of signs (rather than single signs). In particular this means that $\widehat{\mathcal{A}}_{\mathsf{DS}}[\![A[i] + 1]\!]\widehat{\sigma}$ will evaluate to $\{\ \}$ in the case where $\widehat{\sigma}_{\mathsf{V}}(i) = \{-\}$ so the failure observed for $A[i]$ will be propagated to the composite expression.

For the boolean expressions we define:

$$
\begin{array}{rcl}
\widehat{\mathcal{B}}_{\mathsf{DS}}[\![\mathtt{true}]\!]\widehat{\sigma} & = & \{\mathsf{tt}\} \\
\widehat{\mathcal{B}}_{\mathsf{DS}}[\![\mathtt{false}]\!]\widehat{\sigma} & = & \{\mathsf{ff}\} \\
\widehat{\mathcal{B}}_{\mathsf{DS}}[\![a_1\ op_r\ a_2]\!]\widehat{\sigma} & = & \widehat{\mathcal{A}}_{\mathsf{DS}}[\![a_1]\!]\widehat{\sigma}\ \widehat{op}_r\ \widehat{\mathcal{A}}_{\mathsf{DS}}[\![a_2]\!]\widehat{\sigma} \\
\widehat{\mathcal{B}}_{\mathsf{DS}}[\![b_1\ op_b\ b_2]\!]\widehat{\sigma} & = & \widehat{\mathcal{B}}_{\mathsf{DS}}[\![b_1]\!]\widehat{\sigma}\ \widehat{op}_b\ \widehat{\mathcal{B}}_{\mathsf{DS}}[\![b_2]\!]\widehat{\sigma} \\
\widehat{\mathcal{B}}_{\mathsf{DS}}[\![\neg b]\!]\widehat{\sigma} & = & \widehat{\neg}\ \widehat{\mathcal{B}}_{\mathsf{DS}}[\![b]\!]\widehat{\sigma}
\end{array}
$$

Also here we are using the abstract versions $\widehat{op}_r$ of the operators $op_r$ as we are operating on *sets* of signs. The operations $\widehat{\wedge}$, $\widehat{\vee}$ and $\widehat{\neg}$ are the pointwise extension of the operators $\wedge$, $\vee$ and $\neg$ to sets of truth values. As an example, if $\widehat{\sigma}_{\mathsf{V}}(x) = \{0, +\}$ then $\widehat{\mathcal{B}}_{\mathsf{DS}}[\![0 < x]\!]\widehat{\sigma}$ equals $\{\mathsf{tt}, \mathsf{ff}\}$ as both results are possible; however if $\widehat{\sigma}_{\mathsf{V}}(x) = \{-\}$ then it equals $\{\mathsf{ff}\}$ as we know for sure that the test will fail.

Given a memory $\sigma = (\sigma_{\mathsf{V}}, \sigma_{\mathsf{A}}, \sigma_{\mathsf{C}})$ of the semantics we can define the corresponding abstract memory in $\widehat{\mathbf{Mem}}_{\mathsf{DS}}$. It is given by a function $\beta_{\mathsf{DS}} : \mathbf{Mem} \to \widehat{\mathbf{Mem}}$ that will be defined as an extension of the function sign on integers (as illustrated on Figure 5.5). It is defined as follows:

$$
\begin{array}{l}
\beta_{\mathsf{DS}}(\sigma) = (\widehat{\sigma}_{\mathsf{V}}, \widehat{\sigma}_{\mathsf{A}}) \\
\qquad \text{where} \quad \widehat{\sigma}_{\mathsf{V}}(x) \;=\; \{\mathsf{sign}(v)\} \\
\qquad\qquad\qquad\qquad\qquad \text{where } \sigma_{\mathsf{V}}(x) = v \text{ for } x \in \mathbf{Var} \\
\qquad\qquad\quad\; \widehat{\sigma}_{\mathsf{A}}(A) \;=\; \{\mathsf{sign}(v_0), \mathsf{sign}(v_1), \cdots, \mathsf{sign}(v_{k-1})\} \\
\qquad\qquad\qquad\qquad\qquad \text{where } \sigma_{\mathsf{A}}(A) = [v_0, v_1, \cdots, v_{k-1}] \text{ for } A \in \mathbf{Arr}
\end{array}
$$

Thus for a variable we simply give the singleton set with its sign and for an array we collect the set of possible signs of its entries. Figure 5.6 gives an example of a memory and its abstract counterpart.

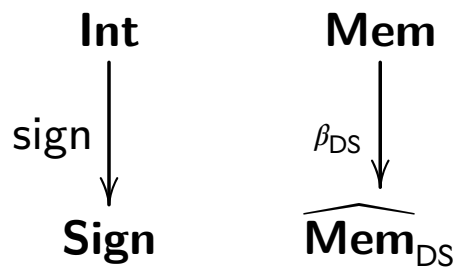

Figure 5.5: Mapping integers to signs and memories to abstract memories.

$\sigma = (\sigma_V, \sigma_A, \sigma_C)$ :

| | |
|---|---|
| x | 0 |
| y | 1 |
| z | −1 |
| i | 3 |
| n | 5 |
| A | $[0, 1, 3, 0, 7]$ |
| out | [ ] |

$\beta_{DS}(\sigma) = (\hat{\sigma}_V, \hat{\sigma}_A)$ :

| | |
|---|---|
| x | $\{0\}$ |
| y | $\{+\}$ |
| z | $\{-\}$ |
| i | $\{+\}$ |
| n | $\{+\}$ |
| A | $\{0, +\}$ |

Figure 5.6: A memory $\sigma$ and its abstract version $\beta_{DS}(\sigma)$.

EXERCISE 5.5: Show that for all arithmetic expressions $a$ we have the following monotonicity result: For all abstract memories $\hat{\sigma}_1$ and $\hat{\sigma}_2$:

$$\hat{\sigma}_1 \sqsubseteq_{DS} \hat{\sigma}_2 \quad \text{implies} \quad \widehat{\mathcal{A}}_{DS}[\![a]\!]\hat{\sigma}_1 \subseteq \widehat{\mathcal{A}}_{DS}[\![a]\!]\hat{\sigma}_2$$

Formulate and show the similar result for boolean expressions. □

ESSENTIAL EXERCISE 5.6: Show that for all arithmetic expressions $a$, for all memories $\sigma$ and values $v$

$$\text{if } \mathcal{A}[\![a]\!]\sigma = v \text{ then } \mathsf{sign}(v) \in \widehat{\mathcal{A}}_{DS}[\![a]\!](\beta_{DS}(\sigma))$$

meaning that the analysis of arithmetic expressions correctly captures their signs. Formulate and show the similar result for boolean expressions.

**The analysis functions** Using these functions we are ready to define the analysis of the actions. Following the approach of Chapter 3 (and Definition 3.30) the functions are of the form $\widehat{\mathcal{S}}_{DS}[\![q_\circ, \alpha, q_\bullet]\!] : \widehat{\mathbf{Mem}}_{DS} \to \widehat{\mathbf{Mem}}_{DS}$; however, our analysis does not make use of the nodes $q_\circ$ and $q_\bullet$ so we shall omit them and simply take

$$\widehat{\mathcal{S}}_{DS}[\![\alpha]\!] : \widehat{\mathbf{Mem}}_{DS} \to \widehat{\mathbf{Mem}}_{DS}$$

In the following the idea is that whenever the analysis is able to determine that the execution of an action is stuck it will return the abstract memory $\bot_{DS}$. And whenever an analysis function is given the argument $\bot_{DS}$ then it will propagate this information about failure and return $\bot_{DS}$.

Recalling that $\hat{\sigma}$ has the form $(\hat{\sigma}_V, \hat{\sigma}_A)$ let us start with the two assignments:

$$\widehat{\mathcal{S}}_{DS}[\![x := a]\!]\hat{\sigma} = \begin{cases} (\hat{\sigma}_V[x \mapsto \widehat{\mathcal{A}}_{DS}[\![a]\!]\hat{\sigma}], \hat{\sigma}_A) \\ \quad \text{if } \widehat{\mathcal{A}}_{DS}[\![a]\!]\hat{\sigma} \neq \{\,\} \text{ and } \hat{\sigma} \neq \bot_{DS} \\ \bot_{DS} \quad \text{otherwise} \end{cases}$$

$$\widehat{\mathcal{S}}_{DS}[\![A[a_1] := a_2]\!]\hat{\sigma} = \begin{cases} (\hat{\sigma}_V, \hat{\sigma}_A[A \mapsto \hat{\sigma}_A(A) \cup \widehat{\mathcal{A}}_{DS}[\![a_2]\!]\hat{\sigma}]) \\ \quad \text{if } \widehat{\mathcal{A}}_{DS}[\![a_1]\!]\hat{\sigma} \cap \{0, +\} \neq \{\,\} \\ \quad \text{and } \widehat{\mathcal{A}}_{DS}[\![a_2]\!]\hat{\sigma} \neq \{\,\} \text{ and } \hat{\sigma} \neq \bot_{DS} \\ \bot_{DS} \quad \text{otherwise} \end{cases}$$

In both cases we ensure that if we cannot determine signs of one of the arithmetic expressions then we return $\bot_{DS}$ as this reflects that the evaluation of the expression failed and hence the execution of the action is stuck. This will for example happen for an assignment as $x := A[-1]$. In the case of updating an array entry we additionally check whether the index could be a valid one; if not, we have yet another reason why the action is stuck and we return $\bot_{\mathsf{DS}}$ as result. An example is an assignment as $A[-1] := x$. When everything looks right we update the information for an array by simply adding the potential new signs; we will not be able to remove any of the previous signs as we do not know which entry has been updated. As an example consider an assignment $A[i] := 0$ and assume that $\widehat{\sigma}_{\mathsf{V}}(i) = \{+\}$ and $\widehat{\sigma}_{\mathsf{A}}(A) = \{+\}$ before the assignment meaning that all entries are positive; after the assignment we can only be sure that all the entries of $A$ are non-negative.

Continuing with the actions involving channels we take

$$\widehat{\mathcal{S}}_{\mathsf{DS}}[\![c?x]\!]\widehat{\sigma} = \begin{cases} (\widehat{\sigma}_{\mathsf{V}}[x \mapsto \{-,0,+\}], \widehat{\sigma}_{\mathsf{A}}) & \text{if } \widehat{\sigma} \neq \bot_{\mathsf{DS}} \\ \bot_{\mathsf{DS}} & \text{otherwise} \end{cases}$$

$$\widehat{\mathcal{S}}_{\mathsf{DS}}[\![c?A[a]]\!]\widehat{\sigma} = \begin{cases} (\widehat{\sigma}_{\mathsf{V}}, \widehat{\sigma}_{\mathsf{A}}[A \mapsto \{-,0,+\}]) & \text{if } \widehat{\mathcal{A}}_{\mathsf{DS}}[\![a]\!]\widehat{\sigma} \cap \{0,+\} \neq \{\,\} \\ & \text{and } \widehat{\sigma} \neq \bot_{\mathsf{DS}} \\ \bot_{\mathsf{DS}} & \text{otherwise} \end{cases}$$

$$\widehat{\mathcal{S}}_{\mathsf{DS}}[\![c!a]\!]\widehat{\sigma} = \begin{cases} \widehat{\sigma} & \text{if } \widehat{\mathcal{A}}_{\mathsf{DS}}[\![a]\!]\widehat{\sigma} \neq \{\,\} \\ \bot_{\mathsf{DS}} & \text{otherwise} \end{cases}$$

For the two input actions we simply record that the input could have any sign – the abstract memory does not contain any information that would help us to do better. As for the assignments we ensure that if the action fails then the analysis function returns $\bot_{\mathsf{DS}}$. For the output action we simply check whether we have proper signs for the arithmetic expressions as otherwise the action will be stuck.

Finally we have the analysis functions for the test and `skip` actions. In the case of a test $b$ we want to determine whether of not it could evaluate to true and only then the target node of the action will be reachable. So a first (and correct) attempt could be

$$\widehat{\mathcal{S}}_{\mathsf{DS}}[\![b]\!]\widehat{\sigma} = \begin{cases} \widehat{\sigma} & \text{if } \mathsf{tt} \in \widehat{\mathcal{B}}_{\mathsf{DS}}[\![b]\!]\widehat{\sigma} \text{ and } \widehat{\sigma} \neq \bot_{\mathsf{DS}} \\ \bot_{\mathsf{DS}} & \text{otherwise} \end{cases}$$

This definition simply reflects that the memory is not modified by passing a test. However, it does not take into account that we might learn something about the values of the variables by passing a test. As an example, if the test is $z > 0$ and we have $\widehat{\sigma}_{\mathsf{V}}(z) = \{0,+\}$ then clearly the test could evaluate to true but once we have passed the test we know that $z$ cannot be 0 so we could improve the abstract memory to have $\widehat{\sigma}_{\mathsf{V}}(z) = \{+\}$. To capture this we shall split $\widehat{\sigma}$ into so-called basic abstract memories and then perform the test on each of those in order to determine which parts of the abstract memory are relevant after the test.

A *basic abstract memory* $\widehat{\sigma}'$ is an abstract memory where for all variables $x$ the set $\widehat{\sigma}'_{\mathsf{V}}(x)$ is a singleton set, that is, it is either $\{-\}$, $\{0\}$ or $\{+\}$. We then define

$$\mathsf{Basic}(\widehat{\sigma}) = \{\widehat{\sigma}' \mid \widehat{\sigma}' \sqsubseteq \widehat{\sigma} \wedge \forall x \in \mathbf{Var} : |\widehat{\sigma}'_{\mathsf{V}}(x)| = 1 \wedge \forall A \in \mathbf{Arr} : \widehat{\sigma}'_{\mathsf{A}}(A) = \widehat{\sigma}_{\mathsf{A}}(A)\}$$

as the set of basic abstract memories that are smaller than $\hat{\sigma}$; here we write $|\cdot|$ for the cardinality of the set given as argument. Note that we do not consider subsets of the information for arrays as this information covers all the entries in the arrays.

Try It Out 5.7: Consider the abstract memory of Figure 5.2; determine the corresponding set of basic abstract memories. How many are there? □

For each of the basic abstract memories in $\mathsf{Basic}(\hat{\sigma})$ we can now determine whether or not the test may evaluate to true – and only if so they will contribute to the abstract memory being the result of the action. This is reflected in the following definition:

$$\begin{array}{rcl}\widehat{\mathcal{S}}_{\mathsf{DS}}[\![b]\!]\hat{\sigma} & = & \bigsqcup\{\hat{\sigma}' \in \mathsf{Basic}(\hat{\sigma}) \mid \mathsf{tt} \in \widehat{\mathcal{B}}_{\mathsf{DS}}[\![b]\!]\hat{\sigma}'\} \\ \widehat{\mathcal{S}}_{\mathsf{DS}}[\![\mathtt{skip}]\!]\hat{\sigma} & = & \hat{\sigma}\end{array}$$

Here $\bigsqcup\{\hat{\sigma}_1, \cdots, \hat{\sigma}_n\} = \bot_{\mathsf{DS}} \sqcup_{\mathsf{DS}} \hat{\sigma}_1 \sqcup_{\mathsf{DS}} \cdots \sqcup_{\mathsf{DS}} \hat{\sigma}_n$ is the generalisation of $\sqcup_{\mathsf{DS}}$ to work on finite sets.

Try It Out 5.8: Determine the result of $\widehat{\mathcal{S}}_{\mathsf{DS}}[\![z > 0]\!]\hat{\sigma}$ in the case where $\hat{\sigma}_{\mathsf{V}}(z) = \{0, +\}$. What is the result of the function in the case where $\hat{\sigma}_{\mathsf{V}}(z) = \{-\}$? □

Exercise 5.9: Show that the analysis functions for the actions defined above are monotone; that is show that for all actions $\alpha$ and abstract memories $\hat{\sigma}_1$ and $\hat{\sigma}_2$ we have

$$\hat{\sigma}_1 \sqsubseteq_{\mathsf{DS}} \hat{\sigma}_2 \quad \text{implies} \quad \widehat{\mathcal{S}}_{\mathsf{DS}}[\![\alpha]\!]\hat{\sigma}_1 \sqsubseteq_{\mathsf{DS}} \widehat{\mathcal{S}}_{\mathsf{DS}}[\![\alpha]\!]\hat{\sigma}_2$$

In the proof you may exploit the results of Exercise 5.5. □

Essential Exercise 5.10: It is important that the analysis functions capture what happens when executing the actions in the semantics. Using the result of Exercise 5.6 show that for all actions $\alpha$, for all memories $\sigma$ and $\sigma'$ it is the case that

$$\text{if } \mathcal{S}[\![\alpha]\!]\sigma = \sigma' \qquad \text{then} \qquad \beta_{\mathsf{DS}}(\sigma') \sqsubseteq_{\mathsf{DS}} \widehat{\mathcal{S}}_{\mathsf{DS}}[\![\alpha]\!](\beta_{\mathsf{DS}}(\sigma))$$

**Analysis of program graphs** Finally we are ready to specify the constraints for the analysis assignment $\mathsf{DS} : \mathbf{Q} \to \widehat{\mathbf{Mem}}_{\mathsf{DS}}$.

For each edge $(q_\circ, \alpha, q_\bullet)$ in the program graph we impose the constraint:

$$\widehat{\mathcal{S}}_{\mathsf{DS}}[\![\alpha]\!](\mathsf{DS}(q_\circ)) \sqsubseteq_{\mathsf{DS}} \mathsf{DS}(q_\bullet)$$

Furthermore, we impose an additional constraint for the entry node:

$$\hat{\sigma}_{\triangleright} \sqsubseteq_{\mathsf{DS}} \mathsf{DS}(q_{\triangleright})$$

Here the abstract memory $\hat{\sigma}_{\triangleright}$ describes the initial signs of the variables and arrays. A natural possibility is to take $\hat{\sigma}_{\triangleright}$ to be $\top_{\mathsf{DS}}$ which is the abstract memory returning the set $\{-, 0, +\}$ for all variables and arrays. (We would generally expect that $\bot_{\mathsf{DS}} \neq \top_{\mathsf{DS}}$ but this fails in the very special case where we have no variables and no arrays.)

EXAMPLE 5.11: Returning to the program graph of Figure 5.1 we obtain one constraint for each of the 10 edges and an additional constraint for the initial node. For the edge $(q_5, \mathtt{x} := \mathtt{x} + \mathtt{z}, q_6)$ we get the constraint

$$\widehat{\mathcal{S}}_{\mathsf{DS}}[\![\mathtt{x} := \mathtt{x} + \mathtt{z}]\!](\mathsf{DS}(q_5)) \sqsubseteq_{\mathsf{DS}} \mathsf{DS}(q_6)$$

where $\widehat{\mathcal{S}}_{\mathsf{DS}}[\![\mathtt{x} := \mathtt{x} + \mathtt{z}]\!](\hat{\sigma}_{\mathsf{V}}, \hat{\sigma}_{\mathsf{A}}) = (\hat{\sigma}_{\mathsf{V}}[\mathtt{x} \mapsto \hat{\sigma}_{\mathsf{V}}(\mathtt{x}) \mathbin{\hat{+}} \hat{\sigma}_{\mathsf{V}}(\mathtt{z})], \hat{\sigma}_{\mathsf{A}})$. For the edge $(q_3, \mathtt{z} > 0, q_5)$ we get the constraint

$$\widehat{\mathcal{S}}_{\mathsf{DS}}[\![\mathtt{z} > 0]\!](\mathsf{DS}(q_3)) \sqsubseteq_{\mathsf{DS}} \mathsf{DS}(q_5)$$

where $\widehat{\mathcal{S}}_{\mathsf{DS}}[\![\mathtt{z} > 0]\!]\hat{\sigma} = \bigsqcup\{(\hat{\sigma}'_{\mathsf{V}}, \hat{\sigma}'_{\mathsf{A}}) \in \mathsf{Basic}(\hat{\sigma}) \mid \mathsf{tt} \in (\hat{\sigma}'_{\mathsf{V}}(\mathtt{z}) \mathbin{\hat{>}} 0)\}$.

The constraint for the initial node expresses an assumption about the initial memory and we may for example take $\hat{\sigma}_{\triangleright}$ to be as in Figure 5.2.

| $q$ | $\mathsf{DS}(q)(\mathtt{x})$ | $\mathsf{DS}(q)(\mathtt{y})$ |
|---|---|---|
| $q_{\triangleright}$ | $\{0,+\}$ | $\{0,+\}$ |
| $q_1$ | $\{0,+\}$ | $\{0,+\}$ |
| $q_2$ | $\{0,+\}$ | $\{0,+\}$ |
| $q_3$ | $\{0,+\}$ | $\{0,+\}$ |
| $q_4$ | $\{0,+\}$ | $\{0,+\}$ |
| $q_5$ | $\{0,+\}$ | $\{0,+\}$ |
| $q_6$ | $\{+\}$ | $\{0,+\}$ |
| $q_7$ | $\{0,+\}$ | $\{0,+\}$ |
| $q_{\blacktriangleleft}$ | $\{0,+\}$ | $\{0,+\}$ |

Figure 5.7: The analysis assignment obtained when $\hat{\sigma}_{\triangleright}(\mathtt{A}) = \{-, 0, +\}$.

Our development of Chapter 3 then suffices for computing the least solution to the constraints. This is because we have ensured that $\widehat{\mathbf{Mem}}_{\mathsf{DS}}$ is indeed an analysis domain and hence satisfies the ascending chain condition, and because Exercise 5.9 established that the analysis functions for the actions are monotone. Hence all conditions for Proposition 3.28 are satisfied.

EXAMPLE 5.12: Figure 5.7 shows (part of) the analysis assignments obtained for the program graph of Figure 5.1 when the signs of the five variables given by $\hat{\sigma}_{\triangleright}$ are as in Figure 5.2 but the signs of the array $\mathtt{A}$ is $\{-, 0, +\}$. We only display the analysis results for the two variables $\mathtt{x}$ and $\mathtt{y}$.

Figure 5.8 shows the similar information in the case where $\hat{\sigma}_{\triangleright}(\mathtt{A}) = \{-, 0\}$. Here the analysis determines that the final node is not reachable and we can see that the reason is that a division by 0 is bound to happen in the output action on the edge from $q_7$ to $q_{\blacktriangleleft}$.

| $q$ | $\mathsf{DS}(q)(\mathtt{x})$ | $\mathsf{DS}(q)(\mathtt{y})$ |
|---|---|---|
| $q_{\triangleright}$ | $\{0\}$ | $\{0\}$ |
| $q_1$ | $\{0\}$ | $\{0\}$ |
| $q_2$ | $\{0\}$ | $\{0\}$ |
| $q_3$ | $\{0\}$ | $\{0\}$ |
| $q_4$ | $\{0\}$ | $\{0\}$ |
| $q_5$ | $\{\ \}$ | $\{\ \}$ |
| $q_6$ | $\{\ \}$ | $\{\ \}$ |
| $q_7$ | $\{0\}$ | $\{0\}$ |
| $q_{\blacktriangleleft}$ | $\{\ \}$ | $\{\ \}$ |

Figure 5.8: The analysis assignment obtained when $\hat{\sigma}_{\triangleright}(\mathtt{A}) = \{-, 0\}$.

EXERCISE 5.13: Compute the analysis result in a scenario where $\hat{\sigma}_{\triangleright}(\mathtt{A}) = \{+\}$ and in a scenario where $\hat{\sigma}_{\triangleright}(\mathtt{A}) = \{0, +\}$. Discuss the information provided by the analysis. □

EXERCISE 5.14: Let us replace the analysis of tests with the simpler definition

$$\widehat{\mathcal{S}}_{\mathsf{DS}}[\![b]\!]\hat{\sigma} = \begin{cases} \hat{\sigma} & \text{if tt} \in \widehat{\mathcal{B}}_{\mathsf{DS}}[\![b]\!]\hat{\sigma} \text{ and } \hat{\sigma} \neq \bot_{\mathsf{DS}} \\ \bot_{\mathsf{DS}} & \text{otherwise} \end{cases}$$

dismissed earlier. How would that modify the analysis results obtained? To answer this you should recompute the analysis result in one of above scenarios (for example the case where $\hat{\sigma}_{\triangleright}(\mathtt{A}) = \{-, 0\}$ or $\hat{\sigma}_{\triangleright}(\mathtt{A}) = \{+\}$). □

The correctness of the constraints can be expressed in a strong way, saying that when the execution reaches the node $q$, then the signs of the variables and arrays will correctly have been predicted by the analysis.

PROPOSITION 5.15: If DS is a solution to the constraints, and if $\pi$ is a path from $q_{\triangleright}$ to $q$, then $\beta_{\mathsf{DS}}(\sigma) \sqsubseteq \mathsf{DS}(q)$ whenever $\sigma = \mathcal{S}[\![\pi]\!]\sigma_{\triangleright}$ and $\beta_{\mathsf{DS}}(\sigma_{\triangleright}) \sqsubseteq \hat{\sigma}_{\triangleright}$.

PROOF: Using Proposition 3.29 we have that

$$\widehat{\mathcal{S}}_{\mathsf{DS}}[\![\pi]\!](\hat{\sigma}_{\triangleright}) \sqsubseteq_{\mathsf{DS}} \mathsf{DS}(q)$$

From Essential Exercise 5.10 it follows by induction on the length of the path $\pi$ that

$$\text{if } \mathcal{S}[\![\pi]\!]\sigma_{\triangleright} = \sigma \quad \text{then} \quad \beta_{\mathsf{DS}}(\sigma) \sqsubseteq_{\mathsf{DS}} \widehat{\mathcal{S}}_{\mathsf{DS}}[\![\pi]\!](\beta_{\mathsf{DS}}(\sigma_{\triangleright}))$$

From $\beta_{\mathsf{DS}}(\sigma_{\triangleright}) \sqsubseteq \hat{\sigma}_{\triangleright}$ and monotonicity of the analysis functions we then get the desired result. □

EXERCISE 5.16: The aim of the *Parity Analysis* is to determine for each node in the program graph the possible parities (even or odd) of the variables and arrays.

The analysis will make use of the set **Parity** = {even, odd} equipped with the ordering ⊆ (as illustrated on Figure 5.9). Construct the analysis domain

$$\widehat{\mathbf{Mem}}_{\mathsf{PA}} = (\mathbf{Var} \to \mathsf{PowerSet}(\mathbf{Parity})) \times (\mathbf{Arr} \to \mathsf{PowerSet}(\mathbf{Parity}))$$

by defining the partial order $\sqsubseteq_{\mathsf{PA}}$ on $\widehat{\mathbf{Mem}}_{\mathsf{PA}}$, by specifying the least element $\bot_{\mathsf{PA}}$ and the join operation $\sqcup_{\mathsf{PA}}$, and by arguing that the ascending chain condition is satisfied.

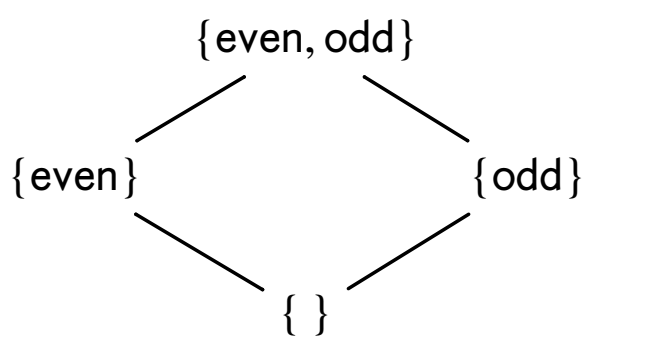

Figure 5.9: The analysis domain PowerSet( **Parity** ).

Next specify the analysis functions for arithmetic and boolean expressions as well as for actions. Show that the functions have the required monotonicity properties. Define a function $\beta_{\mathsf{PA}}$ mapping memories to abstract memories and use it to argue that the functions correctly capture the parities of the values. Finally, specify how to generate constraints from a program graph and illustrate the analysis on an appropriate example. □

EXERCISE 5.17: Continuing Exercise 5.16 consider an improvement of the *Parity Analysis* where the abstract memory keeps the information about the even and odd array entries apart. This can be captured by an analysis domain of the form

$$\begin{aligned} \widehat{\mathbf{Mem}}'_{\mathsf{PA}} &= (\mathbf{Var} \to \mathsf{PowerSet}(\mathbf{Parity})) \\ &\times (\mathbf{Arr} \to (\mathsf{PowerSet}(\mathbf{Parity}) \times \mathsf{PowerSet}(\mathbf{Parity}))) \end{aligned}$$

As an example, we may have $\widehat{\sigma}_{\mathsf{A}}(A) = (\{\mathsf{even}\}, \{\mathsf{even}, \mathsf{odd}\})$ meaning that all the even entries of $A$ (that is, $A[0], A[2], \cdots$) will be even whereas all the odd entries (that is, $A[1], A[3], \cdots$) may be even or odd.

Modify the above analysis to use this analysis domain and discuss the advantages and disadvantages of the new analysis. □

EXERCISE 5.18: We may want to *combine* the Detection of Signs Analysis and the Parity Analysis (of Exercise 5.16). For this we may want for each variable to keep track of the various combinations of signs and parities – and we immediately notice that the combination $(0, \mathsf{odd})$ should never be used. To capture this we may take

$$\begin{array}{rcl}\widehat{\mathbf{Mem}_{\mathsf{SP}}} & = & (\mathbf{Var} \to \mathsf{PowerSet}(\mathbf{Sign} \times \mathbf{Parity})) \\ & \times & (\mathbf{Arr} \to \mathsf{PowerSet}(\mathbf{Sign} \times \mathbf{Parity}))\end{array}$$

Show how $\widehat{\mathbf{Mem}_{\mathsf{SP}}}$ can be turned into an analysis domain and specify analysis functions for the arithmetic and boolean expressions as well as the actions. Note that the abstract versions of the arithmetic, relational and boolean operators combine the previous versions operating on just signs and just parities and in some cases we may be able to do better than a simple combination. Finally show how to generate constraints from a program graph and illustrate the analysis on an appropriate example.

An alternative analysis makes use of

$$\begin{array}{rcl}\widehat{\mathbf{Mem}'_{\mathsf{SP}}} & = & (\mathbf{Var} \to (\mathsf{PowerSet}(\mathbf{Sign}) \times \mathsf{PowerSet}(\mathbf{Parity}))) \\ & \times & (\mathbf{Arr} \to (\mathsf{PowerSet}(\mathbf{Sign}) \times \mathsf{PowerSet}(\mathbf{Parity})))\end{array}$$

Discuss the advantages and disadvantages of developing an analysis based on this analysis domain. □

## 5.2 Constant Propagation Analysis

The aim of the *Constant Propagation* analysis is to determine for each node in the program graph whether or not a variable or array entry has a constant value whenever execution reaches that point. In the case the value is a constant the analysis returns that value, it will return the value $\top$ in case the value is not a constant, and it will return the value $\bot$ if no value is available. This is reflected in the analysis domain of extended values

$$(\mathbf{Int}^{\top}_{\bot}, \sqsubseteq)$$

where the ordering $\sqsubseteq$ is defined by $\bot \sqsubseteq v \sqsubseteq \top$ for all $v \in \mathbf{Int}$; the ordering is illustrated in Figure 5.10. This is indeed an analysis domain; the least element is $\bot$,

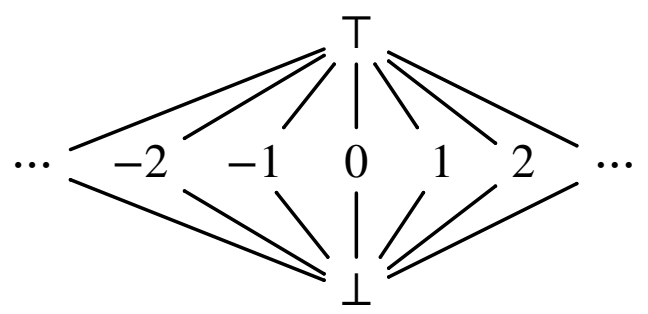

Figure 5.10: The analysis domain $\mathbf{Int}^{\top}_{\bot}$ with the ordering $\sqsubseteq$.

| $z_1 \mathbin{\widehat{+}} z_2$ | $\bot$ | $v_2$ | $\top$ |
|---|---|---|---|
| $\bot$ | $\bot$ | $\bot$ | $\bot$ |
| $v_1$ | $\bot$ | $v_1 + v_2$ | $\top$ |
| $\top$ | $\bot$ | $\top$ | $\top$ |

| $z_1 \mathbin{\widehat{<}} z_2$ | $\bot$ | $v_2$ | $\top$ |
|---|---|---|---|
| $\bot$ | $\{\,\}$ | $\{\,\}$ | $\{\,\}$ |
| $v_1$ | $\{\,\}$ | $\{\,v_1 < v_2\}$ | $\{\mathsf{tt}, \mathsf{ff}\}$ |
| $\top$ | $\{\,\}$ | $\{\mathsf{tt}, \mathsf{ff}\}$ | $\{\mathsf{tt}, \mathsf{ff}\}$ |

Figure 5.11: The abstract operations $\widehat{+}$ and $\widehat{<}$ for Constant Propagation analysis.

the join operation $\sqcup$ is defined by

$$z_1 \sqcup z_2 = \begin{cases} \bot & \text{if } z_1 = \bot \text{ and } z_2 = \bot \\ z_1 & \text{if } z_2 \sqsubseteq z_1 \text{ and } z_1 \in \mathbf{Int} \\ z_2 & \text{if } z_1 \sqsubseteq z_2 \text{ and } z_2 \in \mathbf{Int} \\ \top & \text{otherwise} \end{cases}$$

As an example $0 \sqcup 0 = 0$ whereas $0 \sqcup 1 = \top$. Clearly $\mathbf{Int}_\bot^\top$ satisfies the ascending chain condition.

The analysis will compute with these extended values and for each of the arithmetic and boolean operations ($op$) we shall introduce an abstract version of it (denoted $\widehat{op}$). In the case of addition it is defined by the table in the left part of Figure 5.11 and in the case of the less-than comparison it is defined in the table in the right part of Figure 5.11. In both tables we see that if one of the arguments is not available (that is, it is $\bot$) then the result of the operation is not available either – in the case of addition this is expressed using the abstract value $\bot$ and in the case of the comparison it is expressed using the empty set $\{\,\}$ of boolean values. Otherwise, if one of the arguments is $\top$ meaning that it is not a constant then we cannot determine the precise result of the operation so it is $\top$ in the case of addition and $\{\mathsf{tt}, \mathsf{ff}\}$ in the case of the comparison. Only in the case where both arguments are constants we can determine the precise result of the operations.

Exercise 5.19: Specify abstract versions of the arithmetic operations of subtraction ($-$) and multiplication ($*$), division ($/$) and modulo ($\%$); in particular discuss how you handle division and modulo in the case the second argument is 0. In a similar way specify abstract versions of the relational operators of equality ($=$) and less-than-or-equal ($\leq$). □

**The analysis domain** The analysis domain of the Constant Propagation analysis will for each variable tell us whether or not it is a constant, and for this we can use mappings from $\mathbf{Var} \to \mathbf{Int}_\bot^\top$. For the arrays we could consider using mappings from $\mathbf{Arr} \to \mathbf{Int}_\bot^\top$ but unfortunately, this is not going to be very useful since it is likely to give $\top$ in most cases: only when all the entries of the array have the very same value it will give something useful and this is probably very rare. It is better to exploit that the analysis might give us precise information about the indices and hence use mappings of the form $\mathbf{Arr} \to (\mathbf{Int}_\bot^\top)^*$ so that for an array $A$ of length $k$ we have that, if $\widehat{\sigma}_\mathsf{A}(A) = [z_0, z_1, \cdots, z_{k-1}]$ then $z_i$ (also written as $\widehat{\sigma}_\mathsf{A}(A)_i$) gives us information about whether or not the $i$'th entry of $A$ is a constant. In the construction below we shall ensure that if $z_i = \bot$ for one of the indices $i$ then indeed $z_j = \bot$ for all $j$ – this reflects that if one of the entries of the array is not available then indeed the whole array is not available.

Our analysis therefore uses abstract memories from an analysis domain

$$\widehat{\mathbf{Mem}}_{\mathsf{CP}} \subseteq (\mathbf{Var} \to \mathbf{Int}_{\bot}^{\top}) \times (\mathbf{Arr} \to (\mathbf{Int}_{\bot}^{\top})^{*})$$

that respects the given length of arrays: if $(\hat{\sigma}_{\mathsf{V}}, \hat{\sigma}_{\mathsf{A}}) \in \widehat{\mathbf{Mem}}_{\mathsf{CP}}$ then $\hat{\sigma}_{\mathsf{A}}(A)$ has length $\mathsf{length}(A)$. Thus for each variable $x \in \mathbf{Var}$ we will know whether or not it is a constant and similarly for each array entry $A[i]$ (where $A \in \mathbf{Arr}$ and $0 \leq i < \mathsf{length}(A)$) we will know whether or not it is a constant. Figure 5.12 gives an example of an abstract memory where two of the variables (`x` and `i`) have constant values while the third (`y`) does not; the array `A` has length 4 and two of its elements have constant values whereas the other two do not. The array `B` has length 2 and both entries have constant values. Actually, each memory directly gives rise to an abstract memory as captured by the function $\beta_{\mathsf{CP}} : \mathbf{Mem} \to \widehat{\mathbf{Mem}}_{\mathsf{CP}}$ simply defined by:

$$\left( \begin{array}{|c|c|}\hline \mathtt{x} & 7 \\ \hline \mathtt{y} & \top \\ \hline \mathtt{i} & -3 \\ \hline \end{array}, \begin{array}{|c|c|}\hline \mathtt{A} & [2,\top,\top,0] \\ \hline \mathtt{B} & [0,0] \\ \hline \end{array} \right)$$

Figure 5.12: Example abstract memory for three variables and two arrays.

$$\beta_{\mathsf{CP}}(\sigma) = (\sigma_{\mathsf{V}}, \sigma_{\mathsf{A}}) \qquad (\text{where } \sigma = (\sigma_{\mathsf{V}}, \sigma_{\mathsf{A}}, \sigma_{\mathsf{C}}))$$

In the manner of Essential Exercise 3.20, we shall now provide the additional structure turning $\widehat{\mathbf{Mem}}_{\mathsf{CP}}$ into an analysis domain. Our starting point is that $\mathbf{Int}_{\bot}^{\top}$ with the ordering $\sqsubseteq$ of Figure 5.10 already is an analysis domain.

The ordering $\sqsubseteq_{\mathsf{CP}}$ on $\widehat{\mathbf{Mem}}_{\mathsf{CP}}$ is a pointwise extension of the ordering $\sqsubseteq$ on $\mathbf{Int}_{\bot}^{\top}$:

$$\hat{\sigma}_1 \sqsubseteq_{\mathsf{CP}} \hat{\sigma}_2 \text{ if and only if } \begin{array}{l} \forall x \in \mathbf{Var} : \hat{\sigma}_{1\mathsf{V}}(x) \sqsubseteq \hat{\sigma}_{2\mathsf{V}}(x) \text{ and} \\ \forall A \in \mathbf{Arr} : \forall i \in \{0, \ldots, \mathsf{length}(A)\} : \hat{\sigma}_{1\mathsf{A}}(A)_i \sqsubseteq \hat{\sigma}_{2\mathsf{A}}(A)_i \end{array}$$

The least element $\bot_{\mathsf{CP}}$ is the abstract memory $\hat{\sigma}$ with $\hat{\sigma}_{\mathsf{V}}(x) = \bot$ for all variables $x$ and $\hat{\sigma}_{\mathsf{A}}(A)_i = \bot$ for all arrays $A$ and indices $i < \mathsf{length}(A)$. The join operation $\sqcup_{\mathsf{CP}}$ is the pointwise extension of the join operation $\sqcup$; formally $\hat{\sigma}_1 \sqcup_{\mathsf{CP}} \hat{\sigma}_2$ is the abstract memory $\hat{\sigma}$ given by $\hat{\sigma}_{\mathsf{V}}(x) = \hat{\sigma}_{1\mathsf{V}}(x) \sqcup \hat{\sigma}_{2\mathsf{V}}(x)$ for all variables $x$ and similarly $\hat{\sigma}(A)_i = \hat{\sigma}_{1\mathsf{A}}(A)_i \sqcup \hat{\sigma}_{2\mathsf{A}}(A)_i$ for all arrays $A$ and indices $i < \mathsf{length}(A)$.

Try It Out 5.20: Check that the above definitions of $\sqsubseteq_{\mathsf{CP}}$, $\bot_{\mathsf{CP}}$ and $\sqcup_{\mathsf{CP}}$ satisfy the requirements put forward in Section 3.2 for $\widehat{\mathbf{Mem}}_{\mathsf{CP}}$ being a pointed semi-lattice.□

Exercise 5.21: Argue that $\widehat{\mathbf{Mem}}_{\mathsf{CP}}$ satisfies the ascending chain condition; for this you may assume that there exists an infinite ascending chain $\hat{\sigma}_0 \sqsubset_{\mathsf{CP}} \hat{\sigma}_1 \sqsubset_{\mathsf{CP}} \cdots \sqsubset_{\mathsf{CP}} \hat{\sigma}_n \sqsubset_{\mathsf{CP}} \cdots$ of abstract memories from $\widehat{\mathbf{Mem}}_{\mathsf{CP}}$ and then arrive at a contradiction.□

**Analysing arithmetic and boolean expressions** Before defining the analysis functions for the actions we shall specify how to analyse the arithmetic and boolean expressions. For this we define analysis functions

$$\begin{aligned}\widehat{\mathcal{A}}_{\mathsf{CP}}[\![a]\!] &: \quad \widehat{\mathbf{Mem}}_{\mathsf{CP}} \to \mathbf{Int}_{\perp}^{\top}\\ \widehat{\mathcal{B}}_{\mathsf{CP}}[\![b]\!] &: \quad \widehat{\mathbf{Mem}}_{\mathsf{CP}} \to \mathsf{PowerSet}(\mathbf{Bool})\end{aligned}$$

For arithmetic expressions we obtain an abstract value; this is in line with how we defined the abstract versions of the arithmetic operations. For the boolean expressions we obtain a set of truth values exactly as we did when specifying the abstract versions of the relational operators.

The analysis functions for arithmetic expressions are as follows:

$$\begin{aligned}
\widehat{\mathcal{A}}_{\mathsf{CP}}[\![n]\!]\hat{\sigma} &= n\\
\widehat{\mathcal{A}}_{\mathsf{CP}}[\![x]\!]\hat{\sigma} &= \hat{\sigma}_{\mathsf{V}}(x)\\
\widehat{\mathcal{A}}_{\mathsf{CP}}[\![A[a_0]]\!]\hat{\sigma} &= \begin{cases} \hat{\sigma}_{\mathsf{A}}(A)_i & \\ \quad \text{if } i = \widehat{\mathcal{A}}_{\mathsf{CP}}[\![a_0]\!]\hat{\sigma} \in \mathbf{Int} \text{ and } 0 \le i < \mathsf{length}(A) \\ \bigsqcup\{\hat{\sigma}_{\mathsf{A}}(A)_i \mid 0 \le i < \mathsf{length}(A)\} & \\ \quad \text{if } \widehat{\mathcal{A}}_{\mathsf{CP}}[\![a_0]\!]\hat{\sigma} = \top \\ \perp \quad \text{otherwise} \end{cases}\\
\widehat{\mathcal{A}}_{\mathsf{CP}}[\![a_1 \; op_a \; a_2]\!]\hat{\sigma} &= \widehat{\mathcal{A}}_{\mathsf{CP}}[\![a_1]\!]\hat{\sigma} \; \widehat{op_a} \; \widehat{\mathcal{A}}_{\mathsf{CP}}[\![a_2]\!]\hat{\sigma}\\
\widehat{\mathcal{A}}_{\mathsf{CP}}[\![-a_0]\!]\hat{\sigma} &= \hat{-} \; \widehat{\mathcal{A}}_{\mathsf{CP}}[\![a_0]\!]\hat{\sigma}
\end{aligned}$$

In the case of $A[a_0]$ we first analyse the index $a_0$; if the result is a constant $i$ that is a valid index into $A$ then we consult $\hat{\sigma}$ to determine its abstract value. If the analysis of the index gives $\top$ meaning that we have not been able to determine that it is a constant then we combine all the information available about the entries of the array using the operation $\bigsqcup$ which is a generalisation of the join operation to work on finite sets: $\bigsqcup\{\hat{\sigma}_1, \cdots, \hat{\sigma}_n\} = \perp_{\mathsf{CP}} \sqcup_{\mathsf{CP}} \hat{\sigma}_1 \sqcup_{\mathsf{CP}} \cdots \sqcup_{\mathsf{CP}} \hat{\sigma}_n$. The last case in the clause for $A[a_0]$ captures the situation where the value of the index is not available (that is, when it is $\perp$) and the case where the analysis determines that the index is a constant but it is outside the bounds of the array.

Essential Exercise 5.22: Specify the analysis function $\widehat{\mathcal{B}}_{\mathsf{CP}}[\![b]\!]$ for the boolean expressions.

Exercise 5.23: Show that for all arithmetic expressions $a$ we have the following monotonicity result: For all abstract memories $\hat{\sigma}_1$ and $\hat{\sigma}_2$:

$$\hat{\sigma}_1 \sqsubseteq_{\mathsf{CP}} \hat{\sigma}_2 \qquad \text{implies} \qquad \widehat{\mathcal{A}}_{\mathsf{CP}}[\![a]\!]\hat{\sigma}_1 \sqsubseteq \widehat{\mathcal{A}}_{\mathsf{CP}}[\![a]\!]\hat{\sigma}_2$$

Formulate and show the similar result for boolean expressions. □

Essential Exercise 5.24: Show that for all arithmetic expressions $a$, for all memories $\sigma$ and values $v$

$$\text{if } \mathcal{A}[\![a]\!]\sigma = v \text{ then } v \sqsubseteq \widehat{\mathcal{A}}_{\mathsf{CP}}[\![a]\!](\beta_{\mathsf{CP}}(\sigma))$$

meaning that the analysis of arithmetic expressions correctly captures their value. Formulate and show the similar result for boolean expressions.

**Analysis functions for actions** Using these functions we are ready to define the analysis of the actions. Following the approach of the previous chapters the functions have the form $\widehat{\mathcal{S}}_{\mathsf{CP}}[\![q_\circ, \alpha, q_\bullet]\!] : \widehat{\mathbf{Mem}}_{\mathsf{CP}} \to \widehat{\mathbf{Mem}}_{\mathsf{CP}}$ but as we are not making use of the nodes $q_\circ$ and $q_\bullet$ we shall omit them and simply let

$$\widehat{\mathcal{S}}_{\mathsf{CP}}[\![\alpha]\!] : \widehat{\mathbf{Mem}}_{\mathsf{CP}} \to \widehat{\mathbf{Mem}}_{\mathsf{CP}}$$

For the two assignment actions we take

$$\widehat{\mathcal{S}}_{\mathsf{CP}}[\![x := a]\!]\hat{\sigma} = \begin{cases} (\hat{\sigma}_{\mathsf{V}}[x \mapsto z], \hat{\sigma}_{\mathsf{A}}) \\ \qquad \text{if } z = \widehat{\mathcal{A}}_{\mathsf{CP}}[\![a]\!]\hat{\sigma} \neq \bot \text{ and } \hat{\sigma} \neq \bot_{\mathsf{CP}} \\ \bot_{\mathsf{CP}} \quad \text{otherwise} \end{cases}$$

$$\widehat{\mathcal{S}}_{\mathsf{CP}}[\![A[a_1] := a_2]\!]\hat{\sigma} = \begin{cases} \hat{\sigma}[A[i] \mapsto z] \\ \qquad \text{if } i = \widehat{\mathcal{A}}_{\mathsf{CP}}[\![a_1]\!]\hat{\sigma} \text{ and } 0 \leq i < \mathsf{length}(A) \\ \qquad \text{and } z = \widehat{\mathcal{A}}_{\mathsf{CP}}[\![a_2]\!]\hat{\sigma} \neq \bot \text{ and } \hat{\sigma} \neq \bot_{\mathsf{CP}} \\ \hat{\sigma}[A[0] \mapsto \hat{\sigma}(A[0]) \sqcup z][A[1] \mapsto \hat{\sigma}(A[1]) \sqcup z] \cdots \\ \qquad\qquad [A[k-1] \mapsto \hat{\sigma}(A[k-1]) \sqcup z] \\ \qquad \text{if } \widehat{\mathcal{A}}_{\mathsf{CP}}[\![a_1]\!]\hat{\sigma} = \top \text{ and } k = \mathsf{length}(A) \\ \qquad \text{and } z = \widehat{\mathcal{A}}_{\mathsf{CP}}[\![a_2]\!]\hat{\sigma} \neq \bot \text{ and } \hat{\sigma} \neq \bot_{\mathsf{CP}} \\ \bot_{\mathsf{CP}} \text{ otherwise} \end{cases}$$

In both cases we ensure that if we do not have a value for one of the arithmetic expressions then the resulting abstract memory is $\bot_{\mathsf{CP}}$ thereby reflecting that the semantics will be stuck. In the case of the assignment to an array entry we have two additional cases. If the index is a constant then we can directly update the corresponding entry in the abstract memory with the new information. On the other hand, if the analysis of the index gives $\top$ then we do not know which entry to update. The approach is therefore to update *all* of them with the new information and we use the join operation to combine the old and the new abstract values. As an example, if the analysis of $a_1$ gives $\top$ and the analysis of $a_2$ gives $0$ and if $\hat{\sigma}(A) = [2, \top, 0]$ then the abstract memory will be updated to map $A$ to $[\top, \top, 0]$.

Let us next consider the actions involving channels:

$$\widehat{\mathcal{S}}_{\mathsf{CP}}[\![c?x]\!]\hat{\sigma} = \begin{cases} (\hat{\sigma}_{\mathsf{V}}[x \mapsto \top], \hat{\sigma}_{\mathsf{A}}) & \text{if } \hat{\sigma} \neq \bot_{\mathsf{CP}} \\ \bot_{\mathsf{CP}} & \text{otherwise} \end{cases}$$

$$\widehat{\mathcal{S}}_{\mathsf{CP}}[\![c?A[a]]\!]\hat{\sigma} = \begin{cases} (\hat{\sigma}_{\mathsf{V}}, \hat{\sigma}_{\mathsf{A}}[A[i] \mapsto \top]) \\ \quad \text{if } i = \widehat{\mathcal{A}}_{\mathsf{CP}}[\![a]\!]\hat{\sigma} \text{ and } 0 \leq i < \mathsf{length}(A) \\ \quad \text{and } \hat{\sigma} \neq \bot_{\mathsf{CP}} \\ (\hat{\sigma}_{\mathsf{V}}, \hat{\sigma}_{\mathsf{A}}[A[0] \mapsto \top][A[1] \mapsto \top] \cdots [A[k-1] \mapsto \top]) \\ \quad \text{if } \widehat{\mathcal{A}}_{\mathsf{CP}}[\![a]\!]\hat{\sigma} = \top \text{ and } k = \mathsf{length}(A) \\ \quad \text{and } \hat{\sigma} \neq \bot_{\mathsf{CP}} \\ \bot_{\mathsf{CP}} \quad \text{otherwise} \end{cases}$$

$$\widehat{\mathcal{S}}_{\mathsf{CP}}[\![c!a]\!]\hat{\sigma} = \begin{cases} \hat{\sigma} & \text{if } \widehat{\mathcal{A}}_{\mathsf{CP}}[\![a]\!]\hat{\sigma} \neq \bot \\ \bot_{\mathsf{CP}} & \text{otherwise} \end{cases}$$

For the two input actions we record that we have no constant for the value received on the channel so we use $\top$. As for the assignment to array entries we distinguish between the case where we know exactly which entry is being updated and the case where we do not.

Finally we have the analysis functions for the test and the `skip` action:

$$\widehat{\mathcal{S}}_{\mathsf{CP}}[\![b]\!]\hat{\sigma} = \begin{cases} \hat{\sigma} & \text{if tt} \in \widehat{\mathcal{B}}_{\mathsf{CP}}[\![b]\!]\hat{\sigma} \\ \bot_{\mathsf{CP}} & \text{otherwise} \end{cases}$$

$$\widehat{\mathcal{S}}_{\mathsf{CP}}[\![\texttt{skip}]\!]\hat{\sigma} = \hat{\sigma}$$

Note that for tests we return $\hat{\sigma}$ only if the test has a chance of returning true; otherwise the result is $\bot_{\mathsf{CP}}$ reflecting that the action will never succeed. An alternative definition of $\widehat{\mathcal{S}}_{\mathsf{CP}}[\![b]\!]\hat{\sigma}$ inspired by the Detection of Signs analysis of Section 5.1 is

$$\widehat{\mathcal{S}}_{\mathsf{CP}}[\![b]\!]\hat{\sigma} = \bigsqcup \{\hat{\sigma}' \in \mathsf{Basic}(\hat{\sigma}) \mid \text{tt} \in \widehat{\mathcal{B}}_{\mathsf{CP}}[\![b]\!]\hat{\sigma}'\}$$

where we now define the basic abstract memories to be the ones mapping variables and array entries to proper values:

$$\mathsf{Basic}(\hat{\sigma}) = \{\hat{\sigma}' \mid \hat{\sigma}' \sqsubseteq \hat{\sigma} \wedge \forall x \in \mathbf{Var} : \hat{\sigma}'_{\mathsf{V}}(x) \in \mathbf{Int} \wedge \forall A \in \mathbf{Arr} : \hat{\sigma}'_{\mathsf{A}}(A) \in \mathbf{Int}^{\mathsf{length}(A)}\}$$

Unfortunately this definition is not implementable as it would involve dealing with an infinite number of basic abstract memories.

Exercise 5.25: Show that the analysis functions for the actions as defined above are monotone; that is show that for all actions $\alpha$ and abstract memories $\hat{\sigma}_1$ and $\hat{\sigma}_2$ we have

$$\hat{\sigma}_1 \sqsubseteq_{\mathsf{CP}} \hat{\sigma}_2 \quad \text{implies} \quad \widehat{\mathcal{S}}_{\mathsf{CP}}[\![\alpha]\!]\hat{\sigma}_1 \sqsubseteq_{\mathsf{CP}} \widehat{\mathcal{S}}_{\mathsf{CP}}[\![\alpha]\!]\hat{\sigma}_2$$

In the proof you may exploit the results of Exercise 5.23. □

Essential Exercise 5.26: Use the result of Exercise 5.24 to show that the analysis functions capture what happens in the semantics. That is, show that for all actions $\alpha$, for all memories $\sigma$ and $\sigma'$ it is the case that

$$\text{if } \mathcal{S}[\![\alpha]\!]\sigma = \sigma' \qquad \text{then} \qquad \beta_{\mathsf{CP}}(\sigma') \sqsubseteq_{\mathsf{CP}} \widehat{\mathcal{S}}_{\mathsf{CP}}[\![\alpha]\!](\beta_{\mathsf{CP}}(\sigma))$$

**Constraints** Finally we are ready to specify the constraints for the analysis assignment $\mathsf{CP} : \mathbf{Q} \to \widehat{\mathbf{Mem}}_{\mathsf{CP}}$.

For each edge $(q_\circ, \alpha, q_\bullet)$ in the program graph we impose the constraint:

$$\widehat{\mathcal{S}}_{\mathsf{CP}}[\![\alpha]\!](\mathsf{CP}(q_\circ)) \sqsubseteq_{\mathsf{CP}} \mathsf{CP}(q_\bullet)$$

Furthermore, we impose an additional constraint for the entry node:

$$\hat{\sigma}_\triangleright \sqsubseteq_{\mathsf{CP}} \mathsf{CP}(q_\triangleright)$$

Here the abstract memory $\hat{\sigma}_\triangleright$ describes the initial information about the variables and arrays. A natural possibility is to take $\hat{\sigma}_\triangleright$ to be the abstract memory $\top_{\mathsf{CP}}$ that maps all variables to $\top$ and all arrays to sequences of $\top$'s (of the appropriate length).

Example 5.27: Consider the program graph of Figure 5.13. The edges give rise to six constraints:

$$\begin{array}{ll}
\widehat{\mathcal{S}}_{\mathsf{CP}}[\![\mathtt{x} < 0]\!](\mathsf{CP}(q_\triangleright)) \sqsubseteq_{\mathsf{CP}} \mathsf{CP}(q_1) & \widehat{\mathcal{S}}_{\mathsf{CP}}[\![\mathtt{y} := -1]\!](\mathsf{CP}(q_1)) \sqsubseteq_{\mathsf{CP}} \mathsf{CP}(q_\blacktriangleleft) \\
\widehat{\mathcal{S}}_{\mathsf{CP}}[\![\mathtt{x} = 0]\!](\mathsf{CP}(q_\triangleright)) \sqsubseteq_{\mathsf{CP}} \mathsf{CP}(q_2) & \widehat{\mathcal{S}}_{\mathsf{CP}}[\![\mathtt{y} := 0]\!](\mathsf{CP}(q_2)) \sqsubseteq_{\mathsf{CP}} \mathsf{CP}(q_\blacktriangleleft) \\
\widehat{\mathcal{S}}_{\mathsf{CP}}[\![\mathtt{x} > 0]\!](\mathsf{CP}(q_\triangleright)) \sqsubseteq_{\mathsf{CP}} \mathsf{CP}(q_3) & \widehat{\mathcal{S}}_{\mathsf{CP}}[\![\mathtt{y} := 1]\!](\mathsf{CP}(q_3)) \sqsubseteq_{\mathsf{CP}} \mathsf{CP}(q_\blacktriangleleft)
\end{array}$$

and additionally we have the constraint $\hat{\sigma}_\triangleright \sqsubseteq_{\mathsf{CP}} \mathsf{CP}(q_\triangleright)$ for the initial node.

Let us assume that $\hat{\sigma}_\triangleright(\mathtt{x}) = 7$. Then the analysis gives that $\widehat{\mathcal{S}}_{\mathsf{CP}}[\![\mathtt{x} > 0]\!](\hat{\sigma}_\triangleright) = \hat{\sigma}_\triangleright$ whereas the analysis of the other tests gives $\bot_{\mathsf{CP}}$. The analysis of assignments is such that only $\widehat{\mathcal{S}}_{\mathsf{CP}}[\![\mathtt{y} := 1]\!]$ gives a non-trivial contribution to $\mathsf{CP}(q_\blacktriangleleft)$ so the resulting analysis assignment has $\mathsf{CP}(q_\blacktriangleleft)(\mathtt{y}) = 1$ showing that $\mathtt{y}$ is a constant.

If on the other hand $\hat{\sigma}_\triangleright(\mathtt{x}) = \top$ then the analysis will also conclude that $\mathsf{CP}(q_\blacktriangleleft)(\mathtt{y}) = \top$ as each of the three tests may evaluate to true and hence any of the three assignments might be executed.

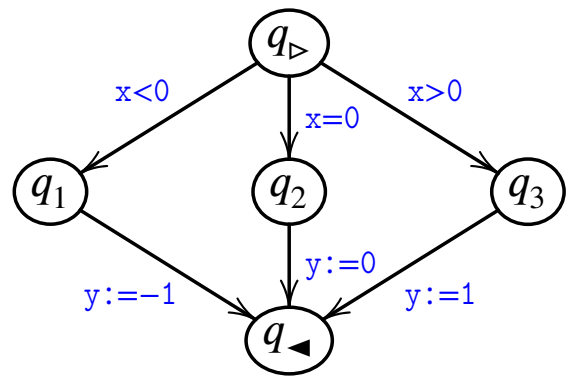

Figure 5.13: Determining the sign of a variable.

Try It Out 5.28: Assume that we replace the analysis of tests with the simpler clause

$$\widehat{\mathcal{S}}_{\mathsf{CP}}[\![b]\!]\hat{\sigma} = \hat{\sigma}$$

How would that modify the analysis results obtained in the above example? □

Exercise 5.29: Argue that our development of Chapter 3 suffices for computing the least solution to the constraints, because all the conditions for Proposition 3.28 are satisifed. □

The correctness of the constraints can be expressed in a strong way, saying that when the execution reaches the node $q$, then the information about the variables and arrays will correctly have been predicted by the analysis.

Proposition 5.30: If CP is a solution to the constraints, and if $\pi$ is a path from $q_{\triangleright}$ to $q$, then $\beta_{\mathsf{CP}}(\sigma) \sqsubseteq \mathsf{CP}(q)$ whenever $\sigma = \mathcal{S}[\![\pi]\!]\sigma_{\triangleright}$ and $\beta_{\mathsf{CP}}(\sigma_{\triangleright}) \sqsubseteq \hat{\sigma}_{\triangleright}$.

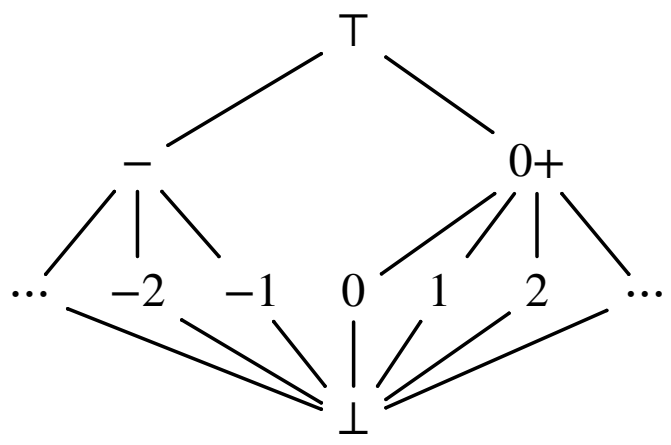

Figure 5.14: The analysis domain $\mathsf{Int}_{\perp}^{\mathsf{Sign}}$ with the ordering $\sqsubseteq$.

Exercise 5.31: Inspired by the Detection of Signs analysis we might want to improve on the precision of the Constant Propagation analysis in the cases where it would return $\top$. The aim of this exercise is to develop a variant of the Constant Propagation analysis based on the analysis domain of Figure 5.14 where we have the additional possibility of distinguishing between non-constant values that are negative (denoted $-$) and non-negative (denoted $0+$).

Modify the Constant Propagation Analysis to make use of this analysis domain. Discuss the advantages and disadvantages of the new analysis. □

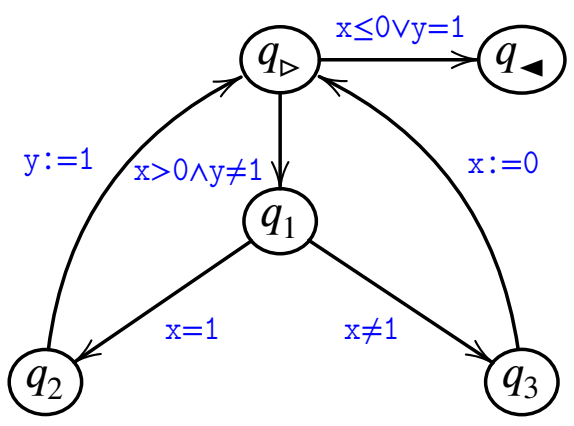

Figure 5.15: Analysing tests.

Teaser 5.32: Consider the program graph of Figure 5.15 and assume that $\hat{\sigma}_{\triangleright}(\mathtt{x}) = 1$ and $\hat{\sigma}_{\triangleright}(\mathtt{y}) = 0$. The analysis presented above will correctly determine that the node $q_3$ is unreachable but it cannot determine that the value of $\mathtt{y}$ always will be 1 at the node $q_{\blacktriangleleft}$ the reason being that the analysis of tests is too imprecise.

To improve on this we shall in this exercise develop a variant of the Constant Propagation analysis that is parametric on a (non-empty) finite set $\mathbf{K} \subseteq \mathbf{Int}$ of constant values of interest. The idea is that we only record constant values from $\mathbf{K}$; all other constant values will be represented by the special symbol †; this is reflected in the definition of the function $\mathsf{repr} : \mathbf{Int} \to \mathbf{K} \cup \{\dagger\}$ defined by

$$\mathsf{repr}(n) = \begin{cases} n & \text{if } n \in \mathbf{K} \\ \dagger & \text{otherwise} \end{cases}$$

The idea is then to use an analysis domain

$$\widehat{\mathbf{Mem}}_{\mathsf{CP}}^{\mathbf{K}} \subseteq (\mathbf{Var} \to (\mathbf{K} \cup \{\dagger\})_{\perp}^{\top}) \times (\mathbf{Arr} \to ((\mathbf{K} \cup \{\dagger\})_{\perp}^{\top})^{*})$$

that respects the length of arrays. We will then have abstract versions of the arithmetic and boolean operators computing with elements of $(\mathbf{K} \cup \{\dagger\})_{\perp}^{\top}$. Modify the Constant Propagation analysis to make use of this analysis domain. Show that the new analysis can give a more precise result for the program graph of Figure 5.15 by an appropriate choice of the set $\mathbf{K}$. (Hint: It is now possible to make $\mathsf{Basic}(\hat{\sigma})$ implementable.) □

## 5.3 Interval Analysis

Our third analysis is an Interval Analysis and it aims to determine an interval over-approximating the set of possible values for each of the variables and arrays whenever execution reaches each node. An interval has the form $[k^-, k^+]$ where $k^-$ is the lower bound and $k^+$ is the upper bound (so it is always the case that $k^- \leq k^+$). Ideally we would like $k^-$ and $k^+$ to be integers but there are cases where we cannot bound the values so we shall add "$-\infty$" as a possible lower bound and "$+\infty$" as a possible upper bound.

For computational reasons we shall impose restrictions on the number of possible endpoints of the intervals. So our development will be parametric on a (non-empty) finite set $\mathbf{K} \subseteq \mathbf{Int}$ of potential endpoints and we shall define the analysis domain from the set

$$\mathbf{Interval_K} \quad = \quad \{\bot\} \cup \{[k^-, k^+] \mid k^- \in \mathbf{K} \cup \{-\infty\}, k^+ \in \mathbf{K} \cup \{+\infty\}, k^- \leq k^+\}$$

where we have extended the ordering $\leq$ on integers to have $-\infty \leq n$ and $n \leq +\infty$ for all $n \in \mathbf{Int}$ in addition to $-\infty \leq +\infty$. As an example, if $\mathbf{K} = \{0\}$ then $\mathbf{Interval}_{\{0\}} = \{\bot, [0,0], [-\infty, 0], [0, +\infty], [-\infty, +\infty]\}$.

The ordering $\sqsubseteq$ on $\mathbf{Interval_K}$ is specifying when one interval is a subinterval of another interval and it is defined as follows:

- $\bot \sqsubseteq int$ for all $int \in \mathbf{Interval_K}$
- $[k_1^-, k_1^+] \sqsubseteq [k_2^-, k_2^+]$ if and only if $k_2^- \leq k_1^- \wedge k_1^+ \leq k_2^+$

Figure 5.16 illustrates the ordering in the case $\mathbf{K} = \{0\}$.

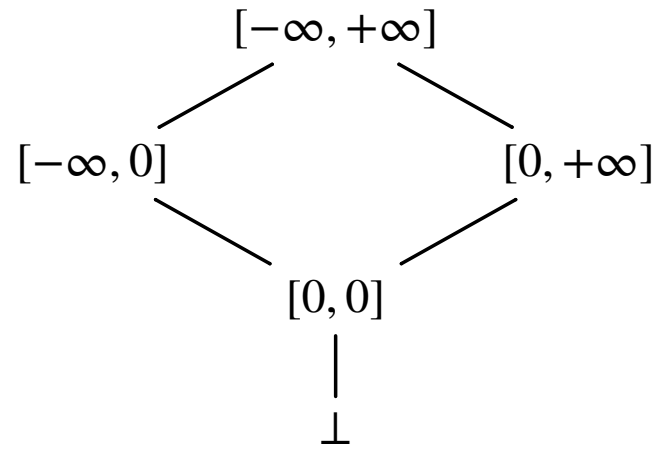

Figure 5.16: The analysis domain $\mathbf{Interval}_{\{0\}}$ with the ordering $\sqsubseteq$.

Try It Out 5.33: Assume that $\mathbf{K} = \{-1, 0, 1\}$ and construct the corresponding analysis domain $\mathbf{Interval_K}$ and its ordering. □

Exercise 5.34: The ordering on intervals reflects the subset ordering on the sets of integers. To see this define the function setof : $\mathbf{Interval_K} \rightarrow$ PowerSet($\mathbf{Int}$) that takes an interval as argument and returns the corresponding set of integers:

$$\mathsf{setof}(int) = \begin{cases} \{\,\} & \text{if } int = \bot \\ \{n \in \mathbf{Int} \mid k^- \leq n \leq k^+\} & \text{if } int = [k^-, k^+] \end{cases}$$

Show that for all intervals $int_1$ and $int_2$ we have that $int_1 \sqsubseteq int_2$ if and only if $\mathsf{setof}(int_1) \subseteq \mathsf{setof}(int_2)$. □

The least element of $\mathbf{Interval_K}$ is $\bot$ and the join operation $\sqcup$ returns the smallest interval containing both its arguments as subintervals and it is given by

$$int_1 \sqcup int_2 = \begin{cases} int_1 & \text{if } int_2 = \bot \\ int_2 & \text{if } int_1 = \bot \\ [k^-, k^+] & \text{if } int_1 = [k_1^-, k_1^+] \text{ and } int_2 = [k_2^-, k_2^+] \\ & \text{and } k^- = \min\{k_1^-, k_2^-\} \\ & \phantom{\text{and }} k^+ = \max\{k_1^+, k_2^+\} \end{cases}$$

EXERCISE 5.35: Show that $\text{setof}(int_1) \cup \text{setof}(int_2) \subseteq \text{setof}(int_1 \sqcup int_2)$ for all intervals $int_1$ and $int_2$. Show that the other inclusion does not hold in general; in particular, give an example of a set **K** and two elements $int_1$ and $int_2$ of $\textbf{Interval}_\textbf{K}$ such that $\text{setof}(int_1 \sqcup int_2) \neq \text{setof}(int_1) \cup \text{setof}(int_2)$. □

When coming to the specification of the analysis we shall make use of the top element $\top$ of the analysis domain and it is $[-\infty, +\infty]$. It is also useful to have the operation $\sqcap$ that is dual to $\sqcup$; it will return the largest interval contained in both of its arguments:

$$int_1 \sqcap int_2 = \begin{cases} [k^-, k^+] & \text{if } int_1 = [k_1^-, k_1^+] \text{ and } int_2 = [k_2^-, k_2^+] \\ & \text{and } k^- = \max\{k_1^-, k_2^-\} \\ & \phantom{\text{and }} k^+ = \min\{k_1^+, k_2^+\} \\ & \text{and } k^- \leq k^+ \\ \bot & \text{otherwise} \end{cases}$$

EXERCISE 5.36: Show that $\text{setof}(int_1) \cap \text{setof}(int_2) = \text{setof}(int_1 \sqcap int_2)$ for all intervals $int_1$ and $int_2$. Give an example of two intervals $[k_1^-, k_1^+]$ and $[k_2^-, k_2^+]$ where the condition $k^- \leq k^+$ of the above definition fails and hence the result of the operation is $\bot$ rather than $[k^-, k^+]$. □

**Computing with intervals** We are now going to present abstract versions of the arithmetic and boolean operations computing with intervals. As a first step we shall introduce the operations $\lfloor \cdot \rfloor : \textbf{Int} \to \textbf{K} \cup \{-\infty\}$ and $\lceil \cdot \rceil : \textbf{Int} \to \textbf{K} \cup \{+\infty\}$ that will be useful when computing the interval boundaries:

$$\begin{aligned} \lfloor n \rfloor &= \sup\{k \in \textbf{K} \mid k \leq n\} \\ \lceil n \rceil &= \inf\{k \in \textbf{K} \mid n \leq k\} \end{aligned}$$

where we recall that $\sup(\{\ \}) = -\infty$ and $\inf(\{\ \}) = +\infty$. Thus $\lfloor n \rfloor$ returns the greatest element in $\textbf{K} \cup \{-\infty\}$ that is smaller than or equal to $n$ whereas $\lceil n \rceil$ returns the smallest element in $\textbf{K} \cup \{+\infty\}$ that is greater than or equal to $n$. Note that if $n \in \textbf{K}$ then $\lfloor n \rfloor = n$ and $\lceil n \rceil = n$.

For the abstract version $\hat{+}$ of addition we take

$$int_1 \mathbin{\hat{+}} int_2 = \begin{cases} \bot & \text{if } int_1 = \bot \text{ or } int_2 = \bot \\ [k^-, k^+] & \text{if } int_1 = [k_1^-, k_1^+] \text{ and } int_2 = [k_2^-, k_2^+] \\ & \text{and } k^- = \lfloor k_1^- + k_2^- \rfloor \\ & \phantom{\text{and }} k^+ = \lceil k_1^+ + k_2^+ \rceil \end{cases}$$

| + | $-\infty$ | $n_2$ | $+\infty$ |
|---|---|---|---|
| $-\infty$ | $-\infty$ | $-\infty$ | undef |
| $n_1$ | $-\infty$ | $n_1 + n_2$ | $+\infty$ |
| $+\infty$ | undef | $+\infty$ | $+\infty$ |

Figure 5.17: Addition ($+$) extended to $-\infty$ and $+\infty$.

Here we have extended the normal addition operator + to operate on the two special values $-\infty$ and $+\infty$ as shown in Figure 5.17; note that the cases where the table gives undefined never will arise when adding intervals.

TRY IT OUT 5.37: Let us assume that $\mathbf{K} = \{-1, 0, 1\}$ as in Try It Out 5.33. Determine the result of adding the intervals [-1,1] and [1,1]. □

To ensure that the definition of $\hat{+}$ correctly captures the addition of the numbers in the intervals we have to check that if $n_1$ is in the interval $int_1$ and $n_2$ is in the interval $int_2$ then indeed $n_1 + n_2$ is in the interval $int_1 \hat{+} int_2$. Using the notation of Exercise 5.34 we can formalise this requirement as follows: for all $n_1, n_2 \in \mathbf{Int}$ it must be the case that

$$n_1 \in \mathsf{setof}(int_1) \wedge n_2 \in \mathsf{setof}(int_2) \text{ implies } n_1 + n_2 \in \mathsf{setof}(int_1 \hat{+} int_2)$$

EXERCISE 5.38: Specify abstract versions of subtraction and multiplication. In each case you will have to extend the concrete operators to operate on $-\infty$ and $+\infty$ along the lines of what we did for addition in Figure 5.17. Argue that the abstract operations correctly capture the meaning of their concrete counterparts. □

TEASER 5.39: Specify abstract versions of division and modulo. In each case you will have to extend the concrete operators to operate on $-\infty$ and $+\infty$ along the lines of what we did for addition in Figure 5.17. Argue that the abstract operations correctly capture the meaning of their concrete counterparts. □

The abstract relational operators will take intervals as arguments and return sets of boolean values as results. In the case of the less-than operation $<$ we define $\hat{<}$ as follows:

$$int_1 \mathbin{\hat{<}} int_2 = \begin{cases} \{\,\} & \text{if } int_1 = \bot \text{ or } int_2 = \bot \\ \{\mathsf{tt}\} & \text{if } int_1 = [k_1^-, k_1^+] \text{ and } int_2 = [k_2^-, k_2^+] \text{ and } k_1^+ < k_2^- \\ \{\mathsf{ff}\} & \text{if } int_1 = [k_1^-, k_1^+] \text{ and } int_2 = [k_2^-, k_2^+] \text{ and } k_1^- \geq k_2^+ \\ \{\mathsf{tt}, \mathsf{ff}\} & \text{otherwise} \end{cases}$$

Again we must verify that the abstract operation correctly captures the meaning of its concrete counterpart so using the notation of Exercise 5.34 we have to ensure that for all $n_1, n_2 \in \mathbf{Int}$:

$$n_1 \in \mathsf{setof}(int_1) \wedge n_2 \in \mathsf{setof}(int_2) \text{ imply } (n_1 < n_2) \in (int_1 \mathbin{\hat{<}} int_2)$$

EXERCISE 5.40: Specify abstract versions of the relational operators = and $\leq$. Argue that they correctly capture the meaning of their concrete counterparts. □

**The analysis domain** In the Interval Analysis the abstract memory will determine an interval for each of the variables specifying a bound on its values. For an array it will determine a single interval bounding the values of *all* the entries of the array – and thus it follows the same pattern as the Detection of Signs Analysis of Section 5.1. The analysis domain has the form

| | |
|---|---|
| x | $[0,0]$ |
| y | $[0,0]$ |
| z | $[-\infty,+\infty]$ |
| i | $[0,0]$ |
| A | $[-\infty,+\infty]$ |

Figure 5.18: An abstract memory.

$$\widehat{\mathbf{Mem}}_{\mathrm{IA}} = (\mathbf{Var} \to \mathbf{Interval}_{\mathbf{K}}) \times (\mathbf{Arr} \to \mathbf{Interval}_{\mathbf{K}})$$

An example is given in Figure 5.18; here we have assumed that $0 \in \mathbf{K}$.

Given a memory $\sigma = (\sigma_{\mathsf{V}}, \sigma_{\mathsf{A}}, \sigma_{\mathsf{C}})$ of the semantics we can define the corresponding abstract memory in $\widehat{\mathbf{Mem}}_{\mathrm{IA}}$ using the function $\beta_{\mathrm{IA}} : \mathbf{Mem} \to \widehat{\mathbf{Mem}}_{\mathrm{IA}}$ defined as follows:

$$\begin{aligned}
&\beta_{\mathrm{IA}}(\sigma) = (\widehat{\sigma}_{\mathsf{V}}, \widehat{\sigma}_{\mathsf{A}}) \\
&\quad \text{where } \widehat{\sigma}_{\mathsf{V}}(x) = [\lfloor v \rfloor, \lceil v \rceil] \\
&\qquad\qquad \text{where } \sigma_{\mathsf{V}}(x) = v \text{ for } x \in \mathbf{Var} \\
&\qquad\quad \widehat{\sigma}_{\mathsf{A}}(A) = [\lfloor v_0 \rfloor, \lceil v_0 \rceil] \sqcup [\lfloor v_1 \rfloor, \lceil v_1 \rceil] \sqcup \cdots \sqcup [\lfloor v_{k-1} \rfloor, \lceil v_{k-1} \rceil] \\
&\qquad\qquad \text{where } \sigma_{\mathsf{A}}(A) = [v_0, v_1, \cdots, v_{k-1}] \text{ for } A \in \mathbf{Arr}
\end{aligned}$$

Thus for a variable we will determine the smallest interval containing its value. For an array we will take the least upper bound of the smallest intervals obtained for each of the entries in the array; this gives us the smallest interval containing all the entries.

$\sigma = (\sigma_{\mathsf{V}}, \sigma_{\mathsf{A}}, \sigma_{\mathsf{C}})$ :

| | |
|---|---|
| x | 0 |
| y | 1 |
| z | −1 |
| i | 3 |
| A | $[0,1,3,0,7]$ |
| out | [ ] |

Figure 5.19: A memory $\sigma$.

Try It Out 5.41: Let $\mathbf{K} = \{-1, 0, 1\}$ as in Try It Out 5.33. Use the above definition to determine the abstract memory $\beta_{\mathrm{IA}}(\sigma)$ corresponding to the memory of Figure 5.19. How does the result change if we take $\mathbf{K} = \{0, 1, 10\}$? □

The ordering $\sqsubseteq_{\mathrm{IA}}$ on $\widehat{\mathbf{Mem}}_{\mathrm{IA}}$ is defined as the pointwise extension of the ordering $\sqsubseteq$ on $\mathbf{Interval}_{\mathbf{K}}$:

$$\hat{\sigma}_1 \sqsubseteq_{\mathrm{IA}} \hat{\sigma}_2 \quad \text{if and only if} \quad \begin{array}{l} \hat{\sigma}_{1\mathsf{V}}(x) \sqsubseteq \hat{\sigma}_{2\mathsf{V}}(x) \text{ for all } x \in \mathbf{Var} \text{ and} \\ \hat{\sigma}_{1\mathsf{A}}(A) \sqsubseteq \hat{\sigma}_{2\mathsf{A}}(A) \text{ for all } A \in \mathbf{Arr} \end{array}$$

The least element $\bot_{\mathrm{IA}}$ is defined to map all variables and array names to the least element $\bot$ of $\mathbf{Interval}_{\mathbf{K}}$. The join operator $\sqcup_{\mathrm{IA}}$ is defined in a pointwise manner from the join operation $\sqcup$ on $\mathbf{Interval}_{\mathbf{K}}$:

Try It Out 5.42: Check that the above definitions of $\sqsubseteq_{\mathrm{IA}}$, $\bot_{\mathrm{IA}}$ and $\sqcup_{\mathrm{IA}}$ satisfy the requirements put forward in Section 3.2 for $\widehat{\mathbf{Mem}}_{\mathrm{IA}}$ being a pointed semi-lattice. □

Exercise 5.43: Argue that $\widehat{\mathbf{Mem}}_{\mathrm{IA}}$ satisfies the ascending chain condition for all finite subsets $\mathbf{K}$ of $\mathbf{Int}$; for this you may assume that there exists an infinite ascending chain $\hat{\sigma}_0 \sqsubset_{\mathrm{IA}} \hat{\sigma}_1 \sqsubset_{\mathrm{IA}} \cdots \sqsubset_{\mathrm{IA}} \hat{\sigma}_n \sqsubset_{\mathrm{IA}} \cdots$ of abstract memories from $\widehat{\mathbf{Mem}}_{\mathrm{IA}}$ and then show that this gives rise to a contradiction. Does the argument also hold when $\mathbf{K}$ is infinite? □

**Analysing arithmetic and boolean expressions** The analysis functions for arithmetic and boolean expressions have the functionality

$$\begin{aligned}\widehat{\mathcal{A}}_{\mathsf{IA}}[\![a]\!] &: \widehat{\mathbf{Mem}}_{\mathsf{IA}} \to \mathbf{Interval}_{\mathbf{K}}\\ \widehat{\mathcal{B}}_{\mathsf{IA}}[\![b]\!] &: \widehat{\mathbf{Mem}}_{\mathsf{IA}} \to \mathsf{PowerSet}(\mathbf{Bool})\end{aligned}$$

The clauses defining the analysis functions follow the same pattern as before and make use of the abstract versions of the operators. For the arithmetic expressions we take

$$\begin{aligned}\widehat{\mathcal{A}}_{\mathsf{IA}}[\![n]\!]\hat{\sigma} &= [\lfloor n\rfloor, \lceil n\rceil]\\ \widehat{\mathcal{A}}_{\mathsf{IA}}[\![x]\!]\hat{\sigma} &= \hat{\sigma}_{\mathsf{V}}(x)\\ \widehat{\mathcal{A}}_{\mathsf{IA}}[\![A[a]]\!]\hat{\sigma} &= \begin{cases}\hat{\sigma}_{\mathsf{A}}(A) & \text{if } \widehat{\mathcal{A}}_{\mathsf{IA}}[\![a]\!]\hat{\sigma} \sqcap [\lfloor 0\rfloor, \lceil \mathsf{length}(A)-1\rceil] \neq \bot\\ \bot & \text{otherwise}\end{cases}\\ \widehat{\mathcal{A}}_{\mathsf{IA}}[\![a_1\ op_a\ a_2]\!]\hat{\sigma} &= \widehat{\mathcal{A}}_{\mathsf{IA}}[\![a_1]\!]\hat{\sigma}\ \widehat{op_a}\ \widehat{\mathcal{A}}_{\mathsf{DS}}[\![a_2]\!]\hat{\sigma}\\ \widehat{\mathcal{A}}_{\mathsf{IA}}[\![-a]\!]\hat{\sigma} &= \hat{-}\ \widehat{\mathcal{A}}_{\mathsf{IA}}[\![a]\!]\hat{\sigma}\end{aligned}$$

Note that in the case of $A[a]$ we first analyse the index $a$ and check whether or not the interval obtained has an overlap with the possible indices of the array $A$. The latter is the interval $[0, \mathsf{length}(A) - 1]$ and the smallest interval of $\mathbf{Interval}_{\mathbf{K}}$ including this is $[\lfloor 0\rfloor, \lceil \mathsf{length}(A) - 1\rceil]$. To check whether the two intervals overlap we use the greatest lower bound operation $\sqcap$ as it returns $\bot$ in case there is no overlap.

Essential Exercise 5.44: Specify the analysis function $\widehat{\mathcal{B}}_{\mathsf{IA}}[\![b]\!]$ for boolean expressions.

Exercise 5.45: Show that for all arithmetic expressions $a$ we have the following monotonicity result: For all abstract memories $\hat{\sigma}_1$ and $\hat{\sigma}_2$:

$$\hat{\sigma}_1 \sqsubseteq_{\mathsf{IA}} \hat{\sigma}_2 \quad \text{implies} \quad \widehat{\mathcal{A}}_{\mathsf{IA}}[\![a]\!]\hat{\sigma}_1 \sqsubseteq \widehat{\mathcal{A}}_{\mathsf{IA}}[\![a]\!]\hat{\sigma}_2$$

Formulate and show the similar result for boolean expressions. □

Essential Exercise 5.46: Show that for all arithmetic expressions $a$, for all memories $\sigma$ and values $v$

$$\text{if } \mathcal{A}[\![a]\!]\sigma = v \text{ then } v \in \mathsf{setof}(\widehat{\mathcal{A}}_{\mathsf{IA}}[\![a]\!](\beta_{\mathsf{IA}}(\sigma)))$$

meaning that the analysis of arithmetic expressions correctly captures their value. Formulate and show the similar result for boolean expressions.

**Analysis functions for actions** Using these functions we can define the analysis of the actions by a function $\widehat{\mathcal{S}}_{\mathsf{IA}}[\![q_\circ, \alpha, q_\bullet]\!] : \widehat{\mathbf{Mem}}_{\mathsf{IA}} \to \widehat{\mathbf{Mem}}_{\mathsf{IA}}$, or omitting the nodes $q_\circ$ and $q_\bullet$ as we are not making use of them:

$$\widehat{\mathcal{S}}_{\mathsf{IA}}[\![\alpha]\!] : \widehat{\mathbf{Mem}}_{\mathsf{IA}} \to \widehat{\mathbf{Mem}}_{\mathsf{IA}}$$

For the assignment actions we take

$$\widehat{\mathcal{S}}_{\mathsf{IA}}[\![x := a]\!]\hat{\sigma} = \begin{cases} (\hat{\sigma}_{\mathsf{V}}[x \mapsto \widehat{\mathcal{A}}_{\mathsf{IA}}[\![a]\!]\hat{\sigma}], \hat{\sigma}_{\mathsf{A}}) \\ \qquad \text{if } \widehat{\mathcal{A}}_{\mathsf{IA}}[\![a]\!]\hat{\sigma} \neq \bot \text{ and } \hat{\sigma} \neq \bot_{\mathsf{IA}} \\ \bot_{\mathsf{IA}} \quad \text{otherwise} \end{cases}$$

$$\widehat{\mathcal{S}}_{\mathsf{IA}}[\![A[a_1] := a_2]\!]\hat{\sigma} = \begin{cases} (\hat{\sigma}_{\mathsf{V}}, \hat{\sigma}_{\mathsf{A}}[A \mapsto \hat{\sigma}_{\mathsf{A}}(A) \sqcup \widehat{\mathcal{A}}_{\mathsf{IA}}[\![a_2]\!]\hat{\sigma}]) \\ \qquad \text{if } \widehat{\mathcal{A}}_{\mathsf{IA}}[\![a_1]\!]\hat{\sigma} \sqcap [\lfloor 0 \rfloor, \lceil \mathsf{length}(A) - 1 \rceil] \neq \bot \\ \qquad \text{and } \widehat{\mathcal{A}}_{\mathsf{IA}}[\![a_2]\!]\hat{\sigma} \neq \bot \text{ and } \hat{\sigma} \neq \bot_{\mathsf{IA}} \\ \bot_{\mathsf{IA}} \quad \text{otherwise} \end{cases}$$

In the case of array update we check whether the interval of the index $a_1$ has an overlap with that of the array $A$ (using the $\sqcap$ operation on $\mathbf{Interval}_{\mathbf{K}}$ as in the case of array lookup above) and only if so we update the abstract memory. As the analysis domain does associate information with the array as a whole (and not the individual indices) we shall use the least upper bound operation $\sqcup$ to construct the new information for the array.

For the actions involving channels we define the analysis functions as follows:

$$\widehat{\mathcal{S}}_{\mathsf{IA}}[\![c?x]\!]\hat{\sigma} = \begin{cases} (\hat{\sigma}_{\mathsf{V}}[x \mapsto [-\infty, +\infty]], \hat{\sigma}_{\mathsf{A}}) \\ \qquad \text{if } \hat{\sigma} \neq \bot_{\mathsf{IA}} \\ \bot_{\mathsf{IA}} \quad \text{otherwise} \end{cases}$$

$$\widehat{\mathcal{S}}_{\mathsf{IA}}[\![c?A[a]]\!]\hat{\sigma} = \begin{cases} (\hat{\sigma}_{\mathsf{V}}, \hat{\sigma}_{\mathsf{A}}[A \mapsto [-\infty, +\infty]]) \\ \qquad \text{if } \widehat{\mathcal{A}}_{\mathsf{IA}}[\![a]\!]\hat{\sigma} \sqcap [\lfloor 0 \rfloor, \lceil \mathsf{length}(A) - 1 \rceil] \neq \bot \\ \qquad \text{and } \hat{\sigma} \neq \bot_{\mathsf{IA}} \\ \bot_{\mathsf{IA}} \quad \text{otherwise} \end{cases}$$

$$\widehat{\mathcal{S}}_{\mathsf{IA}}[\![c!a]\!]\hat{\sigma} = \begin{cases} \hat{\sigma} & \text{if } \widehat{\mathcal{A}}_{\mathsf{IA}}[\![a]\!]\hat{\sigma} \neq \bot \\ \bot_{\mathsf{IA}} & \text{otherwise} \end{cases}$$

For the two input actions we record that any value may be received using the interval $[-\infty, +\infty]$; as was the case in the analysis function for array assignment we ensure that the interval of the actual index overlaps with the interval of the allowed indices.

Finally we have the analysis functions for the test and `skip` actions. In the case of a test $b$ our aim will be to determine whether or not it could evaluate to true. A first attempt would be to take

$$\widehat{\mathcal{S}}_{\mathsf{IA}}[\![b]\!]\hat{\sigma} = \begin{cases} \hat{\sigma} & \text{if } \mathsf{tt} \in \widehat{\mathcal{B}}_{\mathsf{IA}}[\![b]\!]\hat{\sigma} \\ \bot_{\mathsf{IA}} & \text{otherwise} \end{cases}$$

This will indeed be a correct specification but as we discussed for the Detection of Signs Analysis it does not take into account that we might learn something by passing a test. We shall therefore define a *basic abstract memory* $\hat{\sigma}'$ to be a memory where the interval $\hat{\sigma}'(x)$ associated with a variable $x$ is a base interval; a *base interval* is a minimal one, so it has the form $[k^-, k^+]$ (where $k^- \in \mathbf{K} \cup \{-\infty\}$, $k^+ \in \mathbf{K} \cup \{+\infty\}$ and $k^- \leq k^+$) and additionally it satisfies that $k^- + 1 \neq k^+$ and that there are no $k \in \mathbf{K}$ such that $k^- < k < k^+$. In the following we shall write $\mathsf{Base}(\mathbf{K})$ for the set of base intervals.

Try It Out 5.47: Determine $\mathsf{Base}(\mathbf{K})$ in the case where $\mathbf{K} = \{0, 1, 10\}$. □

The set of basic abstract memories for the interval analysis is now defined by

$$\mathsf{Basic}(\hat{\sigma}) = \{\hat{\sigma}' \sqsubseteq \hat{\sigma} \mid \forall x : \hat{\sigma}'_{\mathsf{V}}(x) \in \mathsf{Base}(\mathbf{K}) \wedge \forall A \in \mathbf{Arr} : \hat{\sigma}'_{\mathsf{A}}(A) = \hat{\sigma}_{\mathsf{A}}(A)\}$$

Note that we do not consider subintervals of the information for arrays as this information covers all the entries in the arrays.

For each of the abstract memories of $\mathsf{Basic}(\hat{\sigma})$ we can now determine whether or not a test may evaluate to true and hence whether they contribute to the resulting abstract memory. This is reflected in the following definition:

$$\begin{aligned} \widehat{\mathcal{S}}_{\mathsf{IA}}[\![b]\!]\hat{\sigma} &= \bigsqcup\{\hat{\sigma}' \in \mathsf{Basic}(\hat{\sigma}) \mid \mathsf{tt} \in \widehat{\mathcal{B}}_{\mathsf{IA}}[\![b]\!]\hat{\sigma}'\} \\ \widehat{\mathcal{S}}_{\mathsf{IA}}[\![\mathtt{skip}]\!]\hat{\sigma} &= \hat{\sigma} \end{aligned}$$

Here $\bigsqcup\{\hat{\sigma}_1, \cdots, \hat{\sigma}_n\} = \bot_{\mathsf{IA}} \sqcup_{\mathsf{IA}} \hat{\sigma}_1 \sqcup_{\mathsf{IA}} \cdots \sqcup_{\mathsf{IA}} \hat{\sigma}_n$ is the generalisation of $\sqcup_{\mathsf{IA}}$ to work on finite sets.

Try It Out 5.48: Assume $\mathbf{K} = \{-1, 0, 1\}$ and determine the result of $\widehat{\mathcal{S}}_{\mathsf{IA}}[\![x > 0]\!]\hat{\sigma}$ in the case where $\hat{\sigma}_{\mathsf{V}}(x) = [0, +\infty]$. What is the result of the function in the case where $\hat{\sigma}_{\mathsf{V}}(x) = [-\infty, 1]$? □

Exercise 5.49: Show that the analysis functions for the actions defined above are monotone; that is show that for all actions $\alpha$ and abstract memories $\hat{\sigma}_1$ and $\hat{\sigma}_2$ we have

$$\hat{\sigma}_1 \sqsubseteq_{\mathsf{IA}} \hat{\sigma}_2 \quad \text{implies} \quad \widehat{\mathcal{S}}_{\mathsf{IA}}[\![\alpha]\!]\hat{\sigma}_1 \sqsubseteq_{\mathsf{IA}} \widehat{\mathcal{S}}_{\mathsf{IA}}[\![\alpha]\!]\hat{\sigma}_2$$

In the proof you may exploit the results of Exercise 5.45. □

Essential Exercise 5.50: It is important that the analysis functions capture what happens when executing the actions in the semantics. Using the result of Exercise 5.46 show that for all actions $\alpha$, for all memories $\sigma$ and $\sigma'$ it is the case that

$$\text{if } \mathcal{S}[\![\alpha]\!]\sigma = \sigma' \qquad \text{then} \qquad \beta_{\mathsf{IA}}(\sigma') \sqsubseteq_{\mathsf{IA}} \widehat{\mathcal{S}}_{\mathsf{IA}}[\![\alpha]\!](\beta_{\mathsf{IA}}(\sigma))$$

**Constraints** An analysis assignment is a function $\mathsf{IA} : \mathbf{Q} \to \widehat{\mathbf{Mem}}_{\mathsf{IA}}$. We shall assume that the abstract memory of the initial node is given by $\hat{\sigma}_{\triangleright}$ and we shall derive constraints of the following form:

Whenever we have an edge $(q_\circ, \alpha, q_\bullet)$ in the program graph we will generate a constraint

$$\widehat{\mathcal{S}}_{\mathsf{IA}}[\![\alpha]\!](\mathsf{IA}(q_\circ)) \sqsubseteq_{\mathsf{IA}} \mathsf{IA}(q_\bullet)$$

Furthermore we have the following constraint for the initial node:

$$\hat{\sigma}_{\triangleright} \sqsubseteq_{\mathsf{IA}} \mathsf{IA}(q_{\triangleright})$$

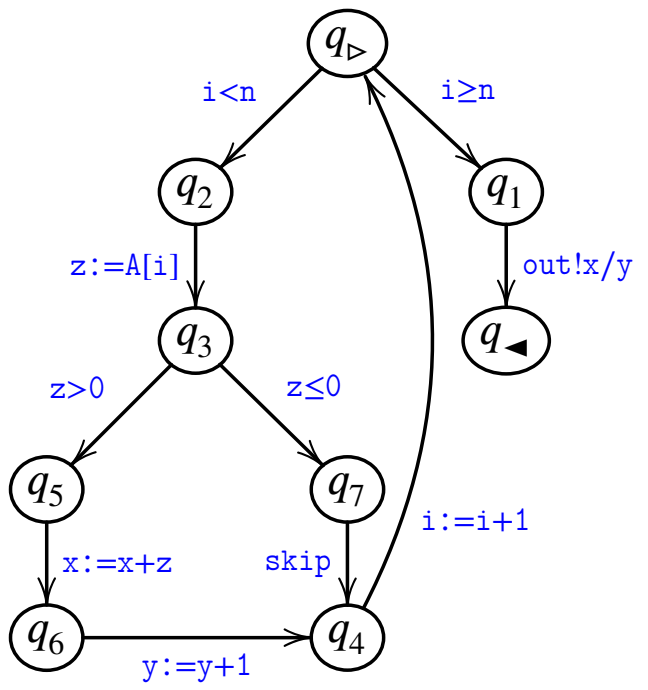

Figure 5.20: Computing the average of some of the elements of an array.

Example 5.51: Consider the program graph of Figure 5.20 computing the average of the positive entries of an array of length 10. Let us assume that $0 \in \mathbf{K}$ and take $\hat{\sigma}_{\triangleright}$ as in Figure 5.18. We can now analyse the program graph for different choices of $\mathbf{K}$; Figure 5.21 lists the analysis result obtained for the variable `i` at the final node $q_\blacktriangleleft$ for various choices of $\mathbf{K}$. Note that the most precise result is obtained when $\mathbf{K}$ contains the integers identifying the borderline of the test on `i` which in this case is 9 and 10.

| $\mathbf{K}$ | $\mathsf{IA}(q_\blacktriangleleft)(\mathtt{i})$ |
|---|---|
| $\{0\}$ | $[0, +\infty]$ |
| $\{0, 1\}$ | $[1, +\infty]$ |
| $\{0, 9, 10\}$ | $[10, +\infty]$ |
| $\{0, 1, 9, 10\}$ | $[10, 10]$ |

Figure 5.21: Analysis results for the variable `i` at the node $q_\blacktriangleleft$ for various choices of $\mathbf{K}$.

Exercise 5.52: Show that the development of Chapter 3 suffices for computing the least solution to the constraints; in particular check that all the conditions for Proposition 3.28 are satisfied. □

The correctness of the constraints amounts to ensuring that whenever the execution reaches the node $q$, then the values of the variables and arrays will be within the intervals predicted by the analysis.

Proposition 5.53: If IA is a solution to the constraints, and if $\pi$ is a path from $q_\triangleright$ to $q$, then $\beta_{\mathsf{IA}}(\sigma) \sqsubseteq \mathsf{IA}(q)$ whenever $\sigma = \mathcal{S}[\![\pi]\!]\sigma_\triangleright$ and $\beta_{\mathsf{IA}}(\sigma_\triangleright) \sqsubseteq \hat{\sigma}_\triangleright$.

# Chapter 6

# Abstract Interpretation

In this chapter we develop the basics of abstract interpretation. Section 6.1 covers *abstraction* and *concretisation* functions for relating analysis domains as well as their intended relationships of being *Galois connections* or *adjunctions*. Section 6.2 introduces a chaotic iteration algorithm based on *widenings* for dealing with analysis domains that do not satisfy the ascending chain condition. Section 6.3 shows how an analysis specification can be transferred to a simpler analysis domain and that correctness of solutions carries over; this is used to develop the Detection of Signs Analysis of Section 5.1 from a *Collecting Semantics*. Finally, Section 6.4 shows how to systematically develop a relational version of the Detection of Signs Analysis.

## 6.1 Abstraction and Concretisation Functions

In this section we shall consider how to relate one analysis domain $\widehat{\mathbf{D}}$ with another $\widehat{\mathbf{D}'}$ that we consider simpler. Throughout we shall merely assume that $\widehat{\mathbf{D}}$ and $\widehat{\mathbf{D}'}$ are pointed semi-lattices and not require that they satisfy the ascending chain condition. We saw that the ascending chain condition is sufficient for ensuring that the Chaotic Iteration algorithm terminates, but while developing an analysis it may be useful to have interim analysis domains that are more complex, and even if an analysis domain satisfies the ascending chain condition there may be a need to replace it with a simpler one.

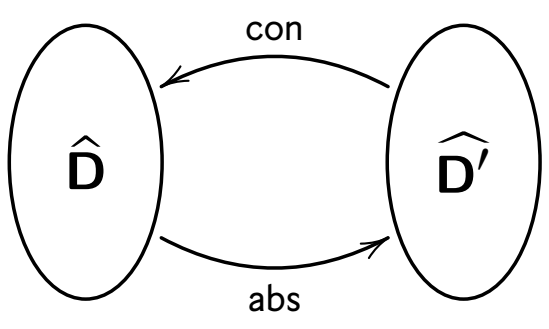

Figure 6.1: Abstraction and concretisation functions.

To relate $\widehat{\mathbf{D}}$ with another $\widehat{\mathbf{D}'}$ it is customary to make use of an *abstraction function* $\mathsf{abs} : \widehat{\mathbf{D}} \to \widehat{\mathbf{D}'}$ and a *concretisation function* $\mathsf{con} : \widehat{\mathbf{D}'} \to \widehat{\mathbf{D}}$ as shown in Figure 6.1. The abstraction function tells us the 'best way' of representing an element of $\widehat{\mathbf{D}}$ in $\widehat{\mathbf{D}'}$ and the concretisation function tells us what an element of $\widehat{\mathbf{D}'}$ means in $\widehat{\mathbf{D}}$.

EXAMPLE 6.1: Borrowing from Section 5.1 let us take $\widehat{\mathbf{D}} = \mathsf{PowerSet}(\mathbf{Int})$ and $\widehat{\mathbf{D}'} = \mathsf{PowerSet}(\{-, 0, +\})$, and recall the function sign given by

$$\mathsf{sign}(z) = \begin{cases} - & \text{if } z < 0 \\ 0 & \text{if } z = 0 \\ + & \text{if } z > 0 \end{cases}$$

for producing the sign of an integer.

Here we may take $\mathsf{abs}(Z) = \{\mathsf{sign}(z) \mid z \in Z\}$ and $\mathsf{con}(S) = \{z \mid \mathsf{sign}(z) \in S\}$.

For this to work the functions need to be suitably related. In Definitions 6.2 and 6.4 we shall consider two ways of doing so and then show in Proposition 6.6 that they are actually equivalent. Having this dual description is very useful for 'getting our intuitions right' and for proofs.

DEFINITION 6.2: A pair $(\mathsf{abs}, \mathsf{con})$ of functions between pointed semi-lattices $\widehat{\mathbf{D}}$ and $\widehat{\mathbf{D}'}$ constitutes a *Galois connection* from $\widehat{\mathbf{D}}$ to $\widehat{\mathbf{D}'}$ when the following properties hold:

- $\mathsf{con} \circ \mathsf{abs} \sqsupseteq \mathsf{id}$
- $\mathsf{abs} \circ \mathsf{con} \sqsubseteq \mathsf{id}$
- $\mathsf{abs} : \widehat{\mathbf{D}} \to \widehat{\mathbf{D}'}$ is monotone
- $\mathsf{con} : \widehat{\mathbf{D}'} \to \widehat{\mathbf{D}}$ is monotone

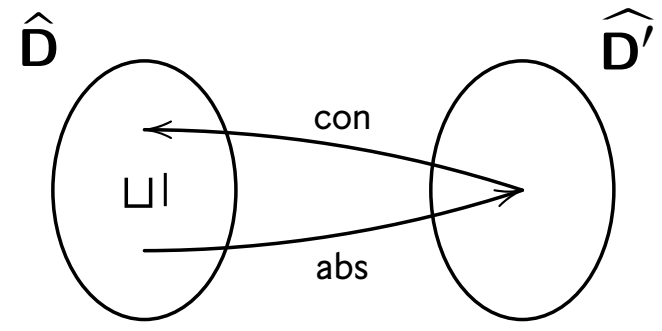

Figure 6.2: $\mathsf{con} \circ \mathsf{abs} \sqsupseteq \mathsf{id}$.

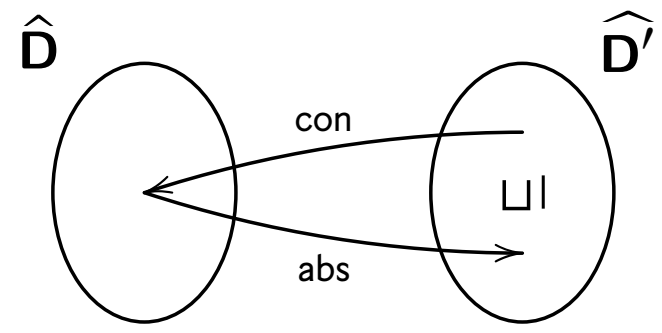

Figure 6.3: $\mathsf{abs} \circ \mathsf{con} \sqsubseteq \mathsf{id}$.

The first condition says that the 'best way' of representing an element of $\widehat{\mathbf{D}}$ in $\widehat{\mathbf{D}'}$ is in fact correct as illustrated in Figure 6.2: if you have an analysis property $\hat{d}$ and take its 'best representation' $\mathsf{abs}(\hat{d})$ then its meaning $\mathsf{con}(\mathsf{abs}(\hat{d}))$ must contain all of $\hat{d}$. The second condition says that the 'best way' actually is best as illustrated in Figure 6.3: if you have an analysis property $\hat{d}'$ and consider its meaning $\mathsf{con}(\hat{d}')$ then the 'best representation' of that is $\mathsf{abs}(\mathsf{con}(\hat{d}'))$ which must be contained in $\hat{d}'$. The third and fourth conditions say that the abstraction and concretisation functions respect our intuitions about the partial orders: if $\hat{d}_1$ is contained in $\hat{d}_2$ then also the 'best representation' $\mathsf{abs}(\hat{d}_1)$ must be contained in the 'best representation' $\mathsf{abs}(\hat{d}_2)$, and similarly if $\hat{d}'_1$ is contained in $\hat{d}'_2$ then also the meaning $\mathsf{con}(\hat{d}'_1)$ must be contained in the meaning $\mathsf{con}(\hat{d}'_2)$.

TRY IT OUT 6.3: Referring back to Example 6.1 show that $(\mathsf{abs}, \mathsf{con})$ is a Galois connection. □

DEFINITION 6.4: A pair (abs, con) of functions between pointed semi-lattices $\widehat{\mathbf{D}}$ and $\widehat{\mathbf{D'}}$ constitutes an *adjunction* from $\widehat{\mathbf{D}}$ to $\widehat{\mathbf{D'}}$ when

$$\mathsf{abs}(\hat{d}) \sqsubseteq \hat{d}' \Leftrightarrow \hat{d} \sqsubseteq \mathsf{con}(\hat{d}')$$

holds for all choices of $\hat{d}$ and $\hat{d}'$.

This is another way of saying that the abstraction and concretisation functions are intimately connected with the partial orders. It says that comparing $\hat{d} \in \widehat{\mathbf{D}}$ with $\hat{d}' \in \widehat{\mathbf{D'}}$ in $\widehat{\mathbf{D}}$ gives the same result as doing it in $\widehat{\mathbf{D'}}$ as illustrated in Figure 6.4: if you want to compare $\hat{d} \in \widehat{\mathbf{D}}$ with $\hat{d}' \in \widehat{\mathbf{D}}'$ you can move $\hat{d}'$ to $\widehat{\mathbf{D}}$ and check if $\hat{d}$ is contained in $\mathsf{con}(\hat{d}')$, or you can move $\hat{d}$ to $\widehat{\mathbf{D}}'$ and check if $\mathsf{abs}(\hat{d})$ is contained in $\hat{d}'$, and since there seems no reason to prefer one over the other we demand that the two checks give the same result.

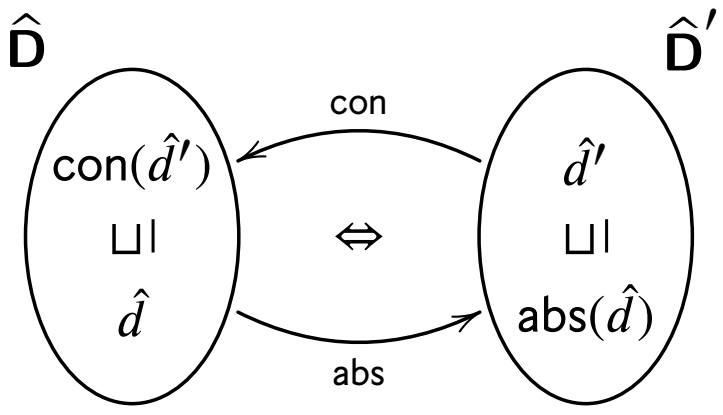

Figure 6.4: $\mathsf{abs}(\hat{d}) \sqsubseteq \hat{d}' \Leftrightarrow \hat{d} \sqsubseteq \mathsf{con}(\hat{d}')$.

TRY IT OUT 6.5: Referring back to Example 6.1 show that (abs, con) is an adjunction. □

PROPOSITION 6.6: A pair (abs, con) of functions between pointed semi-lattices $\widehat{\mathbf{D}}$ and $\widehat{\mathbf{D'}}$ constitutes a Galois connection if and only if it constitutes an adjunction.

PROOF: First suppose that (abs, con) is a Galois connection and make the following calculations (forall $\hat{d}$ and $\hat{d}'$)

$$\begin{array}{lllll} \mathsf{abs}(\hat{d}) \sqsubseteq \hat{d}' & \Rightarrow & \mathsf{con}(\mathsf{abs}(\hat{d})) \sqsubseteq \mathsf{con}(\hat{d}') & \Rightarrow & \hat{d} \sqsubseteq \mathsf{con}(\hat{d}') \\ \hat{d} \sqsubseteq \mathsf{con}(\hat{d}') & \Rightarrow & \mathsf{abs}(\hat{d}) \sqsubseteq \mathsf{abs}(\mathsf{con}(\hat{d}')) & \Rightarrow & \mathsf{abs}(\hat{d}) \sqsubseteq \hat{d}' \end{array}$$

showing that (abs, con) is an adjunction.

Next suppose that (abs, con) is an adjunction and make the following calculations (forall $\hat{d}$ and $\hat{d}'$)

$$\begin{array}{lllllll} & & \mathsf{abs}(\hat{d}) \sqsubseteq \mathsf{abs}(\hat{d}) & \Rightarrow & \hat{d} \sqsubseteq \mathsf{con}(\mathsf{abs}(\hat{d})) & & \text{shows con} \circ \text{abs} \sqsupseteq \text{id} \\ & & \mathsf{con}(\hat{d}') \sqsubseteq \mathsf{con}(\hat{d}') & \Rightarrow & \mathsf{abs}(\mathsf{con}(\hat{d}')) \sqsubseteq \hat{d}' & & \text{shows abs} \circ \text{con} \sqsubseteq \text{id} \\ \hat{d}_1 \sqsubseteq \hat{d}_2 & \Rightarrow & \hat{d}_1 \sqsubseteq \mathsf{con}(\mathsf{abs}(\hat{d}_2)) & \Rightarrow & \mathsf{abs}(\hat{d}_1) \sqsubseteq \mathsf{abs}(\hat{d}_2) & & \text{shows abs monotone} \\ \hat{d}'_1 \sqsubseteq \hat{d}'_2 & \Rightarrow & \mathsf{abs}(\mathsf{con}(\hat{d}'_1)) \sqsubseteq \hat{d}'_2 & \Rightarrow & \mathsf{con}(\hat{d}'_1) \sqsubseteq \mathsf{con}(\hat{d}'_2) & & \text{shows con monotone} \end{array}$$

to show that (abs, con) is a Galois connection. □

EXAMPLE 6.7: (For those who have read Section 5.3.) In Section 5.3 we considered how to perform an interval analysis. It might have been natural to make use of the analysis domain

$$\textbf{Interval} = \{\bot\} \cup \{[z^-, z^+] \mid z^- \in \textbf{Int} \cup \{-\infty\}, z^+ \in \textbf{Int} \cup \{+\infty\}, z^- \leq z^+\}$$

where $\leq$ is extended to work on the set $\textbf{Int} \cup \{-\infty, +\infty\}$ of integers extended with symbols $-\infty$ and $+\infty$. This is a pointed semi-lattice but it does *not* satisfy the ascending chain condition.

So instead we considered the analysis domain

$$\textbf{Interval}_{\textbf{K}} = \{\bot\} \cup \{[k^-, k^+] \mid k^- \in \textbf{K} \cup \{-\infty\}, k^+ \in \textbf{K} \cup \{+\infty\}, k^- \leq k^+\}$$

for some *finite* set $\textbf{K}$ of integers. As discussed in Section 5.3 this is a pointed semi-lattice that satisfies the ascending chain condition.

We can relate these analysis domains by defining a Galois connection $(\mathsf{abs}, \mathsf{con})$ from **Interval** to $\textbf{Interval}_{\textbf{K}}$. The definition of the concretisation function is straightforward:

$$\begin{array}{rcl} \mathsf{con}(\bot) & = & \bot \\ \mathsf{con}([k^-, k^+]) & = & [k^-, k^+] \end{array}$$

The definition of the abstraction function requires a bit more care

$$\begin{array}{rcl} \mathsf{abs}(\bot) & = & \bot \\ \mathsf{abs}([z^-, z^+]) & = & [\lfloor z^- \rfloor, \lceil z^+ \rceil] \end{array}$$

and uses the functions $\lfloor z \rfloor = \sup\{k \in \textbf{K} \mid k \leq z\}$ and $\lceil z \rceil = \inf\{k \in \textbf{K} \mid z \leq k\}$ defined in Section 5.3. It is straightforward to check that this is indeed a Galois connection.

We shall return to this example in Example 6.23.

Abstraction functions and concretisation functions can be seen as 'weak inverses' of one another (see Teaser 6.14) and satisfy a number of useful properties.

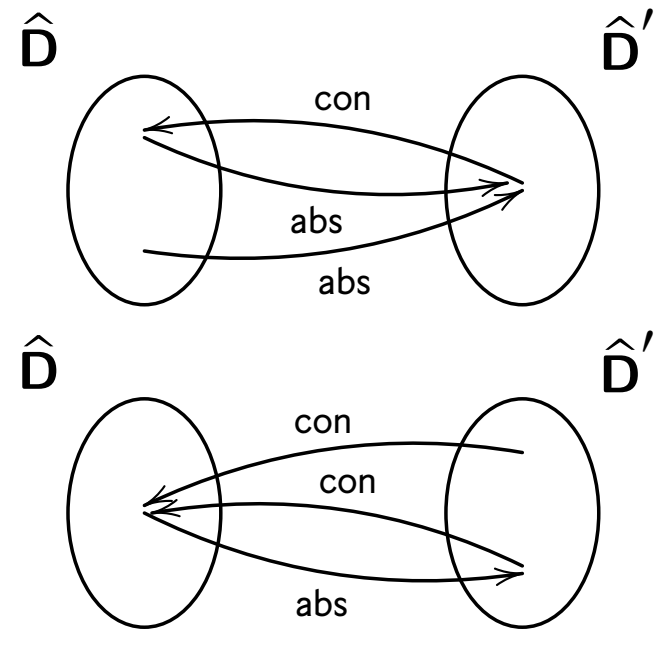

Figure 6.5: Illustrating the equations of Proposition 6.8.

PROPOSITION 6.8: For a Galois connection $(\mathsf{abs}, \mathsf{con})$ we have the equations $\mathsf{abs} \circ \mathsf{con} \circ \mathsf{abs} = \mathsf{abs}$ and $\mathsf{con} \circ \mathsf{abs} \circ \mathsf{con} = \mathsf{con}$.

PROOF: We perform the following calculations

$$\begin{array}{rclclcl} \mathsf{abs} \circ \mathsf{con} \circ \mathsf{abs} & = & (\mathsf{abs} \circ \mathsf{con}) \circ \mathsf{abs} & \sqsubseteq & \mathsf{id} \circ \mathsf{abs} & = & \mathsf{abs} \\ \mathsf{abs} \circ \mathsf{con} \circ \mathsf{abs} & = & \mathsf{abs} \circ (\mathsf{con} \circ \mathsf{abs}) & \sqsupseteq & \mathsf{abs} \circ \mathsf{id} & = & \mathsf{abs} \\ \mathsf{con} \circ \mathsf{abs} \circ \mathsf{con} & = & (\mathsf{con} \circ \mathsf{abs}) \circ \mathsf{con} & \sqsupseteq & \mathsf{id} \circ \mathsf{con} & = & \mathsf{con} \\ \mathsf{con} \circ \mathsf{abs} \circ \mathsf{con} & = & \mathsf{con} \circ (\mathsf{abs} \circ \mathsf{con}) & \sqsubseteq & \mathsf{con} \circ \mathsf{id} & = & \mathsf{con} \end{array}$$

showing that $\mathsf{abs} \circ \mathsf{con} \circ \mathsf{abs} = \mathsf{abs}$ and $\mathsf{con} \circ \mathsf{abs} \circ \mathsf{con} = \mathsf{con}$. □

> PROPOSITION 6.9: For a Galois connection (abs, con) the abstraction function abs preserves finite least upper bounds:
>
> - $\mathsf{abs}(\bot) = \bot$ and
> - $\mathsf{abs}(\hat{d}_1 \sqcup \hat{d}_2) = \mathsf{abs}(\hat{d}_1) \sqcup \mathsf{abs}(\hat{d}_2)$

In fact $\mathsf{abs}(\bigsqcup D) = \bigsqcup\{\mathsf{abs}(\hat{d}) \mid \hat{d} \in D\}$ when both sides exist.

PROOF: We shall feel free to use Proposition 6.6 without explicitly mentioning it. First we calculate

$$\bot \sqsubseteq \mathsf{con}(\bot) \Rightarrow \mathsf{abs}(\bot) \sqsubseteq \bot$$

so as to show that $\mathsf{abs}(\bot) = \bot$. Next we calculate (forall $\hat{d}$ and $\hat{d}'$)

$$\begin{array}{c} \hat{d}_1 \sqsubseteq \mathsf{con}(\mathsf{abs}(\hat{d}_1)) \wedge \hat{d}_2 \sqsubseteq \mathsf{con}(\mathsf{abs}(\hat{d}_2)) \\ \Downarrow \\ \hat{d}_1 \sqsubseteq \mathsf{con}(\mathsf{abs}(\hat{d}_1) \sqcup \mathsf{abs}(\hat{d}_2)) \wedge \hat{d}_2 \sqsubseteq \mathsf{con}(\mathsf{abs}(\hat{d}_1) \sqcup \mathsf{abs}(\hat{d}_2)) \\ \Downarrow \\ \hat{d}_1 \sqcup \hat{d}_2 \sqsubseteq \mathsf{con}(\mathsf{abs}(\hat{d}_1) \sqcup \mathsf{abs}(\hat{d}_2)) \\ \Downarrow \\ \mathsf{abs}(\hat{d}_1 \sqcup \hat{d}_2) \sqsubseteq \mathsf{abs}(\hat{d}_1) \sqcup \mathsf{abs}(\hat{d}_2) \end{array}$$

and

$$\mathsf{abs}(\hat{d}_1) \sqcup \mathsf{abs}(\hat{d}_2) \sqsubseteq \mathsf{abs}(\hat{d}_1 \sqcup \hat{d}_2) \sqcup \mathsf{abs}(\hat{d}_1 \sqcup \hat{d}_2) = \mathsf{abs}(\hat{d}_1 \sqcup \hat{d}_2)$$

so as to show that $\mathsf{abs}(\hat{d}_1 \sqcup \hat{d}_2) = \mathsf{abs}(\hat{d}_1) \sqcup \mathsf{abs}(\hat{d}_2)$. □

EXERCISE 6.10: Suppose that we have Galois connections $(\mathsf{abs}_1, \mathsf{con}_1)$ between $\widehat{\mathbf{D}}_1$ and $\widehat{\mathbf{D}'_1}$ and similarly $(\mathsf{abs}_2, \mathsf{con}_2)$ between $\widehat{\mathbf{D}}_2$ and $\widehat{\mathbf{D}'_2}$. Construct a Galois connection between $\widehat{\mathbf{D}}_1 \times \widehat{\mathbf{D}}_2$ and $\widehat{\mathbf{D}'_1} \times \widehat{\mathbf{D}'_2}$ and show that it indeed is a Galois conncection. □

EXERCISE 6.11: Suppose that we have a Galois connection (abs, con) between $\widehat{\mathbf{D}}$ and $\widehat{\mathbf{D}'}$ and that $\mathbf{V}$ is a (finite and non-empty) set. Construct a Galois connection between $\mathbf{V} \to \widehat{\mathbf{D}}$ and $\mathbf{V} \to \widehat{\mathbf{D}'}$ and show that it indeed is a Galois conncection. □

EXERCISE 6.12: Suppose that we have Galois connections $(\mathsf{abs}_1, \mathsf{con}_1)$ between $\widehat{\mathbf{D}}_1$ and $\widehat{\mathbf{D}}_2$ and similarly $(\mathsf{abs}_2, \mathsf{con}_2)$ between $\widehat{\mathbf{D}}_2$ and $\widehat{\mathbf{D}}_3$. Show that

$$(\mathsf{abs}_2 \circ \mathsf{abs}_1,\ \mathsf{con}_1 \circ \mathsf{con}_2)$$

is a Galois connection. □

TEASER 6.13: Suppose that we have Galois connections $(\mathsf{abs}_1, \mathsf{con}_1)$ between $\widehat{\mathbf{D}}_1$ and $\widehat{\mathbf{D}'_1}$ and similarly $(\mathsf{abs}_2, \mathsf{con}_2)$ between $\widehat{\mathbf{D}}_2$ and $\widehat{\mathbf{D}'_2}$. Let $\widehat{\mathbf{D}}_1 \to_m \widehat{\mathbf{D}}_2$ be the set of monotone functions from $\widehat{\mathbf{D}}_1$ to $\widehat{\mathbf{D}}_2$ and similarly for $\widehat{\mathbf{D}'_1} \to_m \widehat{\mathbf{D}'_2}$. Construct a Galois connection between $\widehat{\mathbf{D}}_1 \to_m \widehat{\mathbf{D}}_2$ and $\widehat{\mathbf{D}'_1} \to_m \widehat{\mathbf{D}'_2}$ and show that it indeed is a Galois conncection. □

TEASER 6.14: Show that the either component of a Galois connection (abs, con) between $\widehat{\mathbf{D}}$ and $\widehat{\mathbf{D}'}$ uniquely determines the other:

- show that if $(\mathsf{abs}_1, \mathsf{con})$ and $(\mathsf{abs}_2, \mathsf{con})$ are Galois connections between $\widehat{\mathbf{D}}$ and $\widehat{\mathbf{D}'}$ then $\mathsf{abs}_1 = \mathsf{abs}_2$, and

- show that if $(\mathsf{abs}, \mathsf{con}_1)$ and $(\mathsf{abs}, \mathsf{con}_2)$ are Galois connections between $\widehat{\mathbf{D}}$ and $\widehat{\mathbf{D}'}$ then $\mathsf{con}_1 = \mathsf{con}_2$. □

In the literature (abs, con) is often written $(\alpha, \gamma)$ – but we already used $\alpha$ for actions in program graphs.

## 6.2 Inducing along the Concretisation Function

In this section we shall show how to use a Galois connection from a pointed semi-lattice $\widehat{\mathbf{D}}$ to another $\widehat{\mathbf{D}'}$ to 'speed up' some of the calculations on $\widehat{\mathbf{D}}$ using properties of $\widehat{\mathbf{D}'}$.

Let us recall the Chaotic Iteration algorithm studied throughout the book for constructing the least solution for a forward analysis problem. It is essentially the algorithm of Figure 6.7 where the operator $\nabla$ is taken to be $\sqcup$ (the least upper bound operation in $\widehat{\mathbf{D}}$). In case $\widehat{\mathbf{D}}$ does not satisfy the ascending chain condition the algorithm might not terminate, and even if $\widehat{\mathbf{D}}$ does satisfy the ascending chain condition its height might be huge so that the algorithm requires more iterations than we would like. The approach taken in this section is to use another operator than $\sqcup$ for $\nabla$.

An operator $\nabla : \widehat{\mathbf{D}} \times \widehat{\mathbf{D}} \to \widehat{\mathbf{D}}$ does not have to be commutative nor associative. We shall therfore assume that it associates to the left, so that $\hat{d}_1 \nabla \hat{d}_2 \nabla \hat{d}_3 = (\hat{d}_1 \nabla \hat{d}_2) \nabla \hat{d}_3$, and similarly for $\hat{d}_1 \nabla \cdots \nabla \hat{d}_n$.

We shall say that the sequence $(\hat{d}_1 \nabla \cdots \nabla \hat{d}_n)_n$ *eventually stabilises* whenever there is a number $N$ such that $\hat{d}_1 \nabla \cdots \nabla \hat{d}_n = \hat{d}_1 \nabla \cdots \nabla \hat{d}_n \nabla \hat{d}_{n+1}$ for all $n > N$.

We are now ready to formulate our requirements on $\nabla$.

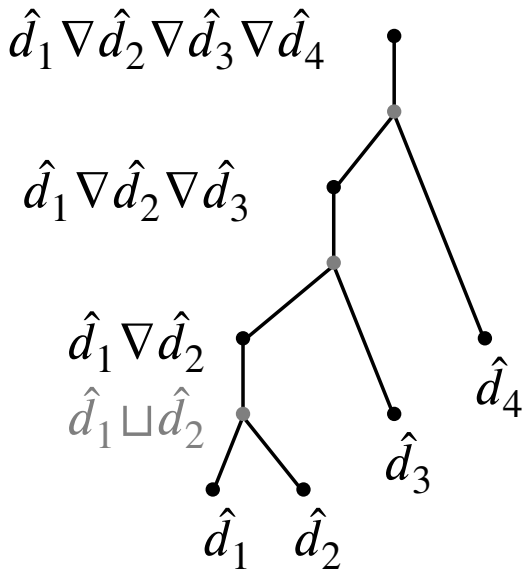

Figure 6.6: Illustrating the first condition of strong widenings.

DEFINITION 6.15: An operator $\nabla : \widehat{\mathbf{D}} \times \widehat{\mathbf{D}} \to \widehat{\mathbf{D}}$ is a *strong widening* whenever

- $\hat{d}_1 \nabla \hat{d}_2 \sqsupseteq \hat{d}_1 \sqcup \hat{d}_2$ holds for all $\hat{d}_1, \hat{d}_2 \in \widehat{\mathbf{D}}$, and

- the sequence $(\hat{d}_1 \nabla \cdots \nabla \hat{d}_n)_n$ eventually stabilises for all choices of sequences $\hat{d}_1, \cdots, \hat{d}_n, \cdots$.

| | |
|---|---|
| INPUT | a program graph $\mathbf{PG} = (\mathbf{Q}, q_{\triangleright}, q_{\blacktriangleleft}, \mathbf{Act}, \mathbf{E})$ |
| | an analysis specification with $\widehat{\mathbf{D}}$, $\widehat{\mathcal{S}}[\![q_\circ, \alpha, q_\bullet]\!]$, $\hat{d}_\diamond$ and $\nabla$ |
| OUTPUT | a solution AA to the forward analysis problem |
| METHOD | forall $q \in \mathbf{Q} \setminus \{q_\triangleright\}$ do $\mathsf{AA}(q) := \bot$ ; |
| | $\mathsf{AA}(q_\triangleright) := \hat{d}_\diamond$ ; |
| | while there exists an edge $(q_\circ, \alpha, q_\bullet) \in \mathbf{E}$ |
| | such that $\widehat{\mathcal{S}}[\![q_\circ, \alpha, q_\bullet]\!]\big(\mathsf{AA}(q_\circ)\big) \not\sqsubseteq \mathsf{AA}(q_\bullet)$ |
| | do $\mathsf{AA}(q_\bullet) := \mathsf{AA}(q_\bullet)\, \nabla\, \widehat{\mathcal{S}}[\![q_\circ, \alpha, q_\bullet]\!]\big(\mathsf{AA}(q_\circ)\big)$ |

Figure 6.7: *Chaotic Iteration with Widening* for forward analyses.

TRY IT OUT 6.16: Show that if $\nabla$ is a strong widening then

$$\hat{d}_1 \nabla \cdots \nabla \hat{d}_n \sqsubseteq \hat{d}_1 \nabla \cdots \nabla \hat{d}_n \nabla \hat{d}_{n+1}$$

for all $n > 0$. □

TRY IT OUT 6.17: Show that if $\widehat{\mathbf{D}}$ satisfies the ascending chain condition then $\sqcup$ is a strong widening. □

PROPOSITION 6.18: If the analysis domain is a pointed semi-lattice and $\nabla$ is a strong widening then the algorithm of Figure 6.7 terminates with an analysis assignment AA that solves the constraints of the analysis problem.

If additionally the analysis functions are monotone then the analysis assignment AA summarises the paths.

PROOF: The first statement is analogous to those of Propositions 3.12 and 3.16. The key idea is that each $\mathsf{AA}(q)$ takes the form $\hat{d}_1 \nabla \cdots \nabla \hat{d}_n$ for suitable choices of $\hat{d}_1, \cdots, \hat{d}_n$. The second result is analogous to that of Proposition 3.29. □

We do not claim that the algorithm computes the least solution nor that the solution produced is independent of the non-deterministic choices made when performing the algorithm. For efficiency, all of the ideas from Chapter 4 may be used. Also the use of strong widening applies equally well to backward analyses, and analyses aiming for the greatest solution, using the ideas of Sections 3.4 and 4.4.

In the literature one more frequently makes use of the notion of *widening* where the chain $(\hat{d}_1 \nabla \cdots \nabla \hat{d}_n)_n$ is only required to eventually stabilise for all sequences $\hat{d}_1, \cdots, \hat{d}_n, \cdots$ satisfying $\hat{d}_1 \sqsubseteq \cdots \sqsubseteq \hat{d}_n \sqsubseteq \cdots$. This is suitable for algorithms computing solutions to equations but less so for algorithms computing solutions to constraints where more than one constraint may influence the same piece of analysis information.

It remains to show how Galois connections can be used to construct strong widenings. So consider a Galois connection (abs, con) from the pointed semi-lattice $\widehat{\mathbf{D}}$ to the pointed semi-lattice $\widehat{\mathbf{D}'}$ and define

$$\hat{d}_1 \nabla \hat{d}_2 = \mathsf{con}\left(\mathsf{abs}(\hat{d}_1) \sqcup \mathsf{abs}(\hat{d}_2)\right)$$

Try It Out 6.19: Referring back to Example 6.1 define the function

$$\nabla : \mathsf{PowerSet}(\mathbf{Int}) \times \mathsf{PowerSet}(\mathbf{Int}) \to \mathsf{PowerSet}(\mathbf{Int})$$

by setting $\hat{d}_1 \nabla \hat{d}_2 = \mathsf{con}\left(\mathsf{abs}(\hat{d}_1) \sqcup \mathsf{abs}(\hat{d}_2)\right)$ as indicated.

Work out the values of $\{-5\}\nabla\{-3\}$ and $\{-5\}\nabla\{+3\}$. □

Using Propositions 6.8 and 6.9 we may calculate $\hat{d}_1 \nabla \hat{d}_2 = \mathsf{con}(\mathsf{abs}(\hat{d}_1 \sqcup \hat{d}_2))$ and $\mathsf{abs}(\hat{d}_1 \nabla \hat{d}_2) = \mathsf{abs}(\hat{d}_1) \sqcup \mathsf{abs}(\hat{d}_2)$.

Essential Exercise 6.20: Show $\hat{d}_1 \nabla \cdots \nabla \hat{d}_n = \mathsf{con}\left(\mathsf{abs}(\hat{d}_1) \sqcup \cdots \sqcup \mathsf{abs}(\hat{d}_n)\right)$ for all $n > 1$ by induction on $n$ (with $n = 2$ as base case).

Proposition 6.21: If (abs, con) is a Galois connection from $\widehat{\mathbf{D}}$ to $\widehat{\mathbf{D}'}$ and $\widehat{\mathbf{D}'}$ satisfies the ascending chain condition then $\nabla$ defined by

$$\hat{d}_1 \nabla \hat{d}_2 = \mathsf{con}\left(\mathsf{abs}(\hat{d}_1) \sqcup \mathsf{abs}(\hat{d}_2)\right)$$

is a strong widening.

Proof: Suppose that there is a sequence $\hat{d}_1, \cdots, \hat{d}_n, \cdots$ such that $(\hat{d}_1 \nabla \cdots \nabla \hat{d}_n)_n$ never eventually stabilises. Then by Essential Exercise 6.20 we have a sequence $\mathsf{abs}(\hat{d}_1), \cdots, \mathsf{abs}(\hat{d}_n), \cdots$ such that $(\mathsf{abs}(\hat{d}_1) \sqcup \cdots \sqcup \mathsf{abs}(\hat{d}_n))_n$ never eventually stabilises and this contradicts the assumption that $\widehat{\mathbf{D}'}$ satisfies the ascending chain condition. □

Teaser 6.22: Suppose that (abs, con) is a Galois connection from the pointed semi-lattice $\widehat{\mathbf{D}}$ to the pointed semi-lattice $\widehat{\mathbf{D}'}$ and that $\nabla'$ is a strong widening on $\widehat{\mathbf{D}'}$. Define $\hat{d}_1 \nabla \hat{d}_2 = \mathsf{con}\left(\mathsf{abs}(\hat{d}_1) \nabla' \mathsf{abs}(\hat{d}_2)\right)$ and investigate what additional requirements on (abs, con) are needed for showing that $\nabla$ is a strong widening. □

Example 6.23: (For those who have read Section 5.3.) Let us continue Example 6.7 where we defined a Galois connection connection (abs, con) from **Interval** to $\mathbf{Interval_K}$. We define $\nabla : \mathbf{Interval} \times \mathbf{Interval} \to \mathbf{Interval}$ from the least upper bound $\sqcup$ : $\mathbf{Interval_K} \times \mathbf{Interval_K} \to \mathbf{Interval_K}$ using the familiar formula

$$int_1 \nabla int_2 = \mathsf{con}\left(\mathsf{abs}(int_1) \sqcup \mathsf{abs}(int_2)\right)$$

This gives rise to a strong widening over **Interval**. Hence we could perform our interval analysis over **Interval** (rather than $\mathbf{Interval_K}$) provided we use Chaotic Iteration with Widening as in Figure 6.7.

We can use our characterisation of $\sqcup$ in Section 5.3 to provide a characterisation of the strong widening as follows:

$$int_1 \nabla int_2 = \begin{cases} \bot & \text{if } int_1 = int_2 = \bot \\ [\lfloor z^- \rfloor, \lceil z^+ \rceil] & \text{if } int_1 = [z^-, z^+] \text{ and } int_2 = \bot \\ [\lfloor z^- \rfloor, \lceil z^+ \rceil] & \text{if } int_1 = \bot \text{ and } int_2 = [z^-, z^+] \\ [\lfloor z^- \rfloor, \lceil z^+ \rceil] & \text{if } int_1 = [z_1^-, z_1^+] \text{ and } int_2 = [z_2^-, z_2^+] \\ & \text{and } z^- = \min\{z_1^-, z_2^-\} \\ & \phantom{\text{and }} z^+ = \max\{z_1^+, z_2^+\} \end{cases}$$

Here we have used that $(\mathsf{con} \circ \mathsf{abs})(\bot) = \bot$ and that $(\mathsf{con} \circ \mathsf{abs})([z^-, z^+]) = [\lfloor z^- \rfloor, \lceil z^+ \rceil]$.

Since $\mathsf{con} \circ \mathsf{abs}$ is not the identity it is often preferable to let the strong widening be more precise in case one of the arguments is $\bot$. To see this note that our Chaotic Iteration with Widening algorithm initialises the analysis information to $\bot$ and applies widening whenever exploration of an edge gives rise to updating the analysis information; this involves applying $\mathsf{con} \circ \mathsf{abs}$ to the analysis information and we would like to avoid doing so the first time that the analysis information is updated. This motivates the following approach.

PROPOSITION 6.24: If $\nabla$ is a strong widening on a pointed semi-lattice $\hat{\mathbf{D}}$ then $\nabla'$ defined by

$$\hat{d}_1 \nabla' \hat{d}_2 = \begin{cases} \hat{d}_1 & \text{if } \hat{d}_2 = \bot \\ \hat{d}_2 & \text{if } \hat{d}_1 = \bot \\ \hat{d}_1 \nabla \hat{d}_2 & \text{otherwise} \end{cases}$$

is also a strong widening on $\hat{\mathbf{D}}$.

PROOF: It is immediate that $\hat{d}_1 \nabla' \hat{d}_2 \sqsupseteq \hat{d}_1 \sqcup \hat{d}_2$.

Next we proceed by way of contradiction and suppose that there is a sequence $\hat{d}_1, \cdots, \hat{d}_n, \cdots$ such that the sequence $(\hat{d}_1 \nabla' \cdots \nabla' \hat{d}_n)_n$ does not eventually stabilise. Then infinitely many of the elements in $\hat{d}_1, \cdots, \hat{d}_n, \cdots$ must be different from $\bot$ and let $\hat{d}'_1, \cdots, \hat{d}'_n, \cdots$ be the infinite subsequence of non-$\bot$ elements in $\hat{d}_1, \cdots, \hat{d}_n, \cdots$. Then also $(\hat{d}'_1 \nabla' \cdots \nabla' \hat{d}'_n)_n$ does not eventually stabilise. But since $\hat{d}'_1 \nabla' \cdots \nabla' \hat{d}'_n = \hat{d}'_1 \nabla \cdots \nabla \hat{d}'_n$ this contradicts our assumption that $\nabla$ is a widening. □

Exercise 6.25: (For those who have read Section 5.3.) Suppose that $\nabla$ : **Interval**× **Interval** → **Interval** is defined as in Example 6.23 and that $\nabla'$ : **Interval** × **Interval** → **Interval** is defined as in Proposition 6.24. Provide a characterisation of $\nabla'$ along the lines of the characterisation of $\nabla$ worked out in Example 6.23. □

Teaser 6.26: (For those who have read Section 5.3.) Define $\nabla''$ by

$$int_1 \nabla'' int_2 = \begin{cases} \bot & \text{if } int_1 = int_2 = \bot \\ [z^-, z^+] & \text{if } int_1 = [z^-, z^+] \text{ and } int_2 = \bot \\ [z^-, z^+] & \text{if } int_1 = \bot \text{ and } int_2 = [z^-, z^+] \\ [z^-, z^+] & \text{if } int_1 = [z_1^-, z_1^+] \text{ and } int_2 = [z_2^-, z_2^+] \\ & \text{and } z^- = \begin{cases} z_1^- & \text{if } z_1^- \leq z_2^- \\ \lfloor z_2^- \rfloor & \text{otherwise} \end{cases} \\ & \quad\;\; z^+ = \begin{cases} z_1^+ & \text{if } z_1^+ \geq z_2^+ \\ \lceil z_2^+ \rceil & \text{otherwise} \end{cases} \end{cases}$$

and show that it is a strong widening. How does it differ from $\nabla$ and $\nabla'$ of Exercise 6.25? □

## 6.3 Inducing along the Abstraction Function

In this section we shall show how to use a Galois connection from a pointed semi-lattice $\widehat{\mathbf{D}}$ to another $\widehat{\mathbf{D}'}$ to transfer an analysis over $\widehat{\mathbf{D}}$ to one over $\widehat{\mathbf{D}'}$. The first part of the section presents the general approach, while the second part introduces an important analysis called the Collecting Semantics, that is used in the third part to obtain the Detection of Signs analysis.

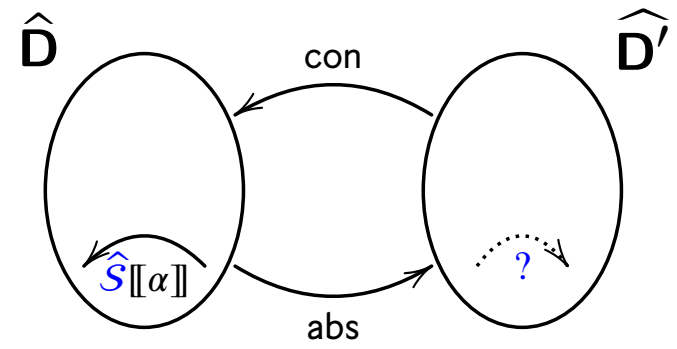

Figure 6.8: How to induce?

**The General Approach** Recall that in this chapter we do not require analysis domains to satisfy the ascending chain condition so that a specification of an analysis is the following modification of Definition 3.30.

Definition 6.27: A *specification* of an analysis in the *extended monotone framework* is given by:

- an analysis domain $\widehat{\mathbf{D}}$ (with ordering $\sqsubseteq$) that is a pointed semi-lattice,
- monotone analysis functions $\widehat{\mathcal{S}}[\![q_\circ, \alpha, q_\bullet]\!] : \widehat{\mathbf{D}} \to \widehat{\mathbf{D}}$ for each choice of action $\alpha$ and nodes $q_\circ$ and $q_\bullet$, and

- an initial element $\hat{d}_\diamond \in \widehat{\mathbf{D}}$.

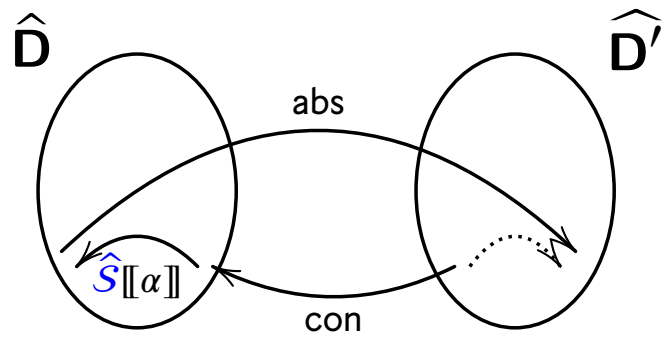

Figure 6.9: Induced analysis functions.

So let (abs, con) be a Galois connection from the pointed semi-lattice $\widehat{\mathbf{D}}$ to the pointed semi-lattice $\widehat{\mathbf{D}'}$. Taking our clue from Figure 6.8 we can see that an analysis function

$$\widehat{\mathcal{S}}[\![q_\circ, \alpha, q_\bullet]\!] : \widehat{\mathbf{D}} \to \widehat{\mathbf{D}}$$

gives rise to an analysis function

$$\mathsf{abs} \circ \widehat{\mathcal{S}}[\![q_\circ, \alpha, q_\bullet]\!] \circ \mathsf{con} : \widehat{\mathbf{D}'} \to \widehat{\mathbf{D}'}$$

as shown in Figure 6.9. This motivates the following definition.

Definition 6.28: A specification of an analysis in the extended monotone framework and a Galois connection (abs, con) from $\widehat{\mathbf{D}}$ to the pointed semi-lattice $\widehat{\mathbf{D}'}$ gives rise to the *induced analysis* given by:

- the analysis domain $\widehat{\mathbf{D}'}$,
- monotone analysis functions $\mathsf{abs} \circ \widehat{\mathcal{S}}[\![q_\circ, \alpha, q_\bullet]\!] \circ \mathsf{con} : \widehat{\mathbf{D}'} \to \widehat{\mathbf{D}'}$ for each choice of action $\alpha$ and nodes $q_\circ$ and $q_\bullet$,
- an initial element $\mathsf{abs}(\hat{d}_\diamond) \in \widehat{\mathbf{D}'}$.

It is straightforward to check that this indeed gives a specification of an analysis in the extended monotone framework.

Exercise 6.29: Referring back to Example 6.1 define the function

$$\mathsf{add} : \mathsf{PowerSet}(\mathbf{Int}) \times \mathsf{PowerSet}(\mathbf{Int}) \to \mathsf{PowerSet}(\mathbf{Int})$$

by the equation

$$\mathsf{add}(Z_1, Z_2) = \{z_1 + z_2 \mid z_1 \in Z_1, z_2 \in Z_2\}$$

and recall the function

$$S_1 \,\hat{+}\, S_2 = \bigcup_{s_1 \in S_1, s_2 \in S_2} s_1 \,\tilde{+}\, s_2$$

defined in Section 5.1 using the function $\tilde{+}$ defined in Figure 5.3.

Show that $S_1 \,\hat{+}\, S_2 = \mathsf{abs}\left(\mathsf{add}\left(\mathsf{con}(S_1), \mathsf{con}(S_2)\right)\right)$. □

In practice it may be too demanding to implement the induced analysis and it will be safe instead to use any analysis function $\widehat{\mathcal{S}'}[\![q_\circ, \alpha, q_\bullet]\!]$ and any initial element $\hat{d}'_\diamond$ provided that they satisfy

$$\mathsf{abs} \circ \widehat{\mathcal{S}}[\![q_\circ, \alpha, q_\bullet]\!] \circ \mathsf{con} \sqsubseteq \widehat{\mathcal{S}'}[\![q_\circ, \alpha, q_\bullet]\!]$$
$$\mathsf{abs}(\hat{d}_\diamond) \sqsubseteq \hat{d}'_\diamond$$

so that they are over-approximations of the induced analysis specification.

We next study to which extent a solution to (an over-approximation of) the induced analysis specification carries over to a solution to the original analysis. For this we let a program graph **PG** with nodes **Q** and initial node $q_\triangleright$ be fixed.

PROPOSITION 6.30: Assume that we have an analysis specification $\widehat{\mathcal{S}'}[\![q_\circ, \alpha, q_\bullet]\!]$ and $\hat{d}'_\diamond$ over $\widehat{\mathbf{D}'}$ satisfying

- $\mathsf{abs} \circ \widehat{\mathcal{S}}[\![q_\circ, \alpha, q_\bullet]\!] \circ \mathsf{con} \sqsubseteq \widehat{\mathcal{S}'}[\![q_\circ, \alpha, q_\bullet]\!]$
- $\mathsf{abs}(\hat{d}_\diamond) \sqsubseteq \hat{d}'_\diamond$

Furthermore, assume that

- $\mathsf{AA} : \mathbf{Q} \to \widehat{\mathbf{D}'}$ solves the constraints obtained from the program graph and the analysis specification $\widehat{\mathcal{S}'}[\![q_\circ, \alpha, q_\bullet]\!]$ and $\hat{d}'_\diamond$ over $\widehat{\mathbf{D}'}$

Then we also have that

- $\mathsf{AA} : \mathbf{Q} \to \widehat{\mathbf{D}'}$ solves the constraints obtained from the analysis specification $\mathsf{abs} \circ \widehat{\mathcal{S}}[\![q_\circ, \alpha, q_\bullet]\!] \circ \mathsf{con}$ and $\mathsf{abs}(\hat{d}_\diamond)$ over $\widehat{\mathbf{D}'}$, and
- $(\mathsf{con} \circ \mathsf{AA}) : \mathbf{Q} \to \widehat{\mathbf{D}}$ solves the constraints obtained from the analysis specification $\widehat{\mathcal{S}}[\![q_\circ, \alpha, q_\bullet]\!]$ and $\hat{d}_\diamond$ over $\widehat{\mathbf{D}}$.

PROOF: Suppose that there is an edge $(q_\circ, \alpha, q_\bullet)$ so that we have

$$\widehat{\mathcal{S}'}[\![q_\circ, \alpha, q_\bullet]\!](\mathsf{AA}(q_\circ)) \sqsubseteq \mathsf{AA}(q_\bullet)$$

Using the assumptions we then get

$$(\mathsf{abs} \circ \widehat{\mathcal{S}}[\![q_\circ, \alpha, q_\bullet]\!] \circ \mathsf{con})(\mathsf{AA}(q_\circ)) \sqsubseteq \mathsf{AA}(q_\bullet)$$

and using that $(\mathsf{abs}, \mathsf{con})$ is a Galois connection we get

$$\widehat{\mathcal{S}}[\![q_\circ, \alpha, q_\bullet]\!]((\mathsf{con} \circ \mathsf{AA})(q_\circ)) \sqsubseteq (\mathsf{con} \circ \mathsf{AA})(q_\bullet)$$

as required.

Next suppose that $\hat{d}'_\diamond \sqsubseteq \mathsf{AA}(q_\triangleright)$; using the assumptions we get $\mathsf{abs}(\hat{d}_\diamond) \sqsubseteq \mathsf{AA}(q_\triangleright)$ and $\hat{d}_\diamond \sqsubseteq (\mathsf{con} \circ \mathsf{AA})(q_\triangleright)$ as required. □

This result shows that it is safe to transfer an analysis to a simpler analysis domain, obtain a solution, and then transfer it back to the original analysis domain. While this presentation has focused on forward analyses, the ideas of Section 3.4 make the approach equally applicable for backward analyses. For analyses where greatest solutions are desired we can still use the ideas of Section 3.4 but should dualise before defining the Galois connections.

**The Collecting Semantics** To prepare for the subsequent development we introduce an important analysis from which analyses like those of Chapter 5 can be obtained using the ideas presented above.

Let us recall the *memory* **Mem** as defined in Section 1.2:

$$\mathbf{Mem} = (\mathbf{Var} \to \mathbf{Int}) \times (\mathbf{Arr} \to \mathbf{Int}^*) \times (\mathbf{Chan} \to \mathbf{Int}^*)$$

A memory $\sigma$ determines the values of the variables ($\sigma_\mathsf{V}$), the values of the entries of the arrays ($\sigma_\mathsf{A}$) and the sequence of values present on the channels ($\sigma_\mathsf{C}$). When we do not want to differentiate between the three parts of the memory we shall simply write $\sigma$; the different components can then be obtained using subscripts V, A and C.

The *Collecting Semantics* essentially lifts the semantics of Chapter 1 to work on sets of states. This gives rise to using 'abstract memories' from the analysis domain $\widehat{\mathbf{Mem}}_\mathsf{CS}$ defined by

$$\widehat{\mathbf{D}} = \widehat{\mathbf{Mem}}_\mathsf{CS} = \mathsf{PowerSet}(\mathbf{Mem})$$

and ordered by $\subseteq$; it is straightforward to check that this is indeed a pointed semi-lattice and that it does *not* satisfy the ascending chain condition. It also gives rise to using analysis functions $\widehat{\mathcal{S}}_\mathsf{CS}[\![q_\circ, \alpha, q_\bullet]\!]$ that apply the semantics $\mathcal{S}[\![\alpha]\!] : \mathbf{Mem} \hookrightarrow \mathbf{Mem}$ in a pointwise manner:

$$\begin{aligned} \widehat{\mathcal{S}}_\mathsf{CS}[\![q_\circ, \alpha, q_\bullet]\!](\widehat{\sigma}_\mathsf{CS}) &= \{\sigma' \mid \exists \sigma \in \widehat{\sigma}_\mathsf{CS} : \sigma' = \mathcal{S}[\![\alpha]\!](\sigma)\} \\ &= \{\mathcal{S}[\![\alpha]\!](\sigma) \mid \sigma \in \widehat{\sigma}_\mathsf{CS}\} \end{aligned}$$

Here we deal with a possible undefined $\mathcal{S}[\![\alpha]\!](\sigma)$ by simply omitting the attempt to include the value in the set.

We can prove a somewhat stronger correctness result for the Collecting Semantics than we did in Propositions 5.15, 5.30 and 5.53 for the analyses of Chapter 5. It shows that the collecting semantics fully incorporates the normal semantics and 'collects' all the states possible at the various nodes of the program graph. For this we let a program graph **PG** with nodes **Q** and initial node $q_\triangleright$ be fixed.

PROPOSITION 6.31: Whenever $\mathsf{CS} : \mathbf{Q} \to \widehat{\mathbf{Mem}}_\mathsf{CS}$ is the *least* solution to the constraints obtained from the program graph and the analysis specification $\widehat{\mathcal{S}}_{\mathsf{CS}}[\![q_\circ, \alpha, q_\bullet]\!]$ and $\widehat{M}_\triangleright$ over $\widehat{\mathbf{Mem}}_{\mathsf{CS}}$ then

$$\mathsf{CS}(q) = \{\mathcal{S}[\![\pi]\!](\sigma_\triangleright) \mid \sigma_\triangleright \in \widehat{M}_\triangleright \text{ and } \pi \text{ is a path from } q_\triangleright \text{ to } q\}$$

where once again we deal with a possible undefined $\mathcal{S}[\![\pi]\!](\sigma_\triangleright)$ by simply omitting the attempt to include the value in the set.

PROOF: That $\supseteq$ holds in the above equation will be the case whenever CS solves the constraints and we shall prove this using the approach of Proposition 5.15. Using Proposition 3.29 (whose proof does not rely on the analysis domains satisfying the ascending chain property) we have that $\widehat{\mathcal{S}}_{\mathsf{CS}}[\![\pi]\!](\widehat{M}_\triangleright) \subseteq \mathsf{CS}(q)$ whenever $\pi$ is a path from $q_\triangleright$ to $q$. It follows by induction on the length of the path $\pi$ that if $\mathcal{S}[\![\pi]\!](\sigma_\triangleright) = \sigma$ then $\sigma \in \widehat{\mathcal{S}}_{\mathsf{CS}}[\![\pi]\!](\{\sigma_\triangleright\})$. From $\sigma_\triangleright \in \widehat{M}_\triangleright$ and monotonicity of the analysis functions we then get the desired result.

That $\subseteq$ holds in the above equation will only be the case when CS is the least solution to the constraints. It is proved by showing that $\mathsf{CS}'$ defined by

$$\mathsf{CS}'(q) = \{\mathcal{S}[\![\pi]\!](\sigma_\triangleright) \mid \sigma_\triangleright \in \widehat{M}_\triangleright \text{ and } \pi \text{ is a path from } q_\triangleright \text{ to } q\}$$

gives rise to a solution to the constraints as then $\mathsf{CS}(q) \subseteq \mathsf{CS}'(q)$ for all $q$. This amounts to checking that $\widehat{M}_\triangleright \subseteq \mathsf{CS}'(q_\triangleright)$ and that $\widehat{\mathcal{S}}_{\mathsf{CS}}[\![q_\circ, \alpha, q_\bullet]\!](\mathsf{CS}'(q_\circ)) \subseteq (\mathsf{CS}'(q_\bullet))$ for all edges $(q_\circ, \alpha, q_\bullet)$ and this is straightforward. □

**Inducing the Detection of Signs analysis from the Collecting Semantics** In the remainder of this section we shall present a worked example of how to develop the Detection of Signs analysis of Section 5.1 from the Collecting Semantic. This means that we will take $\widehat{\mathbf{D}'} = \widehat{\mathbf{Mem}}_{\mathsf{DS}}$ and need to define $\widehat{\mathbf{D}}$ and the Galois connection (abs, con) from $\widehat{\mathbf{D}}$ to $\widehat{\mathbf{D}'} = \widehat{\mathbf{Mem}}_{\mathsf{DS}}$.

Preparing for the abstraction and concretisation functions we first recall the function sign : **Int** → **Sign** used in Section 5.1 and Example 6.1 and define its extension Sign : **Int**$^*$ → **Sign** to sequences of integers:

$$\mathsf{Sign}([v_0, v_1, \cdots, v_{k-1}]) = \{\mathsf{sign}(v_0), \mathsf{sign}(v_1), \cdots, \mathsf{sign}(v_{k-1})\}$$

Next we recall the analysis domain

$$\widehat{\mathbf{Mem}}_{\mathsf{DS}} = (\mathbf{Var} \to \mathsf{PowerSet}(\mathbf{Sign})) \times (\mathbf{Arr} \to \mathsf{PowerSet}(\mathbf{Sign}))$$

used for the Detection of Signs analysis developed in Section 5.1 and where we write $\hat{\sigma} = (\hat{\sigma}_{\mathsf{V}}, \hat{\sigma}_{\mathsf{A}})$ for an abstract memory in $\widehat{\mathbf{Mem}}_{\mathsf{DS}}$.

The abstraction function abs : $\widehat{\mathbf{Mem}}_{\mathsf{CS}} \to \widehat{\mathbf{Mem}}_{\mathsf{DS}}$ from the Collecting Semantics to the Detection of Signs analysis is given by

$$\begin{array}{rl}\mathsf{abs}(\widehat{M}) = & (\widehat{\sigma}_{\mathsf{V}}, \widehat{\sigma}_{\mathsf{A}}) \\ & \text{where } \begin{array}{rcl} \widehat{\sigma}_{\mathsf{V}}(x) & = & \{\mathsf{sign}(\sigma_{\mathsf{V}}(x)) \mid \sigma \in \widehat{M}\} \\ \widehat{\sigma}_{\mathsf{A}}(A) & = & \bigcup\{\mathsf{Sign}(\sigma_{\mathsf{A}}(A)) \mid \sigma \in \widehat{M}\} \end{array}\end{array}$$

and the concretisation function $\mathsf{con} : \widehat{\mathbf{Mem}}_{\mathsf{DS}} \to \widehat{\mathbf{Mem}}_{\mathsf{CS}}$ is given by

$$\mathsf{con}(\widehat{\sigma}_{\mathsf{V}}, \widehat{\sigma}_{\mathsf{A}}) = \left\{ \sigma \in \mathbf{Mem} \;\middle|\; \begin{array}{l} \forall x \in \mathbf{Var} : \mathsf{sign}(x) \in \widehat{\sigma}_{\mathsf{V}}(x), \\ \forall A \in \mathbf{Arr} : \mathsf{Sign}(\sigma_{\mathsf{A}}(A)) \subseteq \widehat{\sigma}_{\mathsf{A}}(A) \end{array} \right\}$$

EXERCISE 6.32: Show that (abs, con) is a Galois connection. (Hint: It may be easier to show that (abs, con) is an adjunction and rely on Proposition 6.6.) □

The *induced analysis* obtained from the Collecting Semantics using the Galois connection (abs, con) is given by using the analysis domain $\widehat{\mathbf{Mem}}_{\mathsf{DS}}$, the analysis functions $\mathsf{abs} \circ \widehat{\mathcal{S}}_{\mathsf{CS}}[\![q_\circ, \alpha, q_\bullet]\!] \circ \mathsf{con}$ and the initial element $\mathsf{abs}(\widehat{M}_\triangleright)$. We have already said that it is safe to use an analysis over $\widehat{\mathbf{Mem}}_{\mathsf{DS}}$ where the analysis functions $\widehat{\mathcal{S}'}[\![q_\circ, \alpha, q_\bullet]\!]$ and the initial element $\widehat{\sigma'_\triangleright}$ satisfy

$$\begin{array}{rcl} \mathsf{abs} \circ \widehat{\mathcal{S}}_{\mathsf{CS}}[\![q_\circ, \alpha, q_\bullet]\!] \circ \mathsf{con} & \sqsubseteq & \widehat{\mathcal{S}'}[\![q_\circ, \alpha, q_\bullet]\!] \\ \mathsf{abs}(\widehat{M}_\triangleright) & \sqsubseteq & \widehat{\sigma'_\triangleright} \end{array}$$

as then both Propositions 6.30 and 6.31 apply.

PROPOSITION 6.33: The analysis functions $\widehat{\mathcal{S}}_{\mathsf{DS}}[\![q_\circ, \alpha, q_\bullet]\!]$ for the Detection of Signs analysis developed in Section 5.1 satisfy

$$\mathsf{abs} \circ \widehat{\mathcal{S}}_{\mathsf{CS}}[\![q_\circ, \alpha, q_\bullet]\!] \circ \mathsf{con} \sqsubseteq \widehat{\mathcal{S}}_{\mathsf{DS}}[\![q_\circ, \alpha, q_\bullet]\!]$$

and hence over-approximate the induced analysis.

SKETCH OF PROOF: We shall prove only two of the cases and leave the remaining cases as an exercise, and to simplify the notation we shall dispense with $q_\circ$ and $q_\bullet$.

For assignments we calculate

$$
\begin{array}{ll}
\multicolumn{2}{l}{(\mathsf{abs} \circ \widehat{\mathcal{S}}_{\mathsf{CS}}[\![x \;{:=}\; a]\!] \circ \mathsf{con})(\hat{\sigma})} \\
= & \mathsf{abs}(\widehat{\mathcal{S}}_{\mathsf{CS}}[\![x \;{:=}\; a]\!](\mathsf{con}(\hat{\sigma})) \\
= & \mathsf{abs}(\{\mathcal{S}[\![x \;{:=}\; a]\!]\sigma \mid \sigma \in \mathsf{con}(\hat{\sigma})\}) \\
= & \mathsf{abs}(\{(\sigma_{\mathsf{V}}[x \mapsto \mathcal{A}[\![a]\!]\sigma], \sigma_{\mathsf{A}}, \sigma_{\mathsf{C}}) \mid \sigma = (\sigma_{\mathsf{V}}, \sigma_{\mathsf{A}}, \sigma_{\mathsf{C}}) \in \mathsf{con}(\hat{\sigma})\}) \\
\sqsubseteq & \begin{cases} (\hat{\sigma}_{\mathsf{V}}[x \mapsto \widehat{\mathcal{A}}_{\mathsf{DS}}[\![a]\!]\hat{\sigma}], \hat{\sigma}_{\mathsf{A}}) \\ \quad \text{if } \widehat{\mathcal{A}}_{\mathsf{DS}}[\![a]\!]\hat{\sigma} \neq \{\,\} \text{ and } \hat{\sigma} \neq \bot_{\mathsf{DS}} \\ \bot_{\mathsf{DS}} \text{ otherwise} \end{cases} \\
= & \widehat{\mathcal{S}}_{\mathsf{DS}}[\![x \;{:=}\; a]\!]\hat{\sigma}
\end{array}
$$

where we have used that $\{\mathsf{sign}(\mathcal{A}[\![a]\!]\sigma) \mid \sigma \in \mathsf{con}(\hat{\sigma})\} \subseteq \widehat{\mathcal{A}}_{\mathsf{DS}}[\![a]\!]\hat{\sigma}$.

For passing a boolean condition we calculate

$$
\begin{array}{ll}
\multicolumn{2}{l}{(\mathsf{abs} \circ \widehat{\mathcal{S}}_{\mathsf{CS}}[\![b]\!] \circ \mathsf{con})(\hat{\sigma})} \\
= & \mathsf{abs}(\widehat{\mathcal{S}}_{\mathsf{CS}}[\![b]\!](\mathsf{con}(\hat{\sigma})) \\
= & \mathsf{abs}(\{\mathcal{S}[\![b]\!]\sigma \mid \sigma \in \mathsf{con}(\hat{\sigma})\}) \\
= & \mathsf{abs}(\{\sigma \mid \mathcal{B}[\![b]\!]\sigma = \mathsf{tt}, \sigma \in \mathsf{con}(\hat{\sigma})\}) \\
\sqsubseteq & \bigsqcup\{\mathsf{abs}(\{\sigma\}) \mid \mathsf{tt} \in \widehat{\mathcal{B}}_{\mathsf{DS}}[\![b]\!](\mathsf{abs}(\{\sigma\})), \mathsf{abs}(\{\sigma\}) \sqsubseteq \hat{\sigma}\} \\
\sqsubseteq & \bigsqcup\{\hat{\sigma}' \in \mathsf{Basic}(\hat{\sigma}) \mid \mathsf{tt} \in \widehat{\mathcal{B}}_{\mathsf{DS}}[\![b]\!]\hat{\sigma}'\} \\
= & \widehat{\mathcal{S}}_{\mathsf{DS}}[\![b]\!]\hat{\sigma}
\end{array}
$$

where we have used that

$$\mathsf{Basic}(\hat{\sigma}) = \{\hat{\sigma}' \mid \hat{\sigma}' \sqsubseteq \hat{\sigma} \wedge \forall x \in \mathbf{Var} : |\hat{\sigma}'_{\mathsf{V}}(x)| = 1 \wedge \forall A \in \mathbf{Arr} : \hat{\sigma}'_{\mathsf{A}}(A) = \hat{\sigma}_{\mathsf{A}}(A)\}$$

and that $\mathsf{tt} \in \widehat{\mathcal{B}}_{\mathsf{DS}}[\![b]\!](\mathsf{abs}(\{\sigma\}))$ whenever $\mathcal{B}[\![b]\!]\sigma = \mathsf{tt}$. □

TEASER 6.34: Investigate the cases where $\mathsf{abs} \circ \widehat{\mathcal{S}}_{\mathsf{CS}}[\![q_{\circ}, \alpha, q_{\bullet}]\!] \circ \mathsf{con} = \widehat{\mathcal{S}}_{\mathsf{DS}}[\![q_{\circ}, \alpha, q_{\bullet}]\!]$ holds. □

## 6.4 Relational Detection of Signs Analysis

The analyses of Chapter 5 are all *independent attribute* analyses meaning that they are not able to capture the relationship between the values of variables and array entries. So as an example, our Detection of Signs analysis in Section 5.1 will not be able to discover that two variables always will have the same sign, in case all signs are possible for each variable. This was different in the Collecting Semantics that was *relational* in that the relationship between values of variables and array entries was fully accounted for. In order to improve on this we now develop a *Relational Detection of Signs* analysis that regains some of the discriminating power of the Collecting Semantics.

To define the analysis domain we shall need an auxiliary set

$$\widehat{\textbf{Mem}}_{\mathsf{R}} = (\textbf{Var} \to \textbf{Sign}) \times (\textbf{Arr} \to \mathsf{PowerSet}(\textbf{Sign}))$$

and write $(\widehat{\rho}_{\mathsf{V}}, \widehat{\rho}_{\mathsf{A}})$ for a typical element; here $\widehat{\rho}_{\mathsf{V}}$ gives the sign of a variable while $\widehat{\rho}_{\mathsf{A}}$ gives the set of signs of an array. Although $\mathsf{PowerSet}(\widehat{\textbf{Mem}}_{\mathsf{R}})$ can be equipped with the subset ordering $\subseteq$, the set $\widehat{\textbf{Mem}}_{\mathsf{R}}$ is not an analysis domain as there is no natural partial order and no way to combine two elements into one.

However, the auxiliary set is useful for defining our analysis domain for the *Relational Detection of Signs* analysis

$$\widehat{\textbf{Mem}}_{\mathsf{RDS}} = \mathsf{PowerSet}(\widehat{\textbf{Mem}}_{\mathsf{R}})$$

and where we write $\widehat{\mathsf{R}}$ for a typical element. This does give rise to an analysis domain when ordered by the subset ordering $\subseteq$, and the least element will be the empty set and the least upper bound is given by set union.

It is helpful to relate this analysis domain to that of the Collecting Semantics by defining a Galois connection from the Collecting Semantics to that of the Relational Detection of Signs analysis defined here. Recall that

$$\begin{array}{rcl} \widehat{\textbf{Mem}}_{\mathsf{CS}} & = & \mathsf{PowerSet}(\textbf{Mem}) \\ & = & \mathsf{PowerSet}\left((\textbf{Var} \to \textbf{Int}) \times (\textbf{Arr} \to \textbf{Int}^*) \times (\textbf{Chan} \to \textbf{Int}^*)\right) \end{array}$$

and also recall the function $\mathsf{sign} : \textbf{Int} \to \textbf{Sign}$ and its extension $\mathsf{Sign} : \textbf{Int}^* \to \textbf{Sign}$ to sequences of integers as defined in Section 6.3.

The abstraction function $\mathsf{abs}' : \widehat{\textbf{Mem}}_{\mathsf{CS}} \to \widehat{\textbf{Mem}}_{\mathsf{RDS}}$ from the Collecting Semantics to the Relational Detection of Signs analysis is given by

$$\mathsf{abs}'(\widehat{\mathsf{M}}) = \{(\mathsf{sign} \circ \sigma_{\mathsf{V}},\ \mathsf{Sign} \circ \sigma_{\mathsf{A}}) \mid (\sigma_{\mathsf{V}}, \sigma_{\mathsf{A}}, \sigma_{\mathsf{C}}) \in \widehat{\mathsf{M}}\}$$

and the concretisation function $\mathsf{con}' : \widehat{\textbf{Mem}}_{\mathsf{RDS}} \to \widehat{\textbf{Mem}}_{\mathsf{CS}}$ is given by

$$\mathsf{con}'(\widehat{\mathsf{R}}) = \{(\sigma_{\mathsf{V}}, \sigma_{\mathsf{A}}, \sigma_{\mathsf{C}}) \in \textbf{Mem} \mid (\mathsf{sign} \circ \sigma_{\mathsf{V}},\ \mathsf{Sign} \circ \sigma_{\mathsf{A}}) \in \widehat{\mathsf{R}}\}$$

EXERCISE 6.35: Show that $(\mathsf{abs}', \mathsf{con}')$ is a Galois connection. (Hint: It may be easier to show that $(\mathsf{abs}', \mathsf{con}')$ is an adjunction and rely on Proposition 6.6.) □

The analysis functions for arithmetic and boolean expressions operate on the auxiliary set and have type

$$\begin{array}{rl} \widehat{\mathcal{A}}_{\mathsf{R}}[\![a]\!] : & \widehat{\mathbf{Mem}}_{\mathsf{R}} \to \mathsf{PowerSet}(\mathbf{Sign}) \\ \widehat{\mathcal{B}}_{\mathsf{R}}[\![b]\!] : & \widehat{\mathbf{Mem}}_{\mathsf{R}} \to \mathsf{PowerSet}(\mathbf{Bool}) \end{array}$$

The function $\widehat{\mathcal{A}}_{\mathsf{R}}$ for analysing arithmetic expressions is defined by:

$$\begin{array}{rcl} \widehat{\mathcal{A}}_{\mathsf{R}}[\![n]\!](\widehat{\rho}_{\mathsf{V}},\widehat{\rho}_{\mathsf{A}}) & = & \{\mathsf{sign}(n)\} \\ \widehat{\mathcal{A}}_{\mathsf{R}}[\![x]\!](\widehat{\rho}_{\mathsf{V}},\widehat{\rho}_{\mathsf{A}}) & = & \{\widehat{\rho}_{\mathsf{V}}(x)\} \\ \widehat{\mathcal{A}}_{\mathsf{R}}[\![a_1\ op_a\ a_2]\!](\widehat{\rho}_{\mathsf{V}},\widehat{\rho}_{\mathsf{A}}) & = & \widehat{\mathcal{A}}_{\mathsf{R}}[\![a_1]\!](\widehat{\rho}_{\mathsf{V}},\widehat{\rho}_{\mathsf{A}})\ \widehat{op_a}\ \widehat{\mathcal{A}}_{\mathsf{R}}[\![a_2]\!](\widehat{\rho}_{\mathsf{V}},\widehat{\rho}_{\mathsf{A}}) \\ \widehat{\mathcal{A}}_{\mathsf{R}}[\![A[a]]\!](\widehat{\rho}_{\mathsf{V}},\widehat{\rho}_{\mathsf{A}}) & = & \begin{cases} \widehat{\rho}_{\mathsf{A}}(A) & \text{if } \widehat{\mathcal{A}}_{\mathsf{R}}[\![a]\!](\widehat{\rho}_{\mathsf{V}},\widehat{\rho}_{\mathsf{A}}) \cap \{0,+\} \neq \{\ \} \\ \{\ \} & \text{otherwise} \end{cases} \end{array}$$

Here we are reusing the functions $\widehat{op_a}$ : PowerSet(**Sign**) × PowerSet(**Sign**) → PowerSet(**Sign**) from Section 5.1.

EXERCISE 6.36: Show that

$$\mathsf{sign}(\mathcal{A}[\![a]\!](\sigma_{\mathsf{V}},\sigma_{\mathsf{A}},\sigma_{\mathsf{C}})) \in \widehat{\mathcal{A}}_{\mathsf{R}}[\![a]\!](\mathsf{sign} \circ \sigma_{\mathsf{V}},\ \mathsf{Sign} \circ \sigma_{\mathsf{A}})$$

for all arithmetic expressions $a$ and states $(\sigma_{\mathsf{V}},\sigma_{\mathsf{A}},\sigma_{\mathsf{C}})$. □

The function $\widehat{\mathcal{B}}_{\mathsf{R}}$ for analysing arithmetic expressions is defined by:

$$\begin{array}{rcl} \widehat{\mathcal{B}}_{\mathsf{R}}[\![\mathtt{true}]\!](\widehat{\rho}_{\mathsf{V}},\widehat{\rho}_{\mathsf{A}}) & = & \{\mathsf{tt}\} \\ \widehat{\mathcal{B}}_{\mathsf{R}}[\![a_1\ op_r\ a_2]\!](\widehat{\rho}_{\mathsf{V}},\widehat{\rho}_{\mathsf{A}}) & = & \widehat{\mathcal{A}}_{\mathsf{R}}[\![a_1]\!](\widehat{\rho}_{\mathsf{V}},\widehat{\rho}_{\mathsf{A}})\ \widehat{op_r}\ \widehat{\mathcal{A}}_{\mathsf{R}}[\![a_2]\!](\widehat{\rho}_{\mathsf{V}},\widehat{\rho}_{\mathsf{A}}) \\ \widehat{\mathcal{B}}_{\mathsf{R}}[\![b_1\ op_b\ b_2]\!](\widehat{\rho}_{\mathsf{V}},\widehat{\rho}_{\mathsf{A}}) & = & \widehat{\mathcal{B}}_{\mathsf{R}}[\![b_1]\!](\widehat{\rho}_{\mathsf{V}},\widehat{\rho}_{\mathsf{A}})\ \widehat{op_b}\ \widehat{\mathcal{B}}_{\mathsf{R}}[\![b_2]\!](\widehat{\rho}_{\mathsf{V}},\widehat{\rho}_{\mathsf{A}}) \end{array}$$

Once more we are reusing the functions $\widehat{op_r}$ : PowerSet(**Sign**) × PowerSet(**Sign**) → PowerSet(**Bool**) and $\widehat{op_b}$ : PowerSet(**Bool**) × PowerSet(**Bool**) → PowerSet(**Bool**) from Section 5.1.

EXERCISE 6.37: Show that

$$\mathcal{B}[\![b]\!](\sigma_{\mathsf{V}},\sigma_{\mathsf{A}},\sigma_{\mathsf{C}}) \in \widehat{\mathcal{A}}_{\mathsf{R}}[\![b]\!](\mathsf{sign} \circ \sigma_{\mathsf{V}},\ \mathsf{Sign} \circ \sigma_{\mathsf{A}})$$

for all boolean expressions $b$ and states $(\sigma_{\mathsf{V}},\sigma_{\mathsf{A}},\sigma_{\mathsf{C}})$. □

The analysis function for actions operates on the analysis domain and has type

$$\widehat{\mathcal{S}}_{\mathsf{R}}[\![\alpha]\!] : \mathsf{PowerSet}(\widehat{\mathbf{Mem}_{\mathsf{R}}}) \to \mathsf{PowerSet}(\widehat{\mathbf{Mem}_{\mathsf{R}}})$$

It is defined by:

$$\begin{array}{rcl}
\widehat{\mathcal{S}}_{\mathsf{R}}[\![b]\!](\widehat{\mathsf{R}}) & = & \{(\widehat{\rho}_{\mathsf{V}}, \widehat{\rho}_{\mathsf{A}}) \in \widehat{\mathsf{R}} \mid \mathsf{tt} \in \widehat{\mathcal{B}}_{\mathsf{R}}[\![b]\!](\widehat{\rho}_{\mathsf{V}}, \widehat{\rho}_{\mathsf{A}})\} \\
\widehat{\mathcal{S}}_{\mathsf{R}}[\![x := a]\!](\widehat{\mathsf{R}}) & = & \{(\widehat{\rho}_{\mathsf{V}}[x \mapsto s], \widehat{\rho}_{\mathsf{A}}) \mid (\widehat{\rho}_{\mathsf{V}}, \widehat{\rho}_{\mathsf{A}}) \in \widehat{\mathsf{R}}, s \in \widehat{\mathcal{A}}_{\mathsf{R}}[\![a]\!](\widehat{\rho}_{\mathsf{V}}, \widehat{\rho}_{\mathsf{A}})] \\
\widehat{\mathcal{S}}_{\mathsf{R}}[\![\mathtt{skip}]\!](\widehat{\mathsf{R}}) & = & \widehat{\mathsf{R}} \\
\widehat{\mathcal{S}}_{\mathsf{R}}[\![c?x]\!](\widehat{\mathsf{R}}) & = & \{(\widehat{\rho}_{\mathsf{V}}[x \mapsto s], \widehat{\rho}_{\mathsf{A}}) \mid (\widehat{\rho}_{\mathsf{V}}, \widehat{\rho}_{\mathsf{A}}) \in \widehat{\mathsf{R}}, s \in \{-, 0, +\}\} \\
\widehat{\mathcal{S}}_{\mathsf{R}}[\![c!a]\!](\widehat{\mathsf{R}}) & = & \{(\widehat{\rho}_{\mathsf{V}}, \widehat{\rho}_{\mathsf{A}}) \in \widehat{\mathsf{R}} \mid \widehat{\mathcal{A}}_{\mathsf{R}}[\![a]\!](\widehat{\rho}_{\mathsf{V}}, \widehat{\rho}_{\mathsf{A}}) \neq \{\ \}\} \\
\widehat{\mathcal{S}}_{\mathsf{R}}[\![A[a_1] := a_2]\!](\widehat{\mathsf{R}}) & = & \{(\widehat{\rho}_{\mathsf{V}}, \widehat{\rho}_{\mathsf{A}}[A \mapsto S]) \mid (\widehat{\rho}_{\mathsf{V}}, \widehat{\rho}_{\mathsf{A}}) \in \widehat{\mathsf{R}}, \\
 & & \quad \widehat{\mathcal{A}}_{\mathsf{R}}[\![a_1]\!](\widehat{\rho}_{\mathsf{V}}, \widehat{\rho}_{\mathsf{A}}) \cap \{0, +\} \neq \{\ \}, \\
 & & \quad \exists s' : \exists s'' \in \widehat{\mathcal{A}}_{\mathsf{R}}[\![a_2]\!](\widehat{\rho}_{\mathsf{V}}, \widehat{\rho}_{\mathsf{A}}) : \\
 & & \qquad (\widehat{\rho}_{\mathsf{A}}(A) \backslash \{s'\}) \cup \{s''\} \subseteq S \subseteq \widehat{\rho}_{\mathsf{A}}(A) \cup \{s''\}\} \\
\widehat{\mathcal{S}}_{\mathsf{R}}[\![c?A[a]]\!](\widehat{\mathsf{R}}) & = & \{(\widehat{\rho}_{\mathsf{V}}, \widehat{\rho}_{\mathsf{A}}[A \mapsto \{-, 0, +\}]) \mid (\widehat{\rho}_{\mathsf{V}}, \widehat{\rho}_{\mathsf{A}}) \in \widehat{\mathsf{R}}, \\
 & & \quad \widehat{\mathcal{A}}_{\mathsf{R}}[\![a]\!](\widehat{\rho}_{\mathsf{V}}, \widehat{\rho}_{\mathsf{A}}) \cap \{0, +\} \neq \{\ \}\}
\end{array}$$

For boolean expressions we examine each of the abstract descriptors $(\widehat{\rho}_{\mathsf{V}}, \widehat{\rho}_{\mathsf{A}})$ in the argument set $\widehat{\mathsf{R}}$ one by one and retain those for which it is possible that the boolean condition could evaluate to true. For assignments we similarly examine each of the abstract descriptors in the argument set one by one and modify the abstract value of the variable assigned as given by the set of signs for the arithmetic expression; this means that each abstract descriptor in the argument set may give rise to more than one abstract descriptor in the resulting set. For input to a variable we treat it as assigning the variable a value that can have any sign whatsoever; once more each abstract descriptor in the argument set may give rise to more than one abstract descriptor in the resulting set. For output we retain those abstract descriptors where the arithmetic expression may evaluate to a value.

For actions modifying arrays we need to be more careful. The simplest case is that of input to an array entry, where we consider each abstract descriptor where the index may evaluate to a non-negative value, and we update the set of signs for the array to be all the possible signs. (Unlike the case for input to a variable this is all handled by a single abstract descriptor.) For assignment to an array entry we consider those abstract descriptors $(\widehat{\rho}_{\mathsf{V}}, \widehat{\rho}_{\mathsf{A}})$ in the argument set $\widehat{\mathsf{R}}$ where the index may evaluate to a non-negative value and we consider each of the possible signs $s''$ of the arithmetic expression to be assigned; the resulting set of signs for the array $A$ will be either $\widehat{\rho}_{\mathsf{A}}(A) \cup \{s''\}$ or $(\widehat{\rho}_{\mathsf{A}}(A) \setminus \{s'\}) \cup \{s''\}$ for some $s'$ (in case the assignment overwrote the last entry in $A$ having the sign $s'$) and we need to produce both possibilities.

PROPOSITION 6.38: The analysis functions $\widehat{\mathcal{S}}_{\mathsf{R}}[\![q_\circ, \alpha, q_\bullet]\!]$ for the Relational Detection of Signs analysis satisfy

$$\mathsf{abs}' \circ \widehat{\mathcal{S}}_{\mathsf{CS}}[\![q_\circ, \alpha, q_\bullet]\!] \circ \mathsf{con}' \sqsubseteq \widehat{\mathcal{S}}_{\mathsf{R}}[\![q_\circ, \alpha, q_\bullet]\!]$$

and hence over-approximate the induced analysis.

SKETCH OF PROOF: The proof is by calculation as in the proof of Proposition 6.33 and we shall prove only two of the cases and leave the remaining cases as an exercise; to simplify the notation we shall dispense with $q_\circ$ and $q_\bullet$ and recall that we write $\sigma = (\sigma_{\mathsf{V}}, \sigma_{\mathsf{A}}, \sigma_{\mathsf{C}})$.

For assignments we calculate

$$\begin{aligned}
&(\mathsf{abs}' \circ \widehat{\mathcal{S}}_{\mathsf{CS}}[\![x := a]\!] \circ \mathsf{con}')(\widehat{R})\\
&\quad = \mathsf{abs}'(\widehat{\mathcal{S}}_{\mathsf{CS}}[\![x := a]\!](\{\sigma \mid (\mathsf{sign} \circ \sigma_{\mathsf{V}}, \mathsf{Sign} \circ \sigma_{\mathsf{A}}) \in \widehat{R}\}))\\
&\quad = \mathsf{abs}'(\{\mathcal{S}[\![x := a]\!]\sigma \mid (\mathsf{sign} \circ \sigma_{\mathsf{V}}, \mathsf{Sign} \circ \sigma_{\mathsf{A}}) \in \widehat{R}\})\\
&\quad = \mathsf{abs}'(\{(\sigma_{\mathsf{V}}[x \mapsto \mathcal{A}[\![a]\!]\sigma], \sigma_{\mathsf{A}}) \mid (\mathsf{sign} \circ \sigma_{\mathsf{V}}, \mathsf{Sign} \circ \sigma_{\mathsf{A}}) \in \widehat{R}\})\\
&\quad = \{((\mathsf{sign}\circ\sigma_{\mathsf{V}})[x \mapsto \mathsf{sign}(\mathcal{A}[\![a]\!]\sigma)], \mathsf{Sign}\circ\sigma_{\mathsf{A}}) \mid (\mathsf{sign}\circ\sigma_{\mathsf{V}}, \mathsf{Sign}\circ\sigma_{\mathsf{A}}) \in \widehat{R}\}\\
&\quad \sqsubseteq \{(\widehat{\rho}_{\mathsf{V}}[x \mapsto \widehat{\mathcal{A}}_{\mathsf{R}}[\![a]\!](\widehat{\rho}_{\mathsf{V}}, \widehat{\rho}_{\mathsf{A}})], \widehat{\rho}_{\mathsf{A}}) \mid (\widehat{\rho}_{\mathsf{V}}, \widehat{\rho}_{\mathsf{A}}) \in \widehat{R}\}
\end{aligned}$$

where we have used that the analysis function $\widehat{\mathcal{A}}_{\mathsf{R}}[\![a]\!] : \widehat{\mathbf{Mem}}_{\mathsf{R}} \to \mathcal{P}(\mathbf{Sign})$ for arithmetic expressions satisfies the condition of Exercise 6.36.

For passing a boolean condition we calculate

$$\begin{aligned}
&(\mathsf{abs}' \circ \widehat{\mathcal{S}}_{\mathsf{CS}}[\![b]\!] \circ \mathsf{con}')\widehat{R}\\
&\quad = \mathsf{abs}'(\widehat{\mathcal{S}}_{\mathsf{CS}}[\![b]\!](\{\sigma \mid (\mathsf{sign} \circ \sigma_{\mathsf{V}}, \mathsf{Sign} \circ \sigma_{\mathsf{A}}) \in \widehat{R}\}))\\
&\quad = \mathsf{abs}'(\{\mathcal{S}[\![b]\!]\sigma \mid (\mathsf{sign} \circ \sigma_{\mathsf{V}}, \mathsf{Sign} \circ \sigma_{\mathsf{A}}) \in \widehat{R}\})\\
&\quad = \mathsf{abs}'(\{\sigma \mid \mathcal{B}[\![b]\!]\sigma = \mathsf{tt}, (\mathsf{sign} \circ \sigma_{\mathsf{V}}, \mathsf{Sign} \circ \sigma_{\mathsf{A}}) \in \widehat{R}\})\\
&\quad = \{(\mathsf{sign} \circ \sigma_{\mathsf{V}}, \mathsf{Sign} \circ \sigma_{\mathsf{A}}) \mid \mathcal{B}[\![b]\!]\sigma = \mathsf{tt}, (\mathsf{sign} \circ \sigma_{\mathsf{V}}, \mathsf{Sign} \circ \sigma_{\mathsf{A}}) \in \widehat{R}\}\\
&\quad \sqsubseteq \{(\widehat{\rho}_{\mathsf{V}}, \widehat{\rho}_{\mathsf{A}}) \in \widehat{R} \mid \mathsf{tt} \in \widehat{\mathcal{B}}_{\mathsf{R}}[\![b]\!](\widehat{\rho}_{\mathsf{V}}, \widehat{\rho}_{\mathsf{A}})\}
\end{aligned}$$

where we have used that the analysis function $\widehat{\mathcal{B}}_{\mathsf{R}}[\![b]\!] : \widehat{\mathbf{Mem}}_{\mathsf{R}} \to \mathcal{P}(\mathbf{Bool})$ for boolean expressions satisfies the condition of Exercise 6.37. □

We shall end this section by illustrating another key idea of abstract interpretation: that it is often preferable to develop analyses in stages, starting with a correct analysis that is too computationally demanding (and that perhaps does not satisfy the ascending chain condition), and obtaining more computationally tractable analyses along thew way. Here we showed how to start with the Collecting Semantics and obtain a Relational Detection of Signs analysis. Next we show how to obtain the Detection of Signs analysis of Section 5.1 using a further step of abstraction.

For this we recall the analysis domain

$$\widehat{\mathbf{Mem}}_{\mathsf{DS}} = (\mathbf{Var} \to \mathsf{PowerSet}(\mathbf{Sign})) \times (\mathbf{Arr} \to \mathsf{PowerSet}(\mathbf{Sign}))$$

used for the Detection of Signs analysis in Sections 5.1 and 6.3 and where we write $\hat{\sigma} = (\widehat{\sigma}_{\mathsf{V}}, \widehat{\sigma}_{\mathsf{A}})$ for an abstract memory in $\widehat{\mathbf{Mem}}_{\mathsf{DS}}$.

The abstraction function $\mathsf{abs}'' : \widehat{\mathbf{Mem}}_{\mathsf{RDS}} \to \widehat{\mathbf{Mem}}_{\mathsf{DS}}$ from the Relational Detection of Signs analysis to the Detection of Signs analysis is given by

$$\begin{array}{lll} \mathsf{abs}''(\widehat{\mathsf{R}}) = & (\widehat{\sigma}_{\mathsf{V}}, \widehat{\sigma}_{\mathsf{A}}) & \\ & \text{where } \widehat{\sigma}_{\mathsf{V}}(x) & = \{\widehat{\rho}_{\mathsf{V}}(x) \mid (\widehat{\rho}_{\mathsf{V}}, \widehat{\rho}_{\mathsf{A}}) \in \widehat{\mathsf{R}}\} \\ & \qquad\quad \widehat{\sigma}_{\mathsf{A}}(A) & = \bigcup\{\widehat{\rho}_{\mathsf{A}}(A) \mid (\widehat{\rho}_{\mathsf{V}}, \widehat{\rho}_{\mathsf{A}}) \in \widehat{\mathsf{R}}\} \end{array}$$

The concretisation function $\mathsf{con}'' : \widehat{\mathbf{Mem}}_{\mathsf{DS}} \to \widehat{\mathbf{Mem}}_{\mathsf{RDS}}$ from the Detection of Signs analysis to the Relational Detection of Signs analysis is given by

$$\mathsf{con}''(\widehat{\sigma}_{\mathsf{V}}, \widehat{\sigma}_{\mathsf{A}}) = \left\{ (\widehat{\rho}_{\mathsf{V}}, \widehat{\rho}_{\mathsf{A}}) \in \widehat{\mathbf{Mem}}_{\mathsf{RDS}} \;\middle|\; \begin{array}{l} \forall x \in \mathbf{Var} : \widehat{\rho}_{\mathsf{V}}(x) \in \widehat{\sigma}_{\mathsf{V}}(x), \\ \forall A \in \mathbf{Arr} : \{\,\} \neq \widehat{\rho}_{\mathsf{A}}(A) \subseteq \widehat{\sigma}_{\mathsf{A}}(A) \end{array} \right\}$$

(The exclusion of those $\widehat{\rho}_{\mathsf{A}}$ with $\widehat{\rho}_{\mathsf{A}}(A) = \{\,\}$ is permitted because we are assuming that all arrays have positive lenghts.)

Exercise 6.39: Show that $(\mathsf{abs}'', \mathsf{con}'')$ is a Galois connection. (Hint: It may be easier to show that $(\mathsf{abs}'', \mathsf{con}'')$ is an adjunction and rely on Proposition 6.6.) □

Exercise 6.40: Recalling the Galois connection $(\mathsf{abs}, \mathsf{con})$ from the Collecting Semantics to the Detection of Signs analysis constructed in Section 6.3 show that $\mathsf{abs} = \mathsf{abs}'' \circ \mathsf{abs}'$ and that $\mathsf{con}'' = \mathsf{con}' \circ \mathsf{con}''$. □

# Chapter 7

# Types for Information Flow Analysis

So far our analyses have been performed directly on program graphs and we have left the construction of program graphs from programs to Appendices A and B. In this chapter we will consider a security analysis and we shall follow tradition and develop it by means of a type system. This requires choosing the programming language and developing the analysis at the level of the syntax of the programming language.

We shall base ourselves on the (probably unfamiliar) language of Guarded Commands introduced by Dijkstra in 1975 (and covered in Appendix A). A key design decision of this language is to admit non-determinism into the language so as to give greater flexibility when implementing the programs. We shall slightly extend the language to support security policies that allow sanitization of data. The required concepts from security will be introduced along the way.

## 7.1 Guarded Commands

In Dijkstra's language of *Guarded Commands* a basic command has one of two forms, either it is an assignment $x$ `:=` $a$ or it is a `skip` command; the latter has no effect but is useful when no memory change is wanted. Commands can be combined in three different ways. Sequencing is written $C_1$ `;` ... `;` $C_k$ and indicates that the commands should be executed in the order they are written. The conditional takes the form `if` $b_1 \rightarrow C_1$ `[]` ... `[]` $b_k \rightarrow C_k$ `fi`; as an example, to express that $C_1$ should be executed when $b$ holds and that otherwise $C_2$ should be executed, we shall write **if** $b \to C_1$ [] $\neg b \to C_2$ **fi**. The iteration construct takes the form **do** $b_1 \to C_1$ [] ... [] $b_k \to C_k$ **od**; as an example, to express that $C$ should be executed as long as $b$ holds, we shall write **do** $b \to C$ **od**.

```
y := 1;
do x > 0 →  y := x * y;
            x := x − 1
od
```

Figure 7.1: Example program for the factorial function.

EXAMPLE 7.1: Figure 7.1 is a program intended to compute the factorial function. In addition to assignments, it makes use of sequencing (as indicated by the semicolons) and the iteration construct do ... od. The body of the construct is a single guarded command consisting of a guard and a command; here the guard is the test x > 0 and the command is the sequence y := x * y; x := x − 1 of assignments. The idea is that as long as the guard is satisfied the associated command will be executed. When the guard fails, the construct terminates.

```
if x ≥ y → z := x
 [] y > x → z := y
fi
```

Figure 7.2: Example program for the maximum function.

EXAMPLE 7.2: Figure 7.2 is a program computing the maximum of two values. It makes use of the conditional if ... fi and contains two guarded commands separated by the choice symbol []. The idea is to select a guard that is satisfied and to execute the corresponding command – in our program exactly one of the guards will be satisfied.

If none of the guards are satisfied the execution stops; this will for example be the case if we change the first guard to x > y and start in a memory where x and y are equal.

It might also be the case that more than one guard is satisfied and then one of the alternatives will be selected non-deterministically for execution; this will for example be the case if we change the second guard to y ≥ x and start in a memory where x and y are equal.

```
i := 0;
do i < A# →  A[i] := A[i] + 27;
             i := i + 1
od
```

Figure 7.3: Incrementation of array.

EXAMPLE 7.3: Figure 7.3 is a program updating all the entries of an array by adding 27 to them. We consider the array A to be indexed from 0 to A# − 1 where A# indicates the number of elements in the array. The program terminates when all entries have been updated and the final value of i will be A#.

We are now ready to introduce the syntax of the simple non-deterministic language to be used in this chapter; as the name indicates it is patterned after Dijkstra's Guarded Commands Language. The arithmetic and boolean expressions are mostly standard but we use && for short-circuit conjunction (where the second component is not evaluated if the first evaluates to false) and san for purposes of security as will be explained in Section 7.2.

DEFINITION 7.4: The syntax of the commands $C$ and guarded commands $GC$ of the *Guarded Commands for Security* language are mutually recursively

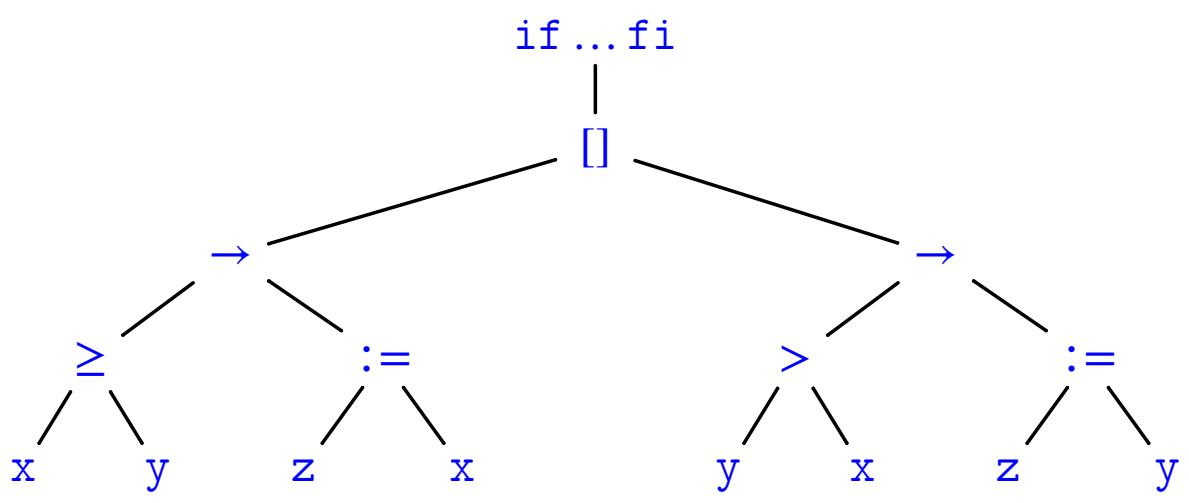

Figure 7.6: Abstract syntax tree for the program in Figure 7.2.

defined using the following *BNF notation*:

$$C \ ::= \ x := a \mid A[a_1] := a_2 \mid \texttt{skip} \mid C_1 \,;\, C_2 \mid \texttt{if}\ GC\ \texttt{fi} \mid \texttt{do}\ GC\ \texttt{od}$$

$$GC \ ::= \ b \rightarrow C \mid GC_1 \,[]\, GC_2$$

We make use of arithmetic expressions $a$ and boolean expressions $b$ given by

$$a \ ::= \ n \mid s \mid x \mid A[a_1] \mid A\# \mid a_1 + a_2 \mid a_1 - a_2 \mid a_1 * a_2 \mid \texttt{san}\ a_1$$

$$b \ ::= \ \texttt{true} \mid a_1 = a_2 \mid a_1 > a_2 \mid a_1 \geq a_2 \mid b_1 \wedge b_2 \mid b_1 \ \&\& \ b_2 \mid \neg b_1$$

The syntax of numbers $n$ (e.g. 27), strings $s$ (e.g. "Jens Madsen"), variables $x$ (e.g. `i`), and arrays $A$ (e.g. `A`) is left unspecified.

The definition specifies an *abstract syntax tree* for commands, guarded commands and arithmetic and boolean expressions. As an example, $x ::= a$ tells us that an arithmetic expression can be a tree consisting of a single node representing the variable $x$, whereas $a ::= a_1 * a_2$ tells us that it can be a binary tree with root $*$ and a subtree corresponding to $a_1$ and another subtree corresponding to $a_2$.

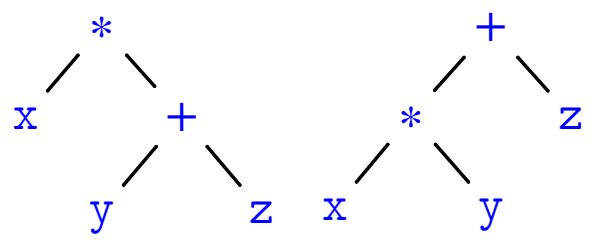

Figure 7.4: Abstract syntax trees for $\texttt{x} * (\texttt{y} + \texttt{z})$ and $(\texttt{x} * \texttt{y}) + \texttt{z}$.

As the definition specifies the *syntactic tree structure* (rather than a linear sequence of characters) we do not need to introduce explicit *brackets* in the syntax, although we shall feel free to use *parentheses* in textual presentations of programs in order to disambiguate the syntax. This is illustrated in Figure 7.4 and when the precedence of operators is clear we allow to write $\texttt{x} * \texttt{y} + \texttt{z}$ for $(\texttt{x} * \texttt{y}) + \texttt{z}$.

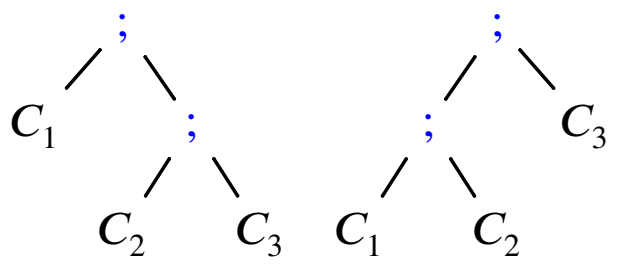

Figure 7.5: Abstract syntax trees for $C_1 \,;\, (C_2 \,;\, C_3)$ and $(C_1 \,;\, C_2) \,;\, C_3$.

Similar comments hold for commands, guarded commands and boolean expressions; see Figure 7.5 for an example. In the case of sequencing we consider the sequencing operator to associate to the right so that $C_1 \,;\, C_2 \,;\, C_3$ is shorthand for $C_1 \,;\, (C_2 \,;\, C_3)$. In the case of choice we consider the choice operator to associate to the right so that $GC_1 \,[]\, GC_2 \,[]\, GC_3$ is shorthand for $GC_1 \,[]\, (GC_2 \,[]\, GC_3)$. Figure 7.6 gives the abstract syntax tree for the program of Figure 7.2.

TRY IT OUT 7.5: Draw the abstract syntax tree for the program of Figure 7.1. □

Program graphs can be constructed much as is done in Appendix A for Guarded Commands as the few additions to the syntax present no obstacles.

## 7.2 Security as Information Flow

The key approach to ensure *security* is to limit the *information flow* allowed by programs. To do so we first explain the notions of *confidentiality* and *integrity* from security, and we then explain how information flow can be used to ensure confidentiality and integrity. Traditionally one distinguishes between *explicit* and *implicit* flows but due to the non-determinism and sanitisation we shall make use of additional kinds of flows. Information flows may be *direct* (happening as one step) or *indirect* (happening in several steps); we shall *not* use the terms direct and indirect flows as synonyms for explicit and implicit flows as done by some authors.

**Confidentiality** What does it mean for a program to maintain *confidentiality* of data? The basic idea is pretty simple: we have a notion of which data is *private* and which data is *public*, and we want to prevent that private data `x` finds its way into public data `y`.

```
if  x < 0 → y := −1
[]  x = 0 → y := 0
[]  x > 0 → y := 1
fi
```

Figure 7.7: Example of implicit flow for confidentiality.

One component of this is that there should be no *explicit* flow of information from `x` to `y`. In particular, an assignment `y := x` would constitute an explicit flow violating the confidentiality of `x` with respect to `y`.

Another component of this is that there should be no *implicit* flow of information from `x` to `y`. In particular, a conditional assignment of the form shown in Figure 7.7 would constitute an implicit flow where the sign of `x` is made visible in `y`.

**Integrity** What does it mean for a program to maintain *integrity* of data? The basic idea is pretty simple: we have a notion of which data is *trusted* and which data is *dubious*, and we want to prevent that trusted data `x` is influenced by dubious data `y`.

```
if  y < 0 → x := −1
[]  y = 0 → x := 0
[]  y > 0 → x := 1
fi
```

Figure 7.8: Example of implicit flow for integrity.

One component of this is that there should be no *explicit* flow of information from `y` to `x`. In particular, an assignment `x := y` would constitute an explicit flow violating the integrity of `x` with respect to `y`.

Another component of this is that there should be no *implicit* flow of information from `y` to `x`. In particular, a conditional assignment of the form shown in Figure 7.8 would constitute an implicit flow where the value of `x` can no longer be trusted if `y` is dubious.

**Introducing information flow** What is common between our description of confidentiality and integrity is that data of a certain security classification should not find its way into data of another security classification. We shall use the notation $x \to y$ to indicate a flow from $x$ to $y$ and may use it both when describing actual flows in programs, and when discussing which flows are permissible with respect to a security policy.

In the confidentiality example of Figure 7.7, where x is private and y is public we would allow y → x but require x $\not\to$ y (meaning that x → y is not allowed). More generally we would allow $y \to x$ whenever $y$ is public or $x$ is private.

In the integrity example of Figure 7.8, where x is trusted and y is dubious we would allow x → y but require y $\not\to$ x. More generally we would allow $x \to y$ whenever $x$ is trusted or $y$ is dubious.

Information flows will not only be between variables but also between array entries and array lengths so we need a concept that encompasses either.

> DEFINITION 7.6: A *data container* $\delta$ is either a variable, an array entry or an array length.

Note that all data containers can be used in arithmetic expressions but only variables and array entries can be modified.

It is useful to extend the flow relation to work on sets of data containers. So if $\Delta_1$ and $\Delta_2$ are two sets of data containers we shall define $\Delta_1 \rightrightarrows \Delta_2$ to mean that each and every data container $\delta_1$ in $\Delta_1$ has a flow to each and every data container $\delta_2$ in $\Delta_2$ – see Figure 7.9 for an example.

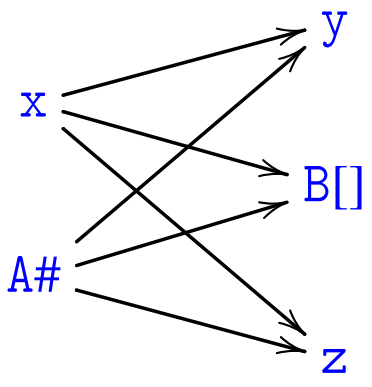

Figure 7.9: {x, A#} $\rightrightarrows$ {y, B[], z}.

EXAMPLE 7.7: Imagine that we have a database with an array A of length A# and an array B of length B#. To increase each entry of A by 27 we may use the program of Figure 7.10; similarly the program of Figure 7.11 increases each entry of B by 12.

In a proper database the execution of these programs might be interleaved in a non-deterministic manner as illustrated by the program

```
i := 0;
j := 0;
do i < A# →  A[i] := A[i] + 27;
             i := i + 1
[] j < B# →  B[j] := B[j] + 12;
             j := j + 1
od
```

To ensure that the operations on A are entirely independent of the operations on B we need to allow {i, A#} $\rightrightarrows$ {i, A[]} and {j, B#} $\rightrightarrows$ {j, B[]} but will need to require A[] $\not\to$ B[] and B[] $\not\to$ A[].

From a *confidentiality* point of view this says that the information in the arrays A and B remain confidential from each other. From an *integrity* point of view this says that the information in the arrays A and B remain unaffected by each other. From an *isolation* point of view this says that the operations on arrays A and B are completely independent. – as they are in Figures 7.10 and 7.11.

```
i := 0;
do i < A# →  A[i] := A[i] + 27;
             i := i + 1
od
```

Figure 7.10: Incrementation of A.

```
j := 0;
do j < B# →  B[j] := B[j] + 12;
             j := j + 1
od
```

Figure 7.11: Incrementation of B.

**Explicit information flows** We now expand on our treatment of explicit and implicit flows and introduce the remaining types of flow that we need in order to deal with non-determinism and sanitisation.

We explained explicit flows using the command

$$\texttt{y} := \texttt{x}$$

and here there is a direct and *explicit* flow from x to y. We write this as $\texttt{x} \to \texttt{y}$ or as

$$\texttt{x} \to^{\mathsf{E}} \texttt{y}$$

to indicate that it is an explicit flow.

```
i:=0;
do i < 128 →
   if true → K[i]:=0
   [] true → K[i]:=1
   fi;
   i:=i+1
od
```

Figure 7.12: Code for non-deterministically creating a secret key.

In general, *explicit* flows arise whenever a data container is used to compute the value of a data container.

A slightly more complex example is the command

$$\texttt{y} := \texttt{x};\ \texttt{z} := \texttt{y}$$

where there are direct explicit flows from x to y and from y to z. The flow from x to z is an *indirect* flow, and in general we use indirect to indicate that we exploit the transitive nature of the flow relation. As for the type of flow we shall say that the indirect flow is also explicit.

```
i:=0;
do i < 128 →
   H[i] := K[i];
   i:=i+1
od
```

Figure 7.13: Hacking the secret key – creating explicit flows.

EXAMPLE 7.8: Consider the code in Figure 7.12 that non-deterministically creates a 128 bit key K. One way for a hacker to steal this key is to execute the code in Figure 7.13 so as to obtain a copy in H. This results in an information flow from K to H that is explicit.

**Implicit information flows** We explained implicit flows using the guarded command

$$\texttt{x} = 0 \to \texttt{y} := 0$$

and here there is a direct and *implicit* flow from x to y. We write $\texttt{x} \to \texttt{y}$ or as

$$\texttt{x} \to^{\mathrm{I}} \texttt{y}$$

to indicate that it is an implicit flow.

In general, *implicit* flows arise whenever a data container is modified inside the body of a command guarded by a boolean condition with a data container.

A slightly more complex example is the command

$$\texttt{if}\ \texttt{x} = 0 \to \texttt{y} := 0\,[]\,\neg(\texttt{x} = 0) \to \texttt{y} := 1\ \texttt{fi};\ \texttt{z} := \texttt{y}$$

where there is a direct implicit flow from x to y and a direct explicit flow from y to z. The flow from x to z is an *indirect* flow and as for the type of flow we shall say that the indirect flow is implicit (since x is not directly copied into z).

EXAMPLE 7.9: Once more consider the code in Figure 7.12 that creates a 128 bit key `K`. Another way for a hacker to steal this key is to excture the code in Figure 7.14 so as to obtain a copy in `H`. This results in an information flow from `K` to `H` that is implicit.

```
i:=0;
do i < 128 →
   if K[i]=0 → H[i]:=0
   [] K[i]=1 → H[i]:=1
   fi;
   i:=i+1
od
```

Figure 7.14: Hacking the secret key – creating implicit flows.

**Bypassing information flows** Let us consider a variation of the program above:

$$\texttt{y} := 0;\ \texttt{if}\ \texttt{x} = 0 \rightarrow \texttt{skip}\,[\,]\,\texttt{true} \rightarrow \texttt{y} := 1\ \texttt{fi}$$

Here there are no explicit flows from `x` to `y` and also no implicit flows. However, it is still the case that the final value of `y` might reveal something about `x` if one is able to run the program many times and observe the different non-determistic outcomes. We write this as $\texttt{x} \rightarrow \texttt{y}$ or as

$$\texttt{x} \rightarrow^{\mathsf{B}} \texttt{y}$$

to indicate that it is a bypassing flow.

In general, *bypassing* flows arise whenever two conditions can be simultaneously true and it is up to the scheduler to determine which branch is taken; in this case there is a bypassing flow from the data containers in the condition of one branch to the data containers modified in the command of the other.

A slightly more complex example is the command

$$\texttt{x} := \texttt{z};\ \texttt{y} := 0;\ \texttt{if}\ \texttt{x} = 0 \rightarrow \texttt{skip}\,[\,]\,\texttt{true} \rightarrow \texttt{y} := 1\ \texttt{fi}$$

where there are is a direct explicit flow from `z` to `x` and a direct bypassing flow from `x` to `y`. The flow from `z` to `y` is an *indirect* flow and as for the type of flow we shall say that the indirect flow is a bypassing one.

```
i:=0;
do i < 128 →
   H[i]:=0;
   i:=i+1
od;
i:=0;
do i < 128 →
   if K[i]=0 → skip
   [] true → H[i]:=1
   fi;
   i:=i+1
od
```

Figure 7.15: Hacking the secret key – creating byassing flows.

EXAMPLE 7.10: Once more consider the code in Figure 7.12 that creates a key `K`. Yet another way for a hacker to steal this key is to excture the code in Figure 7.15 so as to obtain an imperfect copy in `H`; the copy may contain too many 0's but by executing the hacker code several times the number of fake 0's will get reduced. This results in a bypassing flow from `K` to `H`.

**Correlation information flows** Bypassing flows capture some of the power of the scheduler but not all of it. Considering the following program

$$\texttt{if}\ \texttt{true} \rightarrow \texttt{y} := 0\,;\texttt{x} := 0\,[\,]\,\texttt{true} \rightarrow \texttt{y} := 1\,;\texttt{x} := 1\ \texttt{fi}$$

where there are no explicit, implicit or bypassing flows. Yet, if `y` was intended to be a private key (albeit a short one) and `x` is a public variable, then clearly we can learn something about `y` from knowing `x`. We write this as $\texttt{x} \rightarrow \texttt{y}$ and $\texttt{y} \rightarrow \texttt{x}$ and or as

$$\texttt{x} \rightarrow^{\mathsf{C}} \texttt{y} \qquad \texttt{y} \rightarrow^{\mathsf{C}} \texttt{x}$$

to indicate that it is a possible correlation flow between `x` and `y`.

In general, *correlation* flows arise whenever two conditions can be simultaneously true and it is up to the scheduler to determine which branch is taken; in this case there is a correlation flow between the data containers modified in each branch.

A slightly more complex example is the command

$$\texttt{if true} \to \texttt{y} := 0;\texttt{x} := 0\,[]\,\texttt{true} \to \texttt{y} := 1;\texttt{x} := 1\ \texttt{fi};\ \texttt{z} := \texttt{x}$$

where there is a direct explicit flow from x to z and a correlation flow from y to x. The flow from y to z is an *indirect* flow and as for the type of flow we shall say that the indirect flow is a correlation one.

```
i:=0;
do i < 128 →
  if true → K[i]:=0;
            H[i]:=0
  [] true → K[i]:=1;
            H[i]:=1
  fi;
  i:=i+1
od
```

Figure 7.16: Hacking the secret key – creating correlation flows.

Example 7.11: Consider a variation of the code in Figure 7.12 where the hacker is able to get some hacking code executed in the various branches as shown in Figure 7.16. This results in a correlation flow from K to H.

**Sanitised information flows** Returning to the non-deterministic database program in Example 7.7 there are bypassing flows from A# to B[] and similarly from B# to A[]. We might consider these flows to be absolutely unproblematic and a traditional approach is to use *sanitisation* for this; in our case this means using the san construct of Guarded Commands for Security as illustrated in the below program.

```
i := 0;
j := 0;
do san i < san A# →  A[i] := A[i] + 27;
                     i := i + 1
[] san j < san B# →  B[j] := B[j] + 12;
                     j := j + 1
od
```

Rather than neglecting the direct bypassing flows from A# to B[] and from B# to A[] we shall mark these as sanitised flows (that can be disregarded later if so desired) and we write

$$\texttt{A\#} \to^{\mathsf{S}} \texttt{B[]} \qquad \texttt{B\#} \to^{\mathsf{S}} \texttt{A[]}$$

In general, *sanitised* flows arise whenever at least one sanitisation step is involved in the flow.

In line with previous decisions, if a sequence of flows involve a sanitised flow we shall regard the overall flow as a sanitised one.

```
i:=0;
do i < 128 →
  if san(K[i]=0) →
       H[i]:=0
  [] san(K[i]=1) →
       H[i]:=1
  fi;
  i:=i+1
od
```

Figure 7.17: Hacking the secret key – creating sanitisation flows.

Example 7.12: Let us return to the code in Figure 7.12 for the generation of a key K. Suppose that there is a legal requirement to give a security officer access to the key and consider the code in Figure 7.17 as a way to do so. Here we obtain a sanitised flow from `K` to `H`.

**Representation of information flows** We shall take the point of view that some types of information flow are more worrying than others and that whenever there is an information flow between data containers we only record the most worrying one.

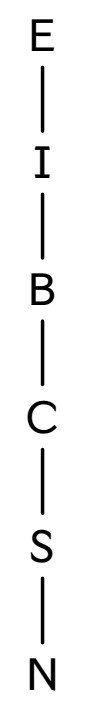

Figure 7.18: Linearly ordering the different types of flows: Explicit, Implicit, Bypassing, Correlation, Sanitised, None.

In Figure 7.18 we present our view on the importance of flows: the higher in the linear order, the more worrying the flow. We use E for explicit flows, I for implicit flows, B for bypassing flows, C for correlation flows, S for sanitised flows, and find it helpful also to include N for when there is no flow. We shall use max and min to choose the most or least worrying of the flows in question.

DEFINITION 7.13: A *flow relation* is a total mapping from pairs of data containers to $\{\mathsf{E}, \mathsf{I}, \mathsf{B}, \mathsf{C}, \mathsf{S}, \mathsf{N}\}$.

We shall use $F$ to range over flow relations, and $\tau$ to range over types of flows.

It might be slightly more intuitive to let a flow relation be a *partial* mapping from pairs of data containers to $\{\mathsf{E}, \mathsf{I}, \mathsf{B}, \mathsf{C}, \mathsf{S}\}$ (omitting the N). This is isomorphic to the choice made here but would give rise to more cumbersome formulations in the sequel.

We shall write

$$\delta_1 \to^{\tau} \delta_2$$

for the flow relation that gives N everywhere, except that the pair $(\delta_1, \delta_2)$ is mapped to $\tau$. This is in keeping with the notation that has already been used above and saying that $\delta_1 \not\to \delta_2$ now is represented by mapping the pair $(\delta_1, \delta_2)$ to N.

Extending this notation to sets of data containers we shall write

$$\Delta_1 \rightrightarrows^{\tau} \Delta_2$$

for the flow relation that gives N everywhere, except that a pair $(\delta_1, \delta_2) \in \Delta_1 \times \Delta_2$ is mapped to $\tau$. As a special case, $\{\,\} \rightrightarrows^{\mathsf{N}} \{\,\}$ denotes the flow relation that gives N everywhere.

EXAMPLE 7.14: Suppose that we want to model a database of records where each record has an integer field for the social security number (ssn), an integer field for the treatment offered (tre), and a text field for the outcome of the treament (res).

In Guarded Commands for Security we may model this by means of three arrays `Assn`, `Atre` and `Ares` such that the entries `Assn[i]`, `Atre[i]` and `Ares[i]` correspond to the same record, and where `Assn[i]` = 0 might indicate an unused entry.

Suppose further that we have an array `Cost` giving the integer cost of each treatment. The program below then traverses the database calculating the resulting price for treating a given patient.

```
price := 0;
i := 0;
do i < Assn# →
    if Assn[i] = patient →
        price:= price + Cost[ Atre [ i ] ]
    [] ¬(Assn[i] = patient) →
        skip
fi;
i := i+1
od
```

Exercise 7.15: Consider the deterministic program in Example 7.14 and determine all direct explicit and implicit flows. □

## 7.3 Multi-Level Security Policies

The purpose of a security policy is to indicate which information flows are acceptable (or secure) and which are unacceptable (or insecure). In this section we define the notion of security policy, give simple examples of policies, and then go on to consider more advanced ways of definining security policies using components and decentralised labels; the latter is not essential to understanding the type systems in the next sections.

**Security policies** The key motivation behind our development is to classify data containers according to a security domain, and to consider it secure to transfer data as expressed by an ordering on the elements of the security domain.

Definition 7.16: A *security policy* consists of a security domain together with a security association.

We now make security domains and security associations precise.

**Definition 7.17:** A *security domain* $\boldsymbol{L}$ is a finite and non-empty set equipped with a preorder $\sqsubseteq$; this is a relation over $\boldsymbol{L}$ that is *reflexive* ($\ell \sqsubseteq \ell$ for all $\ell \in \boldsymbol{L}$) and *transitive* ($\ell_1 \sqsubseteq \ell_3$ whenever $\ell_1 \sqsubseteq \ell_2$ and $\ell_2 \sqsubseteq \ell_3$).

The preorder indicates the direction in which it is secure to move data along; we shall use this approach regardless of whether we deal with confidentiality or integrity or mixtures or modifications of these.

**Definition 7.18:** A *security association* $\mathcal{L}$ is a mapping from the set of data containers of interest into the security domain.

Hence an information flow $\delta_1 \rightarrow^{\tau} \delta_2$ is secure with respect to the security policy whenever $\mathcal{L}(\delta_1) \sqsubseteq \mathcal{L}(\delta_2)$. An information flow $\delta_1 \rightarrow^{\tau} \delta_2$ with $\mathcal{L}(\delta_1) \not\sqsubseteq \mathcal{L}(\delta_2)$ and $\tau \neq \mathsf{N}$ constitutes a security violation at level $\tau$.

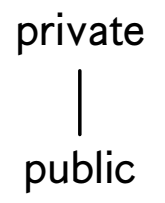

Figure 7.19: A simple lattice for confidentiality.

The simplest way to describe a security domain would be to list its elements and explain the preorder. Often this can be done graphically by means of a *directed acyclic graph* (DAG). So suppose that $(\boldsymbol{L}, \rightarrow)$ is a finite and directed acyclic graph and let $\rightarrow^*$ be the reflexive and transitive closure of $\rightarrow$ (following zero or more edges in $\rightarrow$). Defining $\sqsubseteq$ to be $\rightarrow^*$ then gives rise to a preordered set $(\boldsymbol{L}, \sqsubseteq)$. Recalling *Hasse diagrams* we can draw $(\boldsymbol{L}, \rightarrow)$ in such a way that edges always slant upwards; in particular, no edges slant downwards and no edges are horizontal. It is customary to omit edges that do not contribute to the reflexive and transitive closure of the edge relation. In terms of security this means that we are free to pass data upwards in the Hasse diagram.

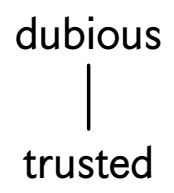

Figure 7.20: A simple lattice for integrity.

It is possible to construct security domains compositionally. The development is based on the fact that if $(\boldsymbol{L}_1, \sqsubseteq_1), \cdots, (\boldsymbol{L}_m, \sqsubseteq_m)$ are security domains then so is the cartesian product $(\boldsymbol{L}_1 \times \cdots \times \boldsymbol{L}_m, \sqsubseteq)$ with the pointwise preorder $((\ell_1, \cdots, \ell_m) \sqsubseteq (\ell'_1, \cdots, \ell'_m)$ if and only if $\ell_1 \sqsubseteq_1 \ell'_1$ and $\cdots$ and $\ell_m \sqsubseteq_m \ell'_m$).

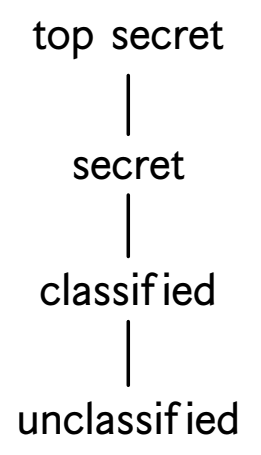

Figure 7.21: Another lattice for confidentiality.

**Simple confidentiality policies** In the confidentiality example of Figure 7.7, where `x` is private and `y` is public, we would like to allow `y` $\rightarrow$ `x` but to have `x` $\not\rightarrow$ `y`. We can achieve this by a security policy that has the security domain of Figure 7.19 and the security association that maps `x` to private and `y` to public.

There are many variations over this theme. A standard textbook example uses a security domain with unclassified, classified, secret and top secret intended to mimic the confidentiality levels of military forces. This is depicted in Figure 7.21.

It is not required that the partially ordered set is totally ordered as has been the case so far. For an example of this consider the security domain PowerSet({financial, medical}) under the subset ordering as illustrated in Figure 7.22. It describes a scenario where access to data may be restricted to those who are able to see financial data, or medical data, or both.

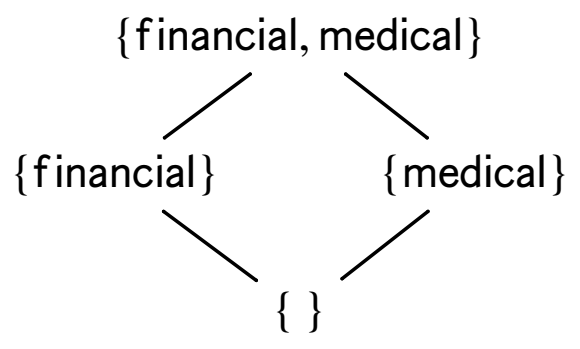

Figure 7.22: Yet another lattice for confidentiality.

Exercise 7.19: Suppose there are three users of a system: Alice, Bob, and Charlie. Define a security lattice PowerSet({Alice, Bob, Charlie}) for modelling readership rights: as information flows it is permitted to *remove* readers but *never to add* readers. □

**Simple integrity policies** In the integrity example of Figure 7.8, where `x` is trusted and `y` is dubious, we would like to allow $x \to y$ but to have $y \not\to x$. We can achieve this by a security policy that has the security domain of Figure 7.20 and the security association that maps `x` to trusted and `y` to dubious.

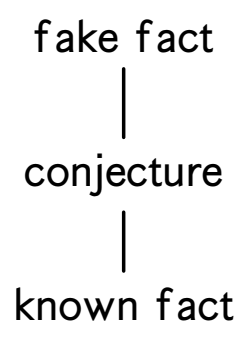

Figure 7.23: Another lattice for integrity.

Once again there are many variations over this theme. As an example consider the security domain of Figure 7.23 intended to record the trustworthiness of information in a scenario where uncorroborated and fake information is abundant.

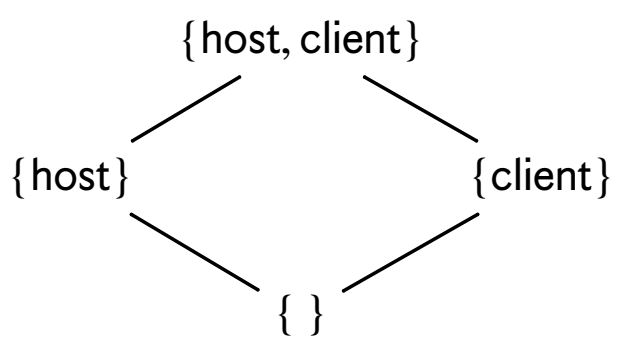

Figure 7.24: Yet another lattice for integrity.

For an example where the partially ordered set is not totally ordered consider the security lattice PowerSet({host, client}) under the subset ordering as illustrated in Figure 7.24. It describes a scenario where the trustworthiness of data may depend on whether the host has not been tampered with, the client has not been tampered with, or both. It is useful for controlling the design of client-server systems so that malicious modification of the (Java or JavaScript) client cannot jeopardise the overall security of the system.

Example 7.20: Returning to Example 7.7 we may obtain the desired security policy as follows.

For the security domain we use clean to indicate neutral data, Microsoft to indicate data proprietary to one client, Google to indicate data proprietary to another client, and Facebook to indicate data proprietary to yet another client. The notion of *proprietary* indicates both confidentiality and integrity and thus we aim at an isolation policy spanning both confidentiality and integrity. Hence we use the preorder shown in Figure 7.25.

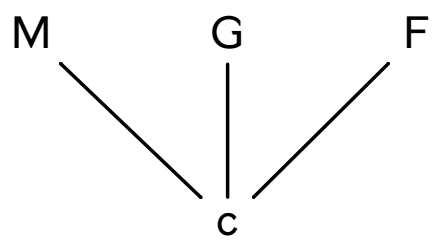

Figure 7.25: A partial order for isolation (abbreviating the security domains).

For the security assocation we might set $\mathcal{L}(\texttt{i}) = \text{clean}$, $\mathcal{L}(\texttt{j}) = \text{clean}$, $\mathcal{L}(\texttt{A\#}) = \text{clean}$, $\mathcal{L}(\texttt{B\#}) = \text{clean}$, $\mathcal{L}(\texttt{A[]}) = \text{Microsoft}$ and $\mathcal{L}(\texttt{B[]}) = \text{Google}$. This would be a more natural way to ensure that $\texttt{A[]} \not\to \texttt{B[]}$ and $\texttt{B[]} \not\to \texttt{A[]}$ than what was done in Example 7.7.

**Security lattices** The preorder given by a Hasse diagram gives rise to a preorder that is in fact a partial order. Recall that a *partial order* on top of being reflexive and transitive also is *antisymmetric* (if $\ell_1 \sqsubseteq \ell_2$ and $\ell_2 \sqsubseteq \ell_1$ then $\ell_1 = \ell_2$). We shall write $\ell_1 \sqsubset \ell_2$ whenever $\ell_1 \sqsubseteq \ell_2$ and $\ell_1 \neq \ell_2$.

Security domains $(\boldsymbol{L}, \sqsubseteq)$ described in the literature are often required to have some additional structure. If for every two elements $\ell_1$ and $\ell_2$ of $L$ we can find an element

$\ell_3$ of $\boldsymbol{L}$ such that

$$\forall \ell \in \boldsymbol{L} : (\ell_1 \sqsubseteq \ell \wedge \ell_2 \sqsubseteq \ell \Leftrightarrow \ell_3 \sqsubseteq \ell)$$

then $\boldsymbol{L}$ is a $\sqcup$-semilattice and we write $\ell_1 \sqcup \ell_2$ for $\ell_3$. (The symbol $\sqcup$ is pronounced *join*.) Similarly, if for every two elements $\ell_1$ and $\ell_2$ of $\boldsymbol{L}$ we can find an element $\ell_3$ of $\boldsymbol{L}$ such that

$$\forall \ell \in \boldsymbol{L} : (\ell \sqsubseteq \ell_1 \wedge \ell \sqsubseteq \ell_2 \Leftrightarrow \ell \sqsubseteq \ell_3)$$

then $\boldsymbol{L}$ is a $\sqcap$-semilattice and we write $\ell_1 \sqcap \ell_2$ for $\ell_3$. (The symbol $\sqcap$ is pronounced *meet*.) When the partially ordered set $\boldsymbol{L}$ is both a $\sqcup$-semilattice and a $\sqcap$-semilattice it is a *lattice*. For this reason $(\boldsymbol{L}, \sqsubseteq)$ is traditionally called a *security lattice*.

Try It Out 7.21: Consider our confidentiality example where $L = \{\text{private}, \text{public}\}$ with public $\sqsubseteq$ private; determine private $\sqcup$ public and private $\sqcap$ public. □

Exercise 7.22: Suppose that $(\boldsymbol{L}, \sqsubseteq)$ is a lattice. Show that the operation $\sqcup$ is commutative (that $l_1 \sqcup l_2 = l_2 \sqcup l_1$) and associative (that $l_1 \sqcup (l_2 \sqcup l_3) = (l_1 \sqcup l_2) \sqcup l_3$). Similarly show that the operation $\sqcap$ is commutative and associate. □

Since a security domain is finite and nonempty, $\boldsymbol{L} = \{\ell_1, \cdots, \ell_N\}$, there also is a *least* security element $\bot = \ell_1 \sqcap \cdots \sqcap \ell_N$ (pronounced *bottom*), and a *greatest* security element $\top = \ell_1 \sqcup \cdots \sqcup \ell_N$ (pronounced *top*).

Teaser 7.23: A partially ordered set $(\boldsymbol{L}, \sqsubseteq)$ is a *complete lattice* when it satisfies the following two properties:

- for all subsets $L' \subseteq \boldsymbol{L}$ we can find an element $\ell'$ of $\boldsymbol{L}$ such that $\forall \ell \in \boldsymbol{L} : (\forall \ell'' \in L' : \ell'' \sqsubseteq \ell) \Leftrightarrow \ell' \sqsubseteq \ell$ and we then write $\bigsqcup L'$ for $\ell'$
- for all subsets $L' \subseteq \boldsymbol{L}$ we can find an element $\ell'$ of $\boldsymbol{L}$ such that $\forall \ell \in \boldsymbol{L} : (\forall \ell'' \in L' : \ell \sqsubseteq \ell'') \Leftrightarrow l \sqsubseteq \ell'$ and we then write $\bigsqcap L'$ for $\ell'$

It is immediate that a complete lattice is also a lattice; argue that a finite and non-empty lattice is also a complete lattice.

Next suppose a flow relation is defined by $\delta_1 \to \delta_2$ if and only if $\mathcal{L}(\delta_1) \sqsubseteq \mathcal{L}(\delta_2)$. Show that $\Delta_1 \rightrightarrows \Delta_2$ is then equivalent to $\bigsqcup\{\mathcal{L}(\delta_1) \mid \delta_1 \in \Delta_1\} \sqsubseteq \bigsqcap\{\mathcal{L}(\delta_2) \mid \delta_2 \in \Delta_2\}$. This links the present development to the formulations often found in research papers on multi-level security and information flow. □

**Components** Describing the security domain as explicitly as done above becomes cumbersome once the security domain grows in size. Above we considered the two security categories financial and medical, that in turn gave rise to the four security components {financial, medical}, {financial}, {medical} and { } that were the elements in the security domain. We used the notation PowerSet({financial, medical}) for the set of these four security components. We now generalise this development to more than two security categories.

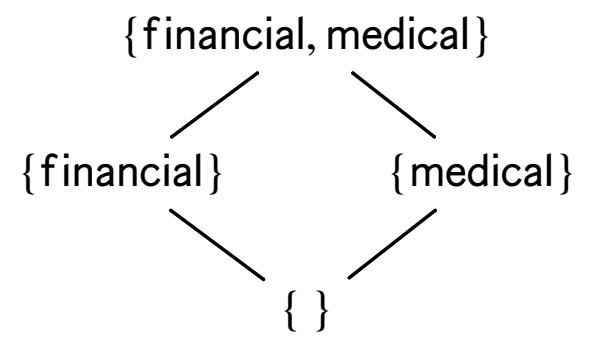

Figure 7.26: Illustrating the restriction ordering ($\sqsubseteq$ = $\subseteq$).

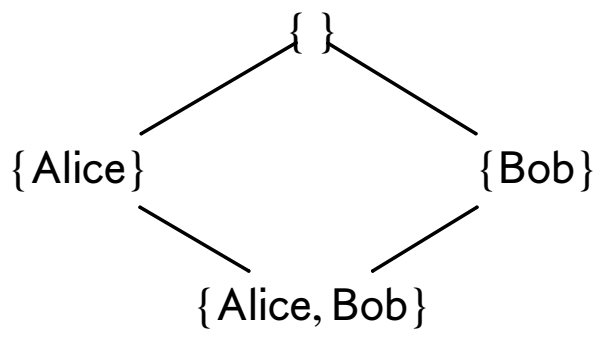

Figure 7.27: Illustrating the permission ordering ($\sqsubseteq$ = $\supseteq$).

DEFINITION 7.24: Define a finite and nonempty set C of *security categories*. A *security component* then is a set of security categories and the *security domain* $\boldsymbol{L} = \mathsf{PowerSet}(\mathsf{C})$ is the set of all such security components.

Our examples so far may have seemed to indicate that data could always *gain* more *restrictions*. This gave rise to the Hasse diagram of Figure 7.22 (redrawn in Figure 7.26) where for example $\{\mathsf{financial}\} \sqsubset \{\mathsf{financial}, \mathsf{medical}\}$. It indicated that is would be permissble to copy a variable requiring only financial trust into a variable requiring both financial and medical trust.

However, this is not the only approach we might take. Instead we might consider that data could always *lose* some of its *permissions*. For this let us consider a version of Exercise 7.19 where $\mathsf{PowerSet}(\{\mathsf{Alice}, \mathsf{Bob}\})$ should be use for modelling readership rights: as data flows it is permitted to *remove* readers but *never to add* readers. In this case we would have $\{\mathsf{Alice}, \mathsf{Bob}\} \sqsubset \{\mathsf{Alice}\}$ indicating that we might lose the ability to give the data to Bob. This motivates the Hasse diagram of Figure 7.27.

DEFINITION 7.25: Whenever the security categories are considered to be *restrictions* that can be gained but cannot be lost, the security domain will be ordered by the subset ordering (taking $\sqsubseteq$ to be $\subseteq$).

Whenever the security categories are considered to be *permissions* that can be lost but cannot be gained, the security domain will be ordered by the superset ordering (taking $\sqsubseteq$ to be $\supseteq$).

The *security domain* is then specified by listing the finite and nonempty set of security categories and indicating whether to use the ordering for restrictions or for permissions.

In both cases we retain the important principle that data may flow upwards in the Hasse diagram (but not downwards). Determining which choice of ordering to go for depends on determining whether or not it is considered to be secure to gain or lose security categories along flows.

EXERCISE 7.26: Argue that we get a security lattice regardless of which choice of ordering is made. □

EXERCISE 7.27: Consider again the program in Example 7.14 and assume that we have security categories doctor and accountant and that these are to be treated as *permissions* to access the data; draw the corresponding Hasse diagram.

Next suppose that we have a security association mapping `Assn#`, `Assn[]`, `patient` and `i` to $\{\mathsf{doctor}, \mathsf{accountant}\}$, mapping `Atre#`, `Atre[]`, `Ares#` and `Ares[]` to $\{\mathsf{doctor}\}$, and mapping `Cost#`, `Cost[]` and `price` to $\{\mathsf{accountant}\}$, Which of the flows characterised in Exercise 7.15 violate the security policy?

How can we sanitise the program to make it permissible? □

We shall allow to write $\{*\}$ for the security component consisting of all security categories. In case of the restriction ordering, the least element then is $\{\,\}$ and the greatest element is $\{*\}$, whereas in the case of the permission ordering, the least element is $\{*\}$ and the greatest element is $\{\,\}$.

**Decentralised Labels** The security perspective of components was of a rather global nature. To accomodate that different security principals might have different views on information flow, that all should be respected, we develop a notion of decentralised labels – motivated by the Decentralised Label Model of Myers and Liskov.

Consider again a version of Exercise 7.19 where $\mathsf{PowerSet}(\{\mathsf{Alice},\mathsf{Bob},\mathsf{Charlie}\})$ should be use for modelling readership rights, However, Alice, Bob and Charlie might have different views on who should be allowed to read the shared data. For example, Alice might want to grant readership permission to everybody whereas Bob and Charlie might want to grant readership permission only to themselves. We could write this as a decentralised label

$$[\mathsf{Alice} \mapsto \{\mathsf{Alice},\mathsf{Bob},\mathsf{Charlie}\}, \mathsf{Bob} \mapsto \{\mathsf{Bob},\mathsf{Charlie}\}, \mathsf{Charlie} \mapsto \{\mathsf{Bob},\mathsf{Charlie}\}]$$

This motivates the following concept.

> DEFINITION 7.28: Define a finite and nonempty set $\mathsf{P}$ of *security principals*. A *decentralised label* then is a total mapping from $\mathsf{P}$ to $\mathsf{PowerSet}(\mathsf{P})$ and the *security domain* $\boldsymbol{L} = (\mathsf{P} \to \mathsf{PowerSet}(\mathsf{P}))$ is the set of all such mappings.

As in the previous section there are two ways of considering the labels: as restrictions that can be gained but cannot be lost, or as *permissions* that can be lost but cannot be gained.

> DEFINITION 7.29: Whenever the labels are considered to be *restrictions* that can be gained but cannot be lost, the security domain will be ordered as $\ell_1 \sqsubseteq \ell_2$ if and only if $\forall P \in \mathsf{P} : \ell_1(P) \subseteq \ell_2(P)$.
>
> Whenever the security categories are considered to be *permissions* that can be lost but cannot be gained, the security domain will be ordered as $\ell_1 \sqsubseteq \ell_2$ if and only if $\forall P \in \mathsf{P} : \ell_1(P) \supseteq \ell_2(P)$.
>
> The *security domain* is then specified by listing the finite and nonempty set of security categories and indicating whether to use the ordering for restrictions or for permissions.

In both cases we retain the important principle that data may flow upwards in the Hasse diagram (but not downwards). Determining which choice of ordering to go for depends on determining whether or not it is considered to be secure to gain or lose security categories along flows.

EXERCISE 7.30: Argue that we get a security lattice regardless of which choice of ordering is made. □

EXERCISE 7.31: Assume that we have security principals Alice (or A) and Bob (or B) and that these are to treated as *restrictions* to access the data; draw the corresponding Hasse diagram. □

We shall allow to write $*$ for the list of all the security principals in P. In case of the restriction ordering, the least element then is $[* \mapsto \{\ \}]$ and the greatest element is $[* \mapsto \{*\}]$, whereas in the case of the permission ordering, the least element is $[* \mapsto \{*\}]$ and the greatest element is $[* \mapsto \{\ \}]$.

Given a decentralised label, if we have omitted writing an entry for a security principal we might take this to mean that the security principal maps to the least element (with respect to the chosen ordering). Similarly, if we are writing more than one entry for a security principal we might take this to mean that the security principal maps to the least upper bound (with respect to the chosen ordering) of the entries provided. Some developments in the literature follow this approach but when mixing permissions and restrictions clarity might be at stake.

## 7.4 Types for Measuring Leakage

In this section we develop a type system that overapproximates the set of flows that may occur in a program. This takes the form of an inference system for defining a judgement $\vdash C : F$ associating the command $C$ with the flow $F$ and similarly for guarded commands. We then subsequently enforce the security policy $(\boldsymbol{L}, \mathcal{L})$ and only retain those flows that are offending with respect to the security policy. This takes the form of defining a judgement $(\boldsymbol{L}, \mathcal{L}) \vdash C : F$. As a preparation we provide auxiliary notation for expressing the set of data containers used or modified in programs.

**Free and sanitised data containers** Given an arithmetic expression $a$ we shall write $\mathbf{fv}(a)$ for the set of data containers (in particular variables) occurring freely in $a$; by this we mean that they do not occur inside any `san` construct. Similarly, given an arithmetic expression $a$ we shall write $\mathbf{sv}(a)$ for the set of data containers (in particular variables) occuring within at least one `san` construct in $a$.

EXERCISE 7.32: To avoid any confusion in the above explanation one may define

**fv**() and **sv**() in a syntax-directed manner. Some of the key cases are the following:

$$\begin{aligned} \mathbf{fv}(n) &= \{\,\} & \mathbf{sv}(n) &= \{\,\} \\ \mathbf{fv}(x) &= \{x\} & \mathbf{sv}(x) &= \{\,\} \\ \mathbf{fv}(A[a_1]) &= \{A[]\} \cup \mathbf{fv}(a_1) & \mathbf{sv}(A[a_1]) &= \mathbf{sv}(a_1) \\ \mathbf{fv}(A\#) &= \{A\#\} & \mathbf{sv}(A\#) &= \{\,\} \\ \mathbf{fv}(a_1 + a_2) &= \mathbf{fv}(a_1) \cup \mathbf{fv}(a_2) & \mathbf{sv}(a_1 + a_2) &= \mathbf{sv}(a_1) \cup \mathbf{sv}(a_2) \\ \mathbf{fv}(\texttt{san}\ a_1) &= \{\,\} & \mathbf{sv}(\texttt{san}\ a_1) &= \mathbf{fv}(a_1) \cup \mathbf{sv}(a_1) \end{aligned}$$

Complete this specification to deal with all arithmetic and boolean expressions of Guarded Commands for Security as defined in Definition 7.4. □

**Modified data containers** Given a command $C$ we shall write **mod**$[\![C]\!]$ for the set of data containers that may be modified in $C$; similarly, given a guarded command $GC$ we write **mod**$[\![GC]\!]$ for the set of data containers that may be modified in $GC$. To avoid any confusion we provide the following formal definition:

$$\begin{aligned} \mathbf{mod}[\![x \texttt{ := } a]\!] &= \{x\} \\ \mathbf{mod}[\![A[a_1] \texttt{ := } a_2]\!] &= \{A[]\} \\ \mathbf{mod}[\![\texttt{skip}]\!] &= \{\,\} \\ \mathbf{mod}[\![C_1 \texttt{;} C_2]\!] &= \mathbf{mod}[\![C_1]\!] \cup \mathbf{mod}[\![C_2]\!] \\ \mathbf{mod}[\![\texttt{if } GC \texttt{ fi}]\!] &= \mathbf{mod}[\![GC]\!] \\ \mathbf{mod}[\![\texttt{do } GC \texttt{ od}]\!] &= \mathbf{mod}[\![GC]\!] \\ \mathbf{mod}[\![b \rightarrow C]\!] &= \mathbf{mod}[\![C]\!] \\ \mathbf{mod}[\![GC_1 \,[]\, GC_2]\!] &= \mathbf{mod}[\![GC_1]\!] \cup \mathbf{mod}[\![GC_2]\!] \end{aligned}$$

This constitutes an overapproximation of the set of data containers that will be modified in any execution: we will not miss any but we may include data containers that (because of conditional choices) are not modified.

**Simple assignments** We are now ready to develop the analysis for overapproximating the set of flows that may occur in a program. For a simple assignment to a variable we need to record a flow from the data containers in the arithmetic expression to the variable being modified. These flows will be explicit unless the data container in question occurs inside at least one `san` construct. This motivates the following axiom scheme.

$$\frac{}{\vdash x \texttt{ := } a : \begin{array}{l}(\mathbf{fv}(a) \rightrightarrows^{\mathsf{E}} \{x\}) \oplus \\ (\mathbf{sv}(a) \rightrightarrows^{\mathsf{S}} \{x\})\end{array}}$$

It makes use of the choice $F_1 \oplus F_2$ between two information flows where we should always choose the most worrying information flow whenever there is a choice. It is given by:

$$(F_1 \oplus F_2)(\delta_1, \delta_2) = \max\{F_1(\delta_1, \delta_2), F_2(\delta_1, \delta_2)\}$$

(As an aside, this definition can also be seen as a pointwise addition of matrices where max plays the role of addition.) It is immediate that both $F \oplus (\{\,\} \rightrightarrows^{\mathsf{N}} \{\,\})$ and $(\{\,\} \rightrightarrows^{\mathsf{N}} \{\,\}) \oplus F$ equal $F$.

EXAMPLE 7.33: For $\texttt{x} := \texttt{y} + \texttt{san}(\texttt{y} + \texttt{z})$ we may use the axiom to obtain that $\vdash \texttt{x} := \texttt{y} + \texttt{san}(\texttt{y} + \texttt{z}) : [\texttt{y} \to^{\mathsf{E}} \texttt{x}, \texttt{z} \to^{\mathsf{S}} \texttt{x}]$.

TRY IT OUT 7.34: Use the axiom scheme to infer the information flows $F_1$ and $F_2$ in $\vdash \texttt{y} := -1 * \texttt{z} : F_1$ and $\vdash \texttt{y} := \texttt{san}\ \texttt{z} : F_2$. □

**Assignments to arrays** For assignment to arrays we take the point of view that the data containers inside the index give rise to implicit rather than explicit flows. This is based on the consideration that for an array A having 5 elements the command $\texttt{if}\ a_1 = 1 \to \texttt{A[1]} := a_2\ [\,]\ \cdots\ [\,]\ a_1 = 5 \to \texttt{A[5]} := a_2\ \texttt{fi}$ is equivalent to the program $\texttt{A}[a_1] := a_2$. This motivates the following axiom scheme.

$$\frac{}{\vdash A[a_1] := a_2 : \begin{array}{l}(\mathbf{fv}(a_2) \rightrightarrows^{\mathsf{E}} \{A[\,]\}) \oplus \\ (\mathbf{fv}(a_1) \rightrightarrows^{\mathsf{I}} \{A[\,]\}) \oplus \\ (\mathbf{sv}(a_1) \cup \mathbf{sv}(a_2) \rightrightarrows^{\mathsf{S}} \{A[\,]\})\end{array}}$$

EXAMPLE 7.35: For $\texttt{A}[\texttt{x} + \texttt{san}(\texttt{y})] := \texttt{san}(\texttt{y}) + \texttt{z}$ we may use the axiom to obtain that $\vdash \texttt{A}[\texttt{x} + \texttt{san}(\texttt{y})] := \texttt{san}(\texttt{y}) + \texttt{z} : [\texttt{x} \to^{\mathsf{I}} \texttt{A}[\,], \texttt{y} \to^{\mathsf{S}} \texttt{A}[\,], \texttt{z} \to^{\mathsf{E}} \texttt{A}[\,]]$.

EXERCISE 7.36: In the assignment A[i] := B[j] the flow from i to A[] is recorded as an implicit flow whereas the flow from j to A[] is recorded as an explicit flow. An alternative would be to record the flow from j to A[] as being implicit.

Modify the axiom schemes for assignment accordingly. To do so, it would be natural to define $\mathbf{iv}(a)$ to be the set of data containers occurring inside some index in $a$ (but outside any san construct), and to define $\mathbf{ov}(a)$ to be the set of data containers occurring outside any index in $a$ (and outside any san construct), such that $\mathbf{fv}(a) = \mathbf{iv}(a) \cup \mathbf{ov}(a)$.

Discuss which choice of action schemes that you find most intuitive. □

**Skip** The axiom for skip produces the least possible flow relation.

$$\vdash \texttt{skip} : (\{\,\} \rightrightarrows^{\mathsf{E}} \{\,\})$$

**Sequencing** For sequential composition $C_1 ; C_2$ we need to compose the flows arising from $C_1$ and $C_2$. However, as our designation of the set of data containers modified represents an overapproximation, and as we have not required information flows to contain explicit flows from a data container to itself, we need to take care to also include the flows from each of the components. This motivates the following axiom scheme.

$$\frac{\vdash C_1 : F_1 \qquad \vdash C_2 : F_2}{\vdash C_1\,;C_2 : (F_1 \otimes F_2) \oplus F_1 \oplus F_2}$$

We now explain the operation $\otimes$ used. It follows from our treatment of the different types of flow, that if we have flows $\delta_1 \to^{\tau_1} \delta_2$ and $\delta_2 \to^{\tau_2} \delta_3$ then we also have a flow $\delta_1 \to^{\tau} \delta_3$ where $\tau$ is the least worrying of the two flows, i.e. $\tau = \min\{\tau_1, \tau_2\}$. If additionally there is another scenario where we have flows $\delta_1 \to^{\tau_1'} \delta_2'$ and $\delta_2' \to^{\tau_2'} \delta_3$ then we want $\tau$ to be the most worrying of the two candidates $\min\{\tau_1, \tau_2\}$ and $\min\{\tau_1', \tau_2'\}$, i.e. $\tau = \max\{\min\{\tau_1, \tau_2\}, \min\{\tau_1', \tau_2'\}\}$. This motivates defining the operation $\otimes$ by:

$$(F_1 \otimes F_2)(\delta_1, \delta_2) = \max\left\{ \min\left\{ \begin{array}{c} F_1(\delta_1, \delta), \\ F_2(\delta, \delta_2) \end{array} \right\} \mid \delta \text{ is a data container} \right\}$$

(As an aside, this definition can also be seen as a matrix multiplication where max plays the role of addition and min plays the role of multiplication.) It is immediate that both $F \otimes (\{\,\} \rightrightarrows^{\tau} \{\,\})$ and $(\{\,\} \rightrightarrows^{\tau} \{\,\}) \otimes F$ equal $\{\,\} \rightrightarrows^{\tau} \{\,\}$.

Example 7.37: For `x := y;z := san(x)` we may use the inference rule and a previous axiom to obtain

$$\frac{\vdash \texttt{x := y} : [\texttt{y} \to^{\mathsf{E}} \texttt{x}] \qquad \vdash \texttt{z := san(x)} : [\texttt{x} \to^{\mathsf{S}} \texttt{z}]}{\vdash \texttt{x := y;z := san(x)} : [\texttt{y} \to^{\mathsf{E}} \texttt{x}, \texttt{x} \to^{\mathsf{S}} \texttt{z}, \texttt{y} \to^{\mathsf{S}} \texttt{z}]}$$

Try It Out 7.38: Continuing Try It Out 7.34 use the rule to infer the information flow $F$ in $\vdash$ `y := -1 * z;y := san z` : $F$. □

**Conditional** For conditional most of the work is left to the analysis of the guarded command inside.

$$\frac{\vdash GC : F}{\vdash \texttt{if } GC \texttt{ fi} : F}$$

**Iteration** For iteration we need to take the transitive closure to reflect the iterative nature.

$$\frac{\vdash GC : F}{\vdash \texttt{do } GC \texttt{ od} : F^{\circledast}}$$

We now explain the operation $\circledast$ used. Generalising our consideration of sequential composition of flows we shall write $F^{\otimes 1} = F$ and $F^{\otimes(n+1)} = F^{\otimes n} \otimes F$. We may observe that

$$F^{\otimes n}(\delta_0, \delta_n) = \max\left\{ \min\left\{ \begin{array}{c} F(\delta_0, \delta_1), \\ \cdots, \\ F(\delta_{n-1}, \delta_n) \end{array} \right\} \mid \delta_1, \cdots, \delta_{n-1} \text{ are data containers} \right\}$$

meaning that $F^{\otimes n}$ summarises paths of length $n$ using the flow relation $F$. To summarise all non-empty paths we then define

$$F^{\circledast} = \bigsqcup_{n>0} F^{\otimes n}$$

(As an aside, this definition can be seen as a form of transitive closure of a matrix.)

EXERCISE 7.39: Argue that $F^{\circledast}$ satisfies the following equation:

$$F^{\circledast}(\delta_{\triangleright}, \delta_{\blacktriangleleft}) = \max\left\{ \min\left\{ \begin{array}{c} F(\delta_0, \delta_1), \\ \cdots, \\ F(\delta_{n-1}, \delta_n) \end{array} \right\} \mid \begin{array}{l} \delta_0, \cdots, \delta_n \text{ are data containers}, \\ n > 0, \delta_0 = \delta_{\triangleright}, \delta_n = \delta_{\blacktriangleleft} \end{array} \right\}$$

TEASER 7.40: Develop an efficient algorithm for computing $F^{\circledast}$. (We shall consider this again in Section 7.6.) □

**Guarded Commands** For a guarded command $b_i \rightarrow C_i$ we need to record the implicit flow from the condition $b_i$ to the command $C_i$ as well as incorporate the flows arising from $C_i$. Some of the implicit flows will actually be sanitised flows in case the data container inside $b$ occurs within at least one `san` construct. In the context of a guarded command $b_1 \rightarrow C_1$ [] $\cdots$ [] $b_n \rightarrow C_n$ with multiple choices these considerations account for the first line of the flow constructed below.

$$\frac{\vdash C_1 : F_1 \qquad \cdots \qquad \vdash C_n : F_n}{\vdash \begin{array}{c} b_1 \rightarrow C_1 \\ [] \cdots [] \\ b_n \rightarrow C_n \end{array} : \bigoplus_{i \leq n} \left( \begin{array}{l} \big(\mathbf{fv}(b_i) \rightrightarrows^{\mathsf{I}} \mathbf{mod}[\![C_i]\!]\big) \oplus \big(\mathbf{sv}(b_i) \rightrightarrows^{\mathsf{S}} \mathbf{mod}[\![C_i]\!]\big) \oplus F_i \oplus \\ \bigoplus_{j\ \mathbf{cosat}\ i} \begin{array}{l} (\mathbf{fv}(b_j) \rightrightarrows^{\mathsf{B}} \mathbf{mod}[\![C_i]\!]) \\ \oplus\ (\mathbf{sv}(b_j) \rightrightarrows^{\mathsf{S}} \mathbf{mod}[\![C_i]\!]) \\ \oplus\ (\mathbf{mod}[\![C_i]\!] \rightrightarrows^{\mathsf{C}} \mathbf{mod}[\![C_i]\!]) \end{array} \end{array} \right)}$$

The remaining three lines take care of the additional complications arising in a non-deterministic language where it is up to the scheduler to make the choice. To express this we shall assume that

$$j\ \mathbf{cosat}\ i$$

overapproximates when $b_j \wedge b_i$ might be satisfiable for *different* choices of $j$ and $i$, so that if $b_j \wedge b_i$ is satisfiable and $j \neq i$ then '$j$ **cosat** $i$' must be true. Whenever

'$j$ **cosat** $i$' we create the bypassing flows possible, taking care of those that will actually be sanitised flows instead, and we create the correlation flows between all data containers modified in either body.

EXAMPLE 7.41: In case of a guarded command $\mathtt{x} > 0 \to \mathtt{x} := \mathtt{y}\,;\mathtt{z} := \mathtt{san}(\mathtt{x})$ we may use the rule to obtain

$$\frac{\vdash \mathtt{x} := \mathtt{y}\,;\mathtt{z} := \mathtt{san}(\mathtt{x}) : [\mathtt{y} \to^{\mathsf{E}} \mathtt{x}, \mathtt{x} \to^{\mathsf{S}} \mathtt{z}, \mathtt{y} \to^{\mathsf{S}} \mathtt{z}]}{\vdash \mathtt{x} > 0 \to \mathtt{x} := \mathtt{y}\,;\mathtt{z} := \mathtt{san}(\mathtt{x}) : [\mathtt{y} \to^{\mathsf{E}} \mathtt{x}, \mathtt{y} \to^{\mathsf{S}} \mathtt{z}, \mathtt{x} \to^{\mathsf{I}} \mathtt{z}, \mathtt{x} \to^{\mathsf{I}} \mathtt{x}]}$$

EXAMPLE 7.42: In case of a multiple guarded command

$$\mathtt{true} \to \mathtt{y} := \mathtt{z}\,[]\,\mathtt{x} > 0 \to \mathtt{x} := \mathtt{y}\,;\mathtt{z} := \mathtt{san}(\mathtt{x})$$

we may use the rule to obtain

$$\frac{\begin{array}{c}\vdash \mathtt{true} \to \mathtt{y} := \mathtt{z} : [\mathtt{z} \to^{\mathsf{E}} \mathtt{y}] \\ \vdash \mathtt{x} > 0 \to \mathtt{x} := \mathtt{y}\,;\mathtt{z} := \mathtt{san}(\mathtt{x}) : [\mathtt{y} \to^{\mathsf{E}} \mathtt{x}, \mathtt{y} \to^{\mathsf{S}} \mathtt{z}, \mathtt{x} \to^{\mathsf{I}} \mathtt{z}, \mathtt{x} \to^{\mathsf{I}} \mathtt{x}]\end{array}}{\vdash \mathtt{true} \to \mathtt{y} := \mathtt{z}\,[]\,\mathtt{x} > 0 \to \mathtt{x} := \mathtt{y}\,;\mathtt{z} := \mathtt{san}(\mathtt{x}) : \begin{array}{c|ccc|} & \mathtt{x} & \mathtt{y} & \mathtt{z} \\ \hline \mathtt{x} & \mathsf{I} & \mathsf{B} & \mathsf{I} \\ \mathtt{y} & \mathsf{E} & \mathsf{C} & \mathsf{S} \\ \mathtt{z} & \mathsf{C} & \mathsf{E} & \mathsf{C} \\ \hline \end{array}}$$

where the resulting flow is presented in a form of a matrix indicating the type of flow from the row entry to the column entry.

```
if  x ≤ 0 → y := −1 ∗ z
[]  san x = 0 → y := 0
[]  x ≥ 0 → y := san z
fi
```

Figure 7.28: Example with many kinds of flows.

TRY IT OUT 7.43: Continuing Try It Out 7.34 consider the program of Figure 7.28 and use the analysis defined here to construct the overapproximation of the set of flows that may arise. □

EXERCISE 7.44: Reconsider the deterministic program in Example 7.14 and use the analysis defined here to construct the overapproximation of the set of flows that may arise. □

TEASER 7.45: Develop a heuristics that given $b_1 \to C_1\,[]\,\cdots\,[]\,b_n \to C_n$ overapproximates '$j$ **cosat** $i$'. (We shall consider this again in Section 7.6.) □

**Offending flows** Given a security policy $(\boldsymbol{L}, \mathcal{L})$ we can now obtain those flows that violate the security policy by means of the following rule.

$$\frac{\vdash C : F}{(\boldsymbol{L}, \mathcal{L}) \vdash C : F'} \text{ where } F'(\delta_1, \delta_2) = \begin{cases} F(\delta_1, \delta_2) & \text{if } \mathcal{L}(\delta_1) \not\sqsubseteq \mathcal{L}(\delta_2) \\ \mathsf{N} & \text{otherwise} \end{cases}$$

A command $C$ is said to be *offending at level* $\tau \neq \mathsf{N}$ with respect to a security policy

$(\boldsymbol{L}, \mathcal{L})$ whenever $(\boldsymbol{L}, \mathcal{L}) \vdash C : F$ and $F(\delta_1, \delta_2) \geq \tau$ for some $(\delta_1, \delta_2)$. In practice, commands offending only at level S should be considered sufficiently secure.

EXAMPLE 7.46: Consider again the program shown in Example 7.7 and suppose that the security domain $\boldsymbol{L} = \{\text{clean}, \text{M}, \text{G}, \text{F}\}$ is ordered by clean $\sqsubset$ M, clean $\sqsubset$ G and clean $\sqsubset$ F as in Example 7.20 and Figure 7.25.

If the security association has $\mathcal{L}(\texttt{i}) = \text{clean}$, $\mathcal{L}(\texttt{j}) = \text{clean}$, $\mathcal{L}(\texttt{A[ ]}) = \text{Microsoft}$, $\mathcal{L}(\texttt{B[ ]}) = \text{Google}$, $\mathcal{L}(\texttt{A\#}) = \text{clean}$ and $\mathcal{L}(\texttt{B\#}) = \text{clean}$ then the only offending flows are correlation flows. So the program is only offending at level C and should be considered sufficiently secure as the only way to get rid of the correlation flows would be to make the program fully deterministic.

If we change the security association to have $\mathcal{L}(\texttt{i}) = \text{Microsoft}$, $\mathcal{L}(\texttt{j}) = \text{Google}$, $\mathcal{L}(\texttt{A[ ]}) = \text{Microsoft}$, $\mathcal{L}(\texttt{E[ ]}) = \text{eeGoogle}$, $\mathcal{L}(\texttt{A\#}) = \text{Microsoft}$ and $\mathcal{L}(\texttt{E\#}) = \text{Google}$ then we also get bypassing flows. So the program is offending at level B and should probably not be considered sufficiently secure.

However, we might additionally change the program to use sanitisation

```
i := 0;
j := 0;
do sani < sanA# →  A[i] := A[i] + 27;
                   i := i + 1
[] sanj < sanB# →  B[j] := B[j] + 12;
                   j := j + 1
od
```

and then the only offending flows are sanitised. So the program is offending at level S and should be considered sufficiently secure.

There are other and more subtle ways in which information may flow than has been covered by our security analysis. The word *covert channel* is used to describe such phenomena.

As an example, the program

```
y := 0 ; x' := x ; do x' > 0 → x' := x' − 1 [] x' < 0 → x' := x' + 1 od
```

always terminates. It has no flows of any kind from `x` to `y` but if we can observe the *execution time* it reveals some information about the absolute value of `x`.

Similar examples can be constructed where the computation on `x` will only terminate successfully for some values of `x` and otherwise enter a loop or a stuck configuration. If we can observe the *non-termination* it also reveals some information about the value of `x`.

## 7.5 Types for Avoiding Leakage

Often we want to accept commands where only sanitised flows may violate the security policy. While this can be expressed using the development of Section 7.4 it is customary to define a type system that enforces the security policy along the way. We perform this development in the present section.

Throughout this section we shall assume that the security domain in question is in fact a complete lattice.

**Arithmetic Expressions** For aritmetic expressions the judgement takes the form

$$\mathcal{L} \vdash a : \ell$$

where $a$ is an arithmetic expression and $\ell$ is an element of the security domain. The formal definition is given by the following inference system.

$$\frac{}{\mathcal{L} \vdash n : \bot} \qquad \frac{}{\mathcal{L} \vdash s : \bot} \qquad \frac{}{\mathcal{L} \vdash x : \ell} \text{ if } \mathcal{L}(x) = \ell$$

$$\frac{\mathcal{L} \vdash a_1 : \ell_1}{\mathcal{L} \vdash A[a_1] : (\ell_1 \sqcup \ell)} \text{ if } \mathcal{L}(A[]) = \ell \qquad \frac{}{\mathcal{L} \vdash A\# : \ell} \text{ if } \mathcal{L}(A\#) = \ell$$

$$\frac{\mathcal{L} \vdash a_1 : \ell_1 \quad \mathcal{L} \vdash a_2 : \ell_2}{\mathcal{L} \vdash a_1 + a_2 : (\ell_1 \sqcup \ell_2)} \qquad \frac{\mathcal{L} \vdash a_1 : \ell_1 \quad \mathcal{L} \vdash a_2 : \ell_2}{\mathcal{L} \vdash a_1 - a_2 : (\ell_1 \sqcup \ell_2)}$$

$$\frac{\mathcal{L} \vdash a_1 : \ell_1 \quad \mathcal{L} \vdash a_2 : \ell_2}{\mathcal{L} \vdash a_1 * e_2 : (\ell_1 \sqcup \ell_2)} \qquad \frac{\mathcal{L} \vdash a_1 : \ell}{\mathcal{L} \vdash \mathtt{san}\ a_1 : \bot}$$

Constants are given the lowest security classification $\bot$, data containers are given the security classification specified by the security assignment, and most of the remaining constructs calculate the least upper bound of the security classifications of their components. The notable exception is in case of sanitisation where the lowest security classification $\bot$ is used.

Exercise 7.47: Recall the definition of free data containers studied in Exercise 7.32. Argue that whenenver $\mathcal{L} \vdash a : \ell$ we have that $\ell = \bigsqcup\{\mathcal{L}(\delta) \mid \delta \in \mathbf{fv}(a)\}$ so that $\ell$ is the least upper bound of the security levels associated with the free data containers in $a$. Note that sanitised data containers are entirely ignored. □

**Boolean Expressions** For boolean expressions the judgement takes the form

$$\mathcal{L} \vdash b : \ell$$

where $b$ is a boolean expression and $\ell$ is an element of the security domain. The formal definition is given by the following inference system.

$$\frac{}{\mathcal{L} \vdash \mathtt{true} : \bot} \qquad \frac{\mathcal{L} \vdash a_1 : \ell_1 \quad \mathcal{L} \vdash a_2 : \ell_2}{\mathcal{L} \vdash a_1 = a_2 : (\ell_1 \sqcup \ell_2)}$$

$$\frac{\mathcal{L} \vdash a_1 : \ell_1 \quad \mathcal{L} \vdash a_2 : \ell_2}{\mathcal{L} \vdash a_1 > a_2 : (\ell_1 \sqcup \ell_2)} \qquad \frac{\mathcal{L} \vdash a_1 : \ell_1 \quad \mathcal{L} \vdash a_2 : \ell_2}{\mathcal{L} \vdash a_1 \geq a_2 : (\ell_1 \sqcup \ell_2)}$$

$$\frac{\mathcal{L} \vdash b_1 : \ell_1 \quad \mathcal{L} \vdash b_2 : \ell_2}{\mathcal{L} \vdash b_1 \wedge b_2 : (\ell_1 \sqcup \ell_2)} \qquad \frac{\mathcal{L} \vdash b_1 : \ell_1 \quad \mathcal{L} \vdash b_2 : \ell_2}{\mathcal{L} \vdash b_1 \,\&\&\, b_2 : (\ell_1 \sqcup \ell_2)} \qquad \frac{\mathcal{L} \vdash b_1 : \ell_1}{\mathcal{L} \vdash \neg b_1 : \ell_1}$$

As in Exercise 7.47 we have that whenenver $\mathcal{L} \vdash b : \ell$ then $\ell = \bigsqcup\{\mathcal{L}(\delta) \mid \delta \in \mathbf{fv}(b)\}$ so that $\ell$ is the least upper bound of the security levels associated with the free data containers in $b$.

**Commands** For commands the judgement takes the form

$$\boldsymbol{L}, \mathcal{L} \vdash C : [\ell_1, \ell_2]$$

where $C$ is a command and $\ell_1$ and $\ell_2$ are elements of the security domain. The component $\ell_1$ would suffice for deterministic commands but $\ell_2$ is needed to also deal with non-deterministic commands.

Recall the definition modified data containers studied in Exercise 7.32. As will become clear when formally defining the judgement the intention is that $\ell_1 = \bigsqcap\{\mathcal{L}(\delta) \mid \delta \in \mathbf{mod}[\![C]\!]\}$ so that $\ell_1$ is the greatest lower bound of the security levels of the data containers modified in $C$. Furthermore, it is the intention that $\ell_2 = \bigsqcup\{\mathcal{L}(\delta) \mid \delta \in \mathbf{mod}[\![C]\!]\}$ so that $\ell_1$ is the least upper bound of the security levels of the data containers modified in $C$. In case $\mathbf{mod}[\![C]\!] \neq \{\,\}$ we then have that $\ell_1 \sqsubseteq \ell_2$.

**Simple assignments** For a simple assignment to a variable we need to record a flow from the data containers in the arithmetic expression to the variable being modified. This motivates the following axiom scheme.

$$\frac{\mathcal{L} \vdash a_1 : \ell_1}{\boldsymbol{L}, \mathcal{L} \vdash x := a_1 : [\ell, \ell]} \text{ if } \begin{cases} \mathcal{L}(x) = \ell \\ \ell_1 \sqsubseteq \ell \end{cases}$$

EXAMPLE 7.48: In case of $\mathtt{x} := \mathtt{y} + \mathtt{san}(\mathtt{y} + \mathtt{z})$ an instance of this takes the form

$$\boldsymbol{L}, \mathcal{L} \vdash \mathtt{x} := \mathtt{y} + \mathtt{san}(\mathtt{y} + \mathtt{z}) : [\ell_{\mathsf{x}}, \ell_{\mathsf{x}}] \quad \text{if} \quad \ell_{\mathsf{y}} \sqsubseteq \ell_{\mathsf{x}}$$

where we write $\mathcal{L}(\mathtt{x}) = \ell_{\mathsf{x}}$ and $\mathcal{L}(\mathtt{y}) = \ell_{\mathsf{y}}$. Note that $\mathtt{z}$ is ignored because it is sanitised.

EXERCISE 7.49: Referring back to the development in Section 7.4, argue that $\boldsymbol{L}, \mathcal{L} \vdash x := a_1 : [\ell, \ell]$ holds if and only if $\ell = \mathcal{L}(x)$ and $(\boldsymbol{L}, \mathcal{L}) \vdash x := a_1 : F$ where $F$ contains no explicit, implicit, bypassing or correlation flows (i.e. $\forall \delta_1, \delta_2 : F(\delta_1, \delta_2) \in \{\mathsf{S}, \mathsf{N}\}$). □

**Assignments to arrays** Assignment to arrays can be handled in the same way as simple assignments.

$$\frac{\mathcal{L} \vdash a_1 : \ell_1 \quad \mathcal{L} \vdash a_2 : \ell_2}{\boldsymbol{L}, \mathcal{L} \vdash A[a_1] := a_2 : [\ell, \ell]} \text{ if } \begin{cases} \mathcal{L}(A[]) = \ell \\ \ell_1 \sqsubseteq \ell \\ \ell_2 \sqsubseteq \ell \end{cases}$$

EXAMPLE 7.50: In case of $\mathtt{A[x + san(y)] := san(y) + z}$ an instance of this takes the form

$$\boldsymbol{L}, \mathcal{L} \vdash \mathtt{A[x + san(y)] := san(y) + z} : [\ell_{\mathsf{A}}, \ell_{\mathsf{A}}] \quad \text{if} \quad \ell_{\mathsf{x}} \sqsubseteq \ell_{\mathsf{A}} \text{ and } \ell_{\mathsf{z}} \sqsubseteq \ell_{\mathsf{A}}$$

where we write $\mathcal{L}(\mathtt{A[\,]}) = \ell_{\mathsf{A}}$, $\mathcal{L}(\mathtt{x}) = \ell_{\mathsf{x}}$ and $\mathcal{L}(\mathtt{z}) = \ell_{\mathsf{z}}$.

**Skip** In the case of `skip` we note that no data containers have been modified, i.e. $\mathbf{mod}[\![C]\!] = \{\,\}$, and that $\bigsqcap\{\,\} = \top$ and $\bigsqcup\{\,\} = \bot$. To ensure that our intentions continue to hold we therefore have the following axiom for `skip`.

$$\frac{}{\boldsymbol{L}, \mathcal{L} \vdash \mathtt{skip} : [\top, \bot]}$$

Note how `skip` differs from x := x.

**Sequencing** For sequential composition $C_1 ; C_2$ we need to compose the flows arising from $C_1$ and $C_2$. This motivates the following axiom scheme.

$$\frac{\boldsymbol{L}, \mathcal{L} \vdash C_1 : [\ell_1^1, \ell_2^1] \qquad \boldsymbol{L}, \mathcal{L} \vdash C_2 : [\ell_1^2, \ell_2^2]}{\boldsymbol{L}, \mathcal{L} \vdash C_1 ; C_2 : [\ell_1^1 \sqcap \ell_1^2, \ell_2^1 \sqcup \ell_2^2]}$$

EXAMPLE 7.51: In case of $\mathtt{x := y ; z := san(x)}$ we obtain

$$\boldsymbol{L}, \mathcal{L} \vdash \mathtt{x := y ; z := san(x)} : [\ell_{\mathsf{x}} \sqcap \ell_{\mathsf{z}}, \ell_{\mathsf{x}} \sqcup \ell_{\mathsf{z}}] \text{ if } \ell_{\mathsf{y}} \sqsubseteq \ell_{\mathsf{x}}$$

where we write $\mathcal{L}(\mathtt{x}) = \ell_{\mathsf{x}}$, $\mathcal{L}(\mathtt{y}) = \ell_{\mathsf{y}}$ and $\mathcal{L}(\mathtt{z}) = \ell_{\mathsf{z}}$.

EXERCISE 7.52: Suppose that $\boldsymbol{L}, \mathcal{L} \vdash C_i : [\ell_1^i, \ell_2^i]$ ensures that $\ell_1^i = \bigsqcap\{\mathcal{L}(\delta) \mid \delta \in \mathbf{mod}[\![C_i]\!]\}$ and $\ell_2^i = \bigsqcup\{\mathcal{L}(\delta) \mid \delta \in \mathbf{mod}[\![C_i]\!]\}$.

Argue that $\boldsymbol{L}, \mathcal{L} \vdash C_1 \,;\, C_2 : [\ell_1^1 \sqcap \ell_1^2, \ell_2^1 \sqcup \ell_2^2]$ ensures that $\ell_1^1 \sqcap \ell_1^2 = \bigsqcap\{\mathcal{L}(\delta) \mid \delta \in \mathbf{mod}[\![C_1 \,;\, C_2]\!]\}$ and $\ell_2^1 \sqcup \ell_2^2 = \bigsqcup\{\mathcal{L}(\delta) \mid \delta \in \mathbf{mod}[\![C_1 \,;\, C_2]\!]\}$. □

**Conditional** For conditional the key consideration is the analysis of the guarded command inside.

$$\frac{\boldsymbol{L}, \mathcal{L} \vdash GC : [\ell_1, \ell_2]}{\boldsymbol{L}, \mathcal{L} \vdash \texttt{if}\ GC\ \texttt{fi} : [\ell_1, \ell_2]}$$

**Iteration** Also for iteration the key consideration is the analysis of the guarded command inside.

$$\frac{\boldsymbol{L}, \mathcal{L} \vdash GC : [\ell_1, \ell_2]}{\boldsymbol{L}, \mathcal{L} \vdash \texttt{do}\ GC\ \texttt{od} : [\ell_1, \ell_2]}$$

**Guarded Commands** For a guarded command $b_1 \to C_1 \,[]\, \cdots \,[]\, b_n \to C_n$ we need to take care of the explicit flows inside each $C_i$; this is taken care of by analysing each $C_i$ in the type system.

Additionally, we need to take care of the implicit flows within each $b_i \to C_i$ and the bypassing and correlation flows between the components of $b_1 \to C_1 \,[]\, \cdots \,[]\, b_n \to C_n$. This motivates the following inference rule to be explained below.

$$\frac{\bigwedge_i \mathcal{L} \vdash b_i : \ell_0^i \quad \bigwedge_i \boldsymbol{L}, \mathcal{L} \vdash C_i : [\ell_1^i, \ell_2^i]}{\boldsymbol{L}, \mathcal{L} \vdash b_1 \to C_1 \,[]\, \cdots \,[]\, b_n \to C_n : [\bigsqcap_i \ell_1^i, \bigsqcup_i \ell_2^i]} \ \text{if} \begin{cases} \bigwedge_i \ell_0^i \sqsubseteq \ell_1^i \\ \bigwedge_{j\ \mathbf{cosat}\ i} \ell_0^j \sqsubseteq \ell_1^i \\ \bigwedge_{j\ \mathbf{cosat}\ i} \ell_1^i \sqsupseteq \ell_2^i \end{cases}$$

The condition $\bigwedge_i \ell_0^i \sqsubseteq \ell_1^i$ takes care of the implicit flow from $b_i$ to $C_i$ within each $b_i \to C_i$. As in Section 7.4 we write $j\ \mathbf{cosat}\ i$ to overapproximate when $b_j \wedge b_i$ might be satisfiable for *different* choices of $j$ and $i$, so that if $b_j \wedge b_i$ is satisfiable and $j \neq i$ then '$j\ \mathbf{cosat}\ i$' must be true. The condition $\ell_0^j \sqsubseteq \ell_1^i$ then takes care of the bypassing flow from $b_j$ to $C_i$. Furthermore, the condition $\ell_1^i \sqsupseteq \ell_2^i$ takes care of the correlation flows within $C_i$ by enforcing that all data containers modifed must have the same security label. (In the case where $C_i$ is not a trivial statement we have $\mathbf{mod}[\![C_i]\!] \neq \{\ \}$ so that the condition $\ell_1^i \sqsupseteq \ell_2^i$ is then equivalent to the more intuitive $\ell_1^i = \ell_2^i$.)

EXAMPLE 7.53: In case of a guarded command $\mathtt{x} > 0 \to \mathtt{x} := \mathtt{y}\,;\mathtt{z} := \mathtt{san}(\mathtt{x})$ we may use the rule to obtain

$$\boldsymbol{L}, \mathcal{L} \vdash \mathtt{x} > 0 \to \mathtt{x} := \mathtt{y}\,;\mathtt{z} := \mathtt{san}(\mathtt{x}) : [\ell_\mathsf{x}, \ell_\mathsf{z}] \text{ if } \ell_\mathsf{y} \sqsubseteq \ell_\mathsf{x} \text{ and } \ell_\mathsf{x} \sqsubseteq \ell_\mathsf{z}$$

where we write $\mathcal{L}(\mathtt{x}) = \ell_\mathsf{x}$, $\mathcal{L}(\mathtt{y}) = \ell_\mathsf{y}$ and $\mathcal{L}(\mathtt{z}) = \ell_\mathsf{z}$.

EXAMPLE 7.54: In case of a multiple guarded command

$$\mathtt{true} \to \mathtt{y} := \mathtt{z} \,[]\, \mathtt{x} > 0 \to \mathtt{x} := \mathtt{y}\,;\mathtt{z} := \mathtt{san}(\mathtt{x})$$

we may use the rule to obtain

$$\boldsymbol{L}, \mathcal{L} \vdash \mathtt{true} \to \mathtt{y} := \mathtt{z} \,[]\, \mathtt{x} > 0 \to \mathtt{x} := \mathtt{y}\,;\mathtt{z} := \mathtt{san}(\mathtt{x}) : [\ell_\mathsf{x}, \ell_\mathsf{x}]$$

$$\text{if } \left(\begin{array}{ll} \ell_\mathsf{z} \sqsubseteq \ell_\mathsf{y} & \text{(leftmost component)} \\ \ell_\mathsf{y} \sqsubseteq \ell_\mathsf{x}, \ell_\mathsf{x} \sqsubseteq \ell_\mathsf{z} & \text{(rightmost component)} \\ \bot \sqsubseteq \ell_\mathsf{x}, \ell_\mathsf{x} \sqsubseteq \ell_\mathsf{y} & \text{(bypassing flows)} \\ \ell_\mathsf{y} \sqsupseteq \ell_\mathsf{y}, \ell_\mathsf{x} \sqsupseteq \ell_\mathsf{x} \sqcup \ell_\mathsf{z} & \text{(correlation flows)} \end{array}\right) \Rightarrow \ell_\mathsf{x} = \ell_\mathsf{y} = \ell_\mathsf{z}$$

where we write $\mathcal{L}(\mathtt{x}) = \ell_\mathsf{x}$, $\mathcal{L}(\mathtt{y}) = \ell_\mathsf{y}$ and $\mathcal{L}(\mathtt{z}) = \ell_\mathsf{z}$.

EXERCISE 7.55: Show that $\boldsymbol{L}, \mathcal{L} \vdash C : [\ell_1, \ell_2]$ ensures that our intentions that $\ell_1 = \bigsqcap\{\mathcal{L}(\delta) \mid \delta \in \mathbf{mod}[\![C]\!]\}$ and $\ell_2 = \bigsqcup\{\mathcal{L}(\delta) \mid \delta \in \mathbf{mod}[\![C]\!]\}$ actually hold. □

TEASER 7.56: Referring back to the development in Section 7.4, argue that

$$\boldsymbol{L}, \mathcal{L} \vdash C : [\ell_1, \ell_2]$$

holds if and only if

$$\begin{array}{c} \ell_1 = \bigsqcap\{\mathcal{L}(\delta) \mid \delta \in \mathbf{mod}[\![C]\!]\} \\ \ell_2 = \bigsqcup\{\mathcal{L}(\delta) \mid \delta \in \mathbf{mod}[\![C]\!]\} \\ (\boldsymbol{L}, \mathcal{L}) \vdash C : F \text{ where } \forall \delta_1, \delta_2 : F(\delta_1, \delta_2) \in \{\mathsf{S}, \mathsf{N}\} \end{array}$$

This shows that the approaches of this and the present section are compatible. □

## 7.6 Algorithmic Issues

We left two algorithmic issues unsolved in the previous sections – namely how to compute transitive closures and how to overapproximate satisfiability of boolean expressions – and we now rectify this.

**Transitive Closure** The key step in ensuring that $F^*$ can be implemented is to provide an upper bound on how many $F^{\otimes n}$ need to be considered.

> Proposition 7.57: If there are at most $N$ data containers in the program considered then $F^{\circledast} = \bigsqcup_{n=1}^{N} F^{\otimes n}$.

Sketch of Proof: It is immediate that $F^{\circledast}(\delta_\circ, \delta_\bullet)$ is greater than or equal to $(\bigsqcup_{n=1}^{N} F^{\otimes n})(\delta_\circ, \delta_\bullet)$.

If they are not equal there must be a sequence of data containers $\delta_\circ = \delta_0 = \cdots = \delta_n = \delta_\bullet$ with $n > N$ such that $\min\{F(\delta_0, \delta_1), \cdots, F(\delta_{n-1}, \delta_n)\}$ is not less than or equal to $(\bigsqcup_{n=1}^{N} F^{\otimes n})(\delta_\circ, \delta_\bullet)$. We proceed by contradiction and without loss of generality we may assume that $n$ is as small as possible.

There must be a data container than occurs more than once in $\delta_\circ = \delta_0 = \cdots = \delta_n = \delta_\bullet$ so consider the reduced sequence obtained by omitting all data containers between the first and the last occurrence and retaining just one occurrence of the data container in question. The reduced sequence will provide a value $\tau$ of the $\min\{\cdots\}$ formula such that $\min\{F(\delta_0, \delta_1), \cdots, F(\delta_{n-1}, \delta_n)\}$ is less than or equal to $\tau$, that is again less than or equal to $(\bigsqcup_{n=1}^{N} F^{\otimes n})(\delta_\circ, \delta_\bullet)$.

This provides the desired contradiction. □

For a more efficient construction using *dynamic programming* let us define

$$F^{[0]} = F \qquad F^{[n+1]} = F \oplus (F^{[n]} \otimes F^{[n]})$$

This is intended to ensure that $F^{[m]}$ correctly summarises the effect of all paths of length between 1 and $2^m$.

Exercise 7.58: Argue that $F^{[m]}$ satisfies the following equation:

$$F^{[m]}(\delta_\circ, \delta_\bullet) = \max\left\{ \min\left\{ \begin{array}{c} F(\delta_0, \delta_1), \\ \cdots, \\ F(\delta_{n-1}, \delta_n) \end{array} \right\} \;\middle|\; \begin{array}{l} \delta_0, \cdots, \delta_n \text{ are data containers,} \\ 1 \le n \le 2^m, \delta_0 = \delta_\circ, \delta_n = \delta_\bullet \end{array} \right\}$$

> Proposition 7.59: If there are at most $N$ data containers in the program considered then $F^{\circledast} = F^{[M]}$ where $M = \lceil \log_2 N \rceil$.

Sketch of Proof: From Proposition 7.57 it follows that we only need to consider paths of length between 1 and $N$ and the result then follows from Exercise 7.58 as $N \le 2^M$. □

**Overapproximating satisfiability** We now develop a heuristics for overapproximating whether or not two boolean expresions might be jointly satisfiable. Recall

```
function sat(b)
  convert b to disjunctive normal form ⋁_i b^i_1∧···∧b^i_{n_i};
  global := false;
  iterating through all i do
    local := true;
    build the ordered DAG for b^i_1∧···∧b^i_{n_i};
    if the DAG contains a marked node ¬t
      where also t is marked then local := false;
    if the DAG contains marked nodes t1 o1 t2 and t1 o2 t2
      with (o1,o2) ∈ ℰ then local := false;
    global := global ∨ local;
  return global
```

Figure 7.29: Algorithm for **sat**$[\![b]\!]$.

that we considered a construct $b_1 \rightarrow C_1 \,[]\, \cdots \,[]\, b_n \rightarrow C_n$ and used the notation '$j$ **cosat** $i$' to overapproximate whether or not $b_i$ and $b_j$ can be jointly satisfied for different choices of $i$ and $j$. We shall define

$$(j\ \mathbf{cosat}\ i) = \left(\ \mathbf{sat}[\![b_i \wedge b_j]\!] \wedge j \neq i\ \right)$$

and now develop a heuristics for **sat**$[\![\cdot]\!]$ as shown in Figure 7.29.

While the syntax only allows $=$, $>$ and $\geq$ we shall introduce $<$ and $\leq$ as part of our development. The syntax allows && but in the construction we shall treat it as $\wedge$, as our analysis will not need to consider the potential undefinedness of boolean expressions. On top of using $\wedge$ we shall make use of $\vee$ in our development. Finally, we shall feel free to disregard the `san` construct as it has no semantic effect.

A boolean expression is a *literal* when it has no occurrences of $\wedge$, &&, or $\vee$ and at most one occurrence of $\neg$. Our heuristics will be based on first constructing an ordered DAG (directed acyclic graph) for conjunctions of literals, and next inspecting the ordered DAG to overapproximate satisfiability.

The first step is of Figure 7.29 is to translate $b$ into disjunctive normal form; this is where $\vee$ may get introduced and && is treated as $\wedge$. The result is an equivalent formula (modulo undefinedness)

$$\bigvee_i\ b^i_1 \wedge \cdots \wedge b^i_{n_i}$$

where each $b^i_j$ is a literal.

Iterating through each conjunction of literals $b^i_1 \wedge \cdots \wedge b^i_{n_i}$ the algorithm of Figure 7.29 first constructs an ordered DAG (directed acyclic graph), and next inspects the ordered DAG to overapproximate satisfiability, as detailed below.

**Constructing the ordered DAG** To increase the amount of sharing in the ordered DAG we need to keep track of the 'transposed variants' of the arithmetic and relational operators:

$$\mathcal{T} = \{(+,+), (*,*), (<,>), (\leq,\geq), (=,=), (\geq,\leq), (<,>)\}$$

This takes care of characterising both those operators that are commutative (like $+$) and those that can be 'transposed' (like $a_1 < a_2$ may be transposed to $a_2 > a_1$). In general, whenever $(o_1, o_2) \in \mathcal{T}$ it must be the case that $t_1\, o_1\, t_2$ is equivalent to $t_2\, o_2\, t_1$.

Given a conjunction of literals $b_1 \wedge \cdots \wedge b_n$ we construct an ordered DAG by a bottom-up traversal over the parse tree. Leaves will be numbers $n$, strings $s$, variables $x$, arrays $A$, and `true`; internal nodes will be $[]$, #, $+$, $-$, $*$, `san`, $<$, $\leq$, $=$, $>$, $\geq$ and $\neg$. Some of the nodes will be marked, and internal nodes will retain the order of their subgraphs.

When we encounter a potential new leaf in the bottom-up traversal over the parse tree of $b_1 \wedge \cdots \wedge b_n$, we reuse the node in the DAG if it is already there, otherwise we construct a new leaf.

When we encounter a potential new internal node $t_1\, o_1\, t_2$, we reuse the node in the DAG if it is already there, otherwise we proceed as follows. If $(o_1, o_2) \in \mathcal{T}$ and there already is a node in the DAG for $t_2\, o_2\, t_1$, we use that node in the DAG, otherwise we construct the node $t_1\, o_1\, t_2$.

Once we encounter the root of one of the $b_i$ we *mark* the node.

**Inspecting the ordered DAG** To detect cases where satisfiability fails we need to keep track of pairs of relational operators that exclude each other:

$$\mathcal{E} = \{(<,=), (<,\geq), (<,>), (\leq,>), (=,<), (=,>), (\geq,<), (>,<), (>,\leq), (>,=)\}$$

In general, whenever $(o_1, o_2) \in \mathcal{E}$ it must be the case that $a_1\, o_1\, a_2$ and $a_1\, o_2\, a_2$ are not jointly satisfiable for any choices of $a_1$ and $a_2$.

We can then establish the overapproximating nature of our heuristics.

PROPOSITION 7.60: If the boolean formula $b$ is satisfiable then the algorithm **sat**$[\![b]\!]$ returns true.

SKETCH OF PROOF: If the ordered DAG for a conjunction of literals contains a marked node $t$ that has an ancestor $\neg t$ that is also marked, then clearly the conjunction of literals is not satisfiable. Similarly, if the ordered DAG for a conjunction of literals contains nodes $t_1$ and $t_2$ that have marked ancestors $t_1\, o_1\, t_2$ and $t_1\, o_2\, t_2$ with $(o_1, o_2) \in \mathcal{E}$, then the conjunction of literals is not satisfiable.

This shows that the resulting value of local for each iteration only reports false when the conjunction of literals is not satisfiable. It follows that the overall algorithm only reports false if none of the conjuncts of the disjunctive normal form are satisfiable. □

We may conclude that '$j$ **cosat** $i$' is a correct overapproximation of joint satisfiability of $b_i$ and $b_j$ (for distinct $i$ and $j$) from $b_1 \rightarrow C_1$ [] $\cdots$ [] $b_n \rightarrow C_n$.

EXERCISE 7.61: We now explore a way to reduce the amount of overapproximation performed by our heuristics. To this end define

$$\mathcal{C} = \{(<, \geq), (\leq, >), (\geq, <), (>, \leq)\}$$

and note that if $(o_1, o_2) \in \mathcal{C}$ then $\neg(a_1\, o_1\, a_2)$ and $a_1\, o_2\, a_2$ are equivalent for all choices of $a_1$ and $a_2$.

Modify the construction of the ordered DAG such that whenever there is a need to construct a node for $\neg(t_1\, o_1\, t_2)$ and $(o_1, o_2) \in \mathcal{C}$ then we construct the node $t_1\, o_2\, t_2$ (instead of $\neg(t_1\, o_1\, t_2)$) making sure to reuse any existing node for $t_1\, o_2\, t_2$ that might already be in the DAG. Argue that the correctness of Proposition 7.60 is maintained, and that it is never undesirable to perform this modification. □

EXERCISE 7.62: Extend the development of the present section by introducing the operator $\neq$ and making the corresponding changes to the sets $\mathcal{T}$, $\mathcal{E}$ and $\mathcal{C}$. Also, consider whether or not to remove `san` wshen constructing the ordered DAG. □

# Chapter 8

# Datalog Based Analyses

So far we have expressed program analyses using constraints and have used the advanced algorithms based on chaotic iteration to solve them. In this chapter we show how to express program analyses using Datalog so as to obtain the following benefits: *(i)* it may be faster to develop an analysis thanks to the uniform format of Datalog clauses, and *(ii)* techniques exist for solving Datalog claues very efficiently even for very large programs.

We begin by recasting the Reaching Definitions analysis from Section 2.1 in Datalog in order to illustrate the approach. Then we show how to perform a Control Flow Analysis for a programming language with recursive functions. Finally we consider a stratified version of Datalog and argue that essentially all analyses in the Monotone Framework based on finite powersets can be expressed using Datalog (including all those of the Bit-Vector Framework).

## 8.1 Reaching Definitions in Datalog

In this section we will be presenting the *Reaching Definitions* analysis of Section 2.1 in a different form.

Recall that an analysis assignment for Reaching Definitions analysis is a mapping RD that maps each node in the program graph to a set of triples of the form $(x, q_\circ, q_\bullet)$ or $(A, q_\circ, q_\bullet)$ where $q_\bullet$ is in **Q** and $q_\circ$ is in either the special symbol '?' or is in **Q**:

$$\mathsf{RD} : \mathbf{Q} \rightarrow \mathsf{PowerSet}(\,(\mathbf{Var} \cup \mathbf{Arr}) \times \mathbf{Q}_? \times \mathbf{Q}\,)$$

where $\mathbf{Q}_? = \{?\} \cup \mathbf{Q}$. Here an element of the *powerset* $\mathsf{PowerSet}(\,(\mathbf{Var} \cup \mathbf{Arr}) \times \mathbf{Q}_? \times \mathbf{Q}\,)$ is merely a subset of $(\mathbf{Var} \cup \mathbf{Arr}) \times \mathbf{Q}_? \times \mathbf{Q}$.

Further recall that an analysis assignment is supposed to solve a set of constraints obtained from a program graph in the following manner.

> For each edge $(q_\circ, \alpha, q_\bullet)$ in the program graph we obtain the constraint
>
> $$\big(\mathsf{RD}(q_\circ) \setminus \mathsf{kill}_{\mathsf{RD}}(q_\circ, \alpha, q_\bullet)\big) \cup \mathsf{gen}_{\mathsf{RD}}(q_\circ, \alpha, q_\bullet) \subseteq \mathsf{RD}(q_\bullet)$$
>
> and for the entry node we additionally obtain the constraint
>
> $$(\mathbf{Var} \cup \mathbf{Arr}) \times \{?\} \times \{q_\triangleright\} \subseteq \mathsf{RD}(q_\triangleright)$$

This can be rephrased into the following equivalent set of constraints.

> For each edge $(q_\circ, \alpha, q_\bullet)$ in the program graph we obtain two constraints
>
> $$\begin{array}{l} \mathsf{RD}(q_\bullet) \supseteq \mathsf{RD}(q_\circ) \setminus \mathsf{kill}_{\mathsf{RD}}(q_\circ, \alpha, q_\bullet) \\ \mathsf{RD}(q_\bullet) \supseteq \mathsf{gen}_{\mathsf{RD}}(q_\circ, \alpha, q_\bullet) \end{array}$$
>
> and for the entry node we additionally obtain the constraint
>
> $$\mathsf{RD}(q_\triangleright) \supseteq (\mathbf{Var} \cup \mathbf{Arr}) \times \{?\} \times \{q_\triangleright\}$$

This can be re-expressed in logical form as follows:

> For each edge $(q_\circ, \alpha, q_\bullet)$ in the program graph we impose two formulae
>
> $$\begin{array}{l} \forall u, v, w : (u, v, w) \in \mathsf{RD}(q_\bullet) \Leftarrow (u, v, w) \in \mathsf{RD}(q_\circ) \wedge (u, v, w) \notin \mathsf{kill}_{\mathsf{RD}}(q_\circ, \alpha, q_\bullet) \\ \forall u, v, w : (u, v, w) \in \mathsf{RD}(q_\bullet) \Leftarrow (u, v, w) \in \mathsf{gen}_{\mathsf{RD}}(q_\circ, \alpha, q_\bullet) \end{array}$$
>
> and for the entry node we additionally impose the two formulae
>
> $$\begin{array}{l} \forall u : (u, ?, q_\triangleright) \in \mathsf{RD}(q_\triangleright) \Leftarrow u \in \mathbf{Var} \\ \forall u : (u, ?, q_\triangleright) \in \mathsf{RD}(q_\triangleright) \Leftarrow u \in \mathbf{Arr} \end{array}$$

Along the lines of the change in functionality conducted in Section 3.1 we should like to represent an analysis assignment in a different form resembling what is sometimes called the 'graph' of the function RD. This leads to representing an analysis assignment in the form

$$\mathsf{RD}' \in \mathsf{PowerSet}(\, \mathbf{Q} \times (\mathbf{Var} \cup \mathbf{Arr}) \times \mathbf{Q}_? \times \mathbf{Q} \,)$$

which is equivalent to $\mathsf{RD}' \subseteq \mathbf{Q} \times (\mathbf{Var} \cup \mathbf{Arr}) \times \mathbf{Q}_? \times \mathbf{Q}$. The two representations are entirely isomorphic, meaning that we can pass freely between them without introducing any errors. Given RD we can construct $\mathsf{RD}'$ by setting

$$\mathsf{RD}' = \{(q, u, v, w) \mid (u, v, w) \in \mathsf{RD}(q)\}$$

and given RD′ we can construct RD by setting

$$\mathsf{RD}(q) = \{(u, v, w) \mid (q, u, v, w) \in \mathsf{RD}'\}$$

Furthermore, going from the mapping-based representation to the set-based representation and back gives us the same set that we started with. Similarly, going from the set-based representation to the mapping-based representation and back again gives us the same mapping that we started with.

We can now re-express the formulae constraining RD into equivalent formulae constraining RD′ as follows:

For each edge $(q_\circ, \alpha, q_\bullet)$ in the program graph we impose two formulae

$$\forall u, v, w : (q_\bullet, u, v, w) \in \mathsf{RD}' \Leftarrow (q_\circ, u, v, w) \in \mathsf{RD}' \land (u, v, w) \notin \mathsf{kill}_{\mathsf{RD}}(q_\circ, \alpha, q_\bullet)$$
$$\forall u, v, w : (q_\bullet, u, v, w) \in \mathsf{RD}' \Leftarrow (u, v, w) \in \mathsf{gen}_{\mathsf{RD}}(q_\circ, \alpha, q_\bullet)$$

and for the entry node we additionally impose the two formulae

$$\forall u : (q_\triangleright, u, ?, q_\triangleright) \in \mathsf{RD}' \Leftarrow u \in \mathbf{Var}$$
$$\forall u : (q_\triangleright, u, ?, q_\triangleright) \in \mathsf{RD}' \Leftarrow u \in \mathbf{Arr}$$

Clearly RD satisfies its constraining formulae exactly when the corresponding RD′ satisfies its constraining formae and vice versa.

At this stage it is useful to expand the definitions of $\mathsf{kill}_{\mathsf{RD}}$ and $\mathsf{gen}_{\mathsf{RD}}$ which gives:

For each edge $(q_\circ, x \mathrel{:=} a, q_\bullet)$ or $(q_\circ, c?x, q_\bullet)$ in the program graph we impose two formulae

$$\forall u, v, w : (q_\bullet, u, v, w) \in \mathsf{RD}' \Leftarrow (q_\circ, u, v, w) \in \mathsf{RD}' \land u \neq x$$
$$(q_\bullet, x, q_\circ, q_\bullet) \in \mathsf{RD}'$$

For each edge $(q_\circ, A[a_1] \mathrel{:=} a_2, q_\bullet)$ or $(q_\circ, c?A[a], q_\bullet)$ in the program graph we impose two formulae

$$\forall u, v, w : (q_\bullet, u, v, w) \in \mathsf{RD}' \Leftarrow (q_\circ, u, v, w) \in \mathsf{RD}'$$
$$(q_\bullet, A, q_\circ, q_\bullet) \in \mathsf{RD}'$$

For each edge $(q_\circ, c!a, q_\bullet)$ or $(q_\circ, b, q_\bullet)$ or $(q_\circ, \texttt{skip}, q_\bullet)$ in the program graph we impose the formula

$$\forall u, v, w : (q_\bullet, u, v, w) \in \mathsf{RD}' \Leftarrow (q_\circ, u, v, w) \in \mathsf{RD}'$$

For the entry node we additionally impose the two formulae

$$\forall u : (q_\triangleright, u, ?, q_\triangleright) \in \mathsf{RD}' \Leftarrow u \in \mathbf{Var}$$
$$\forall u : (q_\triangleright, u, ?, q_\triangleright) \in \mathsf{RD}' \Leftarrow u \in \mathbf{Arr}$$

To translate these formulae into the Datalog notation we shall *(i)* dispense with the universal quantifiers, *(ii)* change the notation for implication from $\Leftarrow$ to $\leftarrow$, *(iii)* change the membership notation $(q_\bullet, u, v, w) \in \mathsf{RD}'$ to predicate notation $\mathsf{RD}'(q_\bullet, u, v, w)$.

Since we have dispensed with the universal quantifiers it makes sense to be clear about which identifiers are supposed to be variables and let us also declare the predicates with their arities (number of arguments); some of the predicates will be initialised outside of the Datalog clauses and we use '/0' to indicate this and '/1' for those only computed by the Datalog clauses.

This leads to the following Datalog clauses for constraining the solution to $\mathsf{RD}'$:

Initially we make the declarations

$$\underline{\mathsf{PRED}}\ \mathbf{Var}(1)/0\ \mathbf{Arr}(1)/0\ \mathsf{RD}'(4)/1$$
$$\underline{\mathsf{VAR}}\ u\ v\ w$$

For each edge $(q_\circ, x := a, q_\bullet)$ or $(q_\circ, c?x, q_\bullet)$ in the program graph we impose two clauses

$$\mathsf{RD}'(q_\bullet, u, v, w) \leftarrow \mathsf{RD}'(q_\circ, u, v, w), u \neq x$$
$$\mathsf{RD}'(q_\bullet, x, q_\circ, q_\bullet) \leftarrow$$

For each edge $(q_\circ, A[a_1] := a_2, q_\bullet)$ or $(q_\circ, c?A[a], q_\bullet)$ in the program graph we impose two clauses

$$\mathsf{RD}'(q_\bullet, u, v, w) \leftarrow \mathsf{RD}'(q_\circ, u, v, w)$$
$$\mathsf{RD}'(q_\bullet, A, q_\circ, q_\bullet) \leftarrow$$

For each edge $(q_\circ, c!a, q_\bullet)$ or $(q_\circ, b, q_\bullet)$ or $(q_\circ, \texttt{skip}, q_\bullet)$ in the program graph we impose the clause

$$\mathsf{RD}'(q_\bullet, u, v, w) \leftarrow \mathsf{RD}'(q_\circ, u, v, w)$$

For the entry node we additionally impose the two clauses

$$\mathsf{RD}'(q_\triangleright, u, ?, q_\triangleright) \leftarrow \mathbf{Var}(u)$$
$$\mathsf{RD}'(q_\triangleright, u, ?, q_\triangleright) \leftarrow \mathbf{Arr}(u)$$

Given the 'usual' semantics of Datalog (to be considered in Section 8.3) the analysis assignment $\mathsf{RD}'$ will satisfy its constraining clauses in Datalog notation exactly when it satisfies the original constraining formulae.

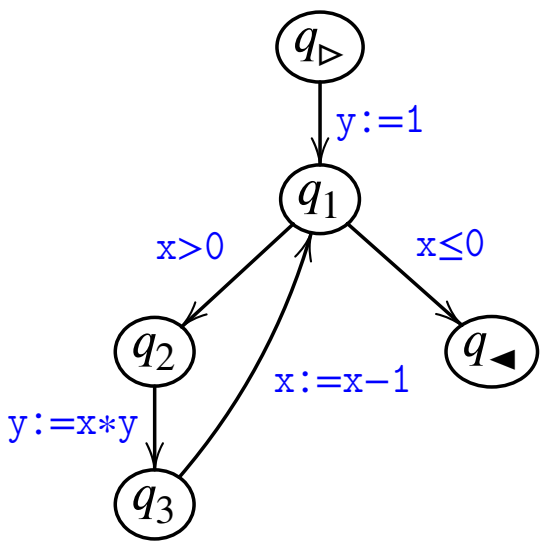

Figure 8.1: Program graph for the factorial function.

TRY IT OUT 8.1: Construct the set of Datalog clauses for the program graph of Figure 8.1 using the new set-based representation. Compare with the constraints produced in Example 2.10 and Try It Out 3.2. □

## 8.2 Control Flow Analysis in Datalog

So far our programs have had a rather simple control structure as could easily be represented by a program graph. In this section we shall consider a simple functional language for which it is non-trivial to determine what functions are called at what point and we shall develop a control flow analysis to provide this information. This can be seen as a way to lift the edges function of Appendices A and B to a functional language.

The expressions $e$ of our functional language is given by the following BNF syntax:

$$\begin{array}{rcl} e & ::= & c \mid x \mid \texttt{fn}\ x\ \texttt{=>}\ e_0 \mid \texttt{fun}\ f\ x\ \texttt{=>}\ e_0 \mid e_1\, e_2 \mid \\ & & \texttt{if}\ e_0\ \texttt{then}\ e_1\ \texttt{else}\ e_2 \mid \texttt{let}\ x = e_1\ \texttt{in}\ e_2 \mid e_1\ op\ e_2 \end{array}$$

The syntax for constants $c$, variables $x \in$ **Var**, function names $f \in$ **Var**, and operators $op$ is left unspecified.

Here `fn` $x$ `=>` $e_0$ defines a function that has $x$ as formal parameter and body $e_0$, and `fun` $f\ x$ `=>` $e_0$ defines a recursive function that has $x$ as formal parameter and body $e_0$ and where the body may invoke a recursive all by means of $f$, and $e_1\, e_2$ denotes a function application of the function resulting from $e_1$ to the argument $e_2$, and finally `let` $x = e_1$ `in` $e_2$ is equivalent to (`fn` $x$ `=>` $e_1$) $e_2$.

EXAMPLE 8.2: An example expression is the following (where we allow to add parantheses for clarity):

```
let g = (fun f x => f (fn y=>y)) in g (fn z=>z)
```

It defines a function `g` that is a recursive function (with local name `f`) that will ignore its argument and call itself recursively on the function `fn y=>y` and hence will loop. It is initally called upon the function `fn z=>z`.

In order to perform our program analysis we should like to have 'pointers' into the abstract syntax tree for an expression. We may achieve this effect by changing the syntax such that an expression is a labelled term: $e = t^\ell$ where $\ell \in$ Pnt should be tought of as the 'pointer' into the abstract syntax trees; these 'pointers' are useful for expressing the analysis but have no importance as far as the semantics of the program is concerned.

$$\begin{array}{rcl} t^\ell & ::= & c^\ell \mid x^\ell \mid (\texttt{fn}\ x\ \texttt{=>}\ t_0^{\ell_0})^\ell \mid (\texttt{fun}\ f\ x\ \texttt{=>}\ t_0^{\ell_0})^\ell \mid (t_1^{\ell_1}\, t_2^{\ell_2})^\ell \mid \\ & & (\texttt{if}\ t_0^{\ell_0}\ \texttt{then}\ t_1^{\ell_1}\ \texttt{else}\ t_2^{\ell_2})^\ell \mid (\texttt{let}\ x = t_1^{\ell_1}\ \texttt{in}\ t_2^{\ell_2})^\ell \mid (t_1^{\ell_1}\ op\ t_2^{\ell_2})^\ell \end{array}$$

EXAMPLE 8.3: Using integers for the 'pointers' into the program we obtain the following labelled term corresponding to the one in Example 8.2:

```
(let g = (fun f x => (f^1 (fn y=>y^2)^3)^4)^5 in (g^6 (fn z=>z^7)^8)^9)^10
```

The *Control Flow Analysis* to be developed will be more precise if we ensure that formal parameters $x$ are not re-used between different functions, recursive functions or `let`-expression; similarly, the analysis will be more precise if we ensure that all 'pointers' are distinct.

It will make use of two analysis domains, one for tracking the values of variables, and one for tracking the value of subexpressions. For tracking variables we use $\mathsf{E} : \mathbf{Var} \to \mathsf{PowerSet}(\mathsf{Terms})$ and for tracking subexpressions we use $\mathsf{C} : \mathsf{Pnt} \to \mathsf{PowerSet}(\mathsf{Terms})$ where Terms is the set of terms.

The analysis can be expressed in logical form as follows:

For each labelled subterm of the form $x^\ell$ we impose the formula

$$\mathsf{E}(x) \subseteq \mathsf{C}(\ell)$$

For each labelled subterm of the form $(\texttt{fn}\ x \texttt{ => } t_0^{\ell_0})^\ell$ we impose the formula

$$(\texttt{fn}\ x \texttt{ => } t_0^{\ell_0}) \in \mathsf{C}(\ell)$$

For each labelled subterm of the form $(\texttt{fun}\ f\ x \texttt{ => } t_0^{\ell_0})^\ell$ we impose the formula

$$(\texttt{fun}\ f\ x \texttt{ => } t_0^{\ell_0}) \in \mathsf{C}(\ell) \ \wedge\ (\texttt{fun}\ f\ x \texttt{ => } t_0^{\ell_0}) \in \mathsf{E}(f)$$

For each labelled subterm of the form $(t_1^{\ell_1}\ t_2^{\ell_2})^\ell$ we impose the formulae

$$\forall(\texttt{fn}\ x \texttt{ => } t_0^{\ell_0}) \in \mathsf{C}(\ell_1) : \mathsf{C}(\ell_2) \subseteq \mathsf{E}(x) \wedge \mathsf{C}(\ell_0) \subseteq \mathsf{C}(\ell)$$
$$\forall(\texttt{fun}\ f\ x \texttt{ => } t_0^{\ell_0}) \in \mathsf{C}(\ell_1) : \mathsf{C}(\ell_2) \subseteq \mathsf{E}(x) \wedge \mathsf{C}(\ell_0) \subseteq \mathsf{C}(\ell)$$

For each labelled subterm of the form $(\texttt{if}\ t_0^{\ell_0}\ \texttt{then}\ t_1^{\ell_1}\ \texttt{else}\ t_2^{\ell_2})^\ell$ we impose the formula

$$\mathsf{C}(\ell_1) \subseteq \mathsf{C}(\ell) \ \wedge\ \mathsf{C}(\ell_2) \subseteq \mathsf{C}(\ell)$$

For each labelled subterm of the form $(\texttt{let}\ x = t_1^{\ell_1}\ \texttt{in}\ t_2^{\ell_2})^\ell$ we impose the formula

$$\mathsf{C}(\ell_1) \subseteq \mathsf{E}(x) \ \wedge\ \mathsf{C}(\ell_2) \subseteq \mathsf{C}(\ell)$$

It is beyond our current aims to prove the correctness of the control flow analysis as this requires defining the formal semantics of the functional language using techniques not otherwise covered here. But to understand what the analysis aims to do the illustrations in Figures 8.2 and 8.3 may be helpful. Figure 8.2 shows for the let clause $(\texttt{let}\ x = t_1^{\ell_1}\ \texttt{in}\ t_2^{\ell_2})^\ell$ how the $t_1^{\ell_1}$ is bound to the $x$ which can then be used as $x^{\ell'}$ inside the $t_2^{\ell_2}$. We already explained that `let` $x = e_1$ `in` $e_2$ is equivalent to (`fn` $x$ `=>` $e_1$) $e_2$. With this in mind Figure 8.3 shows how to analyse a function application $(t_1^{\ell_1}\ t_2^{\ell_2})^{\ell}$ where a result of $t_1^{\ell_1}$ might be the function `fn` $x$ `=>` $t_0^{\ell_0}$. For further explanation and a proof of correctness we refer to Chapter 3 of the book *Principles of Program Analysis* (Springer, 2005).

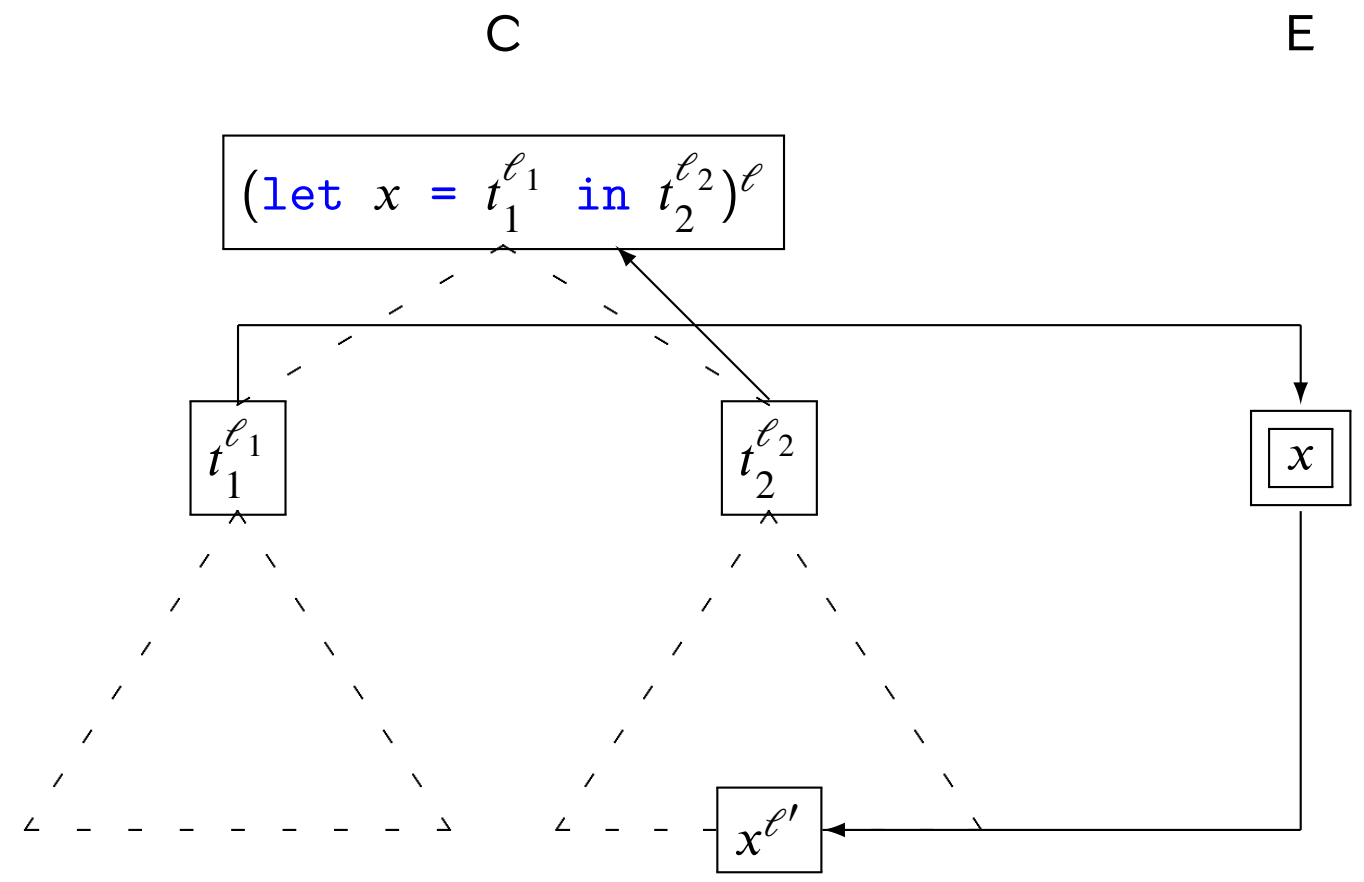

Figure 8.2: Pictorial illustration of the let clause.

TRY IT OUT 8.4: Construct the formulae for the labelled term of Example 8.3. □

EXAMPLE 8.5: To present the result of the control flow analysis for the expression in Example 8.3 it is helpful with a few abbreviations:

$$\begin{aligned} \texttt{F} &= \texttt{fun f x => (f}^1\ \texttt{(fn y=>y}^2\texttt{)}^3\texttt{)}^4 \\ \texttt{Y} &= \texttt{fn y=>y}^2 \\ \texttt{Z} &= \texttt{fn z=>z}^7 \end{aligned}$$

We then obtain

| | | | | | | | | |
|---|---|---|---|---|---|---|---|---|
| C(1) | = | {F} | C(6) | = | {F} | E(f) | = | {F} |
| C(2) | = | { } | C(7) | = | { } | E(g) | = | {F} |
| C(3) | = | {Y} | C(8) | = | {Z} | E(x) | = | {Y, Z} |
| C(4) | = | { } | C(9) | = | { } | E(y) | = | { } |
| C(5) | = | {F} | C(10) | = | { } | E(z) | = | { } |

Since we only have functions in our example expression the fact that C(10) = { } tells us that the expression never terminates. The fact that E(y) = { } and E(z) = { } tells us that the functions `fn y=>y`$^2$ and `fn z=>z`$^7$ are never applied.

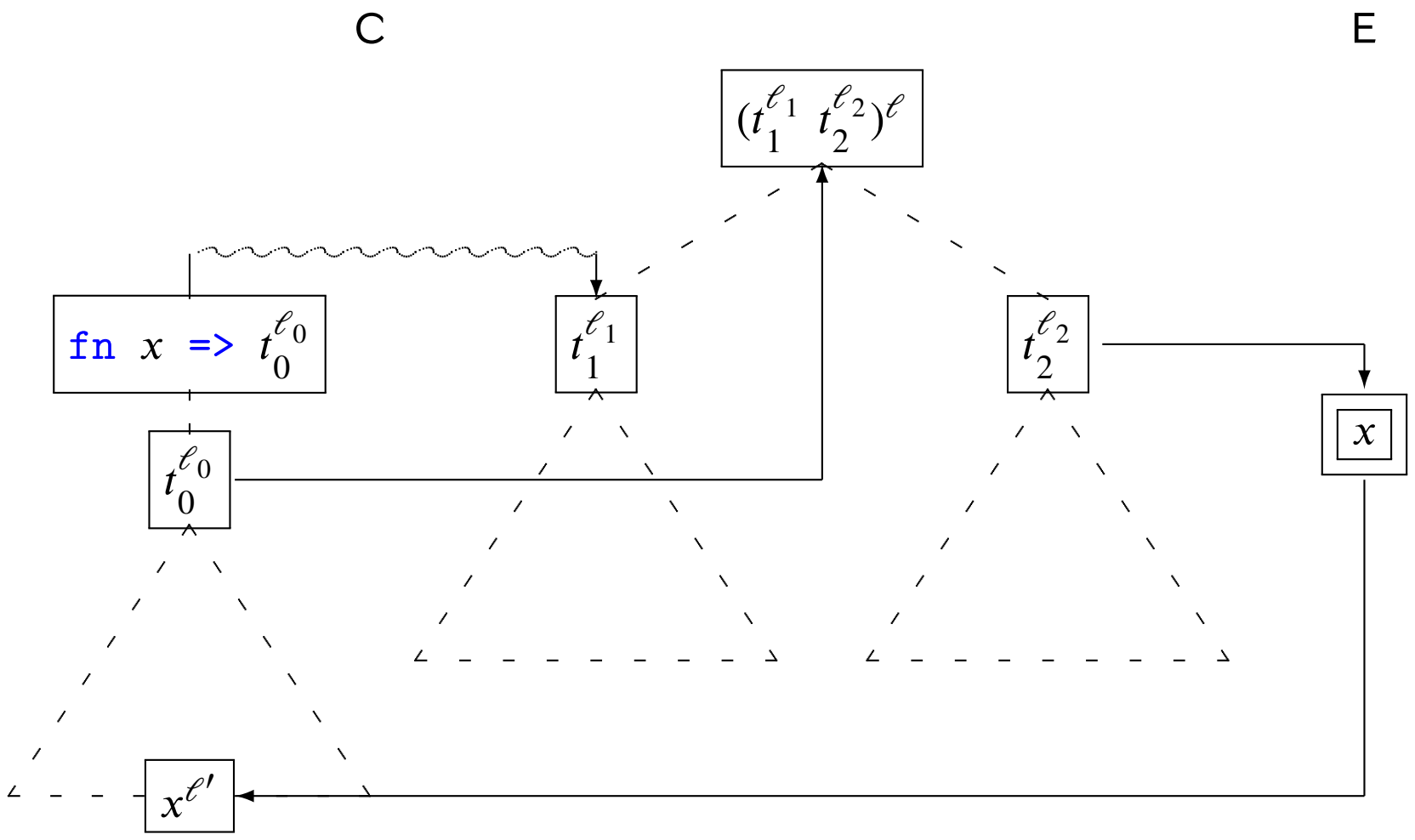

Figure 8.3: Pictorial illustration of function application.

The fact that $\mathsf{E}(\mathtt{x}) = \{\mathtt{Y}, \mathtt{Z}\}$ tells us that the recursive function is applied to both `fn y=>y`$^2$ and `fn z=>z`$^7$.

To avoid complex structured terms inside the domans we shall merely represent functions (`fn` $x$ `=>` $t_0^{\ell_0}$) and (`fun` $f$ $x$ `=>` $t_0^{\ell_0}$) by the pair $(x, \ell_0)$ of formal parameter and 'pointer' to the body. We then obtain the following analysis (reusing the names of the analysis domains):

For each labelled subterm of the form $x^{\ell}$ we impose the formula

$$\mathsf{E}(x) \subseteq \mathsf{C}(\ell)$$

For each labelled subterm of the form (`fn` $x$ `=>` $t_0^{\ell_0})^{\ell}$ we impose the formula

$$(x, \ell_0) \in \mathsf{C}(\ell)$$

For each labelled subterm of the form (`fun` $f$ $x$ `=>` $t_0^{\ell_0})^{\ell}$ we impose the formula

$$(x, \ell_0) \in \mathsf{C}(\ell) \ \wedge \ (x, \ell_0) \in \mathsf{E}(f)$$

For each labelled subterm of the form $(t_1^{\ell_1}\ t_2^{\ell_2})^{\ell}$ we impose the formula

$$\forall (x, \ell_0) \in \mathsf{C}(\ell_1) : \mathsf{C}(\ell_2) \subseteq \mathsf{E}(x) \wedge \mathsf{C}(\ell_0) \subseteq \mathsf{C}(\ell)$$

For each labelled subterm of the form (`if` $t_0^{\ell_0}$ `then` $t_1^{\ell_1}$ `else` $t_2^{\ell_2})^{\ell}$ we impose the formula

$$\mathsf{C}(\ell_1) \subseteq \mathsf{C}(\ell) \ \wedge \ \mathsf{C}(\ell_2) \subseteq \mathsf{C}(\ell)$$

For each labelled subterm of the form $(\texttt{let}\ x = t_1^{\ell_1}\ \texttt{in}\ t_2^{\ell_2})^{\ell}$ we impose the formula

$$\mathsf{C}(\ell_1) \subseteq \mathsf{E}(x) \ \wedge\ \mathsf{C}(\ell_2) \subseteq \mathsf{C}(\ell)$$

TRY IT OUT 8.6: Rewrite the analysis result of Example 8.5 in this notation. □

Turning into Datalog in the manner of the previous section we obtain:

We initially we make the declarations

$$\underline{\mathsf{PRED}}\ \mathsf{E}'(3)/1\ \mathsf{C}'(3)/1$$
$$\underline{\mathsf{VAR}}\ u\ v\ u'\ v'$$

For each labelled subterm of the form $x^{\ell}$ we impose the clause

$$\mathsf{C}(\ell, u, v) \leftarrow \mathsf{E}(x, u, v)$$

For each labelled subterm of the form $(\texttt{fn}\ x\ \texttt{=>}\ t_0^{\ell_0})^{\ell}$ we impose the clause

$$\mathsf{C}(\ell, x, \ell_0) \leftarrow$$

For each labelled subterm of the form $(\texttt{fun}\ f\ x\ \texttt{=>}\ t_0^{\ell_0})^{\ell}$ we impose the clauses

$$\mathsf{C}(\ell, x, \ell_0) \leftarrow$$
$$\mathsf{E}(f, x, \ell_0) \leftarrow$$

For each labelled subterm of the form $(t_1^{\ell_1}\ t_2^{\ell_2})^{\ell}$ we impose the clauses

$$\mathsf{E}(u', u, v) \leftarrow \mathsf{C}(\ell_2, u, v), \mathsf{C}(\ell_1, u', v')$$
$$\mathsf{C}(v', u, v) \leftarrow \mathsf{C}(\ell_0, u, v), \mathsf{C}(\ell_1, u', v')$$

For each labelled subterm of the form $(\texttt{if}\ t_0^{\ell_0}\ \texttt{then}\ t_1^{\ell_1}\ \texttt{else}\ t_2^{\ell_2})^{\ell}$ we impose the clauses

$$\mathsf{C}(\ell, u, v) \leftarrow \mathsf{C}(\ell_1, u, v)$$
$$\mathsf{C}(\ell, u, v) \leftarrow \mathsf{C}(\ell_2, u, v)$$

For each labelled subterm of the form $(\texttt{let}\ x = t_1^{\ell_1}\ \texttt{in}\ t_2^{\ell_2})^{\ell}$ we impose the clauses

$$\mathsf{E}(x, u, v) \leftarrow \mathsf{C}(\ell_1, u, v)$$
$$\mathsf{C}(\ell, u, v) \leftarrow \mathsf{C}(\ell_2, u, v)$$

EXERCISE 8.7: Construct the Datalog program for the labelled term of Example 8.3 and compare with the result of Try It Out 8.4. □

## 8.3 Stratified Datalog

We shall define Datalog programs in stages: first the base syntax and then an important fragment for which least solutions can be computed.

**Datalog Programs and their Solutions** Since well-formedness conditions are essential we shall define Datalog programs by a set of clauses rather than using BNF notation:

DEFINITION 8.8: A *Datalog program* takes the form

$$\underline{\text{PRED}}\ pd^{*}\underline{\text{VAR}}\ id^{*}cl^{*}$$

where the components $pd$, $id$, and $cl$ are given by

- a predicate declaration $pd$ takes the form $p(k)/r$ where $p$ is a predicate, $k \geq 1$ is its *arity* (number of arguments) with which the predicate was declared, and $r \geq 0$ is its *rank* (order of computation)
- we use superscript asterisk to indicate zero or more occurrences and we demand that $pd^{*}$ contains no multiple declarations of the same predicate
- an identifier $id$ is unspecified, but the ones after VAR are Datalog variables that can be instantiated, whereas all other identifers occurring in the Datalog program are elements of the (otherwise unspecified) universe
- $cl$ is a *clause* of the form $p(\vec{a}) \leftarrow rh_1, \cdots, rh_n$ for $n \geq 0$ where $\vec{a}$ is a comma-separated list of identifiers, whose length equals the arity of $p$
- a righthandside $rh$ takes one of the forms $a = a'$, $a \neq a'$, $q(\vec{a}')$, or $\neg q(\vec{a}')$ where $a$ and $a'$ are identifiers, $q$ is a predicate, $\vec{a}'$ is a comma-separated list of identifiers, whose length equals the arity with which $q$ was declared

It is immediate that the example Datalog program of Sections 8.1 and 8.2 adhere to this definition.

The universe $\mathcal{U}$ of a Datalog program consists of all identifiers in the Datalog program except for those declared as variables, together with all the elements in the predicates given rank 0, and who are supposed to be input to the Datalog program.

EXAMPLE 8.9: For the Datalog program

$$\underline{\text{PRED}}\ \text{U}(1)/0\ \text{P}(2)/1$$
$$\underline{\text{VAR}}\ u$$

$$\mathsf{P}(u,c) \leftarrow \mathsf{U}(u)$$
$$\mathsf{P}(c,u) \leftarrow \mathsf{U}(u)$$

and input predicate $\mathsf{U} = \{a,b\}$ we obtain the universe $\{a,b,c\}$.

A *variable valuation* $\sigma$ is a mapping from the set of Datalog variables into the universe. We shall allow to apply a variable valuation $\sigma$ to an arbitrary identifier by letting it act as the identify, and to apply it to lists of identifiers $\vec{a}$ and predicate applications $p(\vec{a})$ in a component-wise manner.

EXAMPLE 8.10: Continuing Example 8.9 we have three possible variable valuations: $\sigma_a$ given by $\sigma_a(u) = a$, $\sigma_b$ given by $\sigma_b(u) = b$, and $\sigma_c$ given by $\sigma_c(u) = a$. Applying the variable valuation $\sigma_b$ to $\mathsf{P}(c,u)$ gives $\mathsf{P}(c,b)$.

A *predicate valuation* $\rho$ is a mapping $\rho$ from the predicates defined in the Datalog progeram to sets of tuples such that $\rho(p) \subseteq \mathcal{U}^k$ whenever $p$ has rank $k$.

Given a predicate valuation $\rho$ and a variable valuation $\sigma$ we can define the meaning of predicate applications and righthandsides as follows:

$$
\begin{aligned}
[\![p(\vec{a})]\!](\rho,\sigma) &= \begin{cases} \mathsf{true} & \text{if } \sigma(\vec{a}) \in \rho(p) \\ \mathsf{false} & \text{otherwise} \end{cases} \\
[\![\neg p(\vec{a})]\!](\rho,\sigma) &= \begin{cases} \mathsf{true} & \text{if } \sigma(\vec{a}) \notin \rho(p) \\ \mathsf{false} & \text{otherwise} \end{cases} \\
[\![a = a']\!](\rho,\sigma) &= \begin{cases} \mathsf{true} & \text{if } \sigma(a) = \sigma(a') \\ \mathsf{false} & \text{otherwise} \end{cases} \\
[\![a \neq a']\!](\rho,\sigma) &= \begin{cases} \mathsf{true} & \text{if } \sigma(a) \neq \sigma(a') \\ \mathsf{false} & \text{otherwise} \end{cases}
\end{aligned}
$$

It may be extended to clauses as follows:

$$
[\![p(\vec{a}) \leftarrow rh_1, \cdots, rh_n]\!](\rho,\sigma) = \begin{cases} \mathsf{false} & \text{if } \bigwedge_i [\![rh_i]\!](\rho,\sigma) \wedge \sigma(\vec{a}) \notin \rho(p) \\ \mathsf{true} & \text{otherwise} \end{cases}
$$

DEFINITION 8.11: A predicate valuation $\rho$ *solves a clause* $p(\vec{a}) \leftarrow rh_1, \cdots, rh_n$ whenever $[\![p(\vec{a}) \leftarrow rh_1, \cdots, rh_n]\!](\rho,\sigma) = \mathsf{true}$ for all variable valuations $\sigma$.

A predicate valuation $\rho$ *solves a Datalog program* whenever it solves all clauses of the Datalog program.

All Datalog programs can be solved, e.g. by the predicate valuation $\rho$ given by $\rho(p) = \mathcal{U}^k$ whenever $p$ has rank $k$, but we shall be more interested in least solutions.

EXAMPLE 8.12: In the presence of negation it is not always possible to obtain a least solution. Consider the Datalog program

$$\underline{\mathsf{PRED}}\ \mathsf{U}(1)/0\ \mathsf{P}(1)/1$$
$$\underline{\mathsf{VAR}}\ u\ v$$
$$\mathsf{P}(u) \leftarrow \mathsf{U}(u), \neg\mathsf{P}(v), u \neq v$$

where the universe $\mathsf{U} = \{a, b\}$ is intended to consist of only two elements. Here there are four possible values of $\rho(\mathsf{P})$ but only three of them will solve the Datalog program: $\{a\}$, $\{b\}$, and $\{a, b\}$. In particular, there are two minimal solutions but no least solution.

EXAMPLE 8.13: The least solution to the Datalog program

$$\underline{\mathsf{PRED}}\ \mathsf{U}(1)/0\ \mathsf{Eq}(2)/1$$
$$\underline{\mathsf{VAR}}\ u$$
$$\mathsf{Eq}(u, u) \leftarrow \mathsf{U}(u)$$

has Eq to be the equality relation (=) over the universe U.

The Datalog program

$$\underline{\mathsf{PRED}}\ \mathsf{U}(1)/0\ \mathsf{Eq}(2)/1\ \mathsf{Neq}(2)/2$$
$$\underline{\mathsf{VAR}}\ u\ v$$
$$\mathsf{Eq}(u, u) \leftarrow \mathsf{U}(u)$$
$$\mathsf{Neq}(u, v) \leftarrow \neg\mathsf{Eq}(u, v)$$

is intended to compute in Eq the equality relation (=) over the universe U and in Neq the in-equality relation (≠) over the universe U. However, it would seem sensible to compute Eq first and NEq next as otherwise elements might be added to Neq that would subsequently have to be removed. This is indicated by using '/1' for Eq and '/2' for NEq.

**Least Solutions** Since we want some kind of 'least solutions' as well the ability to use negation we must carefully constrain the use of negation. Let us write $\mathsf{rnk}(p)$ the rank of $p$ when declared in the Datalog program and let us extend it to predicate applications and righthandsides as follows:

$$\begin{array}{rcl}
\mathsf{rnk}(p(\vec{a})) & = & \mathsf{rnk}(p) \\
\mathsf{rnk}(\neg p(\vec{a})) & = & 1 + \mathsf{rnk}(p) \\
\mathsf{rnk}(a = a') & = & 0 \\
\mathsf{rnk}(a \neq a') & = & 0
\end{array}$$

DEFINITION 8.14: A Datalog program is *stratified* when $\forall i : \mathsf{rnk}(p) \geq \mathsf{rnk}(rh_i)$ holds for every clause $p(\vec{a}) \leftarrow rh_1, \cdots, rh_n$.

EXAMPLE 8.15: The Datalog programs of Sections 8.1 and 8.2 are stratified as are the Datalog programs of Example 8.13. The Datalog program of Example 8.12 is not stratified.

EXERCISE 8.16: In Section 8.1 we dealt with the Reaching Definitions analysis by inlining the definitions of $\mathsf{kill}_{\mathsf{RD}}(q_\circ, \alpha, q_\bullet)$ and $\mathsf{gen}_{\mathsf{RD}}(q_\circ, \alpha, q_\bullet)$.

Suppose instead we have declarations

$$\underline{\mathsf{PRED}}\ \mathbf{Var}(1)/0\ \mathbf{Arr}(1)/0\ \mathsf{RD}'(4)/1\ \mathsf{kill}'_{\mathsf{RD}}(6)/0\ \mathsf{gen}'_{\mathsf{RD}}(6)/0$$
$$\underline{\mathsf{VAR}}\ u\ v\ w\ \cdots$$

We would still impose the following two clauses for the entry node

$$\mathsf{RD}'(q_\triangleright, u, ?, q_\triangleright) \leftarrow \mathbf{Var}(u)$$
$$\mathsf{RD}'(q_\triangleright, u, ?, q_\triangleright) \leftarrow \mathbf{Arr}(u)$$

but for each edge $(q_\circ, \alpha, q_\bullet)$ we can now impose two clauses without performing 'case analysis' on the action $\alpha$.

Specify the two clauses and argue that the resulting Datalog program is stratified.□

For stratified Datalog programs we can then compute a kind of 'least solutions' by computing the values of predicates in increasing order of their rank as briefly discussed in Example 8.13. This is achieved by the chaotic iteration algorithm of Figure 8.4.

The correctness of the algorithm is very much in line with our previous considerations of the chaotic iteration algorithm but because of negation we need to be a bit more careful.

PROPOSITION 8.17: The algorithm of Figure 8.4 terminates and produces a predicate valuation that solves the stratified Datalog program and that contains the initial entries for predicates of rank 0.

PROOF: It is immediate that the algorithm terminates because the universe is finite and after the initialisation phase the only change made to $\rho(p)$ is to enlarge it and this can only happen a finite number of times.

We may prove by inducion that whenever the inner loop terminates with some value of r, all clauses $p(\vec{a}) \leftarrow rh_1, \cdots, rh_n$ with $\mathsf{rnk}(p) \leq \mathsf{r}$ have been solved; this is because the Datalog program is stratified so that no clause $p'(\vec{a}') \leftarrow r'_1, \cdots, r'_n$ with $\mathsf{rnk}(p') < \mathsf{r}$ can become unsolved due to changes on predicates of rank r. □

| | |
|---|---|
| INPUT | a stratified Datalog program<br>for each predicate of rank 0 a finite set of initial entries |
| OUTPUT | the 'least solution' $\rho$ to the Datalog program |
| METHOD | forall declared predicates $p(k)/r$<br>do if $r > 0$ then $\rho(p) := \{\,\}$ else set $\rho(p)$ to the initial entries;<br><br>r := -1;<br>while there exists a clause $p(\vec{a}) \leftarrow rh_1, \cdots, rh_n$<br>such that $\mathsf{rnk}(p) > r$<br>do r := r+1;<br>while there exists a clause $p(\vec{a}) \leftarrow rh_1, \cdots, rh_n$<br>and a variable valuation $\sigma$<br>such that $\mathsf{rnk}(p)$=r and<br>$[\![p(\vec{a}) \leftarrow rh_1, \cdots, rh_n]\!](\rho, \sigma) = \mathsf{false}$<br>do $\rho(p) := \rho(p) \cup \{\sigma(\vec{a})\}$ |

Figure 8.4: Chaotic Iteration for Stratified Datalog.

PROPOSITION 8.18: In case the stratified Datalog program contains no negations the algorithm of Figure 8.4 produces the least predicate valuation that solves the Datalog program and that contains the initial entries for predicates of rank 0.

PROOF: After the initialisation phase it is an invariant of the algorithm that the current value of $\rho$ is contained in every predicate valuation that solves the Datalog program and that contains the initial entries for predicates of rank 0. Upon termination (as guaranteed by Proposition 8.17) the predicate valuation $\rho$ therefore is as stated. □

TEASER 8.19: For a suitable ordering on predicate valuations the algorithm of Figure 8.4 always produces the least predicate valuation that solves the stratified Datalog program and that contains the initial entries for predicates of rank 0.

For this define the relation $\sqsubseteq$ as follows:

$$\rho \sqsubseteq \rho' \Leftrightarrow \left( \rho = \rho' \vee \exists p : \left\{ \begin{array}{l} \rho(p) \subset \rho'(p) \wedge \\ \forall p' : \mathsf{rnk}(p') = \mathsf{rnk}(p) \Rightarrow \rho(p') \subseteq \rho'(p') \wedge \\ \forall p' : \mathsf{rnk}(p') < \mathsf{rnk}(p) \Rightarrow \rho(p') = \rho'(p') \end{array} \right. \right)$$

Show that $\sqsubseteq$ is a partial order. Next show that the algorithm of Figure 8.4 always produces the least predicate valuation (wrt. $\sqsubseteq$) that solves the stratified Datalog program and that contains the initial entries for predicates of rank 0. □

## 8.4 Program Analyses in Stratified Datalog

As stated in the introduction to this chapter the advantage of using stratified Datalog is that (i) it may be faster to develop an analysis thanks to the uniform format of Datalog clauses, and (ii) techniques exist for solving Datalog claues very efficiently even for very large programs.

We now elaborate on the first point. We should like to claim that we can deal with all analyses in the Monotone Framework that operate over powersets but we shall be content with showing through a number of examples and exerices how to deal with all analyses in the Bit-Vector Framework as well as a few analyses outside of the Bit-Vector Framework.

EXAMPLE 8.20: Once more recall from Sections 2.1 and 8.1 that the *Reaching Definitions* analysis was initially described as

For each edge $(q_\circ, \alpha, q_\bullet)$ in the program graph impose

$$\big(\mathsf{RD}(q_\circ) \setminus \mathsf{kill}_{\mathsf{RD}}(q_\circ, \alpha, q_\bullet)\big) \cup \mathsf{gen}_{\mathsf{RD}}(q_\circ, \alpha, q_\bullet) \subseteq \mathsf{RD}(q_\bullet)$$

and for the entry node additionally impose

$$(\mathbf{Var} \cup \mathbf{Arr}) \times \{?\} \times \{q_\triangleright\} \subseteq \mathsf{RD}(q_\triangleright)$$

In Section 8.1 we decided to expand $\mathsf{kill}_{\mathsf{RD}}$ and $\mathsf{gen}_{\mathsf{RD}}$ but for now we shall assume that they are being precomputed.

So let us first make the declarations

$\underline{\mathsf{PRED}}$ $\mathbf{Var}(1)/0$ $\mathbf{Arr}(1)/0$ $\mathsf{gen}_{\mathsf{RD}}(6)/0$ $\mathsf{kill}_{\mathsf{RD}}(6)/0$ $\mathsf{RD}(4)/1$
$\underline{\mathsf{VAR}}$ $u\ v\ w$

Next for each edge $(q_\circ, \alpha, q_\bullet)$ in the program graph impose the clauses

$$\begin{array}{lcl} \mathsf{RD}(q_\bullet, u, v, w) & \leftarrow & \mathsf{RD}(q_\circ.u, v, w), \neg\mathsf{kill}_{\mathsf{RD}}(q_\circ, \alpha, q_\bullet, u, v, w) \\ \mathsf{RD}(q_\bullet, u, v, w) & \leftarrow & \mathsf{gen}_{\mathsf{RD}}(q_\circ, \alpha, q_\bullet, u, v, w) \end{array}$$

and *either* impose additional clauses defining $\mathsf{kill}_{\mathsf{RD}}(q_\circ, \alpha, q_\bullet, u, v, w)$ as well as $\mathsf{gen}_{\mathsf{RD}}(q_\circ, \alpha, q_\bullet, u, v, w)$ based on the action $\alpha$, *or* simply precompute these sets and use them as input to the chaotic iteration algorithm of Figure 8.4.

For the entry node additionally impose the clauses

$$\mathsf{RD}(q_\triangleright, u, ?, q_\triangleright) \leftarrow \mathbf{Var}(u)$$
$$\mathsf{RD}(q_\triangleright, u, ?, q_\triangleright) \leftarrow \mathbf{Arr}(u)$$

and then *either* impose additional clauses defining $\mathbf{Var}(u)$ and $\mathbf{Arr}(u)$ based on the entire program graph, *or* simply precompute these sets and use them as input to the chaotic iteration algorithm of Figure 8.4. The resulting Datalog program will be stratified.

This approach generalises to all *forward* analyses in the Bit Vector Framework for which the *least fixed point* is desired.

EXERCISE 8.21: Perform a development similar to that in Example 8.20 for the *Live Variables* analysis of Section 2.2 and conclude that the approach generalises to all *backward* analyses in the Bit Vector Framework for which the *least fixed point* is desired. □

EXAMPLE 8.22: Once more recall from Section 2.3 that the *Available Expressions* analysis was initially described as

For each edge $(q_\circ, \alpha, q_\bullet)$ in the program graph we impose

$$\big(\mathsf{AE}(q_\circ) \setminus \mathsf{kill}_{\mathsf{AE}}(q_\circ, \alpha, q_\bullet)\big) \cup \mathsf{gen}_{\mathsf{AE}}(q_\circ, \alpha, q_\bullet) \supseteq \mathsf{AE}(q_\bullet)$$

and for the entry node additionally impose

$$\{\ \} \supseteq \mathsf{AE}(q_\triangleright)$$

In the manner of Example 8.20 we shall assume that $\mathsf{kill}_{\mathsf{AE}}$ and $\mathsf{gen}_{\mathsf{AE}}$ are being precomputed. Since we want the greatest solution to the above constraints, and Datalog programs in general compute the least solution, we need to apply the duality ideas of Sections 3.4 and 4.4. To this end we shall need to take the complent with respect to the set **AExp** of non-trivial expressions in the program of interest; let us write for $\complement Y$ for $\mathbf{AExp} \setminus Y$. Using De Morgan's laws we then rephrase the constraints in the following manner

For each edge $(q_\circ, \alpha, q_\bullet)$ in the program graph we impose

$$\complement\mathsf{AE}(q_\bullet) \supseteq \complement\left(\big(\mathsf{AE}(q_\circ) \setminus \mathsf{kill}_{\mathsf{AE}}(q_\circ, \alpha, q_\bullet)\big) \cup \mathsf{gen}_{\mathsf{AE}}(q_\circ, \alpha, q_\bullet)\right)$$

which is equivalent to

$$\begin{array}{rcl} \complement\mathsf{AE}(q_\bullet) & \supseteq & \complement\mathsf{AE}(q_\circ) \cap \complement\mathsf{gen}_{\mathsf{AE}}(q_\circ, \alpha, q_\bullet) \\ \complement\mathsf{AE}(q_\bullet) & \supseteq & \mathsf{kill}_{\mathsf{AE}}(q_\circ, \alpha, q_\bullet) \cap \complement\mathsf{gen}_{\mathsf{AE}}(q_\circ, \alpha, q_\bullet) \end{array}$$

and for the entry node additionally impose

$$\complement\mathsf{AE}(q_\triangleright) \supseteq \complement\{\ \}$$

which is equivalent to

$$\complement\mathsf{AE}(q_\triangleright) \supseteq \mathbf{AExp}$$

where (by De Morgan's laws) we want the least solution for $\complement\mathsf{AE}(\cdot)$.

So let us first make the declarations where the intention is that CAE should be the complement ($\complement$AE) of AE

<u>PRED</u> **AExp**$(1)/0$ gen$_{\mathsf{AE}}(4)/0$ kill$_{\mathsf{AE}}(4)/0$ CAE$(2)/1$ AE$(2)/2$
<u>VAR</u> $u$

Next for each edge $(q_\circ, \alpha, q_\bullet)$ in the program graph impose the clauses

$$\begin{array}{rcl} \mathsf{CAE}(q_\bullet, u) & \leftarrow & \mathsf{CAE}(q_\circ, u), \neg\mathsf{gen}_{\mathsf{AE}}(q_\circ, \alpha, q_\bullet, u) \\ \mathsf{CAE}(q_\bullet, u) & \leftarrow & \mathsf{kill}_{\mathsf{AE}}(q_\circ, \alpha, q_\bullet, u), \neg\mathsf{gen}_{\mathsf{AE}}(q_\circ, \alpha, q_\bullet, u) \end{array}$$

and *either* impose additional clauses for kill$_{\mathsf{AE}}(q_\circ, \alpha, q_\bullet, u)$ and gen$_{\mathsf{AE}}(q_\circ, \alpha, q_\bullet, u)$ based on the action $\alpha$, *or* simply precompute these sets and use them as input to the chaotic iteration algorithm of Figure 8.4.

For the entry node additionally impose

$$\mathsf{CAE}(q_\triangleright, u) \leftarrow \mathbf{AExp}(u)$$

and then *either* impose additional clauses defining **AExp**$(u)$ based on the entire program graph, *or* simply precompute these sets and use them as input to the chaotic iteration algorithm of Figure 8.4.

Finally, for each node $q$ in the program graph we impose

$$\mathsf{AE}(q, u) \leftarrow \neg\mathsf{CAE}(q, u)$$

This ensures that CAE will essentially be the complement (∁AE) of AE and the resulting Datalog program will be stratified because the rank of AE is higher than that of CAE.

This approach generalises to all *forward* analyses in the Bit Vector Framework for which the *greatest fixed point* is desired.

EXERCISE 8.23: Perform a development similar to that in Example 8.22 for the *Very Busy Expressions* analysis of Section 2.4 and conclude that the approach generalises to all *backward* analyses in the Bit Vector Framework for which the *greatest fixed point* is desired. □

EXAMPLE 8.24: We now reconsider the *Faint Variables* analysis of Example 3.38. The analysis functions for computing the Strongly Live Variables (the complement of the Faint Variables were defined as:

| $\alpha$ | $\widehat{S}_{\mathsf{FV}}[\![q_\bullet, \alpha, q_\circ]\!](L)$ |
|---|---|
| $x$ `:=` $a$ | $\begin{cases} (L \setminus \{x\}) \cup \mathbf{fv}(a) & \text{if } x \in L \\ L & \text{if } x \notin L \end{cases}$ |
| $A$`[`$a_1$`]` `:=` $a_2$ | $\begin{cases} L \cup \mathbf{fv}(a_1) \cup \mathbf{fv}(a_2) & \text{if } A \in L \\ L & \text{if } A \notin L \end{cases}$ |
| $c$`?`$x$ | $L \setminus \{x\}$ |
| $c$`?`$A$`[`$a$`]` | $\begin{cases} L \cup \mathbf{fv}(a) & \text{if } A \in L \\ L & \text{if } A \notin L \end{cases}$ |
| $c$`!`$a$ | $L \cup \mathbf{fv}(a)$ |
| $b$ | $L \cup \mathbf{fv}(b)$ |
| `skip` | $L$ |

Turning this into Stratified Datalog we shall use SLV for the Strongly Live Variables, FaintV for the Faint Variables, and FreeV for telling us which variables are free in an (arithmetic or boolean) expression, as well as **Var** and **Arr** for the variables and arrays in the program graph.

The corresponding declarations are:

PRED **Var**$(1)/0$ **Arr**$(1)/0$ **fv**$(()2)/0$ SLV$(2)/1$ FaintV$(2)/2$
VAR $u$

For each edge $(q_\circ, x := a, q_\bullet)$ in the program graph impose

$$\begin{aligned}\mathsf{SLV}(q_\circ, u) &\leftarrow \mathsf{SLV}(q_\bullet, u), u \neq x\\ \mathsf{SLV}(q_\circ, u) &\leftarrow \mathsf{FreeV}(a, u), \mathsf{SLV}(q_\bullet, x)\end{aligned}$$

and *either* impose additional clauses defining $\mathsf{FreeV}(a, u)$ based on the arithmetic expression *a*, *or* simply precompute these sets and use them as input to the chaotic iteration algorithm of Figure 8.4.

For each edge $(q_\circ, A[a_1] := a_2, q_\bullet)$ in the program graph impose

$$\begin{aligned}\mathsf{SLV}(q_\circ, u) &\leftarrow \mathsf{SLV}(q_\bullet, u)\\ \mathsf{SLV}(q_\circ, u) &\leftarrow \mathsf{FreeV}(a_1, u), \mathsf{SLV}(q_\bullet, A)\\ \mathsf{SLV}(q_\circ, u) &\leftarrow \mathsf{FreeV}(a_2, u), \mathsf{SLV}(q_\bullet, A)\end{aligned}$$

with similar remarks concerning $\mathsf{FreeV}(\cdot, u)$ as above.

For each edge $(q_\circ, c?x, q_\bullet)$ in the program graph impose

$$\mathsf{SLV}(q_\circ, u) \leftarrow \mathsf{SLV}(q_\bullet, u), u \neq x$$

For each edge $(q_\circ, c?A[a], q_\bullet)$ in the program graph impose

$$\begin{aligned}\mathsf{SLV}(q_\circ, u) &\leftarrow \mathsf{SLV}(q_\bullet, u)\\ \mathsf{SLV}(q_\circ, u) &\leftarrow \mathsf{FreeV}(a, u), \mathsf{SLV}(q_\bullet, A)\end{aligned}$$

with similar remarks concerning $\mathsf{FreeV}(\cdot, u)$ as above.

For each edge $(q_\circ, c!a, q_\bullet)$ in the program graph impose

$$\begin{aligned}\mathsf{SLV}(q_\circ, u) &\leftarrow \mathsf{SLV}(q_\bullet, u)\\ \mathsf{SLV}(q_\circ, u) &\leftarrow \mathsf{FreeV}(a, u)\end{aligned}$$

with similar remarks concerning $\mathsf{FreeV}(\cdot, u)$ as above.

For each edge $(q_\circ, b, q_\bullet)$ in the program graph impose

$$\begin{aligned}\mathsf{SLV}(q_\circ, u) &\leftarrow \mathsf{SLV}(q_\bullet, u)\\ \mathsf{SLV}(q_\circ, u) &\leftarrow \mathsf{FreeV}(b, u)\end{aligned}$$

with similar remarks concerning $\mathsf{FreeV}(\cdot, u)$ as above.

For each edge $(q_\circ, \texttt{skip}, q_\bullet)$ in the program graph impose

$$\mathsf{SLV}(q_\circ, u) \leftarrow \mathsf{SLV}(q_\bullet, u)$$

Finally, for each node $q$ in the program graph we impose

$$\mathsf{FaintV}(q,u) \leftarrow \neg\mathsf{SLV}(q,u), \mathbf{Var}(u)$$
$$\mathsf{FaintV}(q,u) \leftarrow \neg\mathsf{SLV}(q,u), \mathbf{Arr}(u)$$

and then *either* impose additional clauses defining **Var**$(u)$ and **Arr**$(u)$ based on the entire program graph, *or* simply precompute these sets and use them as input to the chaotic iteration algorithm of Figure 8.4.

The resulting Datalog program will be stratified and will ensure that FaintV will be the desired complement of SLV and hence give us the set of faint variables.

EXERCISE 8.25: Show in the manner of Example 8.24 how to express the *Dangerous Variables* analysis of Example 3.23 in Stratified Datalog. □

**State-of-the-art Solvers for Stratified Datalog** Finally, we elaborate on the advantage stated in the introduction to this chapter that techniques exist for solving Datalog claues very efficiently even for very large programs.

One set of techniques borrows from our chaotic iteration algorithm of Figure 8.4 by using efficient algorithmic techniques to obtain a good running time. The *Succinct Solver* did this for an even more permissive formalim than stratified Datalog where it is possible to encode also model checking problems. The SOUFFLÉ open source programming framework exhibits performance results that are on-par with manually developed state-of-the-art tools.

Another set of techniques translate the stratified Datalog queries into another formalism for which efficient solvers have already been developed. A popular approach is to translate into the primitives of the **bddbddb** package that supports advanced operations on Binary Decision Diagrams (BDDs). An alternative approach is to translate into Boolean Equation Systems.

In summary, program analyses performing at state-of-the-art can be obtained from program analyses expressed as stratified Datalog programs.

# Appendix A

# The Guarded Commands Language

In this appendix we shall present a version of Dijkstra's language of Guarded Commands introduced in 1975. This is a simple imperative language with assignments to variables as the basic command, and constructs for sequencing of commands, choice between commands and iteration of commands. We shall extend the basic commands to include assignments to array entries as well as inputs and outputs over named channels; in this way the basic commands of the language will correspond to the actions on the program graphs introduced in Chapter 1. Following the introduction of the syntax, we shall show how to automatically construct program graphs for the commands.

**Syntax.** In Dijkstra's language of *Guarded Commands* a basic command has one of two forms, either it is an assignment $x$ := $a$ or it is a skip command; the latter has no effect but is useful when no memory change is wanted. Commands can be combined in three different ways. Sequencing is written $C_1$ ; ... ; $C_k$ and indicates that the commands should be executed in the order they are written. The conditional takes the form if $b_1 \rightarrow C_1$ [] ... [] $b_k \rightarrow C_k$ fi. As an example, to express that $C_1$ should be executed when $b$ holds and that otherwise $C_2$ should be executed, we shall write if $b \rightarrow C_1$ [] $\neg b \rightarrow C_2$ fi. To express that $C$ should be executed when $b$ holds and that otherwise we merely continue, it does not suffice to write if $b \rightarrow C$ fi as this construct would not allow us to continue when $b$ is false. The iteration construct takes the form do $b_1 \rightarrow C_1$ [] ... [] $b_k \rightarrow C_k$ od. As an example, to express that $C$ should be executed as long as $b$ holds, we shall write do $b \rightarrow C$ od.

```
y := 1;
do x > 0 → y := x * y;
           x := x − 1
od
```

Figure A.1: Example program for the factorial function.

Example A.1: Figure A.1 is a program intended to compute the factorial function. In addition to assignments, it makes use of sequencing (as indicated by the semicolons) and the iteration construct do . . . od. The body of the construct is a single guarded command consisting of a guard and a command; here the guard is the test x > 0 and the command is the sequence y := x * y; x := x − 1 of assignments. The idea is that as long as the guard is satisfied the associated command will be executed. When the guard fails, the construct terminates.

We shall extend this language with basic commands for array assignment and input and output over channels:

The syntax of the commands $C$ and guarded commands $GC$ of the *Guarded Commands* language are mutually recursively defined using the following *BNF notation*:

$$
\begin{array}{rcl}
C & ::= & x \;\texttt{:=}\; a \mid A[a_1] \;\texttt{:=}\; a_2 \mid c?x \mid c?A[a] \mid c!a \mid \texttt{skip} \\
 & \mid & C_1 \,;\, C_2 \mid \texttt{if}\; GC\; \texttt{fi} \mid \texttt{do}\; GC\; \texttt{od} \\
GC & ::= & b \rightarrow C \mid GC_1 \;[]\; GC_2
\end{array}
$$

We make use of arithmetic expressions $a$ and boolean expressions $b$ given by:

$$
\begin{array}{rcl}
a & ::= & n \mid x \mid A[a_0] \mid a_1 + a_2 \mid a_1 - a_2 \mid a_1 * a_2 \mid A\# \\
b & ::= & \texttt{true} \mid a_1 = a_2 \mid a_1 > a_2 \mid a_1 \geq a_2 \mid b_1 \wedge b_2 \mid b_1 \;\&\&\; b_2 \mid \neg b_0
\end{array}
$$

The syntax of variables $x$, array names $A$, channel names $c$ and numbers $n$ is left unspecified.

The definition specifies an *abstract syntax tree* for commands, guarded commands and arithmetic and boolean expressions. As an example, the notation $a ::= x$ tells us that an arithmetic expression can be a tree consisting of a single node representing the variable $x$, whereas the notation $a ::= a_1 * a_2$ tells us that it can be a binary tree with root labelled "$*$" and two subtrees, one corresponding to $a_1$ and another corresponding to $a_2$.

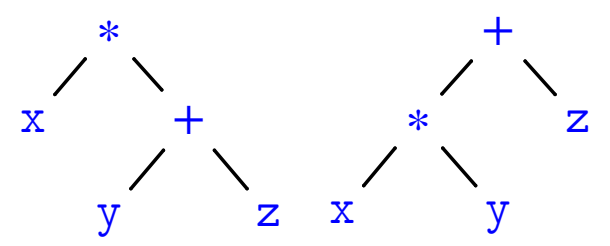

Figure A.2: Abstract syntax trees for x * (y + z) and (x * y) + z.

The definition specifies the *syntactic tree structure* (rather than a linear sequence of characters) and therefore we do not need to introduce explicit *brackets* in the syntax, although we shall feel free to use *parantheses* in textual presentations of programs in order to disambiguate the syntax. This is illustrated in Figure A.2 and when the precedence of operators is clear we allow to write x * y + z for (x * y) + z.

Similar comments holds for commands, guarded commands and boolean expressions. In the case of sequencing we consider the sequencing operator to associate to the right so that $C_1 ; C_2 ; C_3$ is a shorthand for $C_1 ; (C_2 ; C_3)$. In the case of choice we consider the choice operator to associate to the right so that $GC_1 \;[]\; GC_2 \;[]\; GC_3$ is a shorthand for $GC_1 \;[]\; (GC_2 \;[]\; GC_3)$.

Try It Out A.2: Draw the abstract syntax tree for the program of Figure A.1.□

TRY IT OUT A.3: Construct a Guarded Commands program for the power function computing the power $2^n$ of a number $n$; upon termination the variable y should hold the value of the power of the initial value of x. □

EXERCISE A.4: Construct programs in the Guarded Commands language for the following functions:

(a) A function computing the Fibonnaci function; you may assume that it inputs a value from the channel in and outputs the result on the channel out.

(b) A function that sorts an array $A$ with $n$ elements; you may model various sorting algorithms as for example insertion sort and bubble sort. □

Usually we write programs as linear sequences of characters so how do we get them in the form of the abstract syntax trees required here? The answer is that this is the task to be carried out by *parsing* the programs. How to do this, is specified using a context free grammar that differs from our BNF notation in having to deal with issues like precedence of operators. Hence parsing produces parse trees that need to be "simplified" to the abstract syntax trees used here.

**Program Graphs.** We shall now show how to construct program graphs from programs in the Guarded Commands Language. This can be seen as a simple instance of a *Control Flow Analysis*. To this end we will be defining a function **edges** that works on commands and guarded commands and that produces the set of edges of a program graph – given the factorial program of Figure A.1 we expect the function to construct the edges of a program graph as the one of Figure 1.1. We are particularly interested in the initial and final nodes and want to restrict the function **edges** to use the intended initial and final nodes; they are therefore supplied as parameters. So for a command $C$ the result of $\textbf{edges}(q_\circ \rightsquigarrow q_\bullet)[\![C]\!]$ should be the set of edges of a program graph with initial node $q_\circ$ and final node $q_\bullet$. Also we shall ensure that the program graphs constructed do not have edges leaving the final nodes.

For a basic command as an assignment $x := a$ we simply create an edge from $q_\circ$ to $q_\bullet$ that is labelled $x := a$ – this is illustrated in Figure A.3(a). The other basic commands are handled in a similar way:

$$
\begin{array}{rcl}
\textbf{edges}(q_\circ \rightsquigarrow q_\bullet)[\![x := a]\!] & = & \{(q_\circ, x := a, q_\bullet)\} \\
\textbf{edges}(q_\circ \rightsquigarrow q_\bullet)[\![A[a_1] := a_2]\!] & = & \{(q_\circ, A[a_1] := a_2, q_\bullet)\} \\
\textbf{edges}(q_\circ \rightsquigarrow q_\bullet)[\![c?x]\!] & = & \{(q_\circ, c?x, q_\bullet)\} \\
\textbf{edges}(q_\circ \rightsquigarrow q_\bullet)[\![c?A[a]]\!] & = & \{(q_\circ, c?A[a], q_\bullet)\} \\
\textbf{edges}(q_\circ \rightsquigarrow q_\bullet)[\![c!a]\!] & = & \{(q_\circ, c!a, q_\bullet)\} \\
\textbf{edges}(q_\circ \rightsquigarrow q_\bullet)[\![\texttt{skip}]\!] & = & \{(q_\circ, \texttt{skip}, q_\bullet)\}
\end{array}
$$

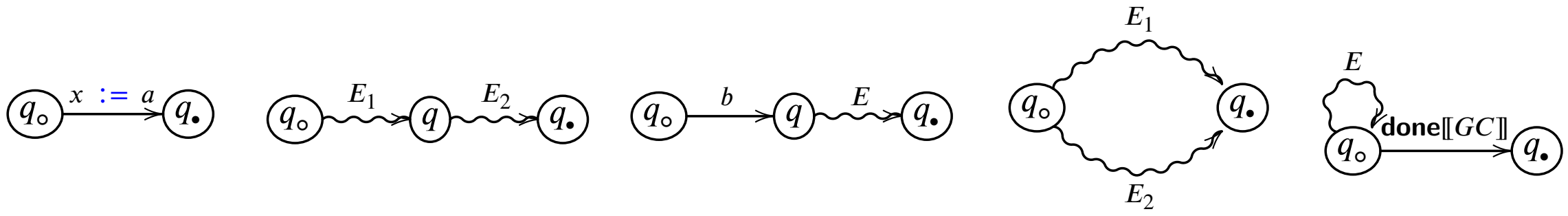

(a) assignment command (b) sequencing of commands (c) single guarded command (d) choice guarded command (e) iteration command

Figure A.3: Figures explaining the construction of program graphs.

During the construction of the program graphs for the composite commands we shall need to create *fresh nodes*. As an example for a sequence of commands $C_1 \,;\, C_2$ we shall create a new node that will glue the edges for the $C_i$'s together; the idea is that the final node of $C_1$ becomes the initial node of $C_2$ – this is illustrated in Figure A.3(b). Here we are exploiting that the program graphs generated for $C_1$ and $C_2$ do not have any edges leaving the final nodes as otherwise the construction might be wrong. For the sake of simplicity we shall refrain from covering the machinery needed to really ensure that fresh nodes are indeed only chosen once.

For the composite commands the edges are constructed as follows:

$$\begin{array}{rcl}
\mathbf{edges}(q_\circ \rightsquigarrow q_\bullet)[\![C_1 \,;\, C_2]\!] & = & \text{let } q \text{ be fresh} \\
 & & \quad E_1 = \mathbf{edges}(q_\circ \rightsquigarrow q)[\![C_1]\!] \\
 & & \quad E_2 = \mathbf{edges}(q \rightsquigarrow q_\bullet)[\![C_2]\!] \\
 & & \text{in } \; E_1 \cup E_2 \\
\mathbf{edges}(q_\circ \rightsquigarrow q_\bullet)[\![\texttt{if } GC \texttt{ fi}]\!] & = & \mathbf{edges}(q_\circ \rightsquigarrow q_\bullet)[\![GC]\!] \\
\mathbf{edges}(q_\circ \rightsquigarrow q_\bullet)[\![\texttt{do } GC \texttt{ od}]\!] & = & \text{let } b = \mathbf{done}[\![GC]\!] \\
 & & \quad E = \mathbf{edges}(q_\circ \rightsquigarrow q_\circ)[\![GC]\!] \\
 & & \text{in } \; E \cup \{(q_\circ, b, q_\bullet)\}
\end{array}$$

Before commenting on the clauses constructing program graphs for the conditional and iteration constructs let us have a look on those for guarded commands.

For guarded commands the edges are constructed as follows:

$$\begin{array}{rcl}
\mathbf{edges}(q_\circ \rightsquigarrow q_\bullet)[\![b \rightarrow C]\!] & = & \text{let } q \text{ be fresh} \\
 & & \quad E = \mathbf{edges}(q \rightsquigarrow q_\bullet)[\![C]\!] \\
 & & \text{in } \; \{(q_\circ, b, q)\} \cup E \\
\mathbf{edges}(q_\circ \rightsquigarrow q_\bullet)[\![GC_1 \;[\,]\; GC_2]\!] & = & \text{let } E_1 = \mathbf{edges}(q_\circ \rightsquigarrow q_\bullet)[\![GC_1]\!] \\
 & & \quad\;\; E_2 = \mathbf{edges}(q_\circ \rightsquigarrow q_\bullet)[\![GC_2]\!] \\
 & & \text{in } \; E_1 \cup E_2
\end{array}$$

Also here we shall need to create fresh nodes. For a single guarded command $b \to C$ we create an edge labelled $b$ to a new node $q$ that is then used as the initial node for the following command – this is illustrated in Figure A.3(c). In case we have a choice between two guarded commands we use the same source and target nodes for both of them as illustrated in Figure A.3(d).

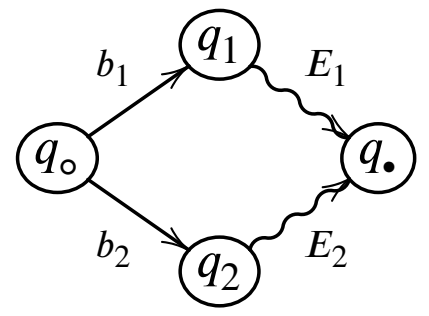

Figure A.4: A conditional with two guarded commands.

Returning to the conditional `if` $GC$ `fi`, we simply construct the edges of the program graph for the embedded guarded command using the given source and target nodes; it will automatically take care of the various tests within the guarded commands. As an example, for a conditional of the form `if` $b_1 \to C_1$ `[]` $b_2 \to C_2$ `fi` we obtain a graph of the form shown in Figure A.4 (where each $E_i$ arises from $C_i$).

The iteration construct `do` $GC$ `od` is slightly more complex as the program graph has to record the looping structure as shown in Figure A.3(e). We construct the edges of the embedded guarded command using $q_\circ$ as source as well as target node thereby reflecting the iterative nature of the construct. When none of the tests of the guarded command are satisfied the iteration should terminate and therefore we add an edge from $q_\circ$ to $q_\bullet$ labelled with the boolean expression **done**$[\![GC]\!]$ expressing this condition. It is defined by:

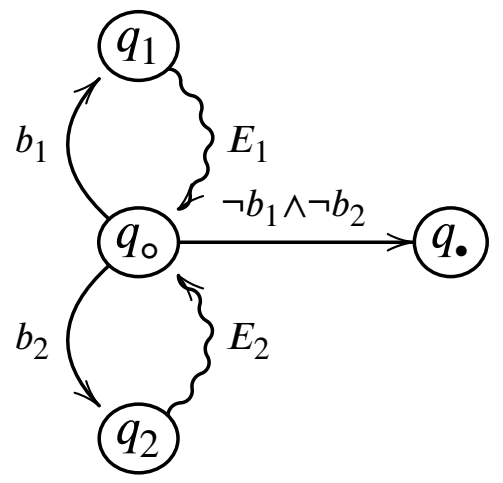

Figure A.5: An iteration construct with two guarded commands.

$$\mathbf{done}[\![b \to C]\!] \quad = \quad \neg b$$
$$\mathbf{done}[\![GC_1 \,[\,]\, GC_2]\!] \quad = \quad \mathbf{done}[\![GC_1]\!] \wedge \mathbf{done}[\![GC_2]\!]$$

As an example, for an iteration construct of the form `do` $b_1 \to C_1$ `[]` $b_2 \to C_2$ `od` we obtain a graph of the form shown in Figure A.5; note that **done**$[\![b_1 \to C_1 \,[\,]\, b_2 \to C_2]\!]$ amounts to $\neg b_1 \wedge \neg b_2$.

EXAMPLE A.5: Let us apply the function $\mathbf{edges}(q_\triangleright \rightsquigarrow q_\blacktriangleleft)[\![\cdot]\!]$ to the program of Figure A.1. We calculate:

$$\mathbf{edges}(q_\triangleright \rightsquigarrow q_\blacktriangleleft)[\![\texttt{y := 1; do x > 0} \to \texttt{y := x * y; x := x - 1 od}]\!]$$
$$= \mathbf{edges}(q_\triangleright \rightsquigarrow q_1)[\![\texttt{y := 1}]\!]$$
$$\cup \ \mathbf{edges}(q_1 \rightsquigarrow q_\blacktriangleleft)[\![\texttt{do x > 0} \to \texttt{y := x * y; x := x - 1 od}]\!]$$
$$= \{(q_\triangleright, \texttt{y := 1}, q_1), (q_1, \neg(\texttt{x > 0}), q_\blacktriangleleft)\}$$
$$\cup \ \mathbf{edges}(q_1 \rightsquigarrow q_1)[\![\texttt{x > 0} \to \texttt{y := x * y; x := x - 1}]\!]$$
$$= \{(q_\triangleright, \texttt{y := 1}, q_1), (q_1, \texttt{x > 0}, q_2), (q_1, \neg(\texttt{x > 0}), q_\blacktriangleleft)\}$$
$$\cup \ \mathbf{edges}(q_2 \rightsquigarrow q_1)[\![\texttt{y := x * y; x := x - 1}]\!]$$
$$= \{(q_\triangleright, \texttt{y := 1}, q_1), (q_1, \texttt{x > 0}, q_2), (q_1, \neg(\texttt{x > 0}), q_\blacktriangleleft)\}$$
$$\cup \ \mathbf{edges}(q_2 \rightsquigarrow q_3)[\![\texttt{y := x * y}]\!] \ \cup \ \mathbf{edges}(q_3 \rightsquigarrow q_1)[\![\texttt{x := x - 1}]\!]$$
$$= \{(q_\triangleright, \texttt{y := 1}, q_1),\ (q_1, \texttt{x > 0}, q_2),\ (q_2, \texttt{y := x * y}, q_3),$$
$$(q_3, \texttt{x := x - 1}, q_1),\ (q_1, \neg(\texttt{x > 0}), q_\blacktriangleleft)\}$$

The resulting program graph is shown in Figure A.6 and coincides with the one of Figure 1.1.

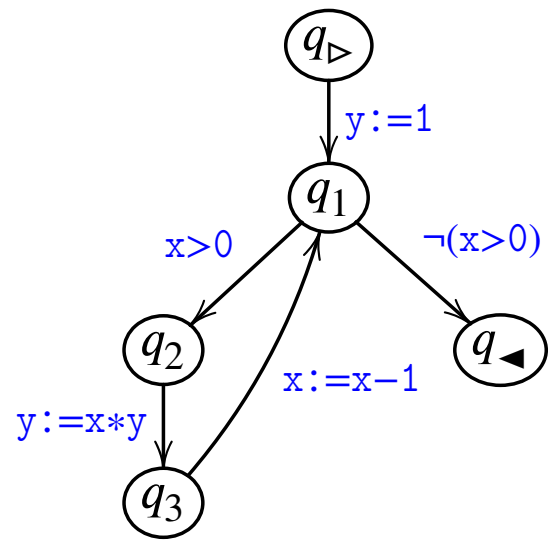

Figure A.6: $\mathbf{edges}(q_\triangleright \rightsquigarrow q_\blacktriangleleft)[\![\cdot]\!]$ applied to the factorial function.

TRY IT OUT A.6: Construct a program graph for the program of Try It Out A.3 using the **edges** function. □

EXERCISE A.7: Construct program graphs for the programs of Exercise A.4 using the **edges** function. □

# Appendix B

# The MicroC Language

In this appendix we shall introduce *MicroC* which is a C-like language. The kernel of the language is a simple imperative language fairly similar to the Guarded Command language of the previous appendix. It is then extended in three stages, first with declarations and constructs for input and output, then with additional control structures and finally with more complex data types.

**The kernel language.** A program in MicroC is simply a *statement* and as in Appendix A we shall specify the form of the syntax trees. We shall feel free to use traditional *round brackets* for expressions and *curly brackets* for statements in order to disambiguate the syntax as shown in the program for the factorial function in Figure B.1.

```
y := 1;
while (x > 0)
  { y := x * y;
    x := x - 1;
  }
```

Figure B.1: Example program for the factorial function.

The syntax of the statements $S$, arithmetic expressions $a$ and boolean expressions $b$ of MicroC are defined using the following BNF notation (where we have included some of the brackets usually found in C) :

$$\begin{array}{rcl} S & ::= & x \texttt{ := } a\,\texttt{;} \mid A\texttt{[}a_1\texttt{]} \texttt{ := } a_2\,\texttt{;} \mid S_1\ S_2 \\ & \mid & \texttt{if}\ (b)\ S_0 \mid \texttt{if}\ (b)\ S_1\ \texttt{else}\ S_2 \mid \texttt{while}\ (b)\ S_0 \\ a & ::= & n \mid x \mid A\texttt{[}a_0\texttt{]} \mid a_1 + a_2 \mid a_1 - a_2 \mid a_1 * a_2 \mid a_1 \,/\, a_2 \mid a_1 \,\%\, a_2 \\ b & ::= & \texttt{true} \mid a_1 = a_2 \mid a_1 > a_2 \mid a_1 \geq a_2 \mid b_1 \wedge b_2 \mid \neg\, b_0 \end{array}$$

The syntax for variables $x$, arrays $A$ and numbers $n$ is left unspecified.

In addition to the two kinds of assignments that are familiar from the Guarded Command language, MicroC has sequencing, conditional and iteration constructs that are inspired by similar constructs in C. As in the Guarded Command language we shall distinguish between arithmetic and boolean expressions (rather than following C where they are merged into one syntactic category).

TASK B.1: [*Control Flow Analysis*] Construct program graphs for the constructs of MicroC in the manner of Appendix A. Illustrate the algorithm on an interesting example program. □

**Declarations and input/output.** Our first extension of the kernel language introduces declarations and constructs for input and output. The declarations may be global or local and in the case of arrays they will specify their length. The extensions are summarised as:

$$\begin{array}{lcl} P & ::= & \{\ D\ S\ \} \\ D & ::= & \texttt{int}\ x\,; \mid \texttt{int}[n]\ A\,; \mid \epsilon \mid D_1\ D_2 \\ S & ::= & \ldots \mid \{\ D\ S_0\ \} \\ & & \quad \mid \texttt{read}\ x\,; \mid \texttt{read}\ A[a]\,; \mid \texttt{write}\ a\,; \end{array}$$

A program $P$ consists of a (possible empty) sequence of declarations followed by a statement.

The declaration of a variable has the form `int` $x$ ; so in this version of the language variables always have type `int`. The declaration of an array has the form `int`$[n]$ $A$ ; where $n$ is a constant giving the length of the array; the entries of the array are then indexed as $A[0], \ldots, A[n-1]$ and they are required to be integers.Variables and arrays must be declared before they are used and in the case of a redeclaration it is the rightmost declaration that is the final one. All variables and array entries are initialised to zero at the point of declaration.

Statements are extended to allow local declarations and they take the form $\{\ D\ S_0\ \}$. The scoping rules are *static* meaning that local variables and arrays are only known inside the block in which they are declarated.

The statements `read` and `write` are only used to read integers and to write integers (akin to `scanf` and `printf` in C). The `read` statement will assign the value being input to either a variable or an array entry whereas `write` will output the value of the arithmetic expression supplied as argument.

TASK B.2: Extend Task B.1 to handle the suggested extensions; in particular discuss how local declarations and redeclarations are handled. Illustrate the algorithm on interesting programs. □

**More Control Structures.** In this paragraph we extend MicroC with additional iteration constructs and loop control constructs from C and with exceptions from C++. The extensions are summarised by:

$$
\begin{array}{rcl}
S & ::= & \ldots \;|\; \texttt{do}\ S_0\ \texttt{while}\ (b) \;|\; \texttt{for}\ (S_1\, b; S_2)\ S_0 \\
 & | & \texttt{break;} \;|\; \texttt{continue;} \;|\; \texttt{skip;} \\
 & | & \texttt{try\{}S_0\texttt{\}}\ H \;|\; \texttt{throw}\ s\,\texttt{;} \\
H & ::= & \texttt{catch(}s\texttt{)\{}S\texttt{\}} \;|\; \texttt{catch(}s\texttt{)\{}S\texttt{\}}\ H
\end{array}
$$

The first line contains the iteration constructs. The construct

$$\texttt{do}\ S_0\ \texttt{while}\ (b)$$

will execute the statement $S_0$ and will continue doing so as long as the test $b$ holds. The construct

$$\texttt{for}\ (S_1\, b; S_2)\ S_0$$

is a looping construct with $S_1$ being the initialisation step, $b$ being the iteration condition for the loop, $S_2$ being the increment/decrement step of the loop and finally $S_0$ being the body of the loop. Figure B.2 shows a program using the `for`-construct; it computes in `y` the `x`'th Fibonacci number.

```
z := 1;
y := 0;
t := 0;
for (i := 0; i < x; i := i + 1;)
  { t := z;
    z := y;
    y := t + z;
  }
```

Figure B.2: Example program for the Fibonacci function.

The second line lists the two loop control statements of C. Here `break` terminates the immediately enclosing loop whereas `continue` skips the remainder of the loop body and evaluates the loop condition once more. It is also convenient to introduce a `skip` action to represent the direct transfer of control from one program point to another (as in the Guarded Command language).

The third line introduces the exception constructs of C++. Here $s$ is the exception name (whose syntax we leave unspecified) and it is defined by a construct of the form

$$\texttt{try\{}S_0\texttt{\}catch(}s_1\texttt{)\{}S_1\texttt{\}} \cdots \texttt{catch(}s_n\texttt{)\{}S_n\texttt{\}}$$

for $n > 0$. If the exception $s_i$ is thrown (using the construct `throw` $s_i$) within $S_0$ then the execution of $S_0$ is terminated and $S_i$ is executed instead. In the case where several constructs are embedded within one another it is always the syntactically innermost exception that is caught.

TASK B.3: Gradually extend the construction from Task B.1 (or B.2) to deal with the additional control structures. For this it may be helpful to extend the **edges** function with additional parameters. Illustrate the algorithm on interesting programs. □

**More Data Structures.** We shall now extend MicroC with more complex data structures; in addition to integers and arrays we also have records. The extension is summarised as follows:

$$
\begin{array}{lcll}
D & ::= & \dots & | \ \{\ \mathtt{int}\ f_1\ ; \cdots ;\ \mathtt{int}\ f_k\ \}\ R \\
S & ::= & \dots & | \ R.f\ := a\ ;\ |\ R\ := (a_1, \cdots, a_k)\ ; \\
a & ::= & \dots & | \ R.f
\end{array}
$$

Here $f_1, \cdots, f_k$ are unique and distinct names for the fields (for $k \geq 1$) of the record and all the fields have type `int`. The fields of a record can be updated individually using the assignment $R.f\ := a$ and here it must be the case that $R$ is declared to have the field $f$. The fields of a record can be updated collectively using the assignment $R\ := (a_1, \cdots, a_k)$; here it must be the case that $R$ has $k$ fields and the intension is that the $i$'th field is updated to have the value of $a_i$ (for $1 \leq i \leq k$).

The arithmetic expressions can extract the value of the individual fields of a record using the construct $R.f$; also here it must be the case that $R$ is declared to have a field named $f$. Figure B.3 is an example of a program written in this version of the language.

```
{ int i;
  {int fst; int snd} R;
  int[10] A;
  while (i < 10)
    { read A[i];
      i := i + 1;
    }
  i := 0;
  while (i < 10)
    { if (A[i] >= 0)
        { R.fst := R.fst + A[i];
          i := i + 1;
        }
      else { i := i + 1;
             break;
           }
      R.snd := R.snd + 1;
    }
  write (R.fst/R.snd);
}
```

Figure B.3: Example program computing the average of some array entries.

A more powerful (and more complex) version of the language is obtained when allowing arrays and records to contain not just integers. As an example one might want to have arrays with elements that are records or one might want records where one or more of the fields are records. Then expressions as $A[a]$ and $R.f$ no longer need to evaluate to integers and we would need to introduce a more involved type checking algorithm in order to ensure that programs are well-formed. We shall not go into further details of this.

Task B.4: Extend the construction from Task B.2 to deal with the additional data structures and the new constructs. Here it would be helpful to introduce a symbol table and supply it as an argument to the **edges** function. Illustrate the algorithm on interesting programs. □

# Appendix C

# Project: An Analysis Module

This appendix introduces a series of tasks for designing, implementing and experimenting with an analysis module constructed using the techniques presented in the chapters of this book. The overall structure is shown in Figure C.1.

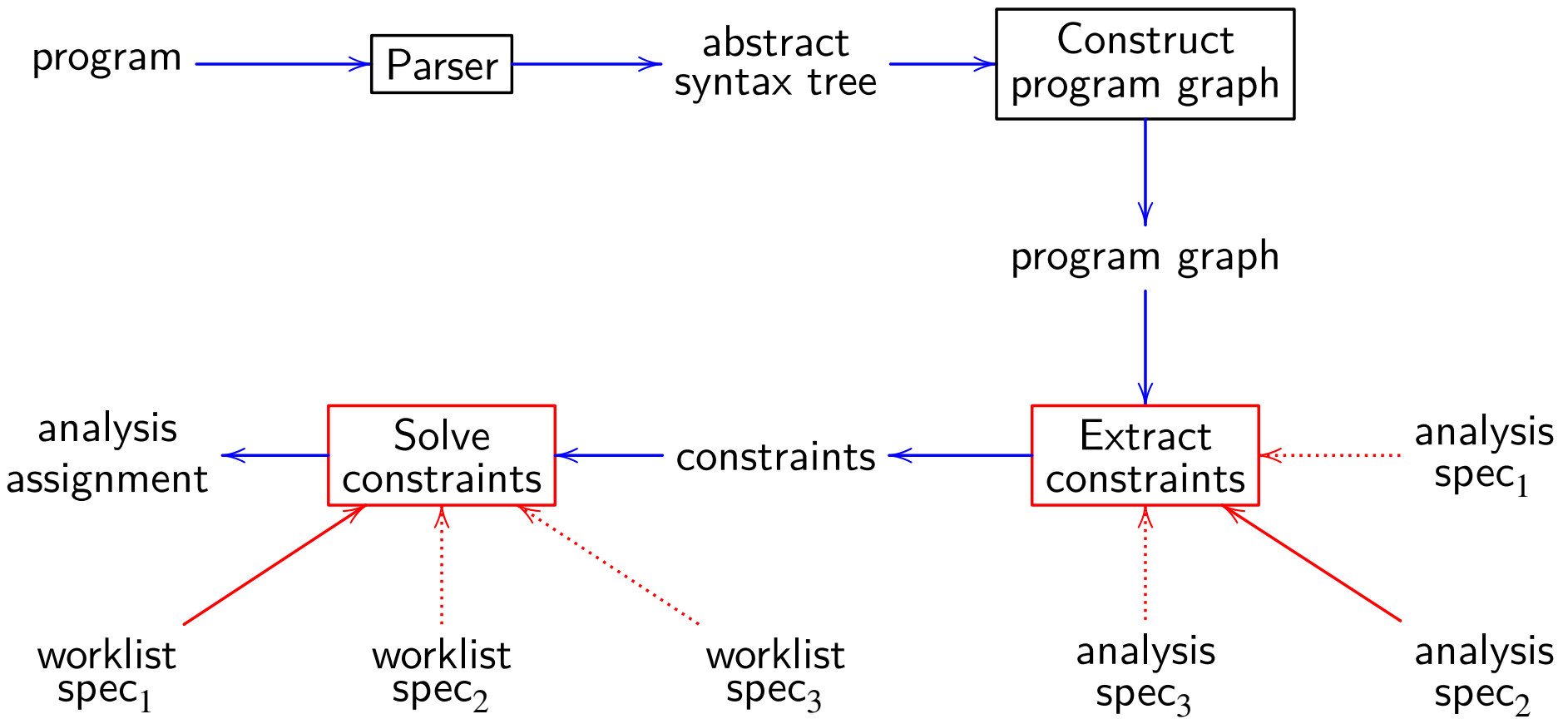

Figure C.1: The analysis module in context.

As illustrated in the upper part of the figure, the system takes as input a program written in some language; this may be the Guarded Commands language of Appendix A, the MicroC language of Appendix B, or some other language. A parser will construct an abstract syntax tree for the program and this is then the basis for the construction of a program graph. This basically amounts to an implementation of functions akin to the ones presented for the Guarded Commands language in Appendix A and developed for MicroC in the tasks of Appendix B.

The bottom part of the figure then gives the structure of the analysis module itself. This part only exploits the form of the actions as present in the program graphs so indeed it will be possible to build a system supporting different programming languages as long as they give rise to the same kind of actions.

The analysis module has two key components, one that can be instantiated to different analyses (to the right) and another (to the left) that can be instantiated to use different data structures for the algorithm solving the analysis problems.

The focus of this project is to get insights in the design and implementation of the analysis module. The implementation itself can be carried out in any programming language although a functional programming style is likely to give rise to fewer lines of code overall. The webpage http://www.formalmethods.dk/pa4fun/ presents a learning environment with much of the functionality of the analysis module being developed in this project.

**Monotone Frameworks.** The analyses of the module are specified as monotone frameworks as presented in Definition 3.30. For each of the analyses this involves representing the analysis domain (including its ordering, bottom element and least upper bound operator), specifying the analyses functions for each of the forms of actions, and defining the initial element associated with the analysis. As part of the design process it must be checked that the analysis domain is indeed a pointed semi-lattice and that it satisfies the ascending chain condition; and it must be checked that the analysis functions are indeed monotone functions.

We shall start with a number of task related to the material of Chapters 2 and 3.

TASK C.1: The aim of this task is to develop the Reaching Definitions analysis of Section 2.1 for the module. For the design stage consider the following subtasks:

(a) For arrays and records discuss the pros and cons of amalgamating the components or dealing with them individually (as discussed in Exercise 2.13). Construct the corresponding analysis domain.

(b) Specify the kill and gen sets for the actions of the program graphs and show how to construct constraints for the programs.

(c) Apply your analysis specification to an illustrative program to ensure that you obtain the expected result; this involves deriving the program graph for the program, constructing the corresponding set of constraints and solving them (by hand at this stage). □

TASK C.2: In this task the aim is to develop the Live Variables analysis of Section 2.2 for the module. As in the previous task, we can indentify a number of subtasks for the design stage:

(a) For arrays and records discuss the pros and cons of amalgamating the components or dealing with them individually (as discussed in Exercise 2.35). Construct the corresponding analysis domain.

(b) Specify the kill and gen sets for the actions of the program graphs and show how to construct constraints for the programs.

(c) Apply your analysis specification to an illustrative program to ensure that you obtain the expected result; this involves deriving the program graph for the program, constructing the corresponding set of constraints and solving them (by hand at this stage). □

The above tasks focus on analyses where we are interested in the least solution to the constraints being constructed. Similar tasks can easily be introduced for analyses where we are interested in the greatest solution. Tasks that develop the Available Expressions analysis of Section 2.3 and the Very Busy Expressions analysis of Section 2.4 will follow the same overall pattern.

Task C.3: In this task we shall extend the previous development with the following two subtasks:

(a) Example 3.23 in Section 3.3 introduces the Dangerous Variables analysis as a modification of the Reaching Definitions analysis; generalise the design of Task C.1 to perform this analysis.

(b) Example 3.38 in Section 3.4 introduces the Faint Variables analysis as a modification of the Live Variables analysis; generalise the design of Task C.2 to perform this analysis. □

**Algorithms.** In Chapter 4 we present a generic algorithm for computing the least solution to the constraints of the analysis problems. The aim is here to obtain an implementation that is parametric on the analysis problem as well as the instantiation of the operations on the worklist. This can then be the basis for a number of experiments providing additional insights in the performance in different settings. The focus in the following two tasks is on the analyses presented in 2 and 3; in Task C.9 we extend this to the analyses of Chapter 5.

Task C.4: In this task we shall study the worklist algorithm and experiment with two of its simple instantitations. We have the following subtasks:

(a) Implement the worklist algorithm of Figure 4.2 making use of an abstract data type for the worklist providing the operations empty, insert and extract; the implementation should also be parametric on the analysis specification.

(b) Instantiate the abstract data type for the worklist to implement the last-in first-out principle as well as the first-in first out principle introduced in Figures 4.6 and 4.7. Ensure that all operations take constant time in both cases.

(c) Experiment with the two versions of the algorithm and the analyses developed in the previous tasks by analysing several program graphs with different characteristica (as for example number of loops). For this it will be interesting to measure the time required for solving the constraints; for simplicity this could just be the number of operations on the worklist.

(d) Based on these experiments discuss whether there are combinations of analyses and worklist representations that have better performance than others. Do the experiments show any difference in the performance of the analyses of Task C.3 compared to those of Tasks C.1 and C.2? □

TASK C.5: Building on the previous task the goal is now to exploit the reverse postorder organisation of the worklist.

(a) The first subtask is to construct the depth first spanning tree from the program graph together with the reverse postorder as described in Section 4.2; this information will be exploited in the following.

(b) Instantiate the abstract data type for the worklist introduced in Task C.4 to implement the Round Robin algorithm (as presented in Example 4.17) together with the algorithms of Section 4.3 exploiting the reverse postorder and strong components. In all cases ensure that the operations on the worklist are reasonably efficient.

(c) Extend the experiments of the previous task to the three new worklist representations. Again discuss whether there are combinations or analyses and worklist representations that have better performance than others. □

**Analysis of Integers.** The following sequence of tasks are concerned with analyses of integers and thus they relate to Chapter 5.

TASK C.6: In this task we shall develop the Detection of Signs analysis of Section 5.1 for the analysis module. For the design stage we have the following subtasks:

(a) Discuss how to handle the arrays and records; in particular, discuss the possibility of letting the analysis capture the relationship between the signs of the various components of the records (as for example, all components are either positive or negative). Construct the analysis domain and argue that it is a pointed semi-lattice satisfying the ascending chain condition.

(b) Specify the analysis functions; this involves specifying how to compute with signs. Argue that the analysis functions correctly capture the semantics and show that they are monotone functions.

(c) Show how to construct the constraints for an illustrative example and present an analysis assignment solving the constraints.

(d) Discuss how to improve the precision of the analysis; this may include modifying the analysis domain as well as the analysis functions. □

A task developing the Parity analysis of Exercise 5.16 (or Exercise5.17) follows the same overall pattern. An interesting and slightly more challenging task is the combined Detection of Signs and Parity analysis of Exercise 5.18; the overall structure of the task is as above.

Task C.7: The Constant Propagation analysis is introduced in Section 5.2 and in this task we shall develop it for the analysis module. The subtasks are as follows:

(a) Discuss how to handle the arrays and records. Construct the analysis domain and argue that it is a pointed semi-lattice satisfying the ascending chain condition.

(b) Specify the analysis functions for the various actions; this involves specifying how to compute with the abstract values of the analysis domain. Argue that the analysis functions correctly capture the semantics and show that they are monotone functions.

(c) Show how to construct the constraints for an illustrative example and present an analysis assignment solving the constraints.

(d) Discuss how to improve the precision of the analysis; this may include modifying the analysis domain as well as the analysis functions. □

Exercise 5.32 suggests a version of the Constant Propagation analysis that is parametric on a (non-empty) finite set of constant values of interest and Exercise 5.31 suggests improving the precision of the Constant Propagation analysis by combining it with the Detection of Signs analysis. Tasks exploring these features follow the same overall structure as above.

Task C.8: Section 5.3 introduces the interval analysis and we shall now develop it for the analysis module. The subtasks are as follows:

(a) Discuss how to handle the arrays and records. Construct the analysis domain and argue that it is a pointed semi-lattice satisfying the ascending chain condition. It may be useful to let the analysis domain be parametric on a finite set of the interval bounds (as in Section 5.3).

(b) Specify the analysis functions for the various actions; this involves specifying how to compute with the intervals of the analysis domain. Argue that the analysis functions correctly capture the semantics and show that they are monotone functions.

(c) Show how to construct the constraints for an illustrative example and present an analysis assignment solving the constraints.

(d) Discuss how to improve the precision of the analysis; this may include modifying the analysis domain as well as the analysis functions. □

TASK C.9: Continuing the experiments of Tasks C.4 and C.5 we shall now study the performance of the analyses of integers. The two tasks provide several implementations of the worklist algorithm and the aim of this task is to obtain insights on their performance for the analyses of integers.

(a) Continuing Task C.6 perform a series of experiments shedding light on the performance of the five worklist algorithms for the Detection of Signs analysis. Experiment with different versions of the analysis (possibly including the combination with Parity analysis) to get insights on the trade off between the performance and the precision of the analyses.

(b) Continuing Task C.7 perform a series of experiments shedding light on the performance of the various worklist algorithms for the Constant Propagation analysis. Experiment with different versions of the analysis to get insights on the trade off between the performance and the precision of the analyses; for the parametric version of the analysis this may include varying the set of constant values of interest.

(c) Continuing Task C.8 perform a series of experiments shedding light on the performance of the various worklist algorithms for the Interval analysis. Experiment with different versions of the analysis to get insights on the trade off between the performance and the precision of the analyses; this may include varying the set of possible interval bounds. □

# Index